\documentclass[floatfix,aps,prc,groupedaddress,10pt,twocolumn,nofootinbib]{revtex4-1}
\usepackage{amsmath}
\usepackage{natbib}
\usepackage{epsfig}
\usepackage{natbib}
\usepackage{color}
\usepackage{longtable}
\usepackage{multirow}
\usepackage{hyperref}
\usepackage{etoolbox}
\usepackage[normalem]{ulem}
\usepackage{color,soul}
\usepackage{xcolor}
\usepackage{pifont}
\setulcolor{blue}
\setstcolor{red}
\sethlcolor{yellow}

\definecolor{darkgreen}{rgb}{0,0.4,0}
\definecolor{dgreen}{rgb}{0.1,0.6,0.7}
\definecolor{violet}{rgb}{0.8, 0.0, 1.0}
\definecolor{bole}{rgb}{0.47, 0.27, 0.23}
\definecolor{airforceblue}{rgb}{0.36, 0.54, 0.66}
\definecolor{amber}{rgb}{1.0, 0.49, 0.0}
\definecolor{darkred}{rgb}{0.55, 0.0, 0.0}
\definecolor{modelref}{rgb}{0.247,0.3647,0.491}
\definecolor{firebrick}{rgb}{0.698,0.1333,0.1333}
\definecolor{seagreen}{rgb}{0.180,0.545,0.341}
\colorlet{notgreen}{orange!40!white}
\sethlcolor{notgreen}
\newcommand{\cmark}{\ding{51}}%
\def\aap{A\&A}%
\def\adsr{Adv. in Space Res.}
\def\aipcs{Am.~Inst.~Phys.~Conf.~Series}
\def\anrps{Ann. Rev. Nucl. Part. Sci.}
\def\ap{Astropart.~Phys.}%
\def\apj{ApJ}%
\def\apjl{ApJ}%
\def\apjs{ApJS}%
\def\araa{ARA\&A}%
\def\cpc{Comput.~Phys.~Commun.}%
\def\icrc{ICRC}%
\def\jcap{J. Cosmology Astropart. Phys.}%
\def\jphg{J.~Phys.~G~Nucl.~Phys.}
\def\nar{New A Rev.}%
\def\nat{Nature}%
\def\nima{Nucl.~Instr.~Meth.~Phys.~Res.~A}%
\def\nimb{Nucl.~Instr.~Meth.~Phys.~Res.~B}%
\def\nphysa{Nucl.~Phys.~A}%
\def\physrep{Phys.~Rep.}%
\def\pr{Phys.~Rev.}%
\def\prc{Phys.~Rev.~C}%
\def\prd{Phys.~Rev.~D}%
\def\prl{Phys.~Rev.~Lett.}%
\begin{document}
\AtEndEnvironment{thebibliography}{\bibitem{6}
Akhababyan N., Baatar Ts., Gasparyan P. et al., JINR Preprint 1-12114,
Dubna, 1979.

\bibitem{8}
Afanas'ev V.N., Borisov V.S., Borodina I.N. et al.,
Yadernaya Fizica, 1988, {\bf 47}, 1656.

\bibitem{13}
Alakoz A.V., Vasil'ev P.S., Vasol'eva L.F. et al., 
Izv. Akad. Nauk SSSR, Ser. Fiz., 1971, {\bf 35}, 2069.

\bibitem{25}
Ashmore A., Jarvis R.G., Mather D.S. et al., Proc. Phys. Soc., 1957,
{\bf A70}, 745.

\bibitem{46}
Jaros J., Wagner A., Anderson L. et al., Phys. Rev., 1978, {\bf C18}, 2273.

\bibitem{48}
Basilova R.N., Grigorov N.L., Kakhidze G.P. et al.,
Izv. Akad. Nauk SSSR, Ser. Fiz., 1966, {\bf 30}, 1610.

\bibitem{53}
Batty C.J., Lock W.O., March P.V., Proc. Phys. Soc., 1958, {\bf 73}, 100.

\bibitem{60}
Belletini G., Cocconi G., Diddens A.N. et al.,
Nucl. Phys., 1966, {\bf 79}, 609.

\bibitem{66}
Bobchenko B.M., Buklei A.E., Vlasov A.V. et al.,
Yadernaya Fizika, 1979, {\bf 30}, 1553.

\bibitem{69}
Booth N.E., Ledley B., Walker D. et al.,
Proc. Phys. Soc., 1957, {\bf A70}, 209.

\bibitem{75}
Bowen T., Nuovo Cimento, 1958, {\bf 9}, 908.

\bibitem{88}
Burg J.P., Chemarin M., Chevallier M. et al.,
Nucl. Phys., 1981, {\bf B187}, 205.


\bibitem{96}
Carlson R.E., Eisenberg R.M., Meyer V.,
University of Minnesota Annual Progress Report, Minnesota, 1965, p.2.


\bibitem{100}
Carroll A.S., Chiang I.H., Kycia T.F. et al.,
Phys. Lett., 1978, {\bf B80}, 319.


\bibitem{102}
Cassels J.M., Lawson J.D.,
Proc. Phys. Soc., 1954, {\bf A67}, 125.

\bibitem{104}
Cester R., Debenedetti A., Garelli C.M. et al.,
Nuovo Cimento, 1958, {\bf 7}, 371.

\bibitem{110}
Chen F.F., Leavitt C.P., Shapiro A.M.,
Phys. Rev., 1955, {\bf 99}, 857.

\bibitem{140}
Denisov S.P., Donskov S.V., Gorin Yu.P. et al.,
Nucl. Phys., 1973, {\bf B61}, 62.


\bibitem{154}
Faldt G., Gilsen L., Lund University Report, Lund, 1975.


\bibitem{165}
Grchurin V.V., Druzhinin B.L., Ezhov B.A. et al.,
Preprint ITEP No. 59, Moscow, 1985.

\bibitem{177}
Goloskie T.J., Strauch K., Nucl. Phys., 1962, {\bf 29}, 474.


\bibitem{187}
Grigorov N.L., Erofeeva I.N., Mishchenko L.G. et al.,
Izv. Akad. Nauk SSSR, Ser. Fiz., 1964, {\bf 28}, 1798.

\bibitem{188}
Grigorov N.L., Nesterov V.E., Rapoport I.D. et al.,
Yadernaya Fizika, 1970, {\bf 11}, 813.

\bibitem{195}
Heckman H.H., Greiner D.E., Lindstrom P.J. et al.,
Phys. Rev., 1978, {\bf C17}, 1735.

\bibitem{207}
Johansson A., Svanberg U., Sundberg O.,
Ark. Phys., 1961, {\bf 19}, 527.

\bibitem{215}
Kirkby P., Link W.T., Can. J. Phys., 1966, {\bf 44}, 1847.

\bibitem{216}
Kirschbaum A.J., Ph.D. Thesis, Report UCRL-1967, Berkeley, 1967.\\
(216a) Klem R., Igo G., Talaga R. et al., Phys. Lett., 1977, {\bf B70}, 155.

\bibitem{224}
Lindstrom P.J., Proc. XXIV Int. Conf. on Cosmic Rays, Munich, 1975, p.2315.

\bibitem{228}
Longo M.L., Ph.D. Thesis, Report UCRL-9497, Berkeley, 1961.




\bibitem{247}
McGill W.F., Carlson R.F., Short T.H. et al., Phys. Rev., 1974, {\bf C10}, 2237.

\bibitem{248}
Measday D.F., Palmiere J.N., Nucl. Phys., 1966, {\bf 85}, 129.


\bibitem{255}
Millburn G.P., Birbaum W., Crandall W.E. et al., Phys Rev., 1954, 
{\bf 95}, 1268.

\bibitem{259}
Moskalev V.N., Gavrilovskii B.V.,
Dokl. Akad. Nauk SSSR, 1956, {\bf 110}, 972.




\bibitem{291}
Renberg P.U., Measday D.F., Pepin M. et al.,
Nucl. Phys., 1972, {\bf A183}, 81.

%
%
\bibitem [{\citenamefont {Dietrich}\ \emph {et~al.}(2002)\citenamefont
  {Dietrich}, \citenamefont {Hartouni}, \citenamefont {Johnson}, \citenamefont
  {Schmid}, \citenamefont {Soltz}, \citenamefont {Abfalterer}, \citenamefont
  {Haight}, \citenamefont {Waters}, \citenamefont {Hanson}, \citenamefont
  {Finlay} \emph {et~al.}}]{341}%
  \BibitemOpen
  \bibfield  {author} {\bibinfo {author} {\bibfnamefont {F.}~\bibnamefont
  {Dietrich}}, \bibinfo {author} {\bibfnamefont {E.}~\bibnamefont {Hartouni}},
  \bibinfo {author} {\bibfnamefont {S.}~\bibnamefont {Johnson}}, \bibinfo
  {author} {\bibfnamefont {G.}~\bibnamefont {Schmid}}, \bibinfo {author}
  {\bibfnamefont {R.}~\bibnamefont {Soltz}}, \bibinfo {author} {\bibfnamefont
  {W.}~\bibnamefont {Abfalterer}}, \bibinfo {author} {\bibfnamefont
  {R.}~\bibnamefont {Haight}}, \bibinfo {author} {\bibfnamefont
  {L.}~\bibnamefont {Waters}}, \bibinfo {author} {\bibfnamefont
  {A.}~\bibnamefont {Hanson}}, \bibinfo {author} {\bibfnamefont
  {R.}~\bibnamefont {Finlay}},  \emph {et~al.},\ }\href@noop {} {\bibfield
  {journal} {\bibinfo  {journal} {Journal of Nuclear Science and Technology}\
  }\textbf {\bibinfo {volume} {39}},\ \bibinfo {pages} {269} (\bibinfo {year}
  {2002})}\BibitemShut {NoStop}%
%
%
%
\bibitem [{\citenamefont {Lantz}\ \emph {et~al.}(2005)\citenamefont {Lantz},
  \citenamefont {Jacobs}, \citenamefont {Auce}, \citenamefont {Carlson},
  \citenamefont {Cowley}, \citenamefont {F{\"o}rtsch}, \citenamefont
  {Hillhouse}, \citenamefont {Ingemarsson}, \citenamefont {Johansson},
  \citenamefont {Lawrie} \emph {et~al.}}]{342}%
  \BibitemOpen
  \bibfield  {author} {\bibinfo {author} {\bibfnamefont {M.}~\bibnamefont
  {Lantz}}, \bibinfo {author} {\bibfnamefont {M.~N.}\ \bibnamefont {Jacobs}},
  \bibinfo {author} {\bibfnamefont {A.}~\bibnamefont {Auce}}, \bibinfo {author}
  {\bibfnamefont {R.~F.}\ \bibnamefont {Carlson}}, \bibinfo {author}
  {\bibfnamefont {A.~A.}\ \bibnamefont {Cowley}}, \bibinfo {author}
  {\bibfnamefont {S.~V.}\ \bibnamefont {F{\"o}rtsch}}, \bibinfo {author}
  {\bibfnamefont {G.~C.}\ \bibnamefont {Hillhouse}}, \bibinfo {author}
  {\bibfnamefont {A.}~\bibnamefont {Ingemarsson}}, \bibinfo {author}
  {\bibfnamefont {R.}~\bibnamefont {Johansson}}, \bibinfo {author}
  {\bibfnamefont {K.~J.}\ \bibnamefont {Lawrie}},  \emph {et~al.},\ }in\
  \href@noop {} {\emph {\bibinfo {booktitle} {AIP Conference Proceedings}}},\
  Vol.\ \bibinfo {volume} {769}\ (\bibinfo {organization} {AIP},\ \bibinfo
  {year} {2005})\ pp.\ \bibinfo {pages} {846--849}\BibitemShut {NoStop}%
%
%
%
%
\bibitem [{\citenamefont {Trzaska}\ \emph {et~al.}(1991)\citenamefont
  {Trzaska}, \citenamefont {Pelte}, \citenamefont {Lemaire}, \citenamefont
  {Alard}, \citenamefont {Augerat}, \citenamefont {Bachelier}, \citenamefont
  {Bastid}, \citenamefont {Boyard}, \citenamefont {Cavata}, \citenamefont
  {Charmensat} \emph {et~al.}}]{343}%
  \BibitemOpen
  \bibfield  {author} {\bibinfo {author} {\bibfnamefont {M.}~\bibnamefont
  {Trzaska}}, \bibinfo {author} {\bibfnamefont {D.}~\bibnamefont {Pelte}},
  \bibinfo {author} {\bibfnamefont {M.-C.}\ \bibnamefont {Lemaire}}, \bibinfo
  {author} {\bibfnamefont {J.}~\bibnamefont {Alard}}, \bibinfo {author}
  {\bibfnamefont {J.}~\bibnamefont {Augerat}}, \bibinfo {author} {\bibfnamefont
  {D.}~\bibnamefont {Bachelier}}, \bibinfo {author} {\bibfnamefont
  {N.}~\bibnamefont {Bastid}}, \bibinfo {author} {\bibfnamefont {J.-L.}\
  \bibnamefont {Boyard}}, \bibinfo {author} {\bibfnamefont {C.}~\bibnamefont
  {Cavata}}, \bibinfo {author} {\bibfnamefont {P.}~\bibnamefont {Charmensat}},
  \emph {et~al.},\ }\href@noop {} {\bibfield  {journal} {\bibinfo  {journal}
  {Zeitschrift f{\"u}r Physik A Hadrons and Nuclei}\ }\textbf {\bibinfo
  {volume} {340}},\ \bibinfo {pages} {325} (\bibinfo {year}
  {1991})}\BibitemShut {NoStop}%
\bibitem[Abgrall et al.(2016)]{344} Abgrall, N., Aduszkiewicz, A., Ali, Y., et al.\ 2016, European Physical Journal C, 76, 84 
%
%

}

\title{Current status and desired accuracy of the isotopic production cross sections relevant to astrophysics of cosmic rays I. Li, Be, B, C, N}

\author{Yoann G\'enolini}
\email[]{yoann.genolini@ulb.ac.be}
\affiliation{Service de Physique Th\'eorique, Universit\'e Libre de Bruxelles, Boulevard du Triomphe, CP225, 1050 Brussels, Belgium}

\author{David Maurin}
\email[]{dmaurin@lpsc.in2p3.fr}
\affiliation{LPSC, Universit\'e Grenoble-Alpes, CNRS/IN2P3, 53 avenue des Martyrs, 38026 Grenoble, France}

\author{Igor V. Moskalenko}
\email[]{imos@stanford.edu}
\affiliation{W. W. Hansen Experimental Physics Laboratory and Kavli Institute for Particle Astrophysics and Cosmology, Stanford University, Stanford, CA 94305, USA}

\author{Michael Unger}
\email[]{michael.unger@kit.edu}
\affiliation{Karlsruhe Institute of Technology, Karlsruhe, Germany}

\date{\today}

\begin{abstract}
The accuracy of the current generation of cosmic-ray (CR) experiments, such as AMS-02, PAMELA, CALET, and ISS-CREAM, is now reaching $\sim$1--3\% in a wide range in energy per nucleon from GeV/n to multi-TeV/n. Their correct interpretation could potentially lead to discoveries of new physics and subtle effects that were unthinkable just a decade ago. However, a major obstacle in doing so is the current uncertainty in the isotopic production cross sections that can be as high as 20--50\% or even larger in some cases. While there is a recently reached consensus in the astrophysics community that new measurements of cross sections are desirable, no attempt to evaluate the importance of particular reaction channels and their required accuracy has been made yet. It is, however, clear that it is a huge work that requires an incremental approach. The goal of this study is to provide the ranking of the isotopic cross sections contributing to the production of the most astrophysically important CR Li, Be, B, C, and N species. In this paper, we (i) rank the reaction channels by their importance for a production of a particular isotope, (ii) provide comparisons plots between the models and data used, and (iii) evaluate a generic beam time necessary to reach a 3\% precision in the production cross-sections pertinent to the AMS-02 experiment. This first roadmap may become a starting point in the planning of new measurement campaigns that could be carried out in several nuclear and/or particle physics facilities around the world. A comprehensive evaluation of other isotopes $Z\leq30$ will be a subject of follow-up studies.
\end{abstract}

\pacs{}
\maketitle

\section{Introduction} \label{sec:intro}

The centennial anniversary of the discovery of CRs (in 2012) was marked by a series of exciting discoveries made a few years before it and during the following years \cite{2009Natur.458..607A,2011Sci...332...69A,2013PhRvL.110n1102A,2015PhRvL.114q1103A,2015PhRvL.115u1101A,2016PhRvL.117w1102A,2016PhRvL.117i1103A,2009PhRvL.102r1101A,2014ApJ...793...64A,2010ApJ...714L..89A}. It became possible due to the superior instrumentation launched to the top of the atmosphere (e.g., BESS-Polar, CREAM) and into space (PAMELA \cite{2007APh....27..296P}, AMS-02 \cite{2013PhRvL.110n1102A}, Fermi-LAT \cite{2009ApJ...697.1071A}) and whose accuracy is now reaching an astonishing level of 1--3\% (see a collection of CR data in \cite{2014A&A...569A..32M}). Not surprisingly, these recent developments raised anticipations that new measurements of composition and spectra of CR species may reveal signatures of yet unknown effects or phenomena and consequently led to the surge of interest in astrophysics and particle physics communities. Meanwhile, achieving this goal demands the appropriate level of accuracy from theoretical models used for interpretation of the data collected by the modern or future experiments. The major obstacle to this is the accuracy of the existing measurements of the nuclear production cross sections \cite{2003ApJ...586.1050M,2005AIPC..769.1612M,2010A&A...516A..67M,2015A&A...580A...9G,2017PhRvD..96j3005T} whose errors are reaching 20--50\% or even worse \cite{2003ApJS..144..153W,2003ICRC....4.1969M,2004AdSpR..34.1288M,2005AIPC..769.1612M, 2011ICRC....6..283M,2013ICRC......0823M} and are unacceptable by nowadays standards.

An accurate calculation of the isotopic production cross sections is a cornerstone of all CR propagation calculations. The cross sections are necessary to calculate the production of secondary isotopes (e.g., isotopes of Li, Be, B) in spallation of CR in the interstellar medium (ISM) and to derive propagation parameters \cite{1998ApJ...509..212S,1998ApJ...506..335W,1998A&A...337..859P,2001ApJ...555..585M} that provide a basis for a number of other studies \cite{2007ARNPS..57..285S}. Even slight excesses or deficits of certain isotopes in CRs relative to expectations from propagation models \cite{2005ApJ...634..351B,2016Sci...352..677B} can be used to pin down the origins of various species, their acceleration mechanisms and propagation history; they also help to locate other deviations \cite{2009Natur.458..607A,1998ApJ...493..694M,1982ApJ...254..391P} that otherwise could remain unnoticed. In turn, such information is necessary for a reliable identification of subtle signatures of the dark matter or new physics \cite{2011ARA&A..49..155P,2012CRPhy..13..740L}, and for accurate predictions of the Galactic diffuse emission and disentangling unexpected features \cite{2010ApJ...724.1044S,2012ApJ...750....3A,2014ApJ...793...64A,2016ApJ...819...44A}. This calls for a dedicated effort to improve on the accuracy of the nuclear production cross sections, especially in the context of recent anomalies seen in CRs, such as, e.g., spectral breaks \cite{2011Sci...332...69A,2014PhR...544..323A,2015PhRvL.114q1103A,2015PhRvL.115u1101A,2017PhRvL.119y1101A,2018PhRvL.120b1101A}. Note that the production cross sections are also the key ingredient for calculations of the production of cosmogenic radionuclides by Galactic CRs in Earth's atmosphere and meteorites \cite{2002RvMG...50..207M,2000M&PS...35..259L,2013NIMPB.294..470R} and for human radiation shielding applications \cite{2016LSSR....9...12M}.

The realization that the correct interpretation of the CR measurements requires a corresponding accuracy of the nuclear cross sections is not entirely new. An evidence can be found in the proceedings of the 16th International Cosmic Ray Conference (ICRC, Kyoto) published back in 1979, as quoted from a talk by \citeauthor{1979ICRC...14..146R} \cite{1979ICRC...14..146R}: \emph{``...this is the first time anyone involved in the experimental determination of nuclear cross sections has been asked to give a rapporteur paper at these meetings. I conclude from this that there is a growing realization of the importance of such measurement for the interpretation of an increasingly abundant and sophisticated body of CR observational data.''}

At the end of the 1980s, several CR and particle physicists gathered in the so-called ``Transport Collaboration'' \cite{1990ICRC....3..428T}, proposing a dedicated program focused on, but not restricted to, data from $Z<26$ beams. A significant effort was made by Bill Webber and his colleagues, who measured a number of isotopic production cross sections using secondary ion beams on liquid hydrogen, carbon, and methylene CH$_2$ targets (and using a CH$_2$ -- C subtraction technique) in the energy range $\sim$400--800 MeV/n \cite{1982ApJ...260..894W,1988PhRvC..37.1490F,1990PhRvC..41..520W,1990PhRvC..41..533W,1990PhRvC..41..547W,1990PhRvC..41..566W,1996PhRvC..53..347K,1997ApJ...479..504C,1997PhRvC..56..398K,1997PhRvC..56.1536C,1998PhRvC..58.3539W,1998ApJ...508..940W,1998ApJ...508..949W}. The cross sections measured by the members of the Transport Collaboration and assembled from the literature along with the existing at that time semi-empirical codes (WNEW and YIELDX \cite{1990PhRvC..41..566W,1998ApJ...501..911S,1998ApJ...501..920T}) were made available to the community through a dedicated web-site.

Besides CRs, extensive efforts for the measurement of production cross sections were driven by the space-flight radiation shielding applications and by the interest in the production of cosmogenic isotopes. The former usually involve heavy targets \cite{2002AIPC..610..285Z,2004AdSpR..34.1383C}, but hydrogen-target cross sections can be derived from C and CH$_2$ target measurements performed by the Zeitlin's group \cite{1997PhRvC..56..388Z,2001PhRvC..64b4902Z,2007PhRvC..76a4911Z,2008PhRvC..77c4605Z,2011PhRvC..83c4909Z}. The latter are focused on the production of radioactive isotopes and involve proton \cite{1998LPI....29.1189S} and neutron \cite{2013NIMPB.294..470R} beams. Nevertheless, extensive measurements by Michel and Leya's group \cite{1989Ana...114..287M,1990NIMPB..52..588D,1993NIMPB..82....9B,1995NIMPB.103..183M,1996NIMPB.114...91S,1997NIMPB.129..153M,1998NIMPB.145..449L,2005NIMPB.229....1L,2006NIMPA.562..760L,2008NIMPB.266....2A} and Sisterson's group \cite{1992LPI....23.1305S,1994NIMPB..92..510S,1997NIMPB.123..324S,1998LPI....29.1234S,2000LPI....31.1432S,2002NIMPB.196..239K,2006NIMPB.251....1S} carried out since 1990s cover many reactions needed for CR studies. Measurements of fragmentation of C \cite{1999ICRC....4..267K,2002JPhG...28.1199K}, Fe \cite{2007PhRvC..75d4603V,2008PhRvC..78c4615T}, and some other nuclei \cite{1997ICRC....4..309F,1988ZPhyA.331..463B,2001ICRC....5.1960F} were also done in the past, but more recent measurements are focused on ultra-heavy species, highly deformed nuclei, and/or short-lived radioactive beams.

Meanwhile, even though many relevant production cross sections have been measured, most of the available data, if exists, is at low energies, below a few tens of MeV/n, and/or between hundreds of MeV/n to a couple of GeV/n with just one or a few data points available. The latter are of interest for astrophysics of CRs, however, the data points published by different groups often differ by a significant factor. Besides, due to the different measurement techniques used by different groups, the published values are not easy to compare, as, e.g., in the case of the individual, direct, cumulative, differential, total, or isobaric cross sections, or reactions with metastable final states, while the target could be a particular isotope, a natural sample with mixed isotopic composition, or a chemical compound. On top of that, many astrophysically important reactions are not measured at all.

To account for the lack of data for many reactions and energies in the early days of CR physics back in the 1960s, efforts were made to establish systematic semi-empirical parametrizations \cite{1966ZNatA..21.1027R}. This approach was refined by Silberberg and Tsao's group from the 1970s till the end of the 1990s \cite{1973ApJS...25..315S,1973ApJS...25..335S,1983ApJS...51..271L,1985ApJS...58..873S,1990PhR...191..351S,1993PhRvC..47.1225S,1993PhRvC..47.1257T,1998ApJ...501..911S,1998ApJ...501..920T}, updating their parametric formula whenever new data became available. As an alternative to the semi-empirical approach, Webber and coworkers developed a data-driven empirical formula that, however, has a very limited validity energy range, with the last update in 2003 \cite{2003ApJS..144..153W}. Unsurprisingly the latter was found to fare better in terms of overall accuracy (for $Z\leq30$) \cite{1990ICRC....3..420C,1998ApJ...501..911S} when compared to the data. Therefore, Webber's WNEW and Silberberg and Tsao's YIELDX codes remained the state-of-the-art cross-section codes used in CR studies for a long time.

An alternative approach has been used by the GALPROP team \cite{2007ARNPS..57..285S} who developed a set of routines called {\tt nuc\_package.cc}\footnote{\url{http://galprop.stanford.edu}}. It is based on a careful inspection of the quality and systematics of various datasets and semi-empirical formulae, and uses the best of parametric formulae (normalized to the data when exists) and results of nuclear codes \cite{2001ICRC....5.1836M,2003ApJ...586.1050M,2003ICRC....4.1969M,2004AdSpR..34.1288M,2005AIPC..769.1612M} or even a direct fit to the data for each particular reaction. The nuclear codes used in this work included a version of the Cascade-Exciton Model (CEM2k) \cite{2004AdSpR..34.1288M} and the ALICE code with the Hybrid Monte Carlo Simulation model (HMS-ALICE) \cite{1996PhRvC..54.1341B,1998PhRvC..57..233B}. The package also includes an extensive nuclear reaction network built using the Nuclear Data Sheets. The total fragmentation cross sections are calculated using CRN6 code by Barashenkov and Polanski \cite{BarPol1994}, or using optional parametrizations \cite{1983ApJS...51..271L} or \cite{1996PhRvC..54.1329W} (with corrections provided by the authors). This was a very laborious work, often with getting into the details of the original measurements to find out which of the conflicting data points is more reliable, but produced probably the most accurate package for massive calculations of the nuclear cross sections so far. More recent attempt to characterize the uncertainties in the calculation of the isotopic production cross sections was made in the framework of the ISOtopic PROduction Cross Sections (ISOPROCS) project \cite{2011ICRC....6..283M,2013ICRC......0823M}.

In a broader outlook, semi-empirical codes are still refined nowadays and systematically evaluated against existing data (NUCFRAG \cite{1986NIMPB..18..225W,1987CoPhC..47..281B,1994NIMPB..94...95W,2012NIMPA.678...21A}, EPACS \cite{1990PhRvC..42.2546S,2000PhRvC..61c4607S,2012PhRvC..86a4601S}, SPACS \cite{2014PhRvC..90f4605S,2016PhRvC..94c9901S}, FRACS \cite{2017PhRvC..95c4608M}). Besides, progress in computing technology have made Monte Carlo simulation codes and event generators more appealing, often motivated by the problem of transport of ions in tissue-equivalent materials (e.g., GEANT4 \cite{2010NIMPB.268..604P}, PHITS \cite{1995PhRvC..52.2620N,2010AdSpR..45..892S,2014cosp...40E3081S}, SHIELD-HIT \cite{2012PMB....57.4369H}, FLUKA \cite{2011NIMPB.269.2850B}). These Monte Carlo codes rely on a combination of different physics implementations incorporating calculations of the nuclear cross sections, and their continuous improvements are of interest for a wide range of applications including Monte Carlo simulations of the design of CR instrumentation. However predictably, the accuracy of the cross section calculations provided by the transport codes still falls behind the dedicated nuclear codes. This is illustrated by the recent results and validations of CEM, LAQGSM, and MCNP6 codes \cite{2011EPJP..126...49M,2014NIMPA.764...59M,2015NIMPB.356..135K,2017PhRvC..95c4613M}.

Despite recent progress and a large variety of currently available nuclear codes, their accuracy does not exceed 5-10\% at best for \emph{some} reaction channels and in a narrow energy range, but those channels are not necessarily the most important for CR studies. In the absence of the reliable measurements for other channels their accuracy remains questionable. Besides, most (if not all) parametrizations listed above assume energy-independent behaviour of the production cross sections above a few GeV/n, whereas a rise in the total and inelastic nucleon-nucleon cross section is reliably observed \cite{2011PhRvL.107u2002B,2015PhRvD..92k4021B}. New measurements in the range from 1 GeV/n to 100 GeV/n would allow a similar rise in the fragmentation cross sections to be tested, which is of great interest for CR studies (e.g., \cite{2017PhRvL.119x1101G}).

The paper is organized as follows: in Sect.~\ref{sec:setup}, we present the propagation setup and cross section parametrizations used. In Sect.~\ref{sec:properties}, we discuss the choice of reactions and criteria we use in this first study. In Sect.~\ref{sec:ranking}, we provide the ranking tables of relevant reactions (projectile, target, and fragments). These tables are used in Sect.~\ref{sec:err_prop} to evaluate how cross section uncertainties propagate to the modelled CR fluxes; they are also used in Sect.~\ref{sec:beamtime} to provide guidelines and recommendations to establish programs to measure cross sections with the accuracy corresponding to that of the AMS-02 data.

For the sake of readability, we moved to the Appendix many tables, plots, and discussions. In particular, the presentation of the dominant production channels (projectile and fragments), which may be of interest mostly to CR physicists, is deferred to App.~\ref{app:channels}. Tables \ref{tab:xs_Li} to \ref{tab:xs_N} for the ranking of the most important reactions are given in App.~\ref{app:table_reactions}, while Fig.~\ref{fig:fabc_LiBeBCN} in App.~\ref{app:plots_fabc} provides a graphical view of the ``flux impact'' coefficients. The evolution of errors on Li to C flux from better measured reactions is shown in App.~\ref{app:plots_errevol}. Plots for the cross section data and models are given in App.~\ref{app:plots_xs_inel} (inelastic) and~\ref{app:plots_xs_prod} (production).

\section{Calculation setup} \label{sec:setup}

\subsection{Propagation}

The key components of CR propagation models include the description of the source spectra (injection spectra and isotopic abundances), a system of transport equations with spatial and momentum diffusion terms (diffusion, convection, reacceleration, energy losses) and their transport coefficients, particle and nuclear production and disintegration cross sections, and the description of the ISM (gas distribution, radiation and magnetic fields). Though the propagation codes may differ by their assumptions, description of their ingredients, geometry, and approaches to the solution of transport equations, the models with similar effective grammages (ISM gas density integrated along the path of a CR particle) yield a similar prediction for the secondary to primary nuclei ratios (e.g., B/C) \cite{2001ApJ...547..264J}. Therefore, the ranking provided below is effectively independent of the specific model implementation.

To facilitate the computations, all calculations in this study are made using a semi-analytical 1D propagation model USINE \cite{2010A&A...516A..66P} incorporating a nuclear reaction network from the heaviest $^{56}$Fe isotope to the lightest $^6$Li (contribution of species heavier than $^{56}$Fe is negligible in the context of this study). Source abundances are normalized to match HEAO-3 elemental abundances at 10.6~GeV/n after the propagation \cite{1990A&A...233...96E}. Recent measurements of spectra of CR species $Z\le8$ by AMS-02 \cite{2017PhRvL.119y1101A,2018PhRvL.120b1101A} are significantly more precise, but the analysis of heavier nuclei is still in progress. The isotopic composition of each element is assumed to match its Solar system values \cite{2003ApJ...591.1220L}. Our results depend on the latter assumption, which may not be valid for all isotopes (e.g., \cite{2008NewAR..52..427B}), but not critically. Besides, the isotopic composition of CRs at 10 GeV/n is unknown, and, therefore, such situation is unavoidable. The injection spectrum is assumed to be a single power law in rigidity ($R=pc/Ze$) without a spectral break\footnote{This is not an oversimplification because the ranking energy, 10~GeV/n (discussed in App.~\ref{sec:10gevn}), is chosen to be close to the normalization energy, 10.6~GeV/n, which makes our results essentially independent of the exact shape of the injection spectrum.}. The fractions of secondary components depend on the transport parameters, but as long as the B/C ratio is recovered (even loosely), the ranking is only mildly affected by their exact value.

To summarize, our results are robust against the choice of the propagation model and its transport parameters and only mildly dependent on the injection spectra and source distribution.

\subsection{Cross section datasets}\label{sec:XS}

To establish the ranking of the cross sections, we rely on several GALPROP and WNEW/YIELDX parametrizations that are used to estimate the cross section uncertainties:
\begin{itemize}

\item {\em WKS93, WKS98, S01, and W03}: the parametrizations developed by Webber and co-workers are based on certain observed properties of nuclear fragmentation. The formulae are fitted to the data at a single energy $\sim$600 MeV/n and take advantage of similar energy dependencies for fragments of similar charge, with three terms: an exponential dependence on the charge difference between the parent nucleus and the fragment, dependence on the width of the mass-yield distribution of the fragment, and the energy dependence with the charge. Both WKS93 and WKS98 are based on the WNEW code, respectively run with initialization files given in \cite{1990PhRvC..41..566W} and \cite{1998PhRvC..58.3539W,1998ApJ...508..940W,1998ApJ...508..949W}. The scaling $\sigma_{\rm He}/\sigma_{p}$ in WNEW is from \cite{1988PhRvC..37.1490F}. The datasets S01 (Aim\'e Soutoul, private communication) and W03 (Bill Webber, private communication) correspond to independently derived updates of WNEW based on new data from Webber and collaborators \cite{2003ApJS..144..153W}, Michel's group \cite{1989Ana...114..287M,1995NIMPB.103..183M,1997NIMPB.129..153M}, and \cite{1988ZPhyA.331..463B,1997ICRC....4..309F,2001ICRC....5.1960F,2002JPhG...28.1199K}.

   \item {\em TS00}: this is the semi-empirical parametrization developed by Tsao and Silberberg in their YIELDX code. It was updated in 1998 \cite{1998ApJ...501..911S,1998ApJ...501..920T}, based on the data from the Transport collaboration, and the last iteration was made available on the internet in 2000. The parametrization relies on regularities observed for certain mass differences between the fragments and the parent nucleus, and the ratio of the number of neutrons and protons in the fragments. There are also parameters related to the nuclear structure, number of stable levels, and pairing factor of neutrons and protons in the fragmentation products.

   \item {\em GALPROP12 and GALPROP22 (GP12, GP22)}: these are based on a careful inspection of the quality and systematics of various datasets and semi-empirical formulae, and use the best of parametric formulae (normalized to the data when exists) and results of nuclear codes \cite{2001ICRC....5.1836M,2003ApJ...586.1050M,2003ICRC....4.1969M,2004AdSpR..34.1288M,2005AIPC..769.1612M} or even a direct fit to the data for each particular reaction, as described in the Introduction. For less important cross sections, WKS93 (option 12 in GALPROP) or TS98 (option 22 in GALPROP) are used, normalized to the data when exists.
\end{itemize}

\section{Properties of $Z=3-7$ fluxes} \label{sec:properties}

In CR studies, it is customary to distinguish between ``primary'' and ``secondary'' species. The primaries are those that are present in the CR sources (e.g., $^1$H, $^4$He, C, O, Fe), whereas those produced mostly in nuclear fragmentation of heavier species in the ISM are called secondaries (e.g., $^2$H, $^3$He, Li, Be, B, sub-Fe). Of course, strictly speaking, there is always some fraction of secondaries present even in species that are mostly ``primary''.

\subsection{Physics case}

Secondary-to-primary ratios are key to CR physics because the source term mostly factors out of the ratio (they only depend on the transport coefficients or grammage). The most studied is the B/C ratio, the easiest one to measure experimentally, compared to sub-Fe/Fe which is less abundant and more difficult to resolve as it has a smaller value of $\Delta Z/A$, or to $^2$H/He and $^3$He/He that require isotopic identification. The B/C ratio is the first secondary-to-primary ratio that have been analysed and published by the AMS-02 collaboration \cite{2016PhRvL.117w1102A}, with an accuracy of a few per cent.

As already emphasized in the Introduction, the scientific case for improving the production cross sections of the secondary species Li, Be, and B, is very strong. Moreover, Be and B nuclei contain imprints of the decay of the so-called radioactive clock $^{10}$Be$\to^{10}$B \cite{1998ApJ...509..212S,1998ApJ...506..335W,1998A&A...337..859P,2002A&A...381..539D}, which is used to break the degeneracy between the normalization of the diffusion coefficient and the diffusion volume of the Galaxy. Lithium is also of great interest, but its measurement is difficult because of the contamination of the much more abundant He nuclei\footnote{See the scarcity of Li data in the cosmic-ray database (CRDB) \cite{2014A&A...569A..32M}: \url{http://lpsc.in2p3.fr/crdb/}}. The interpretation of the recently published Li, Be, and B fluxes by the AMS collaboration \cite{2018PhRvL.120b1101A} will be extremely valuable and can provide complementary tests of the interstellar transport.

C and O are the most abundant CR species after H and He. Their measurements from 2 GV to 3 TV have been recently published by the AMS collaboration \cite{2017PhRvL.119y1101A} together with an updated spectrum of He. Oxygen is so abundant that contribution of heavier species to the production of secondary O is at the level of a few per cent and can be safely neglected for our purposes. Carbon is mostly primary, but has $\sim 20\%$ of secondary contribution at about 1 GeV/n coming mostly from fragmentation of almost entirely primary $^{16}$O. Nitrogen is about 50-50 primary-secondary. An accurate measurement of its isotopic production cross sections may help to unveil the origin of low-energy CRs in the vicinity of the solar system \cite{2003ApJ...586.1050M}, and provide long-awaited clues to the solutions of other current astrophysical puzzles, such as, e.g., the origin of the positron excess \cite{2009Natur.458..607A,2014PhRvL.113l1101A}.

This makes the isotopes of Li-N the highest priority for the first run of the new production cross section measurements.

\subsection{Primary/secondary/radioactive fractions}

The isotopic composition of Li-N elements is compiled in Table~\ref{tab:origin}. The three middle columns indicate the fractions of primary and secondary components in each isotope in CRs, along with the fraction that comes from the radioactive decay (all estimated at 10~GeV/n, see App.~\ref{sec:10gevn}). Pure secondary Li, Be, and B isotopes are also shown in the Table. About 15\% of isotope $^{10}$B is coming from the $\beta^-$-decay of $^{10}$Be. Elements C and N are a mixture of primary and secondary contributions. The Table also lists $^{14}$C isotope, despite its short half-life and very low abundance, because its detection in CRs would shed light on propagation of CRs in the local interstellar medium.

\begingroup
\squeezetable
\begin{table}[t]
\caption{Fractions of primary/fragmentation/radioactive origin (w.r.t. total flux), and contributions of 1-/2-/'more-than-2' step channels (w.r.t. total secondary production) at 10 GeV/n. These numbers are independent of the propagation model if sources have the same spectral index.\label{tab:origin}}
\begin{tabular}{lrcrrrcccc}
\hline\hline
\multicolumn{2}{l}{CR}&~~&\multicolumn{3}{c}{\% of total flux} &~~~~~~~&\multicolumn{3}{c}{\!\!\!\!\!\!\!\!\!\% of multi-step secondaries}\\
\multicolumn{2}{r}{\% isotope} &&~prim.\!\!&frag.&rad.\!&&~~~ 1 ~~~&~~~ 2 ~~~&\!\!\!\!\!\!$>\!2$\\ \hline
Li   &                     &&   0 & 100 &   0&& 66 &  25 &   9\\[-1.0mm]
     & (56\%)~   $^{6}$Li &&   0 & 100 &   0&& 66 &  25 &   9\\[-1.0mm]
     & (44\%)~   $^{7}$Li &&   0 & 100 &   0&& 66 &  26 &   8\\
Be   &                     &&   0 & 100 &   0&& 73 &  20 &   7\\[-1.0mm]
     & (63\%)~   $^{7}$Be &&   0 & 100 &   0&& 78 &  17 &   6\\[-1.0mm]
     & (30\%)~   $^{9}$Be &&   0 & 100 &   0&& 65 &  26 &   9\\[-1.0mm]
     & (6\%)~  $^{10}$Be &&   0 & 100 &   0&& 66 &  26 &   7\\
B    &                     &&   0 &  95 &   5&& 79 &  17 &   5\\[-1.0mm]
     & (33\%)~   $^{10}$B &&   0 &  85 &  15&& 70 &  24 &   6\\[-1.0mm]
     & (67\%)~   $^{11}$B &&   0 & 100 &   0&& 82 &  14 &   4\\
C    &                     &&  79 &  21 &   0&& 77 &  17 &   5\\[-1.0mm]
     & (90\%)~   $^{12}$C &&  88 &  12 &   0&& 72 &  21 &   6\\[-1.0mm]
     & (10\%)~   $^{13}$C &&   7 &  93 &   0&& 83 &  13 &   4\\[-1.0mm]
     & (0.02\%)~   $^{14}$C &&   0 & 100 &   0&& 56 &  35 &   9\\
N    &                     &&  27 &  72 &   2&& 87 &   9 &   4\\[-1.0mm]
     & (54\%)~   $^{14}$N &&  49 &  48 &   3&& 83 &  13 &   4\\[-1.0mm]
     & (46\%)~   $^{15}$N &&   0 & 100 &   0&& 89 &   7 &   3\\
\hline\hline
\end{tabular}
\end{table}
\endgroup

\subsection{Why go beyond 1-step reaction?}

Strictly speaking, if the fluxes of CR species along with the change of their isotopic composition with energy were measured with a good precision, then for estimating the effects of the uncertainties in the production cross sections one would need to account only for direct reactions. However, the fluxes of the majority of CR species are known to 15\%-20\% at best, where uncertainties are steeply increasing with energy. The isotopic abundances were measured at energies below $\sim$500 MeV/n by ACE/CRIS \cite{2013ApJ...770..117L}, Voyager 1, 2 \cite{2002ApJ...568..210W}, Ulysses \cite{1998ApJ...501L..59C}, and by other instruments, but no information of CR isotopic composition is available at higher energies. Therefore, the accuracy of the calculated isotopic composition of each element depends on the accuracy of the production cross sections. Even though the predicted flux of an element often can be normalized to the observations by adjusting the fraction of the primary component, the remaining uncertainty in the isotopic composition propagates to all secondaries produced through fragmentation of this element.

This uncertainty can be accounted for by inclusion of multi-step reactions involving one or several stable or long-lived intermediate nuclei\footnote{This account for all short-lived nuclides decaying into the intermediate or final nuclei, see Sect.~\ref{sec:ghosts}.}. The three right columns in Table~\ref{tab:origin} show the fraction of a contribution to a particular isotope from a 1-step (direct) reaction, from 2-step reactions with a stable or long-lived intermediate nucleus that experiences the second interaction, and from $>$2-step reactions involving more stable intermediate nuclei. They are discussed in the next Section. It can be seen that a contribution from 1-step reaction dominates in all cases, but contributions from reactions involving two or more interactions in the ISM are not negligible.

\section{Ranking of production reactions} \label{sec:ranking}

For the practical purpose of ranking the most important reactions, one must isolate all contributions $X+\{p,\alpha\}\rightarrow F$ (projectile $X$ on H or He target producing a fragment $F$) to the fragment of interest $F$ summed over projectiles $X$ and targets. The CR residence time in the Galaxy is very large, typically a few tens of Myr, as was first hinted at in \cite{1970ApJ...162..837R}. Therefore, all isotopes with a half-life below a few kyr are considered short-lived. Time dilation increases the half-live of such isotopes, but it becomes relevant only at very-high energies (Lorentz factor of $\sim$100-1000), where the CR isotopic composition is not measured yet.

\subsection{Ghost nuclei}\label{sec:ghosts}
Short-lived nuclei produced in the fragmentation of heavier species decay before they can interact with interstellar gas (the ISM can be considered a thin target). Therefore, we are interested only in their stable or long-lived decay products that effectively increase the production cross sections of the corresponding daughter nuclei. The cumulative cross section $\sigma^{\rm c}$ from a projectile X for a given fragment is given by the direct production, plus the production of the short-lived nuclei (decaying into this fragment) weighted by the branching ratio ${\cal B}r$ of the decay channel. For instance, for $^{10}{\rm B}$, there is a single short-lived nucleus,
\begin{equation}
  \sigma^{\rm c}_{{\rm X}\rightarrow ^{10}{\rm B}} =
    \sigma_{{\rm X}\rightarrow {\rm ^{10}B}} + \sigma_{{\rm X}\rightarrow {\rm ^{10}C}} \times {\cal B}r({\rm ^{10}B}\rightarrow {\rm ^{10}C})\;,
    \label{eq:ghosts}
\end{equation}
with ${\cal B}r({\rm ^{10}B}\rightarrow {\rm ^{10}C})=100\%$. We dubbed these short-lived nuclei {\em ghosts}, as they only show up in the cumulative cross section, but do not appear at the propagation stage\footnote{To have a clear picture of which CR reactions receive significant contributions from ghost nuclei, we have reported their contributive fraction in the last column of Tables \ref{tab:xs_Li} to \ref{tab:xs_N} in appendix \ref{app:table_reactions}.}. Ghost nuclei and their branching ratio must be minutely reconstructed from nuclear data tables \cite{2017ChPhC..41c0001A}.

The first comprehensive lists of ghost nuclei and decay networks were compiled by \cite{1984ApJS...56..369L,1987ApJS...64..269G} in 1980s. More recently, the routine {\tt nucdata.dat} calculating the nuclear reaction network was built in 2000 as a part of the GALPROP {\tt nuc\_package.cc}, the package of routines handling isotopic production, nuclear disintegration, and radioactive decay. The routine has a capability to use the network that is built from scratch using the Nuclear Data Sheets or uses the network borrowed from \cite{1987ApJS...64..269G}. Independently, a reconstruction of the ghosts was proposed by \cite{00/04/78/51/PDF/tel-00008773.pdf} in 2001 (based on 1997 NUBASE properties \cite{1997NuPhA.624....1A}). The networks take into account that nuclei in CRs are fully ionized above a few GeV/n and, therefore, their decay via the electronic capture (EC) is blocked and corresponding EC-decay species are stable in CRs. Note that GALPROP can handle H-like ions, electron pick-up from the interstellar gas \cite{WilsonThesis1978}, and electron stripping \cite{1970RSPSA.318....1F,WilsonThesis1978} that can make a difference in the abundances of some EC-decay species at low energies. Meanwhile, accurate experimental determination of the decay mode (EC or $\beta^+$) is often complicated, especially for heavy nuclei. This may lead to over- or under-estimate of their half-life in CRs. On the other hand, accurate determination of the decay modes of the ghost nuclei located far away from the valley of stability is not necessary, given their extremely short life-time and very small production cross sections.

We emphasize that the half-life of a ghost is a crucial input to determine whether it can be measured in the laboratory with a particular experimental setup, or if its contribution is hidden in the cumulative cross sections of the corresponding daughter nuclei.

\subsection{Definition: $f_{abc}$ for reaction $a+b\rightarrow c$\label{sec:f_abs}}

The results obtained in the previous section are based on the cumulative cross sections. Now we have to consider all cross sections separately for two main reasons: first, even the short-lived fragments will most likely fly through the detector before decaying (depending on the experimental setup and nucleus half-life), and, second, at the fundamental level, these cross sections are those that matter for comparisons or validation against cross section calculations.

The light nuclei considered for this study are based on the list compiled in \cite{00/04/78/51/PDF/tel-00008773.pdf}. We have reprocessed the GALPROP GP12 and GP22 cross sections to provide separate contributions for the ghosts in the ranking. We then loop on all CR projectiles $a$, ISM targets $b$ (H and He), and stable and ghosts fragments $c$, to calculate the fraction:
\begin{equation}
   \label{eq:f_abc}
   f_{abc} = \frac{\psi^{\rm sec}({\rm ref})-\psi^{\rm sec}(\sigma^{a+b\rightarrow c}=0)}{\psi^{\rm sec}({\rm ref})}\;.
\end{equation}
This fraction measures the influence of a given cross section on the total flux. In other words, if the corresponding cross section is set to 0, $\psi^{\rm sec}$ would decrease by $f_{abc}$. Note that $\sum_{a,b,c} f_{abc}>100\%$\footnote{It would be exactly 100\% if only 1-step channels were to exist. As seen in the previous section, 2-step channels are not negligible, and they involve 2 cross sections that contribute to the same fraction, hence double counting is unavoidable.}. We emphasize that for each $f_{abc}$ calculation, when switching one cross section off at a time, we always renormalize the elemental fluxes to the observations by re-adjusting the source abundances. This ensures that the provided uncertainties correspond to the standard way the CR propagation calculations are performed.\footnote{The coefficients obtained with renormalization are smaller than those without renormalization and converge to zero faster. For the former, the constraint to match elemental fluxes translates into a readjustment of the elemental source abundance (isotopic source abundances are fixed), whereas it does not for the latter, overestimating the true impact on the final flux.}

\subsection{Ranked reactions}

Tables \ref{tab:xs_Li} to \ref{tab:xs_N} list the ranked reactions at 10 GeV/n for Li to N. The Tables are deliberately cut off when the combined listed $f_{\rm abc}$ reach $70\%$ of the sum of all calculated $f_{\rm abc}$. We rely on the two GALPROP datasets to see the scatter of the current models, the minimum and maximum values. Unsurprisingly, we recover first the reactions involved in the dominant channels of Tables \ref{tab:channels_Li} to \ref{tab:channels_C}, but here the additional information is used:
\begin{itemize}
   \item Targets: using the relative contribution of H (90\%) and He (10\%) in the ISM, and the simple $A^{2/3}$ dependence of cross sections, sufficient for our purposes, we expect the contributions of the reactions on He target to be $\sim0.25$ times of those on H target for similar projectiles and fragments. This is what we observe in the relative ranking of the H and He targets. Applying the same scaling to the next most abundant targets in the ISM, C and O, whose relative abundances are C/H$\approx2.7\cdot10^{-4}$ and O/H$\approx4.9\cdot10^{-4}$ \cite{2009ARA&A..47..481A}, gives a $0.1\%$ contribution for C and $0.3\%$ for O. We conclude that target elements heavier than He in the ISM can be safely discarded.
   \item Ghost nuclei: their contributions (in boldface in Tables \ref{tab:xs_Li} to \ref{tab:xs_N}) can be very important, as much as $\sim 25\%$ for certain isotopes (e.g., $^{11}$C for Be, $^{13}$O for C, and $^{15}$O for N). The most important ghosts are collected in Table~\ref{tab:ghosts} in Appendix~\ref{app:table_reactions}.
   \item Cross section values: the tables also give the corresponding cross sections for these reactions, with the range (minimum and maximum) obtained from the two datasets considered. The cross section values are used in the next section to calculate generic beam time required to reach the AMS-02 precision.
\end{itemize}
We note that individual cross sections are involved in both 1-step and 2-step reactions, so that their contributions do not have a universal energy dependence. Meanwhile, only 1-step reactions matter at high energies and, therefore, the ranking of cross sections would become energy independent. At low energy, a small dependence is still expected.

\section{Error propagation on modelled fluxes}\label{sec:err_prop}

Down to which value do we have to rank the above fractions $f_{abc}$ to ensure an $x\%$ accuracy on the modelled fluxes? As we show in this and the next sections, this question does not have a simple and unique answer, and the best answer may also depend on the way the cross sections are measured.

\subsection{From $f_{abc}$ to error on CR fluxes}

The infinitesimal variation of the secondary flux $\psi^{\rm sec}$ for a fragment, with respect to the reaction $a+b\rightarrow c$ and its cross section $\sigma^{abc}$, can be written in the generic form:
\begin{equation}
   \label{eq5}
d\psi^{\rm sec} = \sum_{a,b,c} \frac{\partial\psi^{\rm sec}}{\partial \sigma^{abc}}\;d\sigma^{abc}\;.
\end{equation}
Using the definition of the flux impact $f_{abc}$ given in Eq.~(\ref{eq:f_abc}), one can rewrite
\begin{equation}
   \label{eq6}
\frac{\partial\psi^{\rm sec}}{\partial \sigma^{abc}} \approx \frac{\Delta\psi^{\rm sec}}{\Delta \sigma^{abc}} \approx f_{abc}\,\frac{\psi^{\rm sec}({\rm ref})}{\sigma^{abc}}\;.
\end{equation}
One can also express
the relative uncertainty in the total flux ($\psi^{\rm tot}$) through the relative uncertainty in the secondary production of the same species ($\psi^{\rm sec}$)
\begin{equation}
   \label{eq7}
\frac{\Delta  \psi^{\rm tot}}{\psi^{\rm tot}} = f^{\rm sec}\, \frac{\Delta  \psi^{\rm sec}}{\psi^{\rm sec}},
\end{equation}
where $f^{\rm sec}$ is the fraction of secondaries in that particular species shown in Table~\ref{tab:origin}. The value $\Delta  \psi^{\rm tot}/\psi^{\rm tot}$ is exactly what we are interested in.  We consider three different assumptions regarding the correlations between the cross section uncertainties, which result in the following formulae for the uncertainties in the total flux:
\begin{itemize}
   \item fully correlated uncertainties:
\begin{eqnarray}
\left(\frac{\Delta  \psi^{\rm tot}}{\psi^{\rm tot}}\right)^{\rm corr} \approx f^{\rm sec}\, \sum_{a,b,c} f_{abc} \frac{\Delta\sigma^{abc}}{\sigma^{abc}}\,;
\label{eq:uncertainty_sumCorr}
\end{eqnarray}
   \item uncorrelated uncertainties:
\begin{eqnarray}
\left(\frac{\Delta  \psi^{\rm tot}}{\psi^{\rm tot}}\right)^{\rm uncorr\!\!\!\!\!} &\approx& f^{\rm sec}\, \sqrt{\sum_{a,b,c} \left(f_{abc} \frac{\Delta\sigma^{abc}}{\sigma^{abc}}\right)^2}\,;
\label{eq:uncertainty_sumUncorr}
\end{eqnarray}
   \item uncorrelated uncertainties for fragments of the same projectile, but correlated for different projectiles:
\begin{eqnarray}
\left(\frac{\Delta  \psi^{\rm tot}}{\psi^{\rm tot}}\right)^{\rm mix\!\!\!\!} \!\!&\approx& f^{\rm sec}\, \sum_a
\sqrt{\sum_{b,c} \left(f_{abc} \frac{\Delta\sigma^{abc}}{\sigma^{abc}}\right)^2}, 
\label{eq:uncertainty_sumMix}
\end{eqnarray}
\end{itemize}
where we used Eqs.~(\ref{eq5})-(\ref{eq7}).

\subsection{Cross-sections to improve: naive approach}
\label{sec:err_evol_naive}

The optimal strategy to reach a desired relative error on the total flux, $\Delta\psi^{\rm tot}_r=\Delta  \psi^{\rm tot}/\psi^{\rm tot}$, is to improve the cross section accuracy for as many reactions as required, starting with the dominating reactions and finishing when the required precision is reached.

Let us assume that we measure all the cross sections whose flux impact is above the threshold, $f_{abc}>f_{\rm thresh}$, and that all new measurements are made with the new relative accuracy $\Delta\sigma_r^{\rm new}=\Delta\sigma^{abc}/\sigma^{abc}$, the same for all newly measured channels, while all other cross sections have a typical $20\%$ uncertainty. The condition that the required accuracy $\Delta\psi^{\rm tot}_r$ is reached can be expressed as:
\begin{equation}
\label{eq11}
f^{\rm sec} \left(\Delta\sigma_r^{\rm new}\sum_{a,b,c} \!f_{abc} + (20\%\!-\!\Delta\sigma_r^{\rm new}) \!\!\sum_{\!\!\!\!\!\!\!f_{abc}<f_{\rm thresh}\!\!\!\!\!\!} \!\!\!\!f_{abc}\!\! \right)\lesssim \Delta\psi^{\rm tot\!\!}_r\;,
\end{equation}
where we used Eq.\,(\ref{eq:uncertainty_sumCorr}) for correlated cross section uncertainties. Defining the total cumulative fraction
\begin{equation}
{\cal C}_{\rm all}\equiv\sum_{a,b,c} f_{abc}\,,
\label{eq:Call_sumCorr}
\end{equation}
we can express the cumulative fraction below the threshold as
\begin{equation}
{\cal C}_{\rm thresh} \equiv \sum_{\!\!\!\!f_{abc}<f_{\rm thresh}\!\!\!\!} f_{abc}\approx \frac{\Delta\psi^{\rm tot}_r/f^{\rm sec}-\Delta\sigma_r^{\rm new}\times {\cal C_{\rm all}}}{(20\%-\Delta\sigma_r^{\rm new})}\,.
\label{eq:Cf_sumCorr}
\end{equation}
In the case of uncorrelated uncertainties, defining
\begin{equation}
\hat{{\cal C}}_{\rm all}\equiv\sqrt{\sum_{a,b,c} f^2_{abc}}\,,
\label{eq:Call_sumUncorr}
\end{equation}
and using Eq.~(\ref{eq:uncertainty_sumUncorr}), one can get the corresponding formula
\begin{align}
\label{eq:Cf_sumUncorr}
\hat{{\cal C}}_{\rm thresh} \equiv& \sqrt{\sum_{f_{abc}<\hat{f}_{\rm thresh}} \!\!\!\!\!\! f^2_{abc}}\\
\approx& \sqrt{\frac{(\Delta\psi^{\rm tot}_r/f^{\rm sec})^2-(\Delta\sigma_r^{\rm new}\times {\hat{\cal C}_{\rm all}})^2}{(20\%)^2-(\Delta\sigma_r^{\rm new})^2}}. \nonumber
\end{align}

\begingroup
\squeezetable
\begin{table}
\caption{The Table provides the number of reactions whose cross section must be measured at relative precision $\Delta\sigma_r^{\rm new}\%$ in order to reach a relative precision $\Delta\psi^{\rm tot}_r=3\%$ for the calculated elemental flux, according to the naive approach discussed in Sect.\ref{sec:err_evol_naive}. The columns below show the element name, its secondary fraction, and the sum (or quadratic sum) over all $f_{abc}$ The remaining sets of columns (for two $\Delta\sigma_r^{\rm new}$ cases) are the number of reactions above threshold, the threshold value, and the sum (or quadratic sum) of $f_{abc}$ above threshold. See text for discussion.\label{tab:thresh_frac}}
\begin{tabular}{lrc @{\hskip 6pt} r|c|l @{\hskip 12pt} r|c|l}
  \multicolumn{9}{c}{Correlated uncertainties: Eqs.~(\ref{eq:uncertainty_sumCorr}), (\ref{eq:Call_sumCorr}), (\ref{eq:Cf_sumCorr})}\\
\hline\hline
     &  &                 &  \multicolumn{6}{c}{$N_{>{\rm thresh}}|f_{\rm thresh}|{\cal C}_{\rm thresh}$}\\
     \cline{4-9}
     &  $f^{\rm sec}$ & ${\cal C}_{\rm all}$ &  \multicolumn{3}{l}{$\Delta\sigma_r^{\rm new}=2\%$} & \multicolumn{3}{l}{$\Delta\sigma_r^{\rm new}=0\%$}\\
 Li  &  100\%  &  1.20 & 356 &  0.01\%  &  0.03    &  67 &  0.16\%  &  0.15   \\[-0.25mm]
 Be  &  100\%  &  1.14 & 236 &  0.02\%  &  0.04    &  65 &  0.20\%  &  0.15   \\[-0.25mm]
  B  &   95\%  &  1.13 &  97 &  0.06\%  &  0.05    &  31 &  0.54\%  &  0.16   \\[-0.25mm]
  C  &   20\%  &  1.08 &   2 &  18.4\%  &  0.70    &   2 &  18.4\%  &  0.73   \\[-0.25mm]
  N  &   73\%  &  1.08 &  21 &  0.43\%  &  0.11    &  11 &  1.47\%  &  0.20   \\[+0.2mm]
\hline\hline
\multicolumn{9}{c}{\phantom{A}}\\
  \multicolumn{9}{c}{Uncorrelated uncertainties: Eqs.~(\ref{eq:uncertainty_sumUncorr}), (\ref{eq:Call_sumUncorr}), (\ref{eq:Cf_sumUncorr})}\\
\hline\hline
     &  &                  &  \multicolumn{6}{c}{$\hat{N}_{>{\rm thresh}}|\hat{f}_{\rm thresh}|\hat{{\cal C}}_{\rm thresh}$}\\
     \cline{4-9}
     &  $f^{\rm sec}$ & $\hat{{\cal C}}_{\rm all}$ &  \multicolumn{3}{l}{$\Delta\sigma_r^{\rm new}=2\%$} & \multicolumn{3}{l}{$\Delta\sigma_r^{\rm new}=0\%$}\\
 Li  &  100\%  &  0.27 &   3 &  11.9\%  &  0.15    &   3 &  11.9\%  &  0.15   \\[-0.25mm]
 Be  &  100\%  &  0.27 &   2 &  15.9\%  &  0.15    &   2 &  15.9\%  &  0.15   \\[-0.25mm]
  B  &   95\%  &  0.30 &   3 &  16.2\%  &  0.16    &   3 &  16.2\%  &  0.16   \\[-0.25mm]
  C  &   20\%  &  0.32 &   0 & $\cdots$ & $\cdots$ &   0 & $\cdots$ & $\cdots$\\[-0.25mm]
  N  &   73\%  &  0.41 &   3 &  20.0\%  &  0.20    &   3 &  20.0\%  &  0.20   \\[+0.2mm]
\hline\hline
\end{tabular}
\end{table}
\endgroup

The values of $f^{\rm sec}$ and ${\cal C}_{\rm all}$ (or $\hat{{\cal C}}_{\rm all}$) for Li to N fluxes at 10~GeV/n are listed in the second and third row of Table~\ref{tab:thresh_frac} for fully correlated (top) or uncorrelated (bottom) errors. If we set the required precision to the level corresponding to the modern CR data $\Delta\psi^{\rm tot}_r=3\%$, then the remaining columns show the number of cross sections $N_{>{\rm thresh}}$ ($\hat{N}_{>{\rm thresh}}$) that have to be measured with the relative accuracy $\Delta\sigma_r^{\rm new}$, the found threshold $f_{\rm thresh}$ ($\hat{f}_{\rm thresh}$), and the threshold cumulative fraction ${\cal C}_{\rm thresh}$ ($\hat{{\cal C}}_{\rm thresh}$). The Table shows very different behaviour for these two scenarios:
\begin{itemize}
   \item In the case of fully correlated errors (top), the number of reactions to measure $N_{>{\rm thresh}}$ strongly depends on the precision of these new measurements, and rapidly increases with $\Delta\sigma_r^{\rm new}$. For species with a subdominant secondary contribution (C), very few measurements are needed, whereas the number of new measurements goes up from 27 for B to more than 60 for Li and Be in the ideal case of an infinite precision $\Delta\sigma_r^{\rm new}=0$.
   \item In the case of uncorrelated errors (bottom), the number of reactions to measure does not depend much on $\Delta\sigma_r^{\rm new}$. This scenario implies that the calculated fluxes are already close to the targeted precision, with only three dominant reactions that require new measurements, and that the precision of the calculated carbon flux is already below $3\%$.
\end{itemize}

\subsection{Cross-sections to improve: wanted reactions}

In reality, cross section uncertainties could be partially correlated and, therefore, the truth is somewhere in between the two extreme scenarios that are described above. For example, measurements made with the same experimental setup are likely to have correlated systematic errors. Instead, the data sets from different groups made with different experimental setups are likely to be uncorrelated. Besides, the degree of correlation may be energy-dependent as many experimental setups can be used in a limited energy range that may be restricted by, e.g., the beam energy, detector efficiency, power of isotope separation, and by many other factors. On top of this, because the data are often scarce and present only in a limited energy range, the cross section calculations rely heavily on various parametrizations (see Sect.~\ref{sec:XS}). In turn, these parametrizations are subject to the same correlations if they are (re-)normalized to the data (e.g., GP12 and GP22), or may introduce additional correlations due to the assumptions made to provide the best average reproduction of a certain collection of data in a particular energy range (e.g., S01, W03).

The regularities observed between various cross-section datasets can hint at the types of correlations between different reactions (or a lack of them).  Fig.~8 in \cite{2015A&A...580A...9G} shows the differences between several parametrizations by Webber and ST used in our analysis. One can see that the amplitudes of the relative differences are large for such combinations of projectiles and fragments when the difference in their nuclear charges is large $\Delta Z_{ac}=Z_a-Z_c\gg1$. Such combinations usually correspond to small absolute values of the production cross sections and often have a few or no experimental data points. Not surprisingly, different parametrizations exhibit significant discrepancies in this case. One can also see that the plots are dominated by one colour (blue in the middle and red at the top and bottom panels in Fig.~8) that implies significant biases. This simple analysis shows that different parametrizations are subjected to errors that correlate with the value of $\Delta Z_{ac}$. Surprisingly that even in the cases when only one or few nucleons are removed $\Delta A_{ac}\sim 1$, the differences between different parametrizations are also considerable. Such cross sections are usually quite large in absolute values and relatively well-measured. Still such discrepancies indicate significant systematic errors between different parametrizations and possibly between different experimental setups.

\begin{figure}[t]
\includegraphics[width=0.48\textwidth]{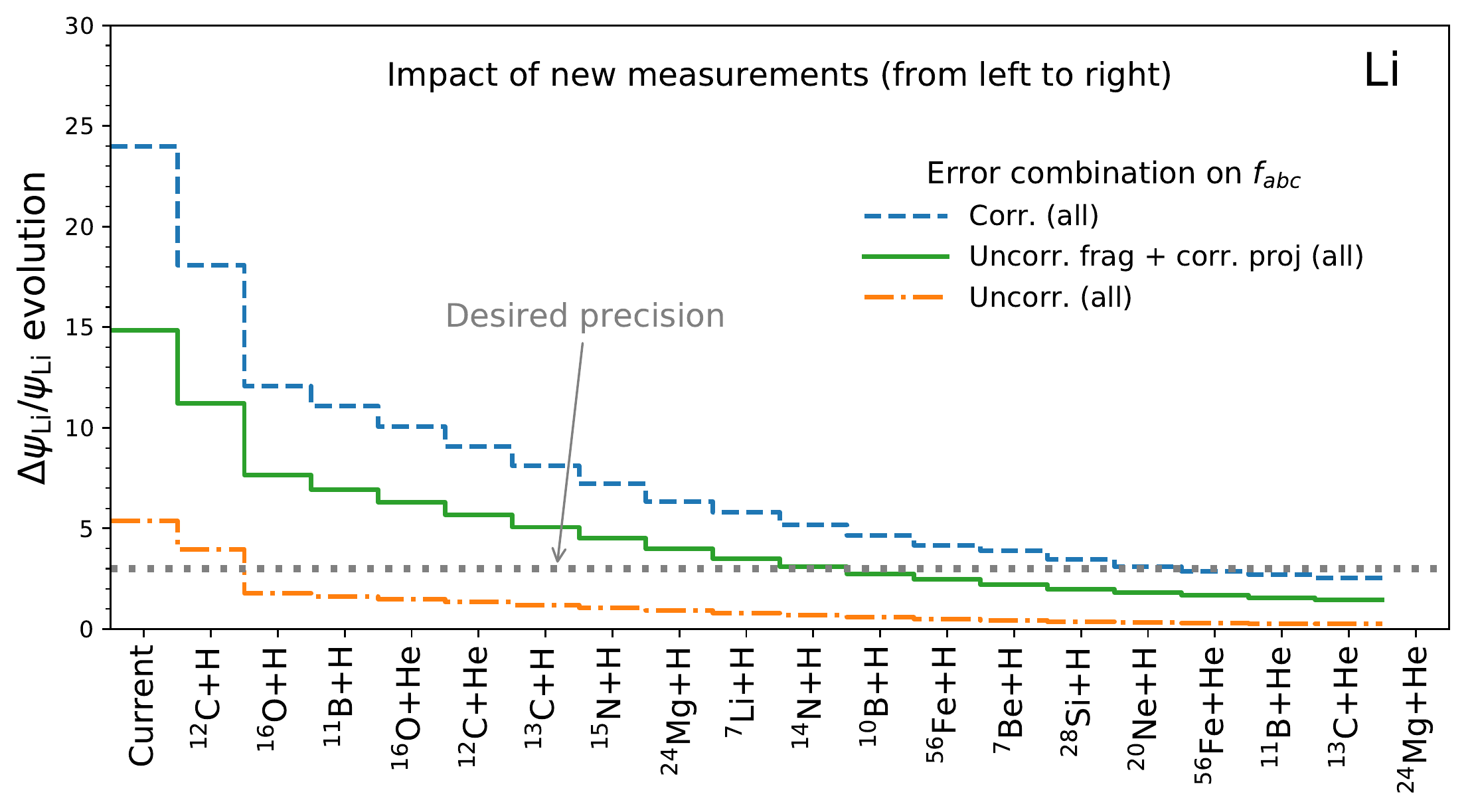}
\includegraphics[width=0.48\textwidth]{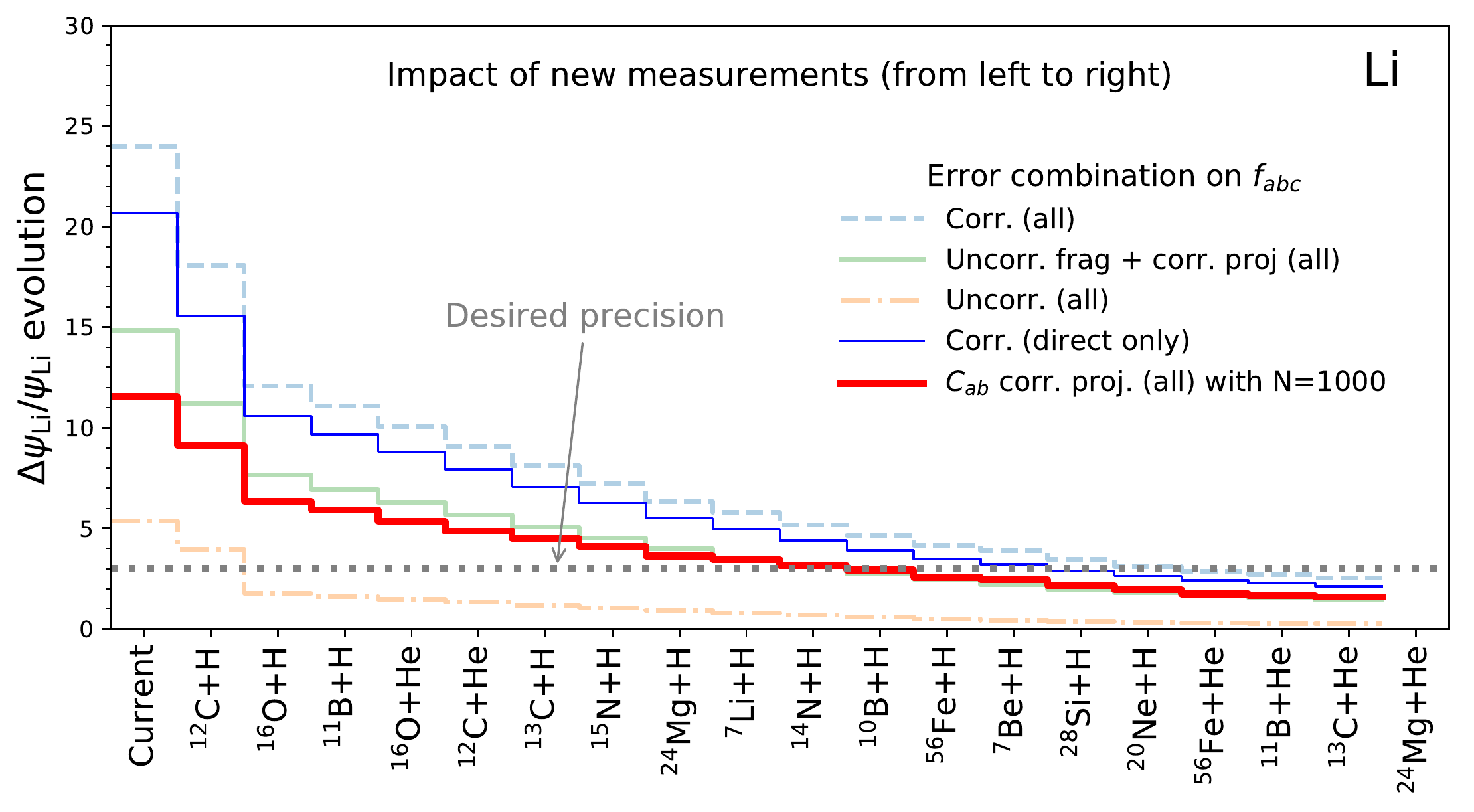}
\caption{Evolution of error on the calculated Li flux as if new reactions are measured with a perfect accuracy. The abscissa labels list the reactions (projectile+target), while the improved accuracy from a new measurement of the corresponding cross section can be read from the respective ordinate value. The plot is read from left to right, with the first bin giving the currently estimated uncertainty. The error of the calculated Li flux is decreasing as more and more reactions are well-measured. All curves assume $\Delta\sigma_r^{\rm current}=20\%$ and $\Delta\sigma_r^{\rm new}=0\%$. {\em Top panel:} three calculations based on various combination of errors, namely correlated, uncorrelated, or a mixture of these two. {\em Bottom panel:} same as in top panel (pale colours), with the additional results from direct production and $C_{ab}$ coefficients. See text for details.\label{fig:err_evol_Li}}
\end{figure}

Improvements in calculations of the nuclear cross sections will certainly remain data driven in the near future, therefore, it is important to stay close to the experimental practice. For this reason, we show in Fig.~\ref{fig:err_evol_Li} the error evolution for the Li flux with new measurements grouped by projectile plus target combinations. In both panels, the starting value of the histogram $\sim$24\% corresponding to the abscissa point marked with ``current'' shows the current estimated uncertainty for Li flux assuming $\Delta\sigma_r^{\rm current}=20\%$. The histogram shows how this uncertainty would decrease if the reactions listed along the abscissa are measured with the absolute accuracy. The projectile plus target combinations to consider for new measurement campaigns can thus be directly read off the abscissa from left to right and the corresponding histogram points then indicate the precision in the flux calculations that can be reached if such measurements are performed (dashed grey horizontal line).

The top panel shows a comparison of the error evolution for three scenarios calculated using Eqs.~(\ref{eq:uncertainty_sumCorr}), (\ref{eq:uncertainty_sumUncorr}), and (\ref{eq:uncertainty_sumMix}): correlated errors (dashed blue line) and uncorrelated errors (dash-dotted orange line) provide extreme cases, whereas a more realistic scenario is provided by the intermediate case (solid green line). Meanwhile, the ranking of the projectiles is mostly insensitive to the exact values of the cross-section uncertainties, and reflects the ranking of individual reactions shown in Table~\ref{tab:xs_Li}. The fragmentation of mostly primary $^{12}$C and $^{16}$O on H and He and of secondary $^{11}$B, $^{15}$N, $^{7}$Li species are among the most important. To have an even more realistic scenario, we have tried (not shown) to use directly the actual errors or the scatter between the data points above 1 GeV/n as the error proxy. At this energy the values of the production cross sections become largely energy independent (see Appendix~\ref{app:plots_xs_prod}). We observed typical errors or error proxies that range from $5\%$ to $20\%$ and the corresponding error evolution plots look less dramatic than in the case of fully correlated errors. One of the main reasons for that is a limiting set of data available above 1 GeV/n.

The bottom panel in Fig.~\ref{fig:err_evol_Li} shows two new histograms superimposed on top of the above-mentioned three scenarios (top panel). The thin blue line accounts only for the direct production of $^{6}$Li and $^7$Li (or through one of the ghosts) assuming correlated errors. This simple calculation involving very few reactions already captures the flux error evolution, giving another view of the fact that at 10 GeV/n most of Li is produced in direct reactions. The thick red curve shows the error evolution based on the $C_{ab}$ coefficients discussed in the next Section (Sect.~\ref{sec:beamtime}). These coefficients are meant to capture realistic multinomial-like statistical uncertainties from measuring all fragments for a given projectile given $N$ interactions in the target. Again, the behaviour is the same, although with a slower convergence.

Finally, we refer the reader to Figs.~\ref{fig:err_evol_LiBeB} and \ref{fig:err_evol_CN} in the Appendix~\ref{app:plots_errevol} for error evolution plots for all elements considered in this study. In these plots, the shaded areas indicate the range of values for the assumed current cross-section uncertainties, namely $\Delta\sigma_r^{\rm new}\in[15\%-25\%]$. The plots are shown for all three scenarios discussed above (fully correlated, uncorrelated, or mixed).

\section{Generic beam time calculation}\label{sec:beamtime}

The purpose of the previous discussion was mostly to illustrate the current flux calculation uncertainties under the assumption of different benchmark scenarios for the uncertainties of currently available cross section parametrizations.  In an experiment dedicated to the measurements of fragmentation cross sections, many fragments associated with a single projectile are measured at once, which leads to a somewhat different arrangement in the error evolution plots.

\subsection{Definition: $C_{ab}$ for reaction $a+b$}

The statistical uncertainty of such an experiment can be estimated using multinomial statistics\footnote{Note that for the validity of multinomial statistics the contributing reactions must be exclusive. This holds strictly true for fragments with a mass larger or equal  half of the projectile mass. For the dominating reactions to produce Li, Be, B, C, or N, this is indeed the case.}. Fragments of type $c$ are produced with probability
\begin{equation}
  p_c = \frac{\sigma^{abc}}{\sigma_{ab}},
\end{equation}
where $\sigma^{ab}$ is the total inelastic cross section for $a+b$ reaction and $\sigma^{abc}$ is the fragmentation cross section to produce a fragment $c$. For a number of $N$ recorded interactions, the covariance of the measured number of fragments, $n_i = p_i\, N$, is
\begin{equation}
 V_{ij}^n \equiv V(n_{c_i},n_{c_j})  =
   \begin{cases}
         N\, p_{i} (1 - p_{i}), &i=j \\
         -N\, p_{i} p_{j}, & i\neq j. 
   \end{cases}
\end{equation}
Furthermore, the Poissonian uncertainty for measuring $N$ interactions has to be taken into account. Defining
\begin{flalign}
  {\cal C}_{ab} \equiv& \left[ \left(\sum_{i=1}^m f_{abc_i}\right)^2 + \right.
          \sum_{i=1}^m f_{abc_i}^2 \left(\frac{\sigma^{ab}}{\sigma^{abc_i}}-1\right)  \nonumber\\
        &\left. - 2 \sum_{i=1}^m \sum_{j=i+1}^m f_{abc_i} f_{abc_j} \right]^\frac{1}{2},
          \label{eq:Cab}
\end{flalign}
this leads to the following expression for the relative (secondary) flux uncertainty resulting from the uncertainties of fragmentation cross sections in $a+b$ interactions:
\begin{flalign}
 \left(\frac{\Delta \psi^{\rm sec}}{\psi^{\rm sec}}\right)_{\!\! ab} =\frac{1}{\sqrt{N}} \; {\cal C}_{ab}.
          \label{eq:Err_Cab}
\end{flalign}

\subsection{Ranked $C_{ab}$}

\begingroup
\squeezetable
\begin{table}[!ht]
\caption{Table of $C_{ab}$ coefficients calculated from Eq.~(\ref{eq:Cab}). Only $C_{ab}>0.05$ are shown.\label{tab:c_ab}}
\begin{minipage}[t]{0.4\linewidth}
\begin{tabular}{cr}\multicolumn{2}{c}{ Li} \\
\multicolumn{2}{c}{$(\sum C_{ab}=5.24)$} \\
\hline\hline
Reaction $(a+b)$ & $C_{ab}$ \\\hline 
$^{16}\text{O} + \text{H}$ & 1.057\\
$^{12}\text{C} + \text{H}$ & 0.773\\
$^{14}\text{N} + \text{He}$ & 0.673\\
$^{16}\text{O} + \text{He}$ & 0.615\\
$^{14}\text{N} + \text{H}$ & 0.410\\
$^{12}\text{C} + \text{He}$ & 0.158\\
$^{24}\text{Mg} + \text{H}$ & 0.152\\
$^{11}\text{B} + \text{H}$ & 0.134\\
$^{15}\text{N} + \text{H}$ & 0.120\\
$^{13}\text{C} + \text{H}$ & 0.115\\
$^{56}\text{Fe} + \text{H}$ & 0.113\\
$^{28}\text{Si} + \text{H}$ & 0.095\\
$^{20}\text{Ne} + \text{H}$ & 0.067\\
$^{10}\text{B} + \text{H}$ & 0.066\\
$^{56}\text{Fe} + \text{He}$ & 0.064\\
$^{7}\text{Li} + \text{H}$ & 0.059\\
\hline\hline
 &\\[-1.2mm] 
 \end{tabular}

\end{minipage}
\begin{minipage}[t]{0.4\linewidth}
\begin{tabular}{cr}\multicolumn{2}{c}{ Be} \\
\multicolumn{2}{c}{$(\sum C_{ab}=6.48)$} \\
\hline\hline
Reaction $(a+b)$ & $C_{ab}$ \\\hline 
$^{16}\text{O} + \text{H}$ & 1.419\\
$^{12}\text{C} + \text{H}$ & 0.986\\
$^{16}\text{O} + \text{He}$ & 0.881\\
$^{14}\text{N} + \text{H}$ & 0.558\\
$^{14}\text{N} + \text{He}$ & 0.536\\
$^{28}\text{Si} + \text{H}$ & 0.202\\
$^{12}\text{C} + \text{He}$ & 0.192\\
$^{24}\text{Mg} + \text{H}$ & 0.192\\
$^{11}\text{B} + \text{H}$ & 0.158\\
$^{20}\text{Ne} + \text{H}$ & 0.130\\
$^{56}\text{Fe} + \text{H}$ & 0.127\\
$^{15}\text{N} + \text{H}$ & 0.121\\
$^{13}\text{C} + \text{H}$ & 0.095\\
$^{10}\text{B} + \text{H}$ & 0.083\\
$^{56}\text{Fe} + \text{He}$ & 0.061\\
\hline\hline
 &\\[-1.2mm] 
 \end{tabular}

\end{minipage}
\begin{minipage}[t]{0.4\linewidth}
\begin{tabular}{cr}\multicolumn{2}{c}{ B} \\
\multicolumn{2}{c}{$(\sum C_{ab}=3.96)$} \\
\hline\hline
Reaction $(a+b)$ & $C_{ab}$ \\\hline 
$^{12}\text{C} + \text{H}$ & 0.808\\
$^{16}\text{O} + \text{H}$ & 0.656\\
$^{16}\text{O} + \text{He}$ & 0.609\\
$^{14}\text{N} + \text{H}$ & 0.574\\
$^{14}\text{N} + \text{He}$ & 0.202\\
$^{12}\text{C} + \text{He}$ & 0.148\\
$^{11}\text{B} + \text{H}$ & 0.108\\
$^{24}\text{Mg} + \text{H}$ & 0.094\\
$^{15}\text{N} + \text{H}$ & 0.088\\
$^{28}\text{Si} + \text{H}$ & 0.080\\
$^{13}\text{C} + \text{H}$ & 0.074\\
$^{20}\text{Ne} + \text{H}$ & 0.073\\
$^{56}\text{Fe} + \text{H}$ & 0.058\\
\hline\hline
 &\\[-1.2mm] 
 \end{tabular}

\end{minipage}
\begin{minipage}[t]{0.4\linewidth}
\begin{tabular}{cr}\multicolumn{2}{c}{ C} \\
\multicolumn{2}{c}{$(\sum C_{ab}=2.40)$} \\
\hline\hline
Reaction $(a+b)$ & $C_{ab}$ \\\hline 
$^{16}\text{O} + \text{H}$ & 1.047\\
$^{16}\text{O} + \text{He}$ & 0.184\\
$^{24}\text{Mg} + \text{H}$ & 0.123\\
$^{15}\text{N} + \text{H}$ & 0.117\\
$^{20}\text{Ne} + \text{H}$ & 0.105\\
$^{14}\text{N} + \text{H}$ & 0.104\\
$^{28}\text{Si} + \text{H}$ & 0.101\\
$^{13}\text{C} + \text{H}$ & 0.084\\
$^{56}\text{Fe} + \text{H}$ & 0.064\\
\hline\hline
 &\\[-1.2mm] 
 \end{tabular}

\end{minipage}
\begin{minipage}[t]{0.4\linewidth}
\begin{tabular}{cr}\multicolumn{2}{c}{ N} \\
\multicolumn{2}{c}{$(\sum C_{ab}=2.32)$} \\
\hline\hline
Reaction $(a+b)$ & $C_{ab}$ \\\hline 
$^{16}\text{O} + \text{H}$ & 1.278\\
$^{16}\text{O} + \text{He}$ & 0.219\\
$^{24}\text{Mg} + \text{H}$ & 0.147\\
$^{20}\text{Ne} + \text{H}$ & 0.138\\
$^{28}\text{Si} + \text{H}$ & 0.131\\
$^{15}\text{N} + \text{H}$ & 0.090\\
\hline\hline
 &\\[-1.2mm] 
 \end{tabular}

\end{minipage}
\end{table}
\endgroup

The constants ${\cal C}_{ab}$ are listed in Tab.~\ref{tab:c_ab}. They are very useful for optimization of the beam requests for future measurements. They also allow the uncertainty of a particular secondary flux due to the statistical uncertainty of the cross section measurements in $a+b$ interactions to be predicted for the given number of recorded interactions. Moreover, they provide clear guidelines on which combinations of projectile and target are the most important ones to measure. For instance, it can be seen that the dominating $C$-values for Boron are ${\cal C}_{\rm Cp}^{\rm B}$ and ${\cal C}_{\rm Op}^{\rm B}$. Their contribution to the relative Boron flux uncertainty is $\frac{1}{\sqrt{N}}\sqrt{ ({\cal C}_{\rm Cp}^{\rm B})^2 + ({\cal C}_{\rm Op}^{\rm B})^2}$, if equal numbers of interactions with Carbon and Oxygen nuclei are recorded.

\subsection{Number of interactions}

If $n$ reactions are to be measured and the aspired combined relative flux uncertainty should be less than $\xi$, then each projectile and target combination needs to be measured until the individual uncertainty from this reaction becomes less than $\xi/\sqrt{n}$. In other words, the number of interactions to be recorded for each reaction is
\begin{equation}
N_{ab}\geq n \,(C_{ab} / \xi)^2.
\label{eq:N_inter}
\end{equation}
For instance, to achieve a combined relative flux uncertainty of $<0.5\%$ for the two aforementioned reactions dominating Boron production ($\rm C+p$ and $\rm O+p$), the required numbers of recorded interactions are
%
$5.2 \times 10^4$ and $3.9 \times 10^4$ for Carbon and Oxygen projectiles respectively.

\section{Conclusions} \label{sec:conclusions}

The main goal of this study is to prioritize the list of cross sections of interest for Galactic CR studies that have to be measured with a higher precision. Indeed, the current generation of CR experiments (AMS-02, CALET, DAMPE, Fermi-LAT, ISS-CREAM, PAMELA) has brought about a revolution in astrophysics of CRs embarking a new high precision era. To fully exploit these data, we need a combined effort of the CR, nuclear, and particle physics communities and their facilities to meet the demand for high precision nuclear fragmentation data.

We have thoroughly discussed how to rank the most important reactions for the production of Li, Be, B, C, and N in CRs (Tables \ref{tab:xs_Li} to \ref{tab:xs_N} in Appendices~\ref{app:table_reactions}). We have also discussed in detail how to propagate the cross section uncertainties to the relevant CR fluxes. Cross section measurements are generally aimed at a certain combination of the projectile ($a$) and target ($b$), measuring as many fragments ($c$) as the experimental setup allows. For this reason, we have also sorted the most important reactions $a+b$ required to reach a given accuracy of the elemental fluxes: this information can be directly read off Figs.~\ref{fig:err_evol_LiBeB} and \ref{fig:err_evol_CN} in Appendix~\ref{app:plots_errevol}. Whereas the exact number of reactions to measure depends somewhat on the degree of correlations between the many cross sections errors, the ranking of these reactions does not. To help planning new experiments and estimate the required beam time in proposals, we have provided a formula to estimate realistically the number of reactions necessary to achieve the given precision in calculations of fluxes of CR species. This is encoded in Eq.~(\ref{eq:N_inter}) and the $C_{ab}$ coefficients (\ref{eq:Cab}) given in Table~\ref{tab:c_ab}.

\begin{figure}[!t]
\includegraphics[width=0.5\textwidth]{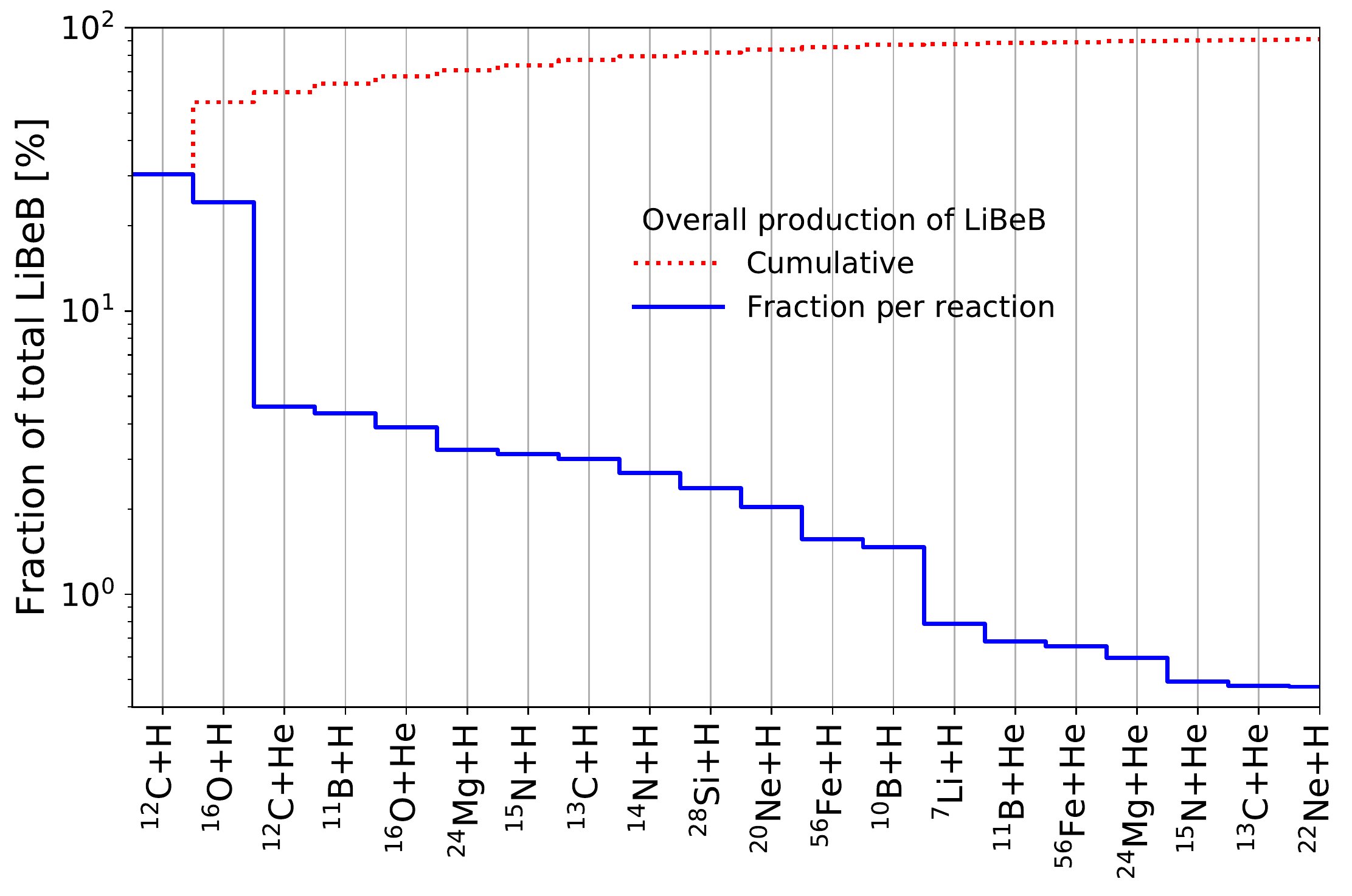}
\caption{Contributive and cumulative fractions of reactions for the overall production of secondary LiBeB in GCRs at 10~GeV/n. The labels on the abscissa give the projectile+target combination considered.\label{fig:frac_LiBeB}}
\end{figure}
Key fluxes for GCR studies are Li, Be, and B. Because of their secondary nature, they give access to the transport mechanisms in the Galaxy and calibrate CR transport to search for possible signatures of new physics; AMS-02 will provide a typical $3\%$ accuracy for all CR measurements. As illustrated in Fig.~\ref{fig:frac_LiBeB}, two reactions, $^{12}$C+H and $^{16}$O+H, provide $50\%$ of all the reactions in LiBeB production; ten more reactions are necessary to have $80\%$ of all the production; the distribution of $a+b$ reactions has then a very large tail above $90\%$. It is worthwhile noting that any improvement on the uncertainties of the most important reactions will also help to refine the cross section modelling and thus improve the accuracy of other reactions including those which are not measured yet.


\acknowledgments

We warmly thank F.\ Donato and P.\ Serpico for organizing the ``XSCRC2017: Cross sections for Cosmic Rays @ CERN''
(\url{https://indico.cern.ch/event/563277/}) where the present collaboration has started -- thanks to the many fruitful discussions we had during the workshop. Y.G.\ and D.M.\ thank P.\ Salati and P.\ Serpico for their support during the preliminary steps of this study. Y.G thank Martin Winkler for useful discussions and for sharing his cross-section data. D.M.\ thanks G.\ Simpson for useful discussions concerning nuclear data. This work has been supported by the ``Investissements d'avenir, Labex ENIGMASS.'' The work of Y.G. is supported by the IISN, the FNRS-FRS and a ULB ARC. I.V.M.\ acknowledges partial support from NASA grant No.\ NNX17AB48G. M.U.\ acknowledges financial support from the EU-funded Marie Curie Outgoing Fellowship, Grant PIOF-GA-2013-624803.

\appendix

\section{Ranking of 1- and 2-step production channels} \label{app:channels}

This Appendix presents the ranking of channels, obtained from the sum over all ghost nuclei and over the chemical composition of the ISM (H, 10\% of He by number). The main benefit of such ranking is that the sum over all channels (involving 1-, 2- or more than 2-steps) of each reaction for the given primary $i$ and secondary $j$ species (e.g., $^{12}$C$\to ^{10}$B) gives the total fraction of the secondary species produced in this reaction, which is not the case when only individual cross sections are considered (see the next Section). An example of the 2-step channels is $^{16}$O$\to ^{12}$C$\to ^{10}$B ($i\to j \to k$).

\subsection{Definition: $f^{\rm 1-step}_{ij}$ and $f^{\rm 2-step}_{ijk}$}

In practice, we calculate the reference flux of secondary fraction of CR species $\psi^{\rm sec}$(ref) in units [m$^2$~s~sr~GeV/n]$^{-1}$ after the propagation. We form the ratio with the flux calculated assigning zero values to all production cross sections, but those involved in the selected reaction channel ($i\to j$). The fraction ratio, or $f$-ratio, for 1-step and 2-step reactions reads:
\begin{eqnarray}
   \label{eq:fij_fijk}
   f^{\rm 1-step}_{ij} &=& \frac{\psi_{ij}^{\rm sec}({\rm only}~\sigma^{ij}\neq 0)}{\psi^{\rm sec}({\rm ref})}\;,\\
   f^{\rm 2-step}_{ijk} &=& \frac{\psi_{ijk}^{\rm sec}({\rm only}~\sigma^{ij}\neq 0, \sigma^{jk}\neq 0)}{\psi^{\rm sec}({\rm ref})}\,.\nonumber
\end{eqnarray}
Here $\sigma^{ij}$ stands for the effective production cross section of species $j$ from fragmentation of species $i$ in the ISM. We do not consider 3-step channels as they are sub-dominant (see Table \ref{tab:origin}).

\subsection{Energy dependence of $f^{\rm 2-step}/f^{\rm 1-step}$\label{sec:10gevn}}

\begin{figure}[t]
\includegraphics[width=\columnwidth]{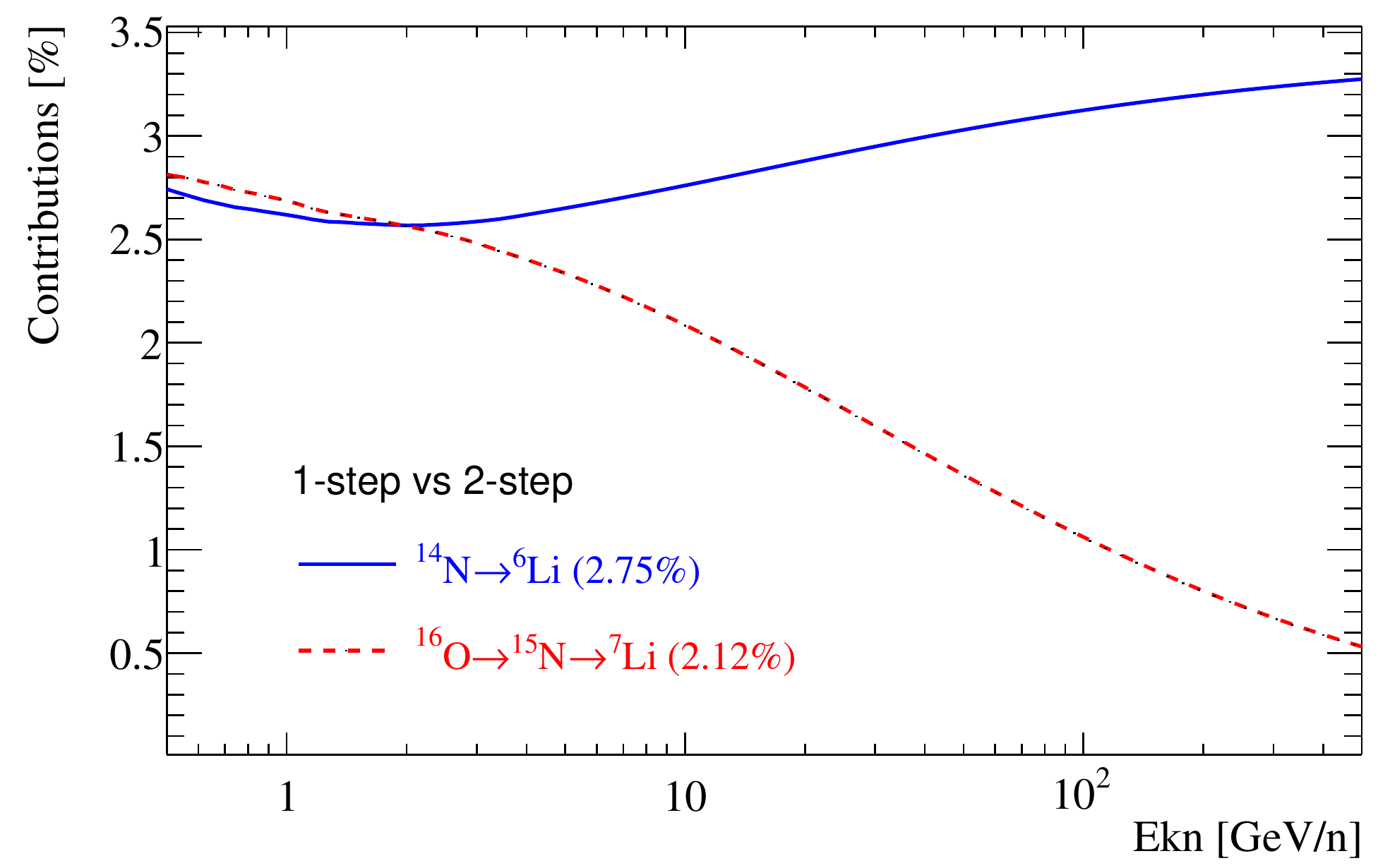}
\caption{Illustration of the different energy dependences of 1-step and 2-step channels. See text for details.\label{fig:edep}}
\end{figure}

The contributions of 1-step and 2-step reactions have different energy dependences, as illustrated in Fig.~\ref{fig:edep}. The origin of these differences is not the energy dependence of the production cross sections, the latter are about constant above a few GeV/n, but the effects of CR propagation. In the first approximation, in the pure diffusive regime with the source term $Q(R)\propto R^{-\alpha}$ and the diffusion coefficient $D(R)\propto R^\delta$, where $R$ is the rigidity, the flux of primary and secondary species produced in 1-step and 2-step reactions can be calculated as:
\begin{eqnarray}
   \label{eq:diff}
   \psi^{\rm prim}(R)= &\displaystyle\frac{Q(R)}{D(R)} &\propto R^{-(\alpha+\delta)\approx 2.8}\,, \\
   \psi^{\rm 1-step}(R)\propto&\displaystyle \frac{\psi^{\rm prim}(R)}{D(R)} &\propto R^{-(\alpha+2\delta)}\,,\nonumber\\
   \psi^{\rm 2-step}(R)\propto&\displaystyle \frac{\psi^{\rm 1-step}(R)}{D(R)} &\propto R^{-(\alpha+3\delta)}\,. \nonumber
\end{eqnarray}
Therefore, the energy-dependences shown in Fig.~\ref{fig:edep} are mostly related to the slope of the rigidity dependence of the diffusion coefficient $f^{\rm 2-step}/f^{\rm 1-step}=\psi^{\rm 2-step}/\psi^{\rm 1-step}\propto R^{-\delta}$. This leads to a higher ranking of 2-step reactions (w.r.t.\ 1-step contributions) at low energies, but conversely to negligible contributions at high energies, in the range of TeV/n.

Since the ranking of the 1-step and 2-step reactions is necessarily energy-dependent, we choose an effective energy of 10~GeV/n for the following reasons:
\begin{itemize}
   \item this energy range encompasses the regime in which 2-step contributions matter: in order not to miss the corresponding cross sections, it must be done at the energy that is low enough;
   \item the ranking depends on composition of CRs in the sources, and intermediate energies are best to mitigate several propagation effects that impact mostly low energies---such as the ionization energy losses, decay of $^{10}$Be, distributed acceleration, convection by the Galactic wind, solar modulation and so on---and the statistical accuracy of the CR measurements that degrades at high energy.
\end{itemize}
We note that in $\sim$1 GeV/n to 10 GeV/n range, the energy dependence of ranking is mild, so that a choice of that particular energy should not significantly affect our conclusions.

\subsection{Ranked channels at 10~GeV/n}
\begingroup
\squeezetable
\begin{table}[!ht]
\caption{Ranking of 1- and 2-step channels for Li at 10 GeV/n, from $f_{ij}^{\rm 1-step}$ and $f_{ijk}^{\rm 2-step}$ coefficients (\ref{eq:fij_fijk}). Channels $<\,$0.1\% and higher-level channels ($>2$-step, contributing to $\sim8.6\,\%$, see Table~\ref{tab:origin}), are not shown.\label{tab:channels_Li}}
\begin{tabular}{rrlllc}
\hline\hline
\multicolumn{3}{c}{\# of channels} & \multicolumn{2}{c}{in range} & {contribution [\%]}\\
\multicolumn{3}{c}{15}& \multicolumn{2}{c}{[1\%,100\%]} & 70.2\\[-0.9mm]
\multicolumn{3}{c}{33}& \multicolumn{2}{c}{[0.1\%,1\%]} & 12.7\\[-0.9mm]
\multicolumn{3}{c}{189}& \multicolumn{2}{c}{[0.01\%,0.1\%]} & 6.7\\[-0.9mm]
\multicolumn{3}{c}{430}& \multicolumn{2}{c}{[0.001\%,0.01\%]} & 1.5\\[-0.9mm]
\multicolumn{3}{c}{618}& \multicolumn{2}{c}{[0.0001\%,0.001\%]} & 0.2\\[-0.9mm]
\multicolumn{3}{c}{2499}& \multicolumn{2}{c}{[0.0\%,0.0001\%]} & 0.0\\[-0.9mm]
\multicolumn{6}{c}{ }\\[-2.5mm]
\multicolumn{5}{l}{Channel}  &  min $|$ {\bf mean} $|$ max \\
\hline
$\rm ^{16}O$&$\rightarrow$ &$ \rm^{6}Li$ & & & 12.8$~|~${\bf15.4}$~|~$17.9\\[-1.20mm]
$\rm ^{12}C$&$\rightarrow$ &$ \rm^{6}Li$ & & & 11.7$~|~${\bf13.9}$~|~$16.1\\[-1.20mm]
$\rm ^{16}O$&$\rightarrow$ &$ \rm^{7}Li$ & & & 9.99$~|~${\bf12.0}$~|~$14.0\\[-1.20mm]
$\rm ^{12}C$&$\rightarrow$ &$ \rm^{7}Li$ & & & 9.50$~|~${\bf11.3}$~|~$13.2\\[-1.20mm]
$\rm ^{24}Mg$&$\rightarrow$ &$ \rm^{6}Li$ & & & 1.99$~|~${\bf2.24}$~|~$2.48\\[-1.20mm]
$\rm ^{16}O$&$\rightarrow$ &$ \rm^{15}N$&$\rightarrow$ &$ \rm^{7}Li$ & 1.55$~|~${\bf1.86}$~|~$2.17\\[-1.20mm]
$\rm ^{56}Fe$&$\rightarrow$ &$ \rm^{6}Li$ & & & 0.00$~|~${\bf1.79}$~|~$3.58\\[-1.20mm]
$\rm ^{16}O$&$\rightarrow$ &$ \rm^{13}C$&$\rightarrow$ &$ \rm^{7}Li$ & 1.48$~|~${\bf1.78}$~|~$2.08\\[-1.20mm]
$\rm ^{12}C$&$\rightarrow$ &$ \rm^{11}B$&$\rightarrow$ &$ \rm^{7}Li$ & 1.47$~|~${\bf1.75}$~|~$2.03\\[-1.20mm]
$\rm ^{24}Mg$&$\rightarrow$ &$ \rm^{7}Li$ & & & 1.45$~|~${\bf1.74}$~|~$2.03\\[-1.20mm]
$\rm ^{16}O$&$\rightarrow$ &$ \rm^{13}C$&$\rightarrow$ &$ \rm^{6}Li$ & 1.41$~|~${\bf1.62}$~|~$1.84\\[-1.20mm]
$\rm ^{16}O$&$\rightarrow$ &$ \rm^{15}N$&$\rightarrow$ &$ \rm^{6}Li$ & 1.30$~|~${\bf1.47}$~|~$1.64\\[-1.20mm]
$\rm ^{56}Fe$&$\rightarrow$ &$ \rm^{7}Li$ & & & 0.00$~|~${\bf1.27}$~|~$2.54\\[-1.20mm]
$\rm ^{28}Si$&$\rightarrow$ &$ \rm^{6}Li$ & & & 0.00$~|~${\bf1.06}$~|~$2.13\\[-1.20mm]
$\rm ^{16}O$&$\rightarrow$ &$ \rm^{11}B$&$\rightarrow$ &$ \rm^{7}Li$ & 0.84$~|~${\bf1.01}$~|~$1.17\\[-1.20mm]
$\rm ^{16}O$&$\rightarrow$ &$ \rm^{12}C$&$\rightarrow$ &$ \rm^{6}Li$ & 0.83$~|~${\bf1.00}$~|~$1.16\\[-1.20mm]
$\rm ^{16}O$&$\rightarrow$ &$ \rm^{7}Li$&$\rightarrow$ &$ \rm^{6}Li$ & 0.86$~|~${\bf0.88}$~|~$0.90\\[-1.20mm]
$\rm ^{12}C$&$\rightarrow$ &$ \rm^{7}Li$&$\rightarrow$ &$ \rm^{6}Li$ & 0.82$~|~${\bf0.83}$~|~$0.84\\[-1.20mm]
$\rm ^{16}O$&$\rightarrow$ &$ \rm^{12}C$&$\rightarrow$ &$ \rm^{7}Li$ & 0.68$~|~${\bf0.81}$~|~$0.95\\[-1.20mm]
$\rm ^{14}N$&$\rightarrow$ &$ \rm^{6}Li$ & & & 0.68$~|~${\bf0.79}$~|~$0.90\\[-1.20mm]
$\rm ^{28}Si$&$\rightarrow$ &$ \rm^{7}Li$ & & & 0.00$~|~${\bf0.76}$~|~$1.52\\[-1.20mm]
$\rm ^{20}Ne$&$\rightarrow$ &$ \rm^{6}Li$ & & & 0.00$~|~${\bf0.69}$~|~$1.38\\[-1.20mm]
$\rm ^{16}O$&$\rightarrow$ &$ \rm^{14}N$&$\rightarrow$ &$ \rm^{6}Li$ & 0.46$~|~${\bf0.64}$~|~$0.82\\[-1.20mm]
$\rm ^{12}C$&$\rightarrow$ &$ \rm^{10}B$&$\rightarrow$ &$ \rm^{6}Li$ & 0.46$~|~${\bf0.53}$~|~$0.61\\[-1.20mm]
$\rm ^{16}O$&$\rightarrow$ &$ \rm^{7}Be$&$\rightarrow$ &$ \rm^{6}Li$ & 0.43$~|~${\bf0.52}$~|~$0.60\\[-1.20mm]
$\rm ^{20}Ne$&$\rightarrow$ &$ \rm^{7}Li$ & & & 0.00$~|~${\bf0.52}$~|~$1.03\\[-1.20mm]
$\rm ^{14}N$&$\rightarrow$ &$ \rm^{7}Li$ & & & 0.44$~|~${\bf0.51}$~|~$0.59\\[-1.20mm]
$\rm ^{12}C$&$\rightarrow$ &$ \rm^{11}B$&$\rightarrow$ &$ \rm^{6}Li$ & 0.50$~|~${\bf0.51}$~|~$0.52\\[-1.20mm]
$\rm ^{16}O$&$\rightarrow$ &$ \rm^{10}B$&$\rightarrow$ &$ \rm^{6}Li$ & 0.39$~|~${\bf0.46}$~|~$0.52\\[-1.20mm]
$\rm ^{12}C$&$\rightarrow$ &$ \rm^{7}Be$&$\rightarrow$ &$ \rm^{6}Li$ & 0.35$~|~${\bf0.42}$~|~$0.49\\[-1.20mm]
$\rm ^{16}O$&$\rightarrow$ &$ \rm^{14}N$&$\rightarrow$ &$ \rm^{7}Li$ & 0.30$~|~${\bf0.42}$~|~$0.53\\[-1.20mm]
$\rm ^{16}O$&$\rightarrow$ &$ \rm^{11}B$&$\rightarrow$ &$ \rm^{6}Li$ & 0.28$~|~${\bf0.29}$~|~$0.30\\[-1.20mm]
$\rm ^{12}C$&$\rightarrow$ &$ \rm^{10}B$&$\rightarrow$ &$ \rm^{7}Li$ & 0.20$~|~${\bf0.24}$~|~$0.28\\[-1.20mm]
$\rm ^{16}O$&$\rightarrow$ &$ \rm^{10}B$&$\rightarrow$ &$ \rm^{7}Li$ & 0.17$~|~${\bf0.20}$~|~$0.24\\[-1.20mm]
$\rm ^{13}C$&$\rightarrow$ &$ \rm^{7}Li$ & & & 0.15$~|~${\bf0.18}$~|~$0.21\\[-1.20mm]
$\rm ^{13}C$&$\rightarrow$ &$ \rm^{6}Li$ & & & 0.14$~|~${\bf0.16}$~|~$0.18\\[-1.20mm]
$\rm ^{32}S$&$\rightarrow$ &$ \rm^{6}Li$ & & & 0.00$~|~${\bf0.14}$~|~$0.27\\[-1.20mm]
$\rm ^{24}Mg$&$\rightarrow$ &$ \rm^{7}Li$&$\rightarrow$ &$ \rm^{6}Li$ & 0.13$~|~${\bf0.13}$~|~$0.13\\[-1.20mm]
$\rm ^{26}Mg$&$\rightarrow$ &$ \rm^{6}Li$ & & & 0.00$~|~${\bf0.12}$~|~$0.25\\[-1.20mm]
$\rm ^{25}Mg$&$\rightarrow$ &$ \rm^{6}Li$ & & & 0.00$~|~${\bf0.12}$~|~$0.24\\[-1.20mm]
$\rm ^{26}Mg$&$\rightarrow$ &$ \rm^{7}Li$ & & & 0.00$~|~${\bf0.11}$~|~$0.23\\[-1.20mm]
$\rm ^{54}Fe$&$\rightarrow$ &$ \rm^{6}Li$ & & & 0.00$~|~${\bf0.11}$~|~$0.23\\[-1.20mm]
$\rm ^{20}Ne$&$\rightarrow$ &$ \rm^{15}N$&$\rightarrow$ &$ \rm^{7}Li$ & 0.09$~|~${\bf0.11}$~|~$0.13\\[-1.20mm]
$\rm ^{56}Fe$&$\rightarrow$ &$ \rm^{7}Li$&$\rightarrow$ &$ \rm^{6}Li$ & 0.00$~|~${\bf0.11}$~|~$0.22\\[-1.20mm]
$\rm ^{24}Mg$&$\rightarrow$ &$ \rm^{16}O$&$\rightarrow$ &$ \rm^{6}Li$ & 0.09$~|~${\bf0.10}$~|~$0.12\\[-1.20mm]
$\rm ^{28}Si$&$\rightarrow$ &$ \rm^{27}Al$&$\rightarrow$ &$ \rm^{6}Li$ & 0.00$~|~${\bf0.10}$~|~$0.21\\[-1.20mm]
$\rm ^{28}Si$&$\rightarrow$ &$ \rm^{24}Mg$&$\rightarrow$ &$ \rm^{6}Li$ & 0.09$~|~${\bf0.10}$~|~$0.11\\[-1.20mm]
$\rm ^{24}Mg$&$\rightarrow$ &$ \rm^{12}C$&$\rightarrow$ &$ \rm^{6}Li$ & 0.08$~|~${\bf0.10}$~|~$0.12\\\hline\hline
\end{tabular}
\end{table}
\endgroup
\begingroup
\squeezetable
\begin{table}[!ht]
\caption{Ranking of 1- and 2-step channels for Be at 10 GeV/n, from $f_{ij}^{\rm 1-step}$ and $f_{ijk}^{\rm 2-step}$ coefficients (\ref{eq:fij_fijk}). Channels $<\,$0.1\% and higher-level channels ($>2$-step, contributing to $\sim6.8\,\%$, see Table~\ref{tab:origin}), are not shown.\label{tab:channels_Be}}
\begin{tabular}{rrlllc}
\hline\hline
\multicolumn{3}{c}{\# of channels} & \multicolumn{2}{c}{in range} & {contribution [\%]}\\
\multicolumn{3}{c}{17}& \multicolumn{2}{c}{[1\%,100\%]} & 71.5\\[-0.9mm]
\multicolumn{3}{c}{46}& \multicolumn{2}{c}{[0.1\%,1\%]} & 13.4\\[-0.9mm]
\multicolumn{3}{c}{207}& \multicolumn{2}{c}{[0.01\%,0.1\%]} & 6.1\\[-0.9mm]
\multicolumn{3}{c}{532}& \multicolumn{2}{c}{[0.001\%,0.01\%]} & 1.8\\[-0.9mm]
\multicolumn{3}{c}{879}& \multicolumn{2}{c}{[0.0001\%,0.001\%]} & 0.3\\[-0.9mm]
\multicolumn{3}{c}{3624}& \multicolumn{2}{c}{[0.0\%,0.0001\%]} & 0.0\\[-0.9mm]
\multicolumn{6}{c}{ }\\[-2.5mm]
\multicolumn{5}{l}{Channel}  &  min $|$ {\bf mean} $|$ max \\
\hline
$\rm ^{16}O$&$\rightarrow$ &$ \rm^{7}Be$ & & & 17.6$~|~${\bf18.9}$~|~$20.9\\[-1.20mm]
$\rm ^{12}C$&$\rightarrow$ &$ \rm^{7}Be$ & & & 15.3$~|~${\bf17.1}$~|~$18.9\\[-1.20mm]
$\rm ^{12}C$&$\rightarrow$ &$ \rm^{9}Be$ & & & 7.12$~|~${\bf8.34}$~|~$9.64\\[-1.20mm]
$\rm ^{16}O$&$\rightarrow$ &$ \rm^{9}Be$ & & & 5.78$~|~${\bf6.18}$~|~$6.48\\[-1.20mm]
$\rm ^{28}Si$&$\rightarrow$ &$ \rm^{7}Be$ & & & 2.70$~|~${\bf3.18}$~|~$3.63\\[-1.20mm]
$\rm ^{24}Mg$&$\rightarrow$ &$ \rm^{7}Be$ & & & 2.53$~|~${\bf2.99}$~|~$3.78\\[-1.20mm]
$\rm ^{20}Ne$&$\rightarrow$ &$ \rm^{7}Be$ & & & 1.63$~|~${\bf2.10}$~|~$2.99\\[-1.20mm]
$\rm ^{56}Fe$&$\rightarrow$ &$ \rm^{7}Be$ & & & 0.16$~|~${\bf1.79}$~|~$3.70\\[-1.20mm]
$\rm ^{12}C$&$\rightarrow$ &$ \rm^{10}Be$ & & & 1.25$~|~${\bf1.72}$~|~$1.99\\[-1.20mm]
$\rm ^{14}N$&$\rightarrow$ &$ \rm^{7}Be$ & & & 1.00$~|~${\bf1.32}$~|~$1.69\\[-1.20mm]
$\rm ^{16}O$&$\rightarrow$ &$ \rm^{10}Be$ & & & 1.17$~|~${\bf1.29}$~|~$1.39\\[-1.20mm]
$\rm ^{12}C$&$\rightarrow$ &$ \rm^{11}B$&$\rightarrow$ &$ \rm^{9}Be$ & 1.21$~|~${\bf1.25}$~|~$1.35\\[-1.20mm]
$\rm ^{28}Si$&$\rightarrow$ &$ \rm^{9}Be$ & & & 1.02$~|~${\bf1.14}$~|~$1.31\\[-1.20mm]
$\rm ^{24}Mg$&$\rightarrow$ &$ \rm^{9}Be$ & & & 0.96$~|~${\bf1.13}$~|~$1.46\\[-1.20mm]
$\rm ^{16}O$&$\rightarrow$ &$ \rm^{12}C$&$\rightarrow$ &$ \rm^{7}Be$ & 0.85$~|~${\bf1.02}$~|~$1.22\\[-1.20mm]
$\rm ^{16}O$&$\rightarrow$ &$ \rm^{15}N$&$\rightarrow$ &$ \rm^{9}Be$ & 0.84$~|~${\bf1.02}$~|~$1.24\\[-1.20mm]
$\rm ^{16}O$&$\rightarrow$ &$ \rm^{15}N$&$\rightarrow$ &$ \rm^{7}Be$ & 0.94$~|~${\bf1.00}$~|~$1.06\\[-1.20mm]
$\rm ^{56}Fe$&$\rightarrow$ &$ \rm^{9}Be$ & & & 0.09$~|~${\bf0.87}$~|~$1.52\\[-1.20mm]
$\rm ^{16}O$&$\rightarrow$ &$ \rm^{14}N$&$\rightarrow$ &$ \rm^{7}Be$ & 0.68$~|~${\bf0.83}$~|~$0.97\\[-1.20mm]
$\rm ^{12}C$&$\rightarrow$ &$ \rm^{11}B$&$\rightarrow$ &$ \rm^{7}Be$ & 0.52$~|~${\bf0.83}$~|~$1.12\\[-1.20mm]
$\rm ^{20}Ne$&$\rightarrow$ &$ \rm^{9}Be$ & & & 0.68$~|~${\bf0.80}$~|~$0.97\\[-1.20mm]
$\rm ^{16}O$&$\rightarrow$ &$ \rm^{11}B$&$\rightarrow$ &$ \rm^{9}Be$ & 0.56$~|~${\bf0.68}$~|~$0.78\\[-1.20mm]
$\rm ^{16}O$&$\rightarrow$ &$ \rm^{13}C$&$\rightarrow$ &$ \rm^{9}Be$ & 0.21$~|~${\bf0.59}$~|~$0.95\\[-1.20mm]
$\rm ^{16}O$&$\rightarrow$ &$ \rm^{13}C$&$\rightarrow$ &$ \rm^{7}Be$ & 0.47$~|~${\bf0.54}$~|~$0.63\\[-1.20mm]
$\rm ^{16}O$&$\rightarrow$ &$ \rm^{12}C$&$\rightarrow$ &$ \rm^{9}Be$ & 0.35$~|~${\bf0.51}$~|~$0.70\\[-1.20mm]
$\rm ^{12}C$&$\rightarrow$ &$ \rm^{10}B$&$\rightarrow$ &$ \rm^{9}Be$ & 0.24$~|~${\bf0.46}$~|~$0.60\\[-1.20mm]
$\rm ^{16}O$&$\rightarrow$ &$ \rm^{11}B$&$\rightarrow$ &$ \rm^{7}Be$ & 0.29$~|~${\bf0.44}$~|~$0.63\\[-1.20mm]
$\rm ^{12}C$&$\rightarrow$ &$ \rm^{11}B$&$\rightarrow$ &$ \rm^{10}Be$ & 0.24$~|~${\bf0.42}$~|~$0.58\\[-1.20mm]
$\rm ^{14}N$&$\rightarrow$ &$ \rm^{9}Be$ & & & 0.19$~|~${\bf0.40}$~|~$0.62\\[-1.20mm]
$\rm ^{16}O$&$\rightarrow$ &$ \rm^{10}B$&$\rightarrow$ &$ \rm^{9}Be$ & 0.18$~|~${\bf0.38}$~|~$0.51\\[-1.20mm]
$\rm ^{12}C$&$\rightarrow$ &$ \rm^{10}B$&$\rightarrow$ &$ \rm^{7}Be$ & 0.29$~|~${\bf0.36}$~|~$0.51\\[-1.20mm]
$\rm ^{32}S$&$\rightarrow$ &$ \rm^{7}Be$ & & & 0.12$~|~${\bf0.29}$~|~$0.53\\[-1.20mm]
$\rm ^{16}O$&$\rightarrow$ &$ \rm^{10}B$&$\rightarrow$ &$ \rm^{7}Be$ & 0.24$~|~${\bf0.29}$~|~$0.38\\[-1.20mm]
$\rm ^{25}Mg$&$\rightarrow$ &$ \rm^{7}Be$ & & & 0.19$~|~${\bf0.26}$~|~$0.42\\[-1.20mm]
$\rm ^{27}Al$&$\rightarrow$ &$ \rm^{7}Be$ & & & 0.15$~|~${\bf0.25}$~|~$0.41\\[-1.20mm]
$\rm ^{16}O$&$\rightarrow$ &$ \rm^{14}N$&$\rightarrow$ &$ \rm^{9}Be$ & 0.13$~|~${\bf0.24}$~|~$0.34\\[-1.20mm]
$\rm ^{24}Mg$&$\rightarrow$ &$ \rm^{10}Be$ & & & 0.18$~|~${\bf0.23}$~|~$0.32\\[-1.20mm]
$\rm ^{16}O$&$\rightarrow$ &$ \rm^{11}B$&$\rightarrow$ &$ \rm^{10}Be$ & 0.11$~|~${\bf0.23}$~|~$0.34\\[-1.20mm]
$\rm ^{26}Mg$&$\rightarrow$ &$ \rm^{7}Be$ & & & 0.15$~|~${\bf0.23}$~|~$0.38\\[-1.20mm]
$\rm ^{28}Si$&$\rightarrow$ &$ \rm^{10}Be$ & & & 0.16$~|~${\bf0.22}$~|~$0.34\\[-1.20mm]
$\rm ^{28}Si$&$\rightarrow$ &$ \rm^{27}Al$&$\rightarrow$ &$ \rm^{7}Be$ & 0.14$~|~${\bf0.19}$~|~$0.25\\[-1.20mm]
$\rm ^{12}C$&$\rightarrow$ &$ \rm^{9}Be$&$\rightarrow$ &$ \rm^{7}Be$ & 0.16$~|~${\bf0.19}$~|~$0.21\\[-1.20mm]
$\rm ^{24}Mg$&$\rightarrow$ &$ \rm^{23}Na$&$\rightarrow$ &$ \rm^{7}Be$ & 0.12$~|~${\bf0.18}$~|~$0.26\\[-1.20mm]
$\rm ^{56}Fe$&$\rightarrow$ &$ \rm^{10}Be$ & & & 0.01$~|~${\bf0.17}$~|~$0.34\\[-1.20mm]
$\rm ^{16}O$&$\rightarrow$ &$ \rm^{15}N$&$\rightarrow$ &$ \rm^{10}Be$ & 0.10$~|~${\bf0.16}$~|~$0.21\\[-1.20mm]
$\rm ^{20}Ne$&$\rightarrow$ &$ \rm^{10}Be$ & & & 0.10$~|~${\bf0.16}$~|~$0.26\\[-1.20mm]
$\rm ^{16}O$&$\rightarrow$ &$ \rm^{13}C$&$\rightarrow$ &$ \rm^{10}Be$ & 0.02$~|~${\bf0.14}$~|~$0.26\\[-1.20mm]
$\rm ^{16}O$&$\rightarrow$ &$ \rm^{9}Be$&$\rightarrow$ &$ \rm^{7}Be$ & 0.13$~|~${\bf0.14}$~|~$0.14\\[-1.20mm]
$\rm ^{26}Mg$&$\rightarrow$ &$ \rm^{9}Be$ & & & 0.09$~|~${\bf0.13}$~|~$0.17\\[-1.20mm]
$\rm ^{24}Mg$&$\rightarrow$ &$ \rm^{16}O$&$\rightarrow$ &$ \rm^{7}Be$ & 0.10$~|~${\bf0.12}$~|~$0.14\\[-1.20mm]
$\rm ^{28}Si$&$\rightarrow$ &$ \rm^{24}Mg$&$\rightarrow$ &$ \rm^{7}Be$ & 0.12$~|~${\bf0.12}$~|~$0.13\\[-1.20mm]
$\rm ^{25}Mg$&$\rightarrow$ &$ \rm^{9}Be$ & & & 0.10$~|~${\bf0.12}$~|~$0.14\\[-1.20mm]
$\rm ^{32}S$&$\rightarrow$ &$ \rm^{9}Be$ & & & 0.06$~|~${\bf0.12}$~|~$0.19\\[-1.20mm]
$\rm ^{24}Mg$&$\rightarrow$ &$ \rm^{12}C$&$\rightarrow$ &$ \rm^{7}Be$ & 0.10$~|~${\bf0.12}$~|~$0.13\\[-1.20mm]
$\rm ^{29}Si$&$\rightarrow$ &$ \rm^{7}Be$ & & & 0.07$~|~${\bf0.11}$~|~$0.17\\[-1.20mm]
$\rm ^{27}Al$&$\rightarrow$ &$ \rm^{9}Be$ & & & 0.08$~|~${\bf0.11}$~|~$0.15\\[-1.20mm]
$\rm ^{20}Ne$&$\rightarrow$ &$ \rm^{16}O$&$\rightarrow$ &$ \rm^{7}Be$ & 0.09$~|~${\bf0.11}$~|~$0.13\\[-1.20mm]
$\rm ^{16}O$&$\rightarrow$ &$ \rm^{12}C$&$\rightarrow$ &$ \rm^{10}Be$ & 0.06$~|~${\bf0.11}$~|~$0.14\\[-1.20mm]
$\rm ^{24}Mg$&$\rightarrow$ &$ \rm^{22}Ne$&$\rightarrow$ &$ \rm^{7}Be$ & 0.08$~|~${\bf0.11}$~|~$0.15\\[-1.20mm]
$\rm ^{20}Ne$&$\rightarrow$ &$ \rm^{12}C$&$\rightarrow$ &$ \rm^{7}Be$ & 0.10$~|~${\bf0.11}$~|~$0.11\\[-1.20mm]
$\rm ^{23}Na$&$\rightarrow$ &$ \rm^{7}Be$ & & & 0.06$~|~${\bf0.10}$~|~$0.21\\[-1.20mm]
$\rm ^{14}N$&$\rightarrow$ &$ \rm^{12}C$&$\rightarrow$ &$ \rm^{7}Be$ & 0.07$~|~${\bf0.10}$~|~$0.14\\[-1.20mm]
$\rm ^{20}Ne$&$\rightarrow$ &$ \rm^{19}F$&$\rightarrow$ &$ \rm^{7}Be$ & 0.07$~|~${\bf0.10}$~|~$0.14\\\hline\hline
\end{tabular}
\end{table}
\endgroup
\begingroup
\squeezetable
\begin{table}[!ht]
\caption{Ranking of 1- and 2-step channels for B at 10 GeV/n, from $f_{ij}^{\rm 1-step}$ and $f_{ijk}^{\rm 2-step}$ coefficients (\ref{eq:fij_fijk}). Channels $<\,$0.1\% and higher-level channels ($>2$-step, contributing to $\sim4.8\,\%$, see Table~\ref{tab:origin}), are not shown.\label{tab:channels_B}}
\begin{tabular}{rrlllc}
\hline\hline
\multicolumn{3}{c}{\# of channels} & \multicolumn{2}{c}{in range} & {contribution [\%]}\\
\multicolumn{3}{c}{13}& \multicolumn{2}{c}{[1\%,100\%]} & 82.2\\[-0.9mm]
\multicolumn{3}{c}{25}& \multicolumn{2}{c}{[0.1\%,1\%]} & 7.7\\[-0.9mm]
\multicolumn{3}{c}{110}& \multicolumn{2}{c}{[0.01\%,0.1\%]} & 3.8\\[-0.9mm]
\multicolumn{3}{c}{346}& \multicolumn{2}{c}{[0.001\%,0.01\%]} & 1.3\\[-0.9mm]
\multicolumn{3}{c}{526}& \multicolumn{2}{c}{[0.0001\%,0.001\%]} & 0.2\\[-0.9mm]
\multicolumn{3}{c}{2340}& \multicolumn{2}{c}{[0.0\%,0.0001\%]} & 0.0\\[-0.9mm]
\multicolumn{6}{c}{ }\\[-2.5mm]
\multicolumn{5}{l}{Channel}  &  min $|$ {\bf mean} $|$ max \\
\hline
$\rm ^{12}C$&$\rightarrow$ &$ \rm^{11}B$ & & & 30.8$~|~${\bf32.7}$~|~$35.3\\[-1.20mm]
$\rm ^{16}O$&$\rightarrow$ &$ \rm^{11}B$ & & & 16.2$~|~${\bf17.7}$~|~$18.8\\[-1.20mm]
$\rm ^{12}C$&$\rightarrow$ &$ \rm^{10}B$ & & & 9.04$~|~${\bf9.95}$~|~$10.9\\[-1.20mm]
$\rm ^{16}O$&$\rightarrow$ &$ \rm^{10}B$ & & & 7.64$~|~${\bf8.17}$~|~$8.68\\[-1.20mm]
$\rm ^{12}C$&$\rightarrow$ &$ \rm^{11}B$&$\rightarrow$ &$ \rm^{10}B$ & 2.07$~|~${\bf2.16}$~|~$2.26\\[-1.20mm]
$\rm ^{16}O$&$\rightarrow$ &$ \rm^{12}C$&$\rightarrow$ &$ \rm^{11}B$ & 1.60$~|~${\bf1.96}$~|~$2.34\\[-1.20mm]
$\rm ^{16}O$&$\rightarrow$ &$ \rm^{15}N$&$\rightarrow$ &$ \rm^{11}B$ & 1.29$~|~${\bf1.69}$~|~$2.04\\[-1.20mm]
$\rm ^{24}Mg$&$\rightarrow$ &$ \rm^{11}B$ & & & 1.51$~|~${\bf1.59}$~|~$1.69\\[-1.20mm]
$\rm ^{20}Ne$&$\rightarrow$ &$ \rm^{11}B$ & & & 1.26$~|~${\bf1.32}$~|~$1.39\\[-1.20mm]
$\rm ^{14}N$&$\rightarrow$ &$ \rm^{11}B$ & & & 1.00$~|~${\bf1.32}$~|~$1.66\\[-1.20mm]
$\rm ^{28}Si$&$\rightarrow$ &$ \rm^{11}B$ & & & 0.85$~|~${\bf1.29}$~|~$1.66\\[-1.20mm]
$\rm ^{16}O$&$\rightarrow$ &$ \rm^{11}B$&$\rightarrow$ &$ \rm^{10}B$ & 1.03$~|~${\bf1.17}$~|~$1.26\\[-1.20mm]
$\rm ^{16}O$&$\rightarrow$ &$ \rm^{13}C$&$\rightarrow$ &$ \rm^{11}B$ & 0.54$~|~${\bf1.15}$~|~$1.62\\[-1.20mm]
$\rm ^{16}O$&$\rightarrow$ &$ \rm^{14}N$&$\rightarrow$ &$ \rm^{11}B$ & 0.68$~|~${\bf0.83}$~|~$0.92\\[-1.20mm]
$\rm ^{24}Mg$&$\rightarrow$ &$ \rm^{10}B$ & & & 0.66$~|~${\bf0.75}$~|~$0.84\\[-1.20mm]
$\rm ^{16}O$&$\rightarrow$ &$ \rm^{12}C$&$\rightarrow$ &$ \rm^{10}B$ & 0.51$~|~${\bf0.59}$~|~$0.69\\[-1.20mm]
$\rm ^{16}O$&$\rightarrow$ &$ \rm^{15}N$&$\rightarrow$ &$ \rm^{10}B$ & 0.50$~|~${\bf0.59}$~|~$0.68\\[-1.20mm]
$\rm ^{20}Ne$&$\rightarrow$ &$ \rm^{10}B$ & & & 0.47$~|~${\bf0.54}$~|~$0.63\\[-1.20mm]
$\rm ^{28}Si$&$\rightarrow$ &$ \rm^{10}B$ & & & 0.32$~|~${\bf0.53}$~|~$0.67\\[-1.20mm]
$\rm ^{14}N$&$\rightarrow$ &$ \rm^{10}B$ & & & 0.39$~|~${\bf0.50}$~|~$0.65\\[-1.20mm]
$\rm ^{56}Fe$&$\rightarrow$ &$ \rm^{11}B$ & & & 0.11$~|~${\bf0.49}$~|~$1.10\\[-1.20mm]
$\rm ^{16}O$&$\rightarrow$ &$ \rm^{13}C$&$\rightarrow$ &$ \rm^{10}B$ & 0.12$~|~${\bf0.32}$~|~$0.50\\[-1.20mm]
$\rm ^{16}O$&$\rightarrow$ &$ \rm^{14}N$&$\rightarrow$ &$ \rm^{10}B$ & 0.26$~|~${\bf0.31}$~|~$0.36\\[-1.20mm]
$\rm ^{24}Mg$&$\rightarrow$ &$ \rm^{12}C$&$\rightarrow$ &$ \rm^{11}B$ & 0.21$~|~${\bf0.22}$~|~$0.25\\[-1.20mm]
$\rm ^{56}Fe$&$\rightarrow$ &$ \rm^{10}B$ & & & 0.00$~|~${\bf0.21}$~|~$0.71\\[-1.20mm]
$\rm ^{20}Ne$&$\rightarrow$ &$ \rm^{12}C$&$\rightarrow$ &$ \rm^{11}B$ & 0.19$~|~${\bf0.20}$~|~$0.22\\[-1.20mm]
$\rm ^{14}N$&$\rightarrow$ &$ \rm^{12}C$&$\rightarrow$ &$ \rm^{11}B$ & 0.14$~|~${\bf0.20}$~|~$0.25\\[-1.20mm]
$\rm ^{13}C$&$\rightarrow$ &$ \rm^{11}B$ & & & 0.15$~|~${\bf0.18}$~|~$0.24\\[-1.20mm]
$\rm ^{28}Si$&$\rightarrow$ &$ \rm^{12}C$&$\rightarrow$ &$ \rm^{11}B$ & 0.10$~|~${\bf0.18}$~|~$0.21\\[-1.20mm]
$\rm ^{25}Mg$&$\rightarrow$ &$ \rm^{11}B$ & & & 0.14$~|~${\bf0.17}$~|~$0.19\\[-1.20mm]
$\rm ^{32}S$&$\rightarrow$ &$ \rm^{11}B$ & & & 0.09$~|~${\bf0.14}$~|~$0.17\\[-1.20mm]
$\rm ^{26}Mg$&$\rightarrow$ &$ \rm^{11}B$ & & & 0.11$~|~${\bf0.13}$~|~$0.14\\[-1.20mm]
$\rm ^{27}Al$&$\rightarrow$ &$ \rm^{11}B$ & & & 0.08$~|~${\bf0.12}$~|~$0.16\\[-1.20mm]
$\rm ^{24}Mg$&$\rightarrow$ &$ \rm^{16}O$&$\rightarrow$ &$ \rm^{11}B$ & 0.10$~|~${\bf0.12}$~|~$0.13\\[-1.20mm]
$\rm ^{24}Mg$&$\rightarrow$ &$ \rm^{23}Na$&$\rightarrow$ &$ \rm^{11}B$ & 0.10$~|~${\bf0.11}$~|~$0.14\\[-1.20mm]
$\rm ^{20}Ne$&$\rightarrow$ &$ \rm^{15}N$&$\rightarrow$ &$ \rm^{11}B$ & 0.09$~|~${\bf0.11}$~|~$0.12\\[-1.20mm]
$\rm ^{24}Mg$&$\rightarrow$ &$ \rm^{11}B$&$\rightarrow$ &$ \rm^{10}B$ & 0.10$~|~${\bf0.11}$~|~$0.11\\[-1.20mm]
$\rm ^{20}Ne$&$\rightarrow$ &$ \rm^{16}O$&$\rightarrow$ &$ \rm^{11}B$ & 0.09$~|~${\bf0.10}$~|~$0.12\\\hline\hline
\end{tabular}
\end{table}
\endgroup
\begingroup
\squeezetable
\begin{table}[!ht]
\caption{Ranking of 1- and 2-step channels for C at 10 GeV/n, from $f_{ij}^{\rm 1-step}$ and $f_{ijk}^{\rm 2-step}$ coefficients (\ref{eq:fij_fijk}). Channels $<\,$1.0\% and higher-level channels ($>2$-step, contributing to $\sim5.2\,\%$, see Table~\ref{tab:origin}), are not shown.\label{tab:channels_C}}
\begin{tabular}{rrlllc}
\hline\hline
\multicolumn{3}{c}{\# of channels} & \multicolumn{2}{c}{in range} & {contribution [\%]}\\
\multicolumn{3}{c}{12}& \multicolumn{2}{c}{[1\%,100\%]} & 81.5\\[-0.9mm]
\multicolumn{3}{c}{35}& \multicolumn{2}{c}{[0.1\%,1\%]} & 7.5\\[-0.9mm]
\multicolumn{3}{c}{139}& \multicolumn{2}{c}{[0.01\%,0.1\%]} & 4.2\\[-0.9mm]
\multicolumn{3}{c}{346}& \multicolumn{2}{c}{[0.001\%,0.01\%]} & 1.4\\[-0.9mm]
\multicolumn{3}{c}{535}& \multicolumn{2}{c}{[0.0001\%,0.001\%]} & 0.2\\[-0.9mm]
\multicolumn{3}{c}{3450}& \multicolumn{2}{c}{[0.0\%,0.0001\%]} & 0.0\\[-0.9mm]
\multicolumn{6}{c}{ }\\[-2.5mm]
\multicolumn{5}{l}{Channel}  &  min $|$ {\bf mean} $|$ max \\
\hline
$\rm ^{16}O$&$\rightarrow$ &$ \rm^{13}C$ & & & 33.1$~|~${\bf33.8}$~|~$34.6\\[-1.20mm]
$\rm ^{16}O$&$\rightarrow$ &$ \rm^{12}C$ & & & 26.7$~|~${\bf27.3}$~|~$28.0\\[-1.20mm]
$\rm ^{16}O$&$\rightarrow$ &$ \rm^{13}C$&$\rightarrow$ &$ \rm^{12}C$ & 2.68$~|~${\bf2.87}$~|~$3.05\\[-1.20mm]
$\rm ^{24}Mg$&$\rightarrow$ &$ \rm^{12}C$ & & & 2.62$~|~${\bf2.72}$~|~$2.83\\[-1.20mm]
$\rm ^{16}O$&$\rightarrow$ &$ \rm^{15}N$&$\rightarrow$ &$ \rm^{13}C$ & 2.43$~|~${\bf2.47}$~|~$2.50\\[-1.20mm]
$\rm ^{20}Ne$&$\rightarrow$ &$ \rm^{12}C$ & & & 2.45$~|~${\bf2.46}$~|~$2.46\\[-1.20mm]
$\rm ^{16}O$&$\rightarrow$ &$ \rm^{15}N$&$\rightarrow$ &$ \rm^{12}C$ & 1.95$~|~${\bf2.18}$~|~$2.42\\[-1.20mm]
$\rm ^{14}N$&$\rightarrow$ &$ \rm^{12}C$ & & & 1.73$~|~${\bf1.84}$~|~$1.96\\[-1.20mm]
$\rm ^{28}Si$&$\rightarrow$ &$ \rm^{12}C$ & & & 1.25$~|~${\bf1.80}$~|~$2.34\\[-1.20mm]
$\rm ^{16}O$&$\rightarrow$ &$ \rm^{14}N$&$\rightarrow$ &$ \rm^{12}C$ & 1.18$~|~${\bf1.48}$~|~$1.78\\[-1.20mm]
$\rm ^{20}Ne$&$\rightarrow$ &$ \rm^{13}C$ & & & 1.34$~|~${\bf1.38}$~|~$1.41\\[-1.20mm]
$\rm ^{24}Mg$&$\rightarrow$ &$ \rm^{13}C$ & & & 1.05$~|~${\bf1.16}$~|~$1.27\\\hline\hline
\end{tabular}
\end{table}
\endgroup
\begingroup
\squeezetable
\begin{table}[!ht]
\caption{Ranking of 1- and 2-step channels for N at 10 GeV/n, from $f_{ij}^{\rm 1-step}$ and $f_{ijk}^{\rm 2-step}$ coefficients (\ref{eq:fij_fijk}). Channels $<\,$0.1\% and higher-level channels ($>2$-step, contributing to $\sim3.5\,\%$, see Table~\ref{tab:origin}), are not shown.\label{tab:channels_N}}
\begin{tabular}{rrlllc}
\hline\hline
\multicolumn{3}{c}{\# of channels} & \multicolumn{2}{c}{in range} & {contribution [\%]}\\
\multicolumn{3}{c}{9}& \multicolumn{2}{c}{[1\%,100\%]} & 85.6\\[-0.9mm]
\multicolumn{3}{c}{28}& \multicolumn{2}{c}{[0.1\%,1\%]} & 5.5\\[-0.9mm]
\multicolumn{3}{c}{140}& \multicolumn{2}{c}{[0.01\%,0.1\%]} & 4.0\\[-0.9mm]
\multicolumn{3}{c}{312}& \multicolumn{2}{c}{[0.001\%,0.01\%]} & 1.2\\[-0.9mm]
\multicolumn{3}{c}{495}& \multicolumn{2}{c}{[0.0001\%,0.001\%]} & 0.2\\[-0.9mm]
\multicolumn{3}{c}{1858}& \multicolumn{2}{c}{[0.0\%,0.0001\%]} & 0.0\\[-0.9mm]
\multicolumn{6}{c}{ }\\[-2.5mm]
\multicolumn{5}{l}{Channel}  &  min $|$ {\bf mean} $|$ max \\
\hline
$\rm ^{16}O$&$\rightarrow$ &$ \rm^{15}N$ & & & 43.3$~|~${\bf47.1}$~|~$50.4\\[-1.20mm]
$\rm ^{16}O$&$\rightarrow$ &$ \rm^{14}N$ & & & 19.6$~|~${\bf23.4}$~|~$26.3\\[-1.20mm]
$\rm ^{20}Ne$&$\rightarrow$ &$ \rm^{15}N$ & & & 2.95$~|~${\bf3.09}$~|~$3.38\\[-1.20mm]
$\rm ^{24}Mg$&$\rightarrow$ &$ \rm^{15}N$ & & & 2.40$~|~${\bf2.73}$~|~$3.05\\[-1.20mm]
$\rm ^{20}Ne$&$\rightarrow$ &$ \rm^{14}N$ & & & 2.02$~|~${\bf2.23}$~|~$2.72\\[-1.20mm]
$\rm ^{28}Si$&$\rightarrow$ &$ \rm^{15}N$ & & & 1.84$~|~${\bf2.14}$~|~$2.39\\[-1.20mm]
$\rm ^{16}O$&$\rightarrow$ &$ \rm^{15}N$&$\rightarrow$ &$ \rm^{14}N$ & 1.81$~|~${\bf2.04}$~|~$2.36\\[-1.20mm]
$\rm ^{24}Mg$&$\rightarrow$ &$ \rm^{14}N$ & & & 1.50$~|~${\bf1.70}$~|~$2.02\\[-1.20mm]
$\rm ^{28}Si$&$\rightarrow$ &$ \rm^{14}N$ & & & 0.98$~|~${\bf1.14}$~|~$1.40\\[-1.20mm]
$\rm ^{56}Fe$&$\rightarrow$ &$ \rm^{15}N$ & & & 0.36$~|~${\bf0.52}$~|~$0.83\\[-1.20mm]
$\rm ^{26}Mg$&$\rightarrow$ &$ \rm^{15}N$ & & & 0.24$~|~${\bf0.32}$~|~$0.38\\[-1.20mm]
$\rm ^{25}Mg$&$\rightarrow$ &$ \rm^{15}N$ & & & 0.28$~|~${\bf0.31}$~|~$0.34\\[-1.20mm]
$\rm ^{24}Mg$&$\rightarrow$ &$ \rm^{16}O$&$\rightarrow$ &$ \rm^{15}N$ & 0.27$~|~${\bf0.31}$~|~$0.35\\[-1.20mm]
$\rm ^{32}S$&$\rightarrow$ &$ \rm^{15}N$ & & & 0.21$~|~${\bf0.27}$~|~$0.33\\[-1.20mm]
$\rm ^{20}Ne$&$\rightarrow$ &$ \rm^{16}O$&$\rightarrow$ &$ \rm^{15}N$ & 0.24$~|~${\bf0.27}$~|~$0.30\\[-1.20mm]
$\rm ^{24}Mg$&$\rightarrow$ &$ \rm^{23}Na$&$\rightarrow$ &$ \rm^{15}N$ & 0.24$~|~${\bf0.27}$~|~$0.31\\[-1.20mm]
$\rm ^{56}Fe$&$\rightarrow$ &$ \rm^{14}N$ & & & 0.14$~|~${\bf0.26}$~|~$0.52\\[-1.20mm]
$\rm ^{27}Al$&$\rightarrow$ &$ \rm^{15}N$ & & & 0.21$~|~${\bf0.25}$~|~$0.31\\[-1.20mm]
$\rm ^{28}Si$&$\rightarrow$ &$ \rm^{16}O$&$\rightarrow$ &$ \rm^{15}N$ & 0.19$~|~${\bf0.21}$~|~$0.23\\[-1.20mm]
$\rm ^{24}Mg$&$\rightarrow$ &$ \rm^{22}Ne$&$\rightarrow$ &$ \rm^{15}N$ & 0.15$~|~${\bf0.20}$~|~$0.23\\[-1.20mm]
$\rm ^{28}Si$&$\rightarrow$ &$ \rm^{27}Al$&$\rightarrow$ &$ \rm^{15}N$ & 0.17$~|~${\bf0.20}$~|~$0.23\\[-1.20mm]
$\rm ^{32}S$&$\rightarrow$ &$ \rm^{14}N$ & & & 0.12$~|~${\bf0.16}$~|~$0.21\\[-1.20mm]
$\rm ^{22}Ne$&$\rightarrow$ &$ \rm^{15}N$ & & & 0.15$~|~${\bf0.15}$~|~$0.17\\[-1.20mm]
$\rm ^{24}Mg$&$\rightarrow$ &$ \rm^{16}O$&$\rightarrow$ &$ \rm^{14}N$ & 0.14$~|~${\bf0.15}$~|~$0.16\\[-1.20mm]
$\rm ^{23}Na$&$\rightarrow$ &$ \rm^{15}N$ & & & 0.09$~|~${\bf0.15}$~|~$0.20\\[-1.20mm]
$\rm ^{20}Ne$&$\rightarrow$ &$ \rm^{19}F$&$\rightarrow$ &$ \rm^{15}N$ & 0.11$~|~${\bf0.14}$~|~$0.19\\[-1.20mm]
$\rm ^{20}Ne$&$\rightarrow$ &$ \rm^{16}O$&$\rightarrow$ &$ \rm^{14}N$ & 0.10$~|~${\bf0.14}$~|~$0.16\\[-1.20mm]
$\rm ^{20}Ne$&$\rightarrow$ &$ \rm^{15}N$&$\rightarrow$ &$ \rm^{14}N$ & 0.11$~|~${\bf0.14}$~|~$0.18\\[-1.20mm]
$\rm ^{26}Mg$&$\rightarrow$ &$ \rm^{14}N$ & & & 0.08$~|~${\bf0.13}$~|~$0.21\\[-1.20mm]
$\rm ^{25}Mg$&$\rightarrow$ &$ \rm^{14}N$ & & & 0.11$~|~${\bf0.13}$~|~$0.16\\[-1.20mm]
$\rm ^{27}Al$&$\rightarrow$ &$ \rm^{14}N$ & & & 0.08$~|~${\bf0.12}$~|~$0.16\\[-1.20mm]
$\rm ^{24}Mg$&$\rightarrow$ &$ \rm^{15}N$&$\rightarrow$ &$ \rm^{14}N$ & 0.09$~|~${\bf0.12}$~|~$0.15\\[-1.20mm]
$\rm ^{24}Mg$&$\rightarrow$ &$ \rm^{21}Ne$&$\rightarrow$ &$ \rm^{15}N$ & 0.08$~|~${\bf0.12}$~|~$0.15\\[-1.20mm]
$\rm ^{24}Mg$&$\rightarrow$ &$ \rm^{20}Ne$&$\rightarrow$ &$ \rm^{15}N$ & 0.08$~|~${\bf0.12}$~|~$0.15\\[-1.20mm]
$\rm ^{28}Si$&$\rightarrow$ &$ \rm^{24}Mg$&$\rightarrow$ &$ \rm^{15}N$ & 0.10$~|~${\bf0.11}$~|~$0.13\\[-1.20mm]
$\rm ^{28}Si$&$\rightarrow$ &$ \rm^{16}O$&$\rightarrow$ &$ \rm^{14}N$ & 0.09$~|~${\bf0.11}$~|~$0.12\\[-1.20mm]
$\rm ^{24}Mg$&$\rightarrow$ &$ \rm^{23}Na$&$\rightarrow$ &$ \rm^{14}N$ & 0.10$~|~${\bf0.10}$~|~$0.10\\\hline\hline
\end{tabular}
\end{table}
\endgroup

The $f$-ratios (Eq.~[\ref{eq:fij_fijk}]) calculated for Li through C species for 1-step and 2-step channels at 10 GeV/n are listed in Tables \ref{tab:channels_Li} to \ref{tab:channels_C}). The top portion of each Table provides an estimate of the total number of channels whose percentage contribution to the production of each species falls into one of the equally spaced logarithmic intervals:  0-0.0001\%, \dots, 0.1\%-1\%, 1\%-100\%. The bottom portion is the actual ranking starting from the largest contributor down to the channels whose relative contribution exceeds $\sim$0.1\%.

To ensure their robustness, the $f$-ratios are calculated using several available cross section parametrizations: Table \ref{tab:channels_Li} is based on GP12 and GP22 since WNEW parametrization does not provide Li production cross sections, whereas for other Tables GP12, GP22, S01, and W03 parametrizations are used. We report the minimum, median, and maximum $f$-ratio values derived from those parametrizations. The results are explicitly checked to be robust against acceptable choices of the injection indices and transport parameters.

Abundances of CR species depend on the isotopic composition of the ISM, which reflects the properties of stellar nucleosynthesis \cite{1999ARA&A..37..239B}, and acceleration selectivity in the CR acceleration sites \cite{1997ApJ...487..182M,2003ApJ...591.1220L}. In turn, the dominant channels in 1-step reactions can be found by forming a product of the relative abundance of CR species and the associated production cross sections \cite{2013ICRC......0823M}. Such simple estimate can help to understand the main results of our ranking.

Though the modern experiments, such as, e.g., AMS-02, provide an unmatched precision (see Introduction), they are still in the process of data acquisition and their published results are limited by the spectra of  light species ($Z\le8$). The best measurement of CR abundances from Be to Ni in the energy range from 0.62--35 GeV/n so far was done by the HEAO-3 instrument launched in 1979 \cite{1990A&A...233...96E}. Table 2 in the HEAO-3 paper \cite{1990A&A...233...96E} provides CR abundances at 10 GeV/n normalized to Oxygen (=1000):
\begin{center}
\begin{tabular}[c]{c|c|c|c|c|c|c|c|c|c|c|c}
Element & C & N & O & Ne & Na & Mg & Al & Si & S & Ca & Fe\\ 
\hline
Abund. & 986 & 219 & 1000 & 152 & 26 & 197 & 31 & 163 & 30 & 18 & 110\smallskip
\end{tabular}
\end{center}

Combining these abundance values with the typical $A^{2/3}$ dependence for the nuclear cross sections, one can see that $^{16}$O and $^{12}$C are (well-known) dominant species for production of Li, Be, and B. Sub-dominant channels also follow the same trend with most prominent being $^{24}$Mg, $^{20}$Ne, $^{28}$Si, and $^{56}$Fe. Despite its abundance, Nitrogen is not one of the dominant species because it has only a $\sim$30\% primary contribution (see Table~\ref{tab:origin}), but appears in the 2-step reactions. In fact, $^{15}$N is ranked higher than $^{14}$N because of its larger production cross section ($^{16}$O$\to ^{14,15}$N). Note that the accurate cross section values mostly matter for the relative ranking of isotopes produced in fragmentation of the same species (e.g., relative production of $^6$Li and $^7$Li), or when the abundances of parent nuclei are similar (e.g., $^{20}$Ne and $^{28}$Si). Meanwhile, the accuracy of the isotopic production cross sections and especially their precise values are what we need to know.

\clearpage
\section{Tables of ranked reactions (and ghosts) at 10~GeV/n\label{app:table_reactions}}

Tables~\ref{tab:xs_Li} to \ref{tab:xs_N} show ranked $f_{abc}$ coefficients, as calculated from Eq.~(\ref{eq:f_abc}) and discussed in Sect.~\ref{sec:ranking}, along with their cross section values (extreme value and average). The next-to-last column indicates whether any data were found for this reaction (see App.~\ref{app:plots_xs_prod}). The last column shows the ratio of the cumulative cross section $\sigma^{\rm c}$ to the direct production $\sigma$; only values $\sigma^{\rm c}/\sigma>1.05$ are shown (reactions involving ghosts, in boldface, have no cumulative).

Full ASCII files from which the tables are extracted are available upon request.

\begingroup
\squeezetable
\begin{table}[!ht]
\caption{Reactions and associated cross sections important for calculations of Li flux at 10 GeV/n, sorted according to the flux impact $f_{abc}$, Eq.~(\ref{eq:f_abc}), until the cumulative of the flux impact $>0.8\times f_{\rm sec} \times \sum f_{abc}$, with $f_{\rm sec}=100\%$ and $\sum f_{abc}=1.20$ (see Sect.~\ref{sec:f_abs}). Reactions in {\bf bold} highlight short-lived fragments (see Sect.~\ref{sec:ghosts}), whose properties are gathered in Table~\ref{tab:ghosts}.\label{tab:xs_Li}}
\begin{tabular}{lr|c|lc@{\hskip 0.4cm}c@{\hskip 0.2cm}c}
\hline\hline
Reaction $a+b\rightarrow c$ &  \multicolumn{3}{c}{Flux impact $f_{abc}$ [\%]} & $\sigma$ [mb]& Data & $\sigma^{\rm c}\!\!/\!\sigma$\\
\cline{2-4}
  &  min & {\bf mean} & max &  range & & \\
\hline
$\rm \sigma(^{12}C+H\rightarrow^{6}$$\rm Li)$ & 11.0 & {\bf13.6} & 16.0 & 14.0 & \cmark & \\[-1.20mm]
$\rm \sigma(^{16}O+H\rightarrow^{6}$$\rm Li)$ & 11.0 & {\bf13.5} & 16.0 & 13.0 & \cmark & \\[-1.20mm]
$\rm \sigma(^{12}C+H\rightarrow^{7}$$\rm Li)$ & 10.0 & {\bf11.9} & 14.0 & 12.6 & \cmark & \\[-1.20mm]
$\rm \sigma(^{16}O+H\rightarrow^{7}$$\rm Li)$ & 9.6 & {\bf11.3} & 13.0 & 11.2 & \cmark & \\[-1.20mm]
$\rm \sigma(^{11}B+H\rightarrow^{7}$$\rm Li)$ & 3.00 & {\bf3.52} & 4.00 & 21.5 & \cmark & \\[-1.20mm]
$\rm \sigma(^{13}C+H\rightarrow^{7}$$\rm Li)$ & 2.00 & {\bf2.39} & 2.80 & 22.1 &  & \\[-1.20mm]
$\rm \sigma(^{16}O+He\rightarrow^{6}$$\rm Li)$ & 2.00 & {\bf2.38} & 2.80 & 20.6 &  & \\[-1.20mm]
$\rm \sigma(^{7}Li+H\rightarrow^{6}$$\rm Li)$ & 2.30 & {\bf2.35} & 2.40 & 31.5 & \cmark & \\[-1.20mm]
$\rm \sigma(^{12}C+He\rightarrow^{6}$$\rm Li)$ & 1.90 & {\bf2.33} & 2.70 & 21.6 &  & \\[-1.20mm]
$\rm \sigma(^{15}N+H\rightarrow^{7}$$\rm Li)$ & 1.90 & {\bf2.27} & 2.60 & 18.6 & \cmark & \\[-1.20mm]
$\rm \sigma(^{12}C+He\rightarrow^{7}$$\rm Li)$ & 1.70 & {\bf2.04} & 2.40 & 19.4 &  & \\[-1.20mm]
$\rm \sigma(^{16}O+He\rightarrow^{7}$$\rm Li)$ & 1.70 & {\bf2.00} & 2.30 & 17.8 &  & \\[-1.20mm]
$\rm \sigma(^{24}Mg+H\rightarrow^{6}$$\rm Li)$ & 1.70 & {\bf1.98} & 2.30 & 12.6 &  & \\[-1.20mm]
$\rm \sigma(^{13}C+H\rightarrow^{6}$$\rm Li)$ & 1.60 & {\bf1.97} & 2.30 & 17.8 &  & \\[-1.20mm]
$\rm \sigma(^{24}Mg+H\rightarrow^{7}$$\rm Li)$ & 1.50 & {\bf1.74} & 2.00 & 11.4 &  & \\[-1.20mm]
$\rm \sigma(^{10}B+H\rightarrow^{6}$$\rm Li)$ & 1.40 & {\bf1.64} & 1.90 & 20.0 &  & \\[-1.20mm]
$\rm \sigma(^{14}N+H\rightarrow^{6}$$\rm Li)$ & 1.40 & {\bf1.62} & 1.90 & 13.0 & \cmark & \\[-1.20mm]
$\rm \sigma(^{15}N+H\rightarrow^{6}$$\rm Li)$ & 1.30 & {\bf1.60} & 1.90 & 12.8 & \cmark & \\[-1.20mm]
$\rm \sigma(^{12}C+H\rightarrow^{11}$$\rm B)$ & 1.20 & {\bf1.38} & 1.60 & 30.0 & \cmark & 1.8\\[-1.20mm]
$\rm \sigma(^{7}Be+H\rightarrow^{6}$$\rm Li)$ & 1.20 & {\bf1.34} & 1.50 & 21.0 &  & \\[-1.20mm]
$\rm \sigma(\mathbf{^{12}C+H\rightarrow^{11}}$${\rm\mathbf C})$ & 1.10 & {\bf1.24} & 1.40 & 26.9 & \cmark & {\bf n\!/\!a} \\[-1.20mm]
$\rm \sigma(^{14}N+H\rightarrow^{7}$$\rm Li)$ & 0.95 & {\bf1.13} & 1.30 & 9.3 & \cmark & \\[-1.20mm]
$\rm \sigma(^{56}Fe+H\rightarrow^{7}$$\rm Li)$ & 0.00 & {\bf0.94} & 1.90 &  [0.0, 23.0] &  & \\[-1.20mm]
$\rm \sigma(^{56}Fe+H\rightarrow^{6}$$\rm Li)$ & 0.00 & {\bf0.94} & 1.90 &  [0.0, 22.0] &  & \\[-1.20mm]
$\rm \sigma(^{16}O+H\rightarrow^{11}$$\rm B)$ & 0.80 & {\bf0.90} & 1.00 & 18.2 & \cmark & 1.5\\[-1.20mm]
$\rm \sigma(^{11}B+H\rightarrow^{6}$$\rm Li)$ & 0.71 & {\bf0.84} & 0.97 & 5.0 & \cmark & \\[-1.20mm]
$\rm \sigma(^{28}Si+H\rightarrow^{6}$$\rm Li)$ & 0.00 & {\bf0.80} & 1.60 &  [0.0, 13.0] &  & \\[-1.20mm]
$\rm \sigma(^{10}B+H\rightarrow^{7}$$\rm Li)$ & 0.70 & {\bf0.80} & 0.90 & 10.0 &  & \\[-1.20mm]
$\rm \sigma(^{28}Si+H\rightarrow^{7}$$\rm Li)$ & 0.00 & {\bf0.71} & 1.40 &  [0.0, 11.0] &  & \\[-1.20mm]
$\rm \sigma(^{16}O+H\rightarrow^{15}$$\rm N)$ & 0.57 & {\bf0.64} & 0.71 & 34.3 & \cmark & 1.8\\[-1.20mm]
$\rm \sigma(^{12}C+H\rightarrow^{10}$$\rm B)$ & 0.53 & {\bf0.64} & 0.74 & 12.3 & \cmark & 1.1\\[-1.20mm]
$\rm \sigma(^{20}Ne+H\rightarrow^{6}$$\rm Li)$ & 0.00 & {\bf0.63} & 1.30 &  [0.0, 13.0] &  & \\[-1.20mm]
$\rm \sigma(\mathbf{^{16}O+H\rightarrow^{13}}$${\rm\mathbf O})$ & 0.55 & {\bf0.63} & 0.71 & 30.5 & \cmark & {\bf n\!/\!a} \\[-1.20mm]
$\rm \sigma(^{16}O+H\rightarrow^{10}$$\rm B)$ & 0.50 & {\bf0.60} & 0.70 & 10.9 & \cmark & \\[-1.20mm]
$\rm \sigma(^{11}B+He\rightarrow^{7}$$\rm Li)$ & 0.52 & {\bf0.60} & 0.69 & 33.2 &  & \\[-1.20mm]
$\rm \sigma(\mathbf{^{16}O+H\rightarrow^{15}}$${\rm\mathbf O})$ & 0.51 & {\bf0.57} & 0.63 & 30.5 & \cmark & {\bf n\!/\!a} \\[-1.20mm]
$\rm \sigma(^{20}Ne+H\rightarrow^{7}$$\rm Li)$ & 0.00 & {\bf0.56} & 1.10 &  [0.0, 11.0] &  & \\[-1.20mm]
$\rm \sigma(^{16}O+H\rightarrow^{7}$$\rm Be)$ & 0.37 & {\bf0.45} & 0.54 & 10.0 & \cmark & \\[-1.20mm]
$\rm \sigma(\mathbf{^{16}O+H\rightarrow^{11}}$${\rm\mathbf C})$ & 0.40 & {\bf0.45} & 0.50 & 9.1 &  & {\bf n\!/\!a} \\[-1.20mm]
$\rm \sigma(^{56}Fe+He\rightarrow^{7}$$\rm Li)$ & 0.00 & {\bf0.44} & 0.88 &  [0.0, 97.0] &  & \\[-1.20mm]
$\rm \sigma(^{56}Fe+He\rightarrow^{6}$$\rm Li)$ & 0.00 & {\bf0.44} & 0.88 &  [0.0, 95.0] &  & \\[-1.20mm]
$\rm \sigma(^{7}Li+He\rightarrow^{6}$$\rm Li)$ & 0.42 & {\bf0.43} & 0.45 & 52.2 &  & \\[-1.20mm]
$\rm \sigma(^{13}C+He\rightarrow^{7}$$\rm Li)$ & 0.34 & {\bf0.41} & 0.48 & 34.2 &  & \\[-1.20mm]
$\rm \sigma(^{12}C+H\rightarrow^{7}$$\rm Be)$ & 0.34 & {\bf0.41} & 0.48 & 9.7 & \cmark & \\[-1.20mm]
$\rm \sigma(^{16}O+H\rightarrow^{13}$$\rm C)$ & 0.36 & {\bf0.41} & 0.46 & 17.5 & \cmark & 1.2\\[-1.20mm]
$\rm \sigma(^{24}Mg+He\rightarrow^{6}$$\rm Li)$ & 0.33 & {\bf0.39} & 0.46 & 22.5 &  & \\[-1.20mm]
$\rm \sigma(^{15}N+He\rightarrow^{7}$$\rm Li)$ & 0.33 & {\bf0.39} & 0.45 & 28.6 &  & \\[-1.20mm]
$\rm \sigma(\mathbf{^{7}Li+H\rightarrow^{6}}$${\rm\mathbf He})$ & 0.00 & {\bf0.38} & 0.76 &  [0.0, 10.0] &  & {\bf n\!/\!a} \\[-1.20mm]
$\rm \sigma(^{11}B+H\rightarrow^{10}$$\rm B)$ & 0.29 & {\bf0.35} & 0.40 & 38.9 & \cmark & \\[-1.20mm]
$\rm \sigma(^{24}Mg+He\rightarrow^{7}$$\rm Li)$ & 0.29 & {\bf0.34} & 0.40 & 20.3 &  & \\[-1.20mm]
$\rm \sigma(^{13}C+He\rightarrow^{6}$$\rm Li)$ & 0.28 & {\bf0.34} & 0.40 & 27.5 &  & \\[-1.20mm]
$\rm \sigma(\mathbf{^{56}Fe+H\rightarrow^{6}}$${\rm\mathbf He})$ & 0.00 & {\bf0.29} & 0.57 &  [0.0, 6.9] &  & {\bf n\!/\!a} \\
\hline\hline
 \end{tabular}
\end{table}
\endgroup
\begingroup
\squeezetable
\begin{table}[!ht]
\caption{Reactions and associated cross sections important for calculations of Be flux at 10 GeV/n, sorted according to the flux impact $f_{abc}$, Eq.~(\ref{eq:f_abc}), until the cumulative of the flux impact $>0.8\times f_{\rm sec} \times \sum f_{abc}$, with $f_{\rm sec}=100\%$ and $\sum f_{abc}=1.14$ (see Sect.~\ref{sec:f_abs}). Reactions in {\bf bold} highlight short-lived fragments (see Sect.~\ref{sec:ghosts}), whose properties are gathered in Table~\ref{tab:ghosts}.\label{tab:xs_Be}}
\begin{tabular}{lr|c|lc@{\hskip 0.4cm}c@{\hskip 0.2cm}c}
\hline\hline
Reaction $a+b\rightarrow c$ &  \multicolumn{3}{c}{Flux impact $f_{abc}$ [\%]} & $\sigma$ [mb]& Data & $\sigma^{\rm c}\!\!/\!\sigma$\\
\cline{2-4}
  &  min & {\bf mean} & max &  range & & \\
\hline
$\rm \sigma(^{16}O+H\rightarrow^{7}$$\rm Be)$ & 17.0 & {\bf17.6} & 19.0 & 10.0 & \cmark & \\[-1.20mm]
$\rm \sigma(^{12}C+H\rightarrow^{7}$$\rm Be)$ & 15.0 & {\bf15.9} & 17.0 & 9.7 & \cmark & \\[-1.20mm]
$\rm \sigma(^{12}C+H\rightarrow^{9}$$\rm Be)$ & 8.80 & {\bf9.27} & 9.80 & 6.8 & \cmark & \\[-1.20mm]
$\rm \sigma(^{16}O+H\rightarrow^{9}$$\rm Be)$ & 5.00 & {\bf5.34} & 5.60 & 3.7 & \cmark & \\[-1.20mm]
$\rm \sigma(^{16}O+He\rightarrow^{7}$$\rm Be)$ & 2.70 & {\bf2.87} & 3.00 & 14.7 &  & \\[-1.20mm]
$\rm \sigma(^{28}Si+H\rightarrow^{7}$$\rm Be)$ & 2.60 & {\bf2.77} & 2.90 & 10.8 &  & \\[-1.20mm]
$\rm \sigma(^{24}Mg+H\rightarrow^{7}$$\rm Be)$ & 2.50 & {\bf2.65} & 2.80 & 10.0 &  & \\[-1.20mm]
$\rm \sigma(^{12}C+He\rightarrow^{7}$$\rm Be)$ & 2.30 & {\bf2.48} & 2.60 & 13.7 &  & \\[-1.20mm]
$\rm \sigma(^{11}B+H\rightarrow^{9}$$\rm Be)$ & 2.30 & {\bf2.36} & 2.50 & 10.0 & \cmark & \\[-1.20mm]
$\rm \sigma(^{12}C+H\rightarrow^{10}$$\rm Be)$ & 2.00 & {\bf2.16} & 2.30 & 4.0 & \cmark & \\[-1.20mm]
$\rm \sigma(^{14}N+H\rightarrow^{7}$$\rm Be)$ & 2.00 & {\bf2.12} & 2.20 & 10.1 & \cmark & \\[-1.20mm]
$\rm \sigma(^{20}Ne+H\rightarrow^{7}$$\rm Be)$ & 1.60 & {\bf1.73} & 1.90 &  [7.4, 9.7] &  & \\[-1.20mm]
$\rm \sigma(^{10}B+H\rightarrow^{9}$$\rm Be)$ & 1.60 & {\bf1.62} & 1.70 & 13.9 &  & \\[-1.20mm]
$\rm \sigma(^{12}C+He\rightarrow^{9}$$\rm Be)$ & 1.40 & {\bf1.45} & 1.50 & 9.6 &  & \\[-1.20mm]
$\rm \sigma(^{12}C+H\rightarrow^{11}$$\rm B)$ & 1.30 & {\bf1.43} & 1.60 & 30.0 & \cmark & 1.8\\[-1.20mm]
$\rm \sigma(^{15}N+H\rightarrow^{9}$$\rm Be)$ & 1.20 & {\bf1.29} & 1.40 & 7.3 & \cmark & \\[-1.20mm]
$\rm \sigma(\mathbf{^{12}C+H\rightarrow^{11}}$${\rm\mathbf C})$ & 1.20 & {\bf1.28} & 1.40 & 26.9 & \cmark & {\bf n\!/\!a} \\[-1.20mm]
$\rm \sigma(^{16}O+H\rightarrow^{10}$$\rm Be)$ & 1.20 & {\bf1.27} & 1.40 & 2.2 & \cmark & \\[-1.20mm]
$\rm \sigma(^{11}B+H\rightarrow^{10}$$\rm Be)$ & 1.10 & {\bf1.21} & 1.30 & 12.9 & \cmark & \\[-1.20mm]
$\rm \sigma(^{11}B+H\rightarrow^{7}$$\rm Be)$ & 0.99 & {\bf1.16} & 1.30 &  [3.6, 4.5] & \cmark & \\[-1.20mm]
$\rm \sigma(^{15}N+H\rightarrow^{7}$$\rm Be)$ & 1.10 & {\bf1.15} & 1.20 & 5.4 & \cmark & \\[-1.20mm]
$\rm \sigma(^{13}C+H\rightarrow^{9}$$\rm Be)$ & 0.96 & {\bf1.03} & 1.10 & 6.7 & \cmark & \\[-1.20mm]
$\rm \sigma(^{28}Si+H\rightarrow^{9}$$\rm Be)$ & 0.91 & {\bf0.96} & 1.00 & 4.5 & \cmark & \\[-1.20mm]
$\rm \sigma(^{10}B+H\rightarrow^{7}$$\rm Be)$ & 0.93 & {\bf0.95} & 0.98 & 6.9 & \cmark & \\[-1.20mm]
$\rm \sigma(^{24}Mg+H\rightarrow^{9}$$\rm Be)$ & 0.89 & {\bf0.94} & 0.99 & 4.3 &  & \\[-1.20mm]
$\rm \sigma(^{16}O+H\rightarrow^{11}$$\rm B)$ & 0.87 & {\bf0.94} & 1.00 & 18.2 & \cmark & 1.5\\[-1.20mm]
$\rm \sigma(^{56}Fe+H\rightarrow^{7}$$\rm Be)$ & 0.11 & {\bf0.92} & 1.70 &  [0.6, 11.0] &  & \\[-1.20mm]
$\rm \sigma(^{16}O+He\rightarrow^{9}$$\rm Be)$ & 0.82 & {\bf0.87} & 0.92 & 5.4 &  & \\[-1.20mm]
$\rm \sigma(^{13}C+H\rightarrow^{7}$$\rm Be)$ & 0.71 & {\bf0.76} & 0.81 & 4.1 & \cmark & \\[-1.20mm]
$\rm \sigma(^{20}Ne+H\rightarrow^{9}$$\rm Be)$ & 0.68 & {\bf0.72} & 0.76 & 4.3 &  & \\[-1.20mm]
$\rm \sigma(^{12}C+H\rightarrow^{10}$$\rm B)$ & 0.59 & {\bf0.64} & 0.68 & 12.3 & \cmark & 1.1\\[-1.20mm]
$\rm \sigma(^{16}O+H\rightarrow^{10}$$\rm B)$ & 0.56 & {\bf0.60} & 0.65 & 10.9 & \cmark & \\[-1.20mm]
$\rm \sigma(^{9}Be+H\rightarrow^{7}$$\rm Be)$ & 0.59 & {\bf0.59} & 0.60 & 10.6 & \cmark & \\[-1.20mm]
$\rm \sigma(^{28}Si+He\rightarrow^{7}$$\rm Be)$ & 0.53 & {\bf0.56} & 0.60 & 19.8 &  & \\[-1.20mm]
$\rm \sigma(^{56}Fe+H\rightarrow^{9}$$\rm Be)$ & 0.06 & {\bf0.53} & 1.00 &  [0.4, 7.5] &  & \\[-1.20mm]
$\rm \sigma(^{24}Mg+He\rightarrow^{7}$$\rm Be)$ & 0.47 & {\bf0.50} & 0.52 & 16.8 &  & \\[-1.20mm]
$\rm \sigma(\mathbf{^{16}O+H\rightarrow^{11}}$${\rm\mathbf C})$ & 0.43 & {\bf0.47} & 0.50 & 9.1 &  & {\bf n\!/\!a} \\[-1.20mm]
$\rm \sigma(^{16}O+H\rightarrow^{15}$$\rm N)$ & 0.41 & {\bf0.44} & 0.47 & 34.3 & \cmark & 1.8\\[-1.20mm]
$\rm \sigma(^{56}Fe+He\rightarrow^{7}$$\rm Be)$ & 0.05 & {\bf0.41} & 0.77 &  [2.4, 43.0] &  & \\[-1.20mm]
$\rm \sigma(\mathbf{^{16}O+H\rightarrow^{15}}$${\rm\mathbf O})$ & 0.37 & {\bf0.39} & 0.42 & 30.5 & \cmark & {\bf n\!/\!a} \\[-1.20mm]
$\rm \sigma(^{27}Al+H\rightarrow^{7}$$\rm Be)$ & 0.30 & {\bf0.38} & 0.45 &  [5.3, 8.9] &  & \\[-1.20mm]
$\rm \sigma(^{14}N+H\rightarrow^{9}$$\rm Be)$ & 0.35 & {\bf0.37} & 0.39 & 2.1 & \cmark & \\[-1.20mm]
$\rm \sigma(^{11}B+He\rightarrow^{9}$$\rm Be)$ & 0.35 & {\bf0.37} & 0.38 & 14.0 &  & \\[-1.20mm]
$\rm \sigma(^{13}C+H\rightarrow^{10}$$\rm Be)$ & 0.33 & {\bf0.37} & 0.40 & 5.9 & \cmark & \\[-1.20mm]
$\rm \sigma(^{23}Na+H\rightarrow^{7}$$\rm Be)$ & 0.30 & {\bf0.35} & 0.41 &  [5.8, 8.6] &  & \\[-1.20mm]
$\rm \sigma(^{11}B+H\rightarrow^{10}$$\rm B)$ & 0.33 & {\bf0.35} & 0.37 & 38.9 & \cmark & \\[-1.20mm]
$\rm \sigma(^{25}Mg+H\rightarrow^{7}$$\rm Be)$ & 0.29 & {\bf0.34} & 0.40 &  [5.6, 8.8] &  & \\[-1.20mm]
$\rm \sigma(^{12}C+He\rightarrow^{10}$$\rm Be)$ & 0.31 & {\bf0.34} & 0.36 & 5.6 &  & \\[-1.20mm]
$\rm \sigma(^{14}N+He\rightarrow^{7}$$\rm Be)$ & 0.32 & {\bf0.34} & 0.36 & 14.4 &  & \\[-1.20mm]
$\rm \sigma(^{20}Ne+He\rightarrow^{7}$$\rm Be)$ & 0.28 & {\bf0.30} & 0.32 &  [12.0, 15.0] &  & \\[-1.20mm]
$\rm \sigma(^{22}Ne+H\rightarrow^{7}$$\rm Be)$ & 0.22 & {\bf0.25} & 0.28 &  [4.7, 6.4] &  & \\[-1.20mm]
$\rm \sigma(^{10}B+He\rightarrow^{9}$$\rm Be)$ & 0.25 & {\bf0.25} & 0.26 & 19.6 &  & \\[-1.20mm]
$\rm \sigma(^{26}Mg+H\rightarrow^{7}$$\rm Be)$ & 0.21 & {\bf0.25} & 0.29 &  [4.7, 7.2] &  & \\[-1.20mm]
$\rm \sigma(\mathbf{^{16}O+H\rightarrow^{9}}$${\rm\mathbf Li})$ & 0.23 & {\bf0.24} & 0.26 & 0.3 & \cmark & {\bf n\!/\!a} \\
\hline\hline
 \end{tabular}
\end{table}
\endgroup
\begingroup
\squeezetable
\begin{table}[!ht]
\caption{Reactions and associated cross sections important for calculations of B flux at 10 GeV/n, sorted according to the flux impact $f_{abc}$, Eq.~(\ref{eq:f_abc}), until the cumulative of the flux impact $>0.9\times f_{\rm sec} \times \sum f_{abc}$, with $f_{\rm sec}=95\%$ and $\sum f_{abc}=1.13$ (see Sect.~\ref{sec:f_abs}). Reactions in {\bf bold} highlight short-lived fragments (see Sect.~\ref{sec:ghosts}), whose properties are gathered in Table~\ref{tab:ghosts}.\label{tab:xs_B}}
\begin{tabular}{lr|c|lc@{\hskip 0.4cm}c@{\hskip 0.2cm}c}
\hline\hline
Reaction $a+b\rightarrow c$ &  \multicolumn{3}{c}{Flux impact $f_{abc}$ [\%]} & $\sigma$ [mb]& Data & $\sigma^{\rm c}\!\!/\!\sigma$\\
\cline{2-4}
  &  min & {\bf mean} & max &  range & & \\
\hline
$\rm \sigma(^{12}C+H\rightarrow^{11}$$\rm B)$ & 18.0 & {\bf18.1} & 19.0 & 30.0 & \cmark & 1.8\\[-1.20mm]
$\rm \sigma(\mathbf{^{12}C+H\rightarrow^{11}}$${\rm\mathbf C})$ & 16.0 & {\bf16.2} & 17.0 & 26.9 & \cmark & {\bf n\!/\!a} \\[-1.20mm]
$\rm \sigma(^{16}O+H\rightarrow^{11}$$\rm B)$ & 11.3 & {\bf11.8} & 12.0 & 18.2 & \cmark & 1.5\\[-1.20mm]
$\rm \sigma(^{12}C+H\rightarrow^{10}$$\rm B)$ & 7.20 & {\bf7.41} & 7.60 & 12.3 & \cmark & 1.1\\[-1.20mm]
$\rm \sigma(^{16}O+H\rightarrow^{10}$$\rm B)$ & 6.82 & {\bf7.03} & 7.21 & 10.9 & \cmark & \\[-1.20mm]
$\rm \sigma(\mathbf{^{16}O+H\rightarrow^{11}}$${\rm\mathbf C})$ & 5.67 & {\bf5.89} & 6.00 & 9.1 &  & {\bf n\!/\!a} \\[-1.20mm]
$\rm \sigma(^{11}B+H\rightarrow^{10}$$\rm B)$ & 4.00 & {\bf4.07} & 4.20 & 38.9 & \cmark & \\[-1.20mm]
$\rm \sigma(^{12}C+He\rightarrow^{11}$$\rm B)$ & 2.50 & {\bf2.59} & 2.70 & 38.6 &  & 1.8\\[-1.20mm]
$\rm \sigma(\mathbf{^{12}C+He\rightarrow^{11}}$${\rm\mathbf C})$ & 2.10 & {\bf2.14} & 2.20 & 32.0 &  & {\bf n\!/\!a} \\[-1.20mm]
$\rm \sigma(^{15}N+H\rightarrow^{11}$$\rm B)$ & 2.00 & {\bf2.03} & 2.10 & 26.1 & \cmark & 1.2\\[-1.20mm]
$\rm \sigma(\mathbf{^{12}C+H\rightarrow^{10}}$${\rm\mathbf C})$ & 1.80 & {\bf1.87} & 1.90 & 3.1 & \cmark & {\bf n\!/\!a} \\[-1.20mm]
$\rm \sigma(^{16}O+He\rightarrow^{11}$$\rm B)$ & 1.67 & {\bf1.75} & 1.80 & 24.4 &  & 1.5\\[-1.20mm]
$\rm \sigma(^{13}C+H\rightarrow^{11}$$\rm B)$ & 1.50 & {\bf1.53} & 1.60 & 22.2 &  & 1.7\\[-1.20mm]
$\rm \sigma(^{12}C+H\rightarrow^{10}$$\rm Be)$ & 1.40 & {\bf1.48} & 1.50 & 4.0 & \cmark & \\[-1.20mm]
$\rm \sigma(^{14}N+H\rightarrow^{11}$$\rm B)$ & 1.30 & {\bf1.34} & 1.36 & 17.3 & \cmark & 1.7\\[-1.20mm]
$\rm \sigma(^{12}C+He\rightarrow^{10}$$\rm B)$ & 1.00 & {\bf1.06} & 1.10 & 15.8 &  & 1.1\\[-1.20mm]
$\rm \sigma(^{16}O+He\rightarrow^{10}$$\rm B)$ & 0.99 & {\bf1.05} & 1.09 & 14.6 &  & \\[-1.20mm]
$\rm \sigma(^{24}Mg+H\rightarrow^{11}$$\rm B)$ & 0.98 & {\bf1.01} & 1.00 & 10.4 &  & 1.6\\[-1.20mm]
$\rm \sigma(\mathbf{^{14}N+H\rightarrow^{11}}$${\rm\mathbf C})$ & 0.90 & {\bf0.92} & 0.94 & 11.9 &  & {\bf n\!/\!a} \\[-1.20mm]
$\rm \sigma(^{20}Ne+H\rightarrow^{11}$$\rm B)$ & 0.87 & {\bf0.90} & 0.93 & 12.0 &  & 1.7\\[-1.20mm]
$\rm \sigma(\mathbf{^{16}O+He\rightarrow^{11}}$${\rm\mathbf C})$ & 0.83 & {\bf0.88} & 0.90 & 12.2 &  & {\bf n\!/\!a} \\[-1.20mm]
$\rm \sigma(^{16}O+H\rightarrow^{10}$$\rm Be)$ & 0.84 & {\bf0.87} & 0.91 & 2.2 & \cmark & \\[-1.20mm]
$\rm \sigma(^{11}B+H\rightarrow^{10}$$\rm Be)$ & 0.81 & {\bf0.83} & 0.85 & 12.9 & \cmark & \\[-1.20mm]
$\rm \sigma(^{14}N+H\rightarrow^{10}$$\rm B)$ & 0.77 & {\bf0.79} & 0.82 & 10.3 & \cmark & \\[-1.20mm]
$\rm \sigma(^{15}N+H\rightarrow^{10}$$\rm B)$ & 0.72 & {\bf0.74} & 0.77 & 9.6 & \cmark & \\[-1.20mm]
$\rm \sigma(^{28}Si+H\rightarrow^{11}$$\rm B)$ & 0.39 & {\bf0.63} & 0.87 &  [4.0, 9.5] &  & 2.1\\[-1.20mm]
$\rm \sigma(^{13}C+H\rightarrow^{10}$$\rm B)$ & 0.59 & {\bf0.62} & 0.65 & 9.0 &  & 1.6\\[-1.20mm]
$\rm \sigma(^{24}Mg+H\rightarrow^{10}$$\rm B)$ & 0.58 & {\bf0.60} & 0.62 & 6.2 &  & \\[-1.20mm]
$\rm \sigma(^{11}B+He\rightarrow^{10}$$\rm B)$ & 0.57 & {\bf0.58} & 0.59 & 50.0 &  & \\[-1.20mm]
$\rm \sigma(\mathbf{^{13}C+H\rightarrow^{11}}$${\rm\mathbf C})$ & 0.54 & {\bf0.56} & 0.59 & 8.2 &  & {\bf n\!/\!a} \\[-1.20mm]
$\rm \sigma(\mathbf{^{20}Ne+H\rightarrow^{11}}$${\rm\mathbf C})$ & 0.52 & {\bf0.54} & 0.56 & 7.2 & \cmark & {\bf n\!/\!a} \\[-1.20mm]
$\rm \sigma(\mathbf{^{24}Mg+H\rightarrow^{11}}$${\rm\mathbf C})$ & 0.51 & {\bf0.53} & 0.56 &  [5.1, 5.9] &  & {\bf n\!/\!a} \\[-1.20mm]
$\rm \sigma(^{20}Ne+H\rightarrow^{10}$$\rm B)$ & 0.49 & {\bf0.51} & 0.52 &  [6.4, 7.1] &  & \\[-1.20mm]
$\rm \sigma(\mathbf{^{28}Si+H\rightarrow^{11}}$${\rm\mathbf C})$ & 0.42 & {\bf0.44} & 0.46 &  [4.3, 5.0] &  & {\bf n\!/\!a} \\[-1.20mm]
$\rm \sigma(\mathbf{^{15}N+H\rightarrow^{11}}$${\rm\mathbf C})$ & 0.40 & {\bf0.41} & 0.43 & 5.3 & \cmark & {\bf n\!/\!a} \\[-1.20mm]
$\rm \sigma(^{28}Si+H\rightarrow^{10}$$\rm B)$ & 0.27 & {\bf0.39} & 0.52 &  [2.8, 5.7] &  & \\[-1.20mm]
$\rm \sigma(^{56}Fe+H\rightarrow^{11}$$\rm B)$ & 0.03 & {\bf0.35} & 0.67 &  [0.4, 11.0] &  & 3.3\\[-1.20mm]
$\rm \sigma(^{15}N+He\rightarrow^{11}$$\rm B)$ & 0.29 & {\bf0.29} & 0.30 & 34.1 &  & 1.2\\[-1.20mm]
$\rm \sigma(^{22}Ne+H\rightarrow^{11}$$\rm B)$ & 0.27 & {\bf0.28} & 0.30 &  [16.0, 18.0] & \cmark & 1.2\\[-1.20mm]
$\rm \sigma(^{13}C+H\rightarrow^{10}$$\rm Be)$ & 0.24 & {\bf0.25} & 0.26 & 5.9 & \cmark & \\[-1.20mm]
$\rm \sigma(\mathbf{^{12}C+He\rightarrow^{10}}$${\rm\mathbf C})$ & 0.24 & {\bf0.25} & 0.25 & 3.7 &  & {\bf n\!/\!a} \\[-1.20mm]
$\rm \sigma(^{56}Fe+H\rightarrow^{10}$$\rm B)$ & 0.01 & {\bf0.24} & 0.47 &  [0.2, 7.8] &  & 1.1\\[-1.20mm]
$\rm \sigma(^{12}C+He\rightarrow^{10}$$\rm Be)$ & 0.22 & {\bf0.23} & 0.24 & 5.6 &  & \\
\hline\hline
 \end{tabular}
\end{table}
\endgroup
\begingroup
\squeezetable
\begin{table}[!ht]
\caption{Reactions and associated cross sections important for calculations of C flux at 10 GeV/n, sorted according to the flux impact $f_{abc}$, Eq.~(\ref{eq:f_abc}), until the cumulative of the flux impact $>3.5\times f_{\rm sec} \times \sum f_{abc}$, with $f_{\rm sec}=20\%$ and $\sum f_{abc}=1.08$ (see Sect.~\ref{sec:f_abs}). Reactions in {\bf bold} highlight short-lived fragments (see Sect.~\ref{sec:ghosts}), whose properties are gathered in Table~\ref{tab:ghosts}.\label{tab:xs_C}}
\begin{tabular}{lr|c|lc@{\hskip 0.4cm}c@{\hskip 0.2cm}c}
\hline\hline
Reaction $a+b\rightarrow c$ &  \multicolumn{3}{c}{Flux impact $f_{abc}$ [\%]} & $\sigma$ [mb]& Data & $\sigma^{\rm c}\!\!/\!\sigma$\\
\cline{2-4}
  &  min & {\bf mean} & max &  range & & \\
\hline
$\rm \sigma(^{16}O+H\rightarrow^{12}$$\rm C)$ & 21.0 & {\bf21.5} & 22.0 & 32.3 & \cmark & \\[-1.20mm]
$\rm \sigma(\mathbf{^{16}O+H\rightarrow^{13}}$${\rm\mathbf O})$ & 18.0 & {\bf18.4} & 19.0 & 30.5 & \cmark & {\bf n\!/\!a} \\[-1.20mm]
$\rm \sigma(^{16}O+H\rightarrow^{13}$$\rm C)$ & & {\bf11.8} &  & 17.5 & \cmark & 1.2\\[-1.20mm]
$\rm \sigma(^{14}N+H\rightarrow^{12}$$\rm C)$ & 3.30 & {\bf3.65} & 4.00 &  [40.0, 52.0] & \cmark & 1.1\\[-1.20mm]
$\rm \sigma(\mathbf{^{16}O+H\rightarrow^{13}}$${\rm\mathbf N})$ & 3.40 & {\bf3.45} & 3.50 & 5.1 & \cmark & {\bf n\!/\!a} \\[-1.20mm]
$\rm \sigma(^{16}O+He\rightarrow^{12}$$\rm C)$ & 2.90 & {\bf2.95} & 3.00 & 39.8 &  & \\[-1.20mm]
$\rm \sigma(^{15}N+H\rightarrow^{13}$$\rm C)$ & 2.70 & {\bf2.72} & 2.80 & 33.4 & \cmark & 1.5\\[-1.20mm]
$\rm \sigma(^{20}Ne+H\rightarrow^{12}$$\rm C)$ & & {\bf2.38} &  &  [30.0, 31.0] & \cmark & \\[-1.20mm]
$\rm \sigma(^{24}Mg+H\rightarrow^{12}$$\rm C)$ & 2.10 & {\bf2.28} & 2.50 &  [20.0, 25.0] &  & \\[-1.20mm]
$\rm \sigma(\mathbf{^{16}O+H\rightarrow^{13}}$${\rm\mathbf O})$ & 2.20 & {\bf2.22} & 2.30 & 30.5 & \cmark & {\bf n\!/\!a} \\[-1.20mm]
$\rm \sigma(\mathbf{^{16}O+He\rightarrow^{13}}$${\rm\mathbf O})$ & & {\bf2.20} &  & 32.8 &  & {\bf n\!/\!a} \\[-1.20mm]
$\rm \sigma(^{15}N+H\rightarrow^{12}$$\rm C)$ & 1.90 & {\bf2.05} & 2.20 &  [23.0, 28.0] & \cmark & 1.2\\[-1.20mm]
$\rm \sigma(^{13}C+H\rightarrow^{12}$$\rm C)$ & 1.90 & {\bf1.97} & 2.00 &  [27.0, 29.0] &  & 2.0\\
\hline\hline
 \end{tabular}
\end{table}
\endgroup
\begingroup
\squeezetable
\begin{table}[!ht]
\caption{Reactions and associated cross sections important for calculations of N flux at 10 GeV/n, sorted according to the flux impact $f_{abc}$, Eq.~(\ref{eq:f_abc}), until the cumulative of the flux impact $>1.3\times f_{\rm sec} \times \sum f_{abc}$, with $f_{\rm sec}=73\%$ and $\sum f_{abc}=1.08$ (see Sect.~\ref{sec:f_abs}). Reactions in {\bf bold} highlight short-lived fragments (see Sect.~\ref{sec:ghosts}), whose properties are gathered in Table~\ref{tab:ghosts}.\label{tab:xs_N}}
\begin{tabular}{lr|c|lc@{\hskip 0.4cm}c@{\hskip 0.2cm}c}
\hline\hline
Reaction $a+b\rightarrow c$ &  \multicolumn{3}{c}{Flux impact $f_{abc}$ [\%]} & $\sigma$ [mb]& Data & $\sigma^{\rm c}\!\!/\!\sigma$\\
\cline{2-4}
  &  min & {\bf mean} & max &  range & & \\
\hline
$\rm \sigma(^{16}O+H\rightarrow^{15}$$\rm N)$ & 26.0 & {\bf26.3} & 27.0 & 34.3 & \cmark & 1.8\\[-1.20mm]
$\rm \sigma(\mathbf{^{16}O+H\rightarrow^{15}}$${\rm\mathbf O})$ & 23.0 & {\bf23.4} & 24.0 & 30.5 & \cmark & {\bf n\!/\!a} \\[-1.20mm]
$\rm \sigma(^{16}O+H\rightarrow^{14}$$\rm N)$ & 18.0 & {\bf20.0} & 22.0 &  [23.0, 29.0] & \cmark & 1.1\\[-1.20mm]
$\rm \sigma(^{16}O+He\rightarrow^{15}$$\rm N)$ & 3.30 & {\bf3.34} & 3.40 & 39.3 &  & 1.8\\[-1.20mm]
$\rm \sigma(\mathbf{^{16}O+He\rightarrow^{15}}$${\rm\mathbf O})$ & 2.70 & {\bf2.79} & 2.90 & 32.8 &  & {\bf n\!/\!a} \\[-1.20mm]
$\rm \sigma(^{16}O+He\rightarrow^{14}$$\rm N)$ & 2.30 & {\bf2.55} & 2.80 &  [26.0, 33.0] &  & 1.1\\[-1.20mm]
$\rm \sigma(^{15}N+H\rightarrow^{14}$$\rm N)$ & 2.10 & {\bf2.18} & 2.20 & 24.3 & \cmark & \\[-1.20mm]
$\rm \sigma(^{20}Ne+H\rightarrow^{14}$$\rm N)$ & & {\bf2.18} &  &  [23.0, 24.0] & \cmark & \\[-1.20mm]
$\rm \sigma(^{20}Ne+H\rightarrow^{15}$$\rm N)$ & & {\bf2.09} &  &  [22.0, 23.0] & \cmark & 1.6\\[-1.20mm]
$\rm \sigma(^{24}Mg+H\rightarrow^{15}$$\rm N)$ & 1.60 & {\bf1.65} & 1.70 & 13.9 & \cmark & 1.7\\[-1.20mm]
$\rm \sigma(^{24}Mg+H\rightarrow^{14}$$\rm N)$ & & {\bf1.47} &  & 12.4 & \cmark & \\[-1.20mm]
$\rm \sigma(^{28}Si+H\rightarrow^{15}$$\rm N)$ & 1.10 & {\bf1.33} & 1.50 &  [9.9, 13.0] &  & 1.8\\[-1.20mm]
$\rm \sigma(\mathbf{^{20}Ne+H\rightarrow^{15}}$${\rm\mathbf O})$ & 1.30 & {\bf1.32} & 1.40 &  [14.0, 15.0] & \cmark & {\bf n\!/\!a} \\[-1.20mm]
$\rm \sigma(^{16}O+H\rightarrow^{14}$$\rm C)$ & 1.20 & {\bf1.27} & 1.30 & 1.7 & \cmark & \\[-1.20mm]
$\rm \sigma(\mathbf{^{24}Mg+H\rightarrow^{15}}$${\rm\mathbf O})$ & 0.98 & {\bf1.08} & 1.20 &  [8.5, 9.8] & \cmark & {\bf n\!/\!a} \\[-1.20mm]
$\rm \sigma(^{28}Si+H\rightarrow^{14}$$\rm N)$ & 0.83 & {\bf0.95} & 1.10 &  [6.9, 9.2] &  & \\[-1.20mm]
$\rm \sigma(^{15}N+H\rightarrow^{14}$$\rm C)$ & 0.83 & {\bf0.85} & 0.87 &  [9.7, 9.8] & \cmark & \\[-1.20mm]
$\rm \sigma(\mathbf{^{28}Si+H\rightarrow^{15}}$${\rm\mathbf O})$ & 0.74 & {\bf0.79} & 0.83 &  [6.5, 6.9] & \cmark & {\bf n\!/\!a} \\[-1.20mm]
$\rm \sigma(\mathbf{^{16}O+H\rightarrow^{14}}$${\rm\mathbf O})$ & 0.61 & {\bf0.64} & 0.67 &  [0.8, 0.8] & \cmark & {\bf n\!/\!a} \\[-1.20mm]
$\rm \sigma(^{22}Ne+H\rightarrow^{15}$$\rm N)$ & 0.56 & {\bf0.57} & 0.57 &  [31.0, 32.0] & \cmark & 1.1\\[-1.20mm]
$\rm \sigma(^{23}Na+H\rightarrow^{15}$$\rm N)$ & 0.41 & {\bf0.43} & 0.44 &  [19.0, 21.0] &  & 1.3\\[-1.20mm]
$\rm \sigma(^{26}Mg+H\rightarrow^{15}$$\rm N)$ & & {\bf0.41} &  &  [24.0, 25.0] & \cmark & 1.2\\[-1.20mm]
$\rm \sigma(^{27}Al+H\rightarrow^{15}$$\rm N)$ & 0.33 & {\bf0.34} & 0.35 &  [14.0, 15.0] &  & 1.4\\[-1.20mm]
$\rm \sigma(^{17}O+H\rightarrow^{15}$$\rm N)$ & 0.29 & {\bf0.33} & 0.38 &  [31.0, 41.0] &  & 1.4\\[-1.20mm]
$\rm \sigma(^{25}Mg+H\rightarrow^{15}$$\rm N)$ & 0.30 & {\bf0.32} & 0.33 &  [16.0, 18.0] &  & 1.3\\[-1.20mm]
$\rm \sigma(^{19}F+H\rightarrow^{15}$$\rm N)$ & 0.26 & {\bf0.30} & 0.35 &  [20.0, 26.0] &  & 1.3\\[-1.20mm]
$\rm \sigma(^{20}Ne+He\rightarrow^{14}$$\rm N)$ & 0.29 & {\bf0.29} & 0.30 &  [28.0, 30.0] &  & \\[-1.20mm]
$\rm \sigma(^{21}Ne+H\rightarrow^{15}$$\rm N)$ & 0.24 & {\bf0.29} & 0.34 &  [22.0, 32.0] &  & 1.2\\[-1.20mm]
$\rm \sigma(^{18}O+H\rightarrow^{15}$$\rm N)$ & 0.23 & {\bf0.28} & 0.34 &  [21.0, 30.0] &  & 1.2\\[-1.20mm]
$\rm \sigma(^{20}Ne+He\rightarrow^{15}$$\rm N)$ & & {\bf0.28} &  &  [27.0, 28.0] &  & 1.6\\[-1.20mm]
$\rm \sigma(^{15}N+He\rightarrow^{14}$$\rm N)$ & 0.27 & {\bf0.27} & 0.28 & 27.2 &  & \\[-1.20mm]
$\rm \sigma(^{24}Mg+He\rightarrow^{15}$$\rm N)$ & 0.24 & {\bf0.24} & 0.25 & 18.4 &  & 1.7\\[-1.20mm]
$\rm \sigma(\mathbf{^{23}Na+H\rightarrow^{15}}$${\rm\mathbf O})$ & 0.21 & {\bf0.24} & 0.27 &  [9.9, 12.0] & \cmark & {\bf n\!/\!a} \\
\hline\hline
 \end{tabular}
\end{table}
\endgroup

\begingroup
\squeezetable
\begin{table}[!th]
\caption{List of ghost nuclei with significant contributions to Li-C fluxes from Tables \ref{tab:xs_Li} to \ref{tab:xs_N}: the half-life, decay channel, and branching ratio are taken from \cite{2017ChPhC..41c0001A}.\label{tab:ghosts}}
\begin{tabular}{cll}
\hline\hline
Nucleus &$T_{1/2}$~~~~~~~~~& Daughter (decay mode)\\
\hline
$^{6}$He & 806.92 ms& $^{6}$Li ($\beta^-$, 100\%)  \\
$^{9}$Li & 178.3 ms & $^{9}$Be ($\beta^-$, 49.2\%, $^4$He ($\beta^-n$, 50.8\%) \\
$^{10}$C & 19.3009 s& $^{10}$B ($\beta^+$, 100\%)  \\
$^{11}$C & 20.364 m & $^{11}$B ($\beta^+$, 100\%)  \\
$^{12}$B & 20.20 ms & $^{12}$C ($\beta^-$, 98.4\%), $^4$He ($\beta^-3\alpha$, 1.6\%) \\
$^{13}$N & 9.965 m  & $^{13}$C ($\beta^+$, 100\%)  \\
$^{13}$O & 8.58 ms  & $^{13}$C ($\beta^+$, 89.1\%), $^{12}$C ($\beta^+p$, 10.9\%)\\
$^{14}$O & 70.620 s & $^{14}$N ($\beta^+$, 100\%)  \\
$^{15}$O & 122.24 s & $^{15}$N ($\beta^+$, 100\%)  \\
\hline\hline
\end{tabular}
\end{table}
\endgroup
\clearpage
\onecolumngrid
\section{Graphical view of $f_{aHc}$ coefficients\label{app:plots_fabc}}

We provide here a complementary view of the $f_{abc}$ coefficients (see App.~\ref{app:table_reactions}) for $b=$H, i.e. hydrogen target.

\begin{figure}[!h]
\centering
\def\figw{0.41}
\includegraphics[width=\figw\textwidth]{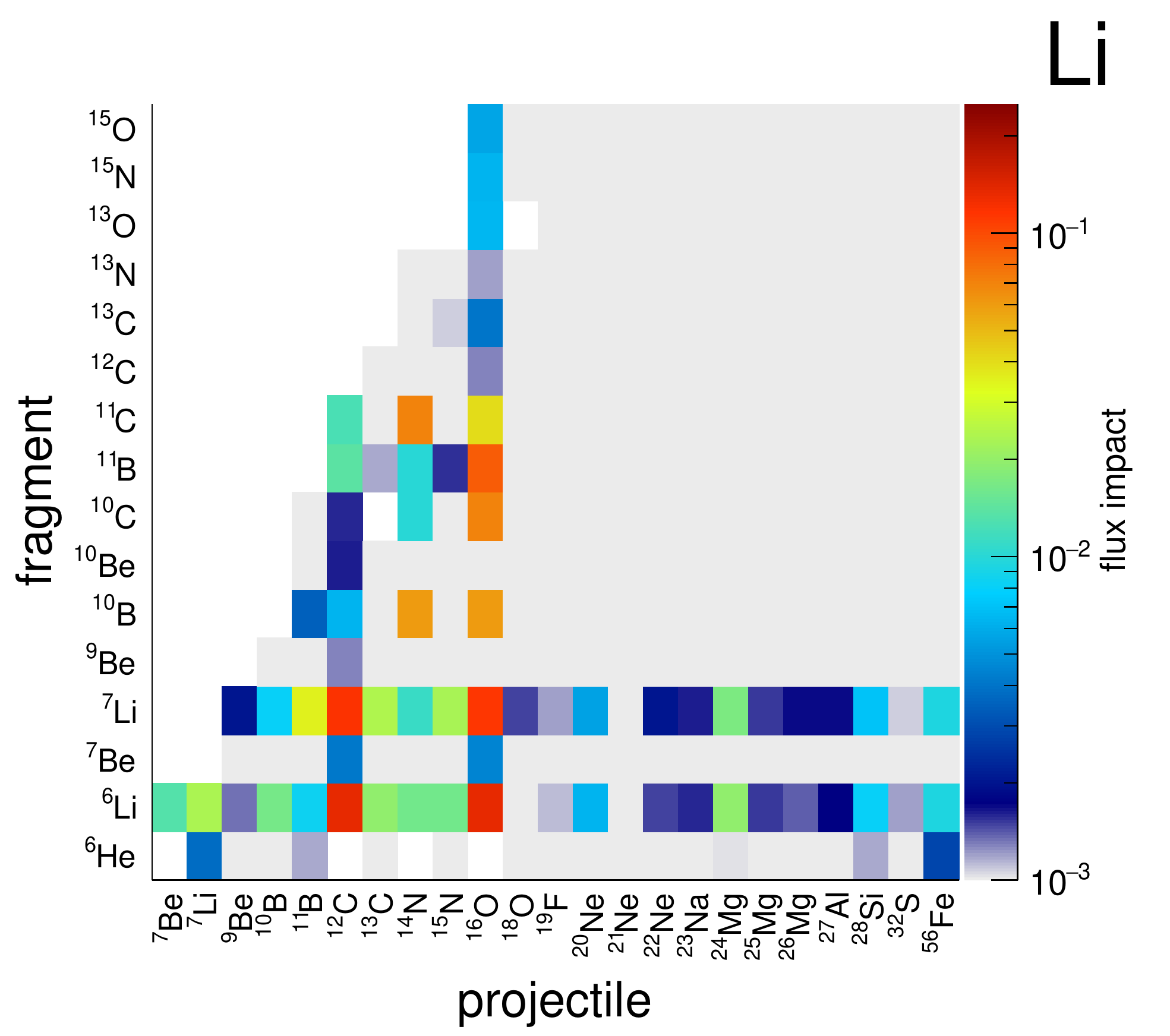}
\includegraphics[width=\figw\textwidth]{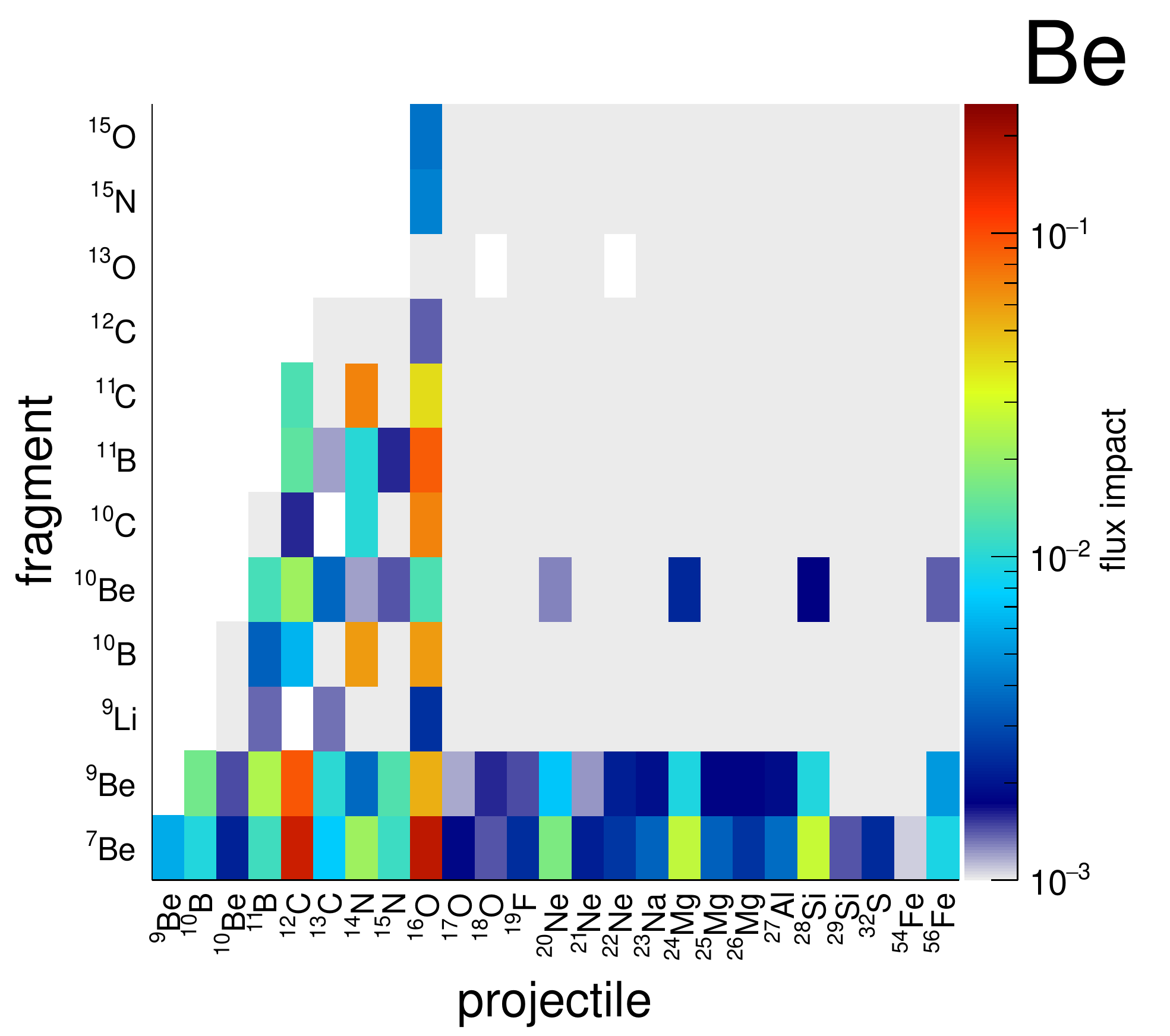}\\
\includegraphics[width=\figw\textwidth]{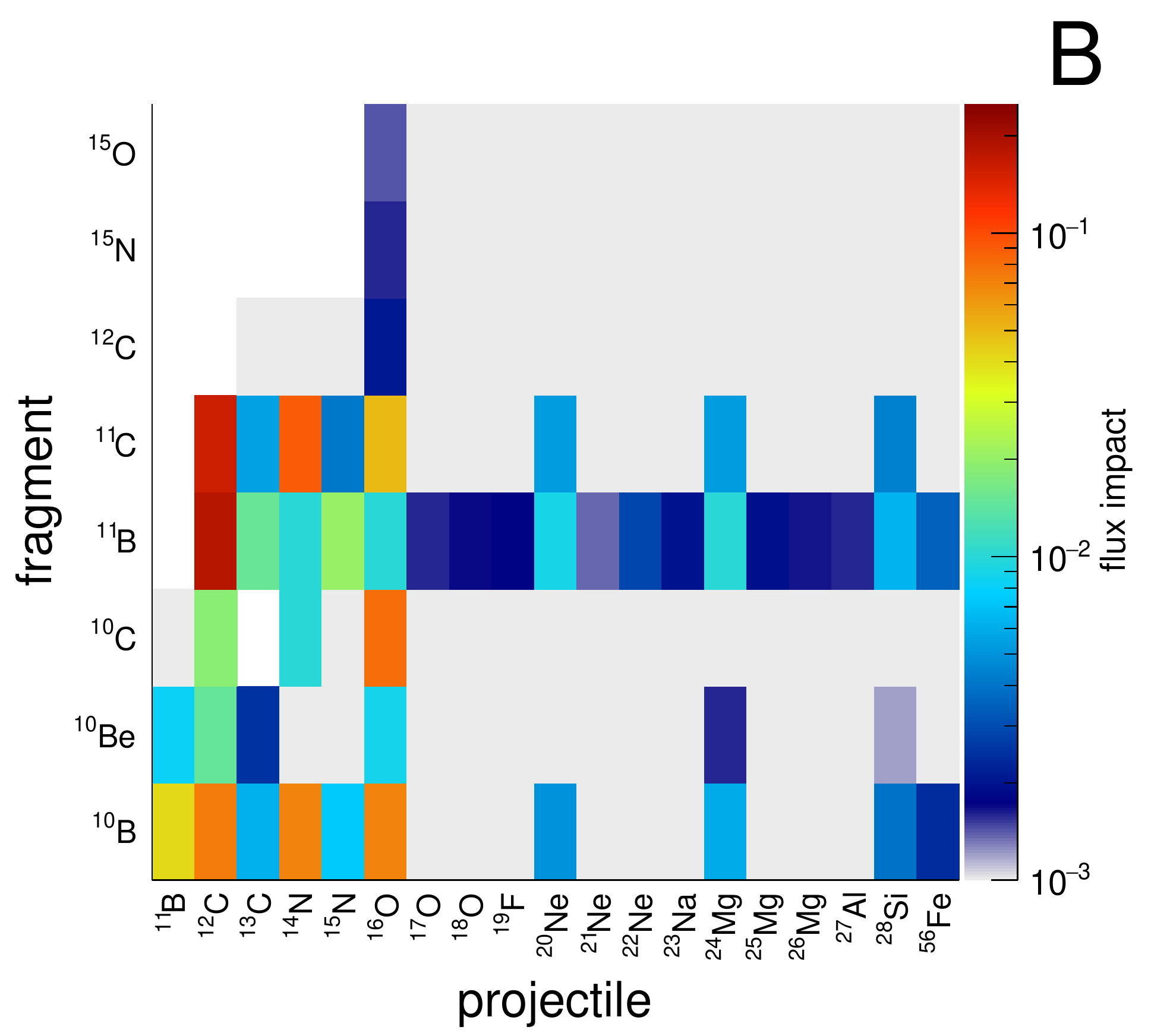}
\includegraphics[width=\figw\textwidth]{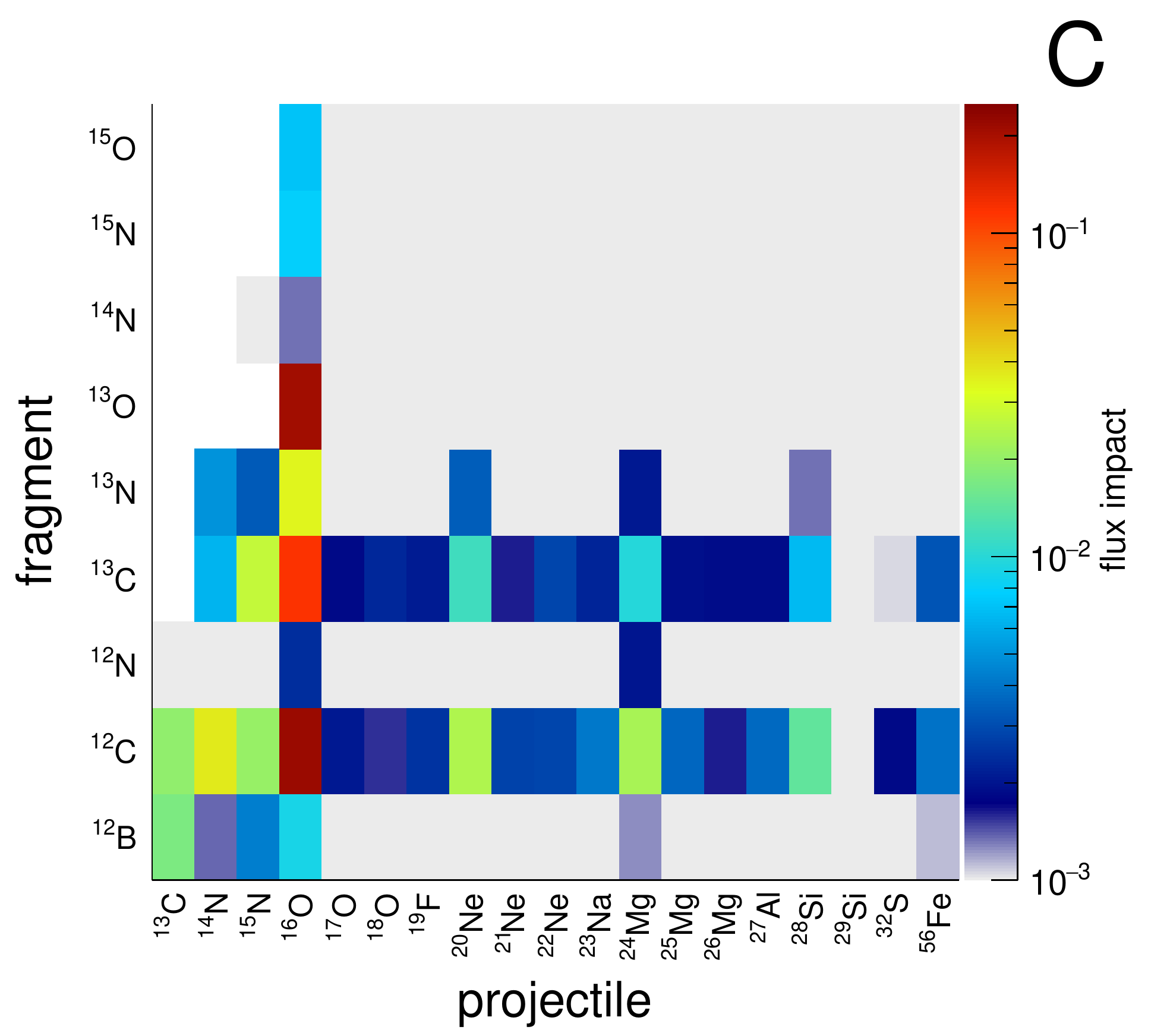}\\
\includegraphics[width=\figw\textwidth]{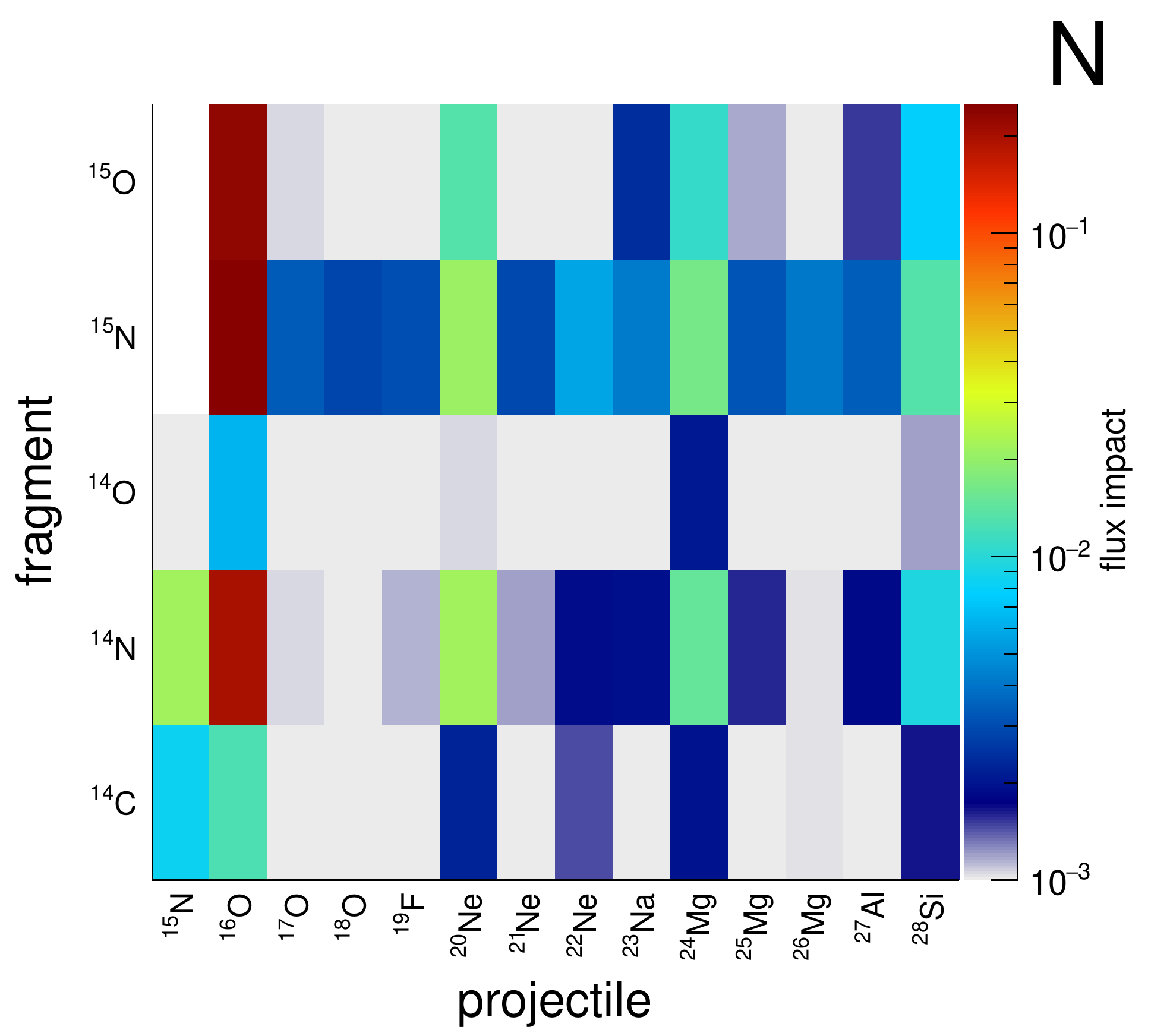}
\caption{Flux impact of reactions for Li, Be, B, C and N, as encoded in the $f_{a{\rm H}c}$ coefficients. The figure reads as follows: projectiles (ordinate) interacting on H lead to fragments (abscissa), whose impact on the flux is given by the colour scale from $10^{-3}$ to 0.25 as indicated on the right-hand side of each plot. Over- and underflows are set to the maximum and minimum colour scale respectively.\label{fig:fabc_LiBeBCN}}
\end{figure}

\section{Error evolution on fluxes at 10~GeV/n\label{app:plots_errevol}}

Here we summarize the error evolution plots for production of Li, Be, B, C, and N.

\begin{figure*}[!th]
\includegraphics[width=0.65\textwidth]{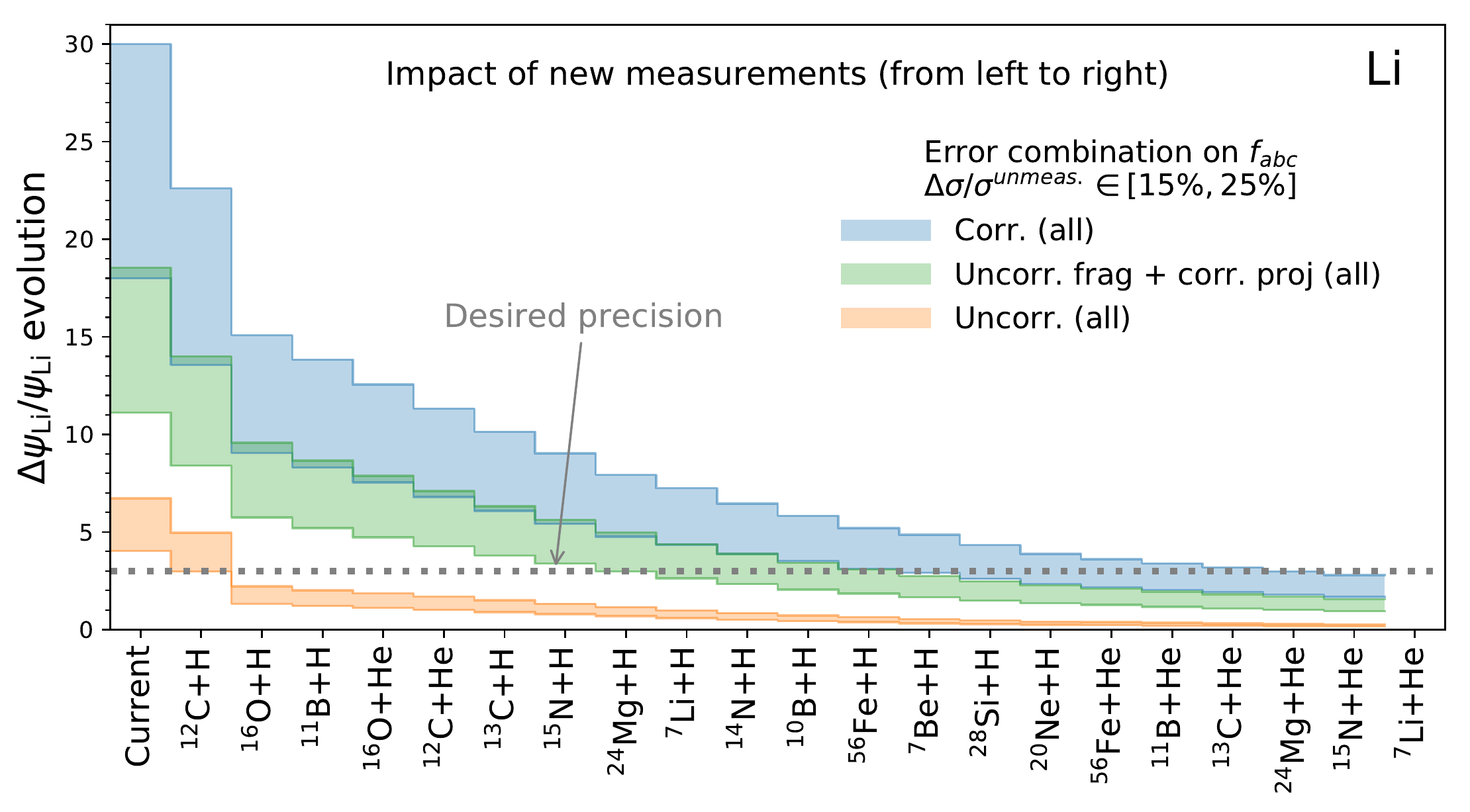}
\includegraphics[width=0.65\textwidth]{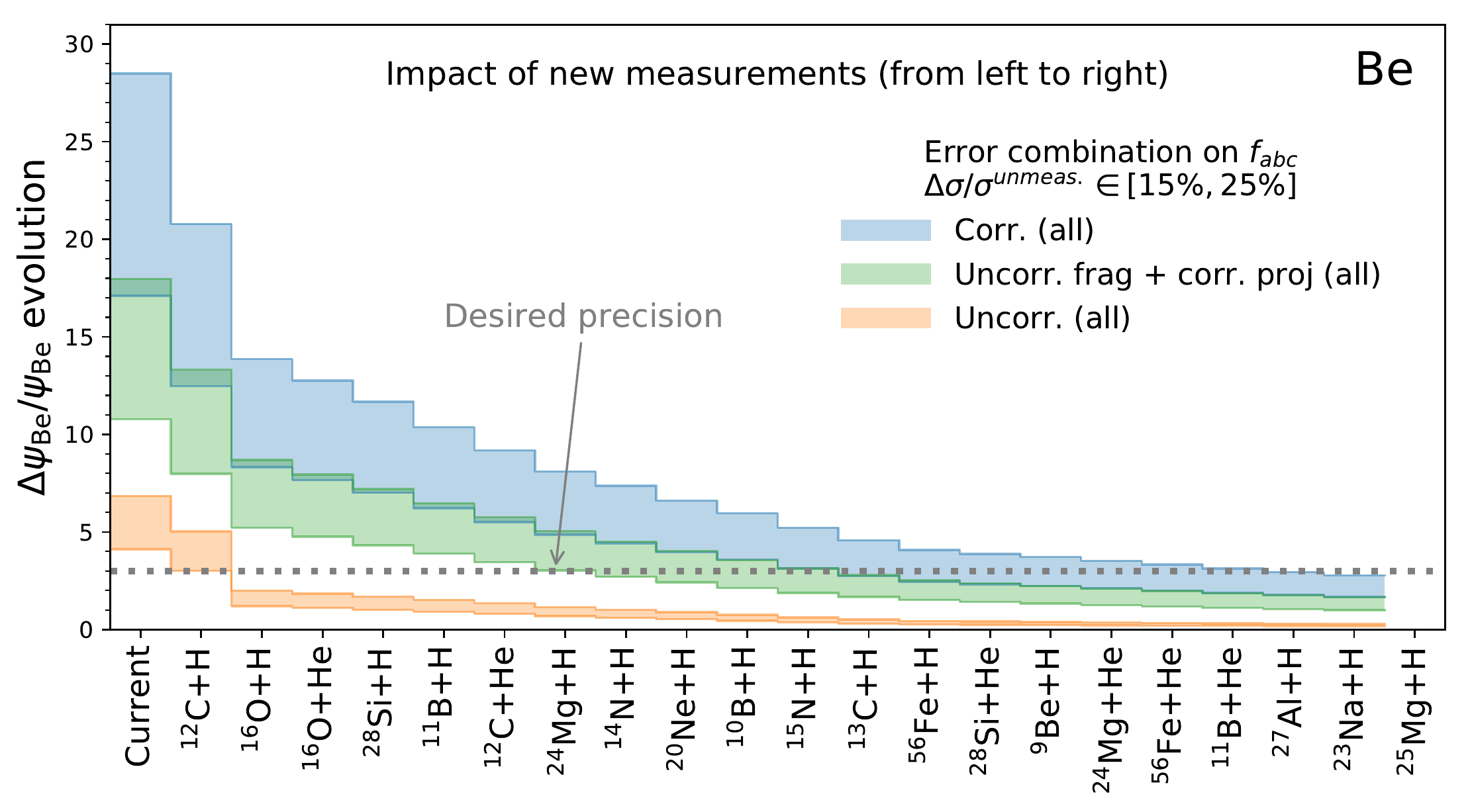}
\includegraphics[width=0.65\textwidth]{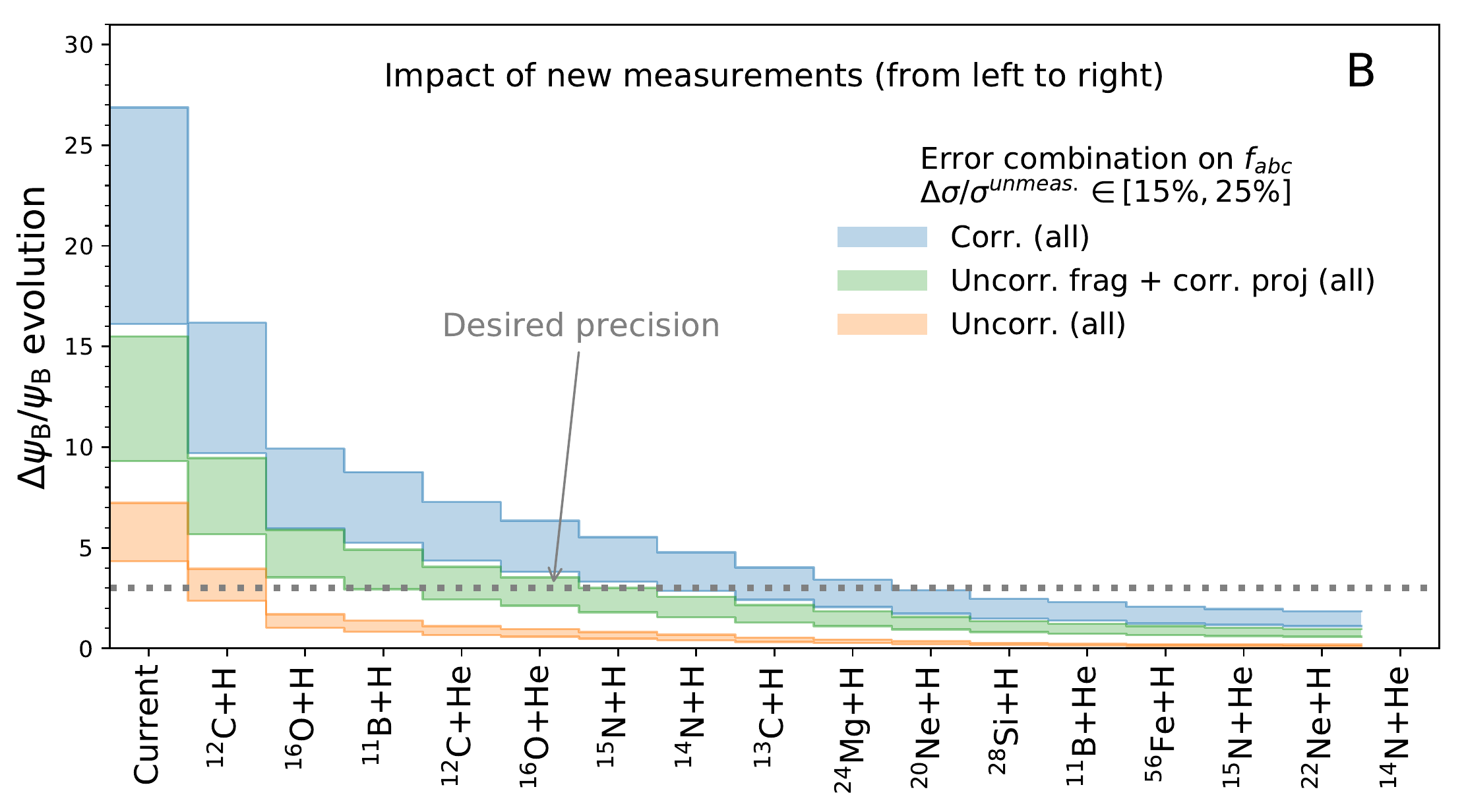}
\caption{Evolution of error on the calculated Li, Be, and B fluxes as if new reactions (from left to right) are measured with a perfect accuracy. The plot is read from left to right, with the first bin giving the currently estimated uncertainty on the flux (no new cross section measurement). The three sets of curves correspond to three different assumptions made on the cross-section errors, namely correlated Eq.~(\ref{eq:uncertainty_sumCorr}), uncorrelated Eq.~(\ref{eq:uncertainty_sumUncorr}), or a mixture of these two Eq.~(\ref{eq:uncertainty_sumMix}). The shaded areas is obtained by varying the assumption made on the current uncertainty on all cross section, $\Delta\sigma_r^{\rm current}$, between $15\%$ and $25\%$.\label{fig:err_evol_LiBeB}}
\end{figure*}

\begin{figure*}[!th]
\includegraphics[width=0.65\textwidth]{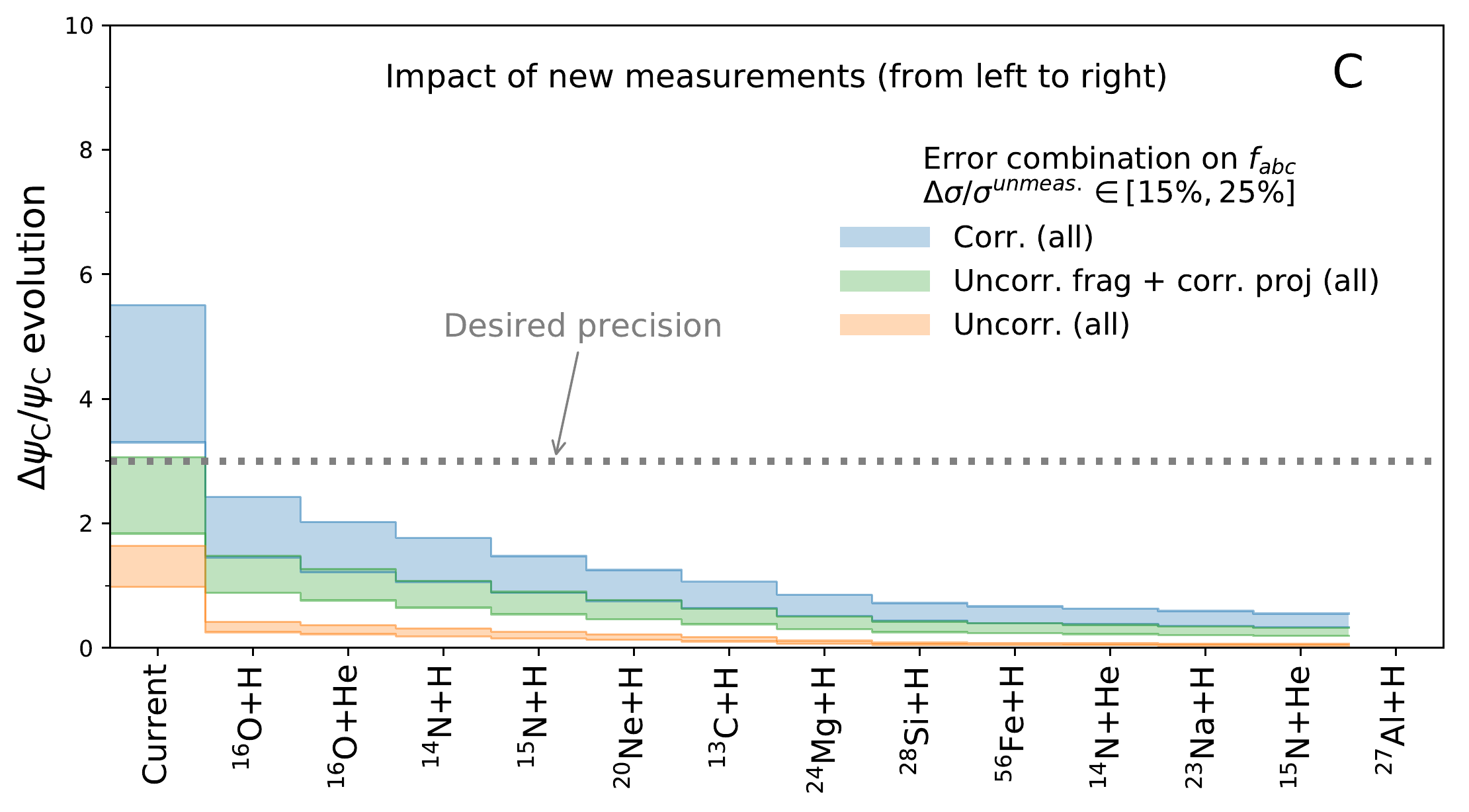}
\includegraphics[width=0.65\textwidth]{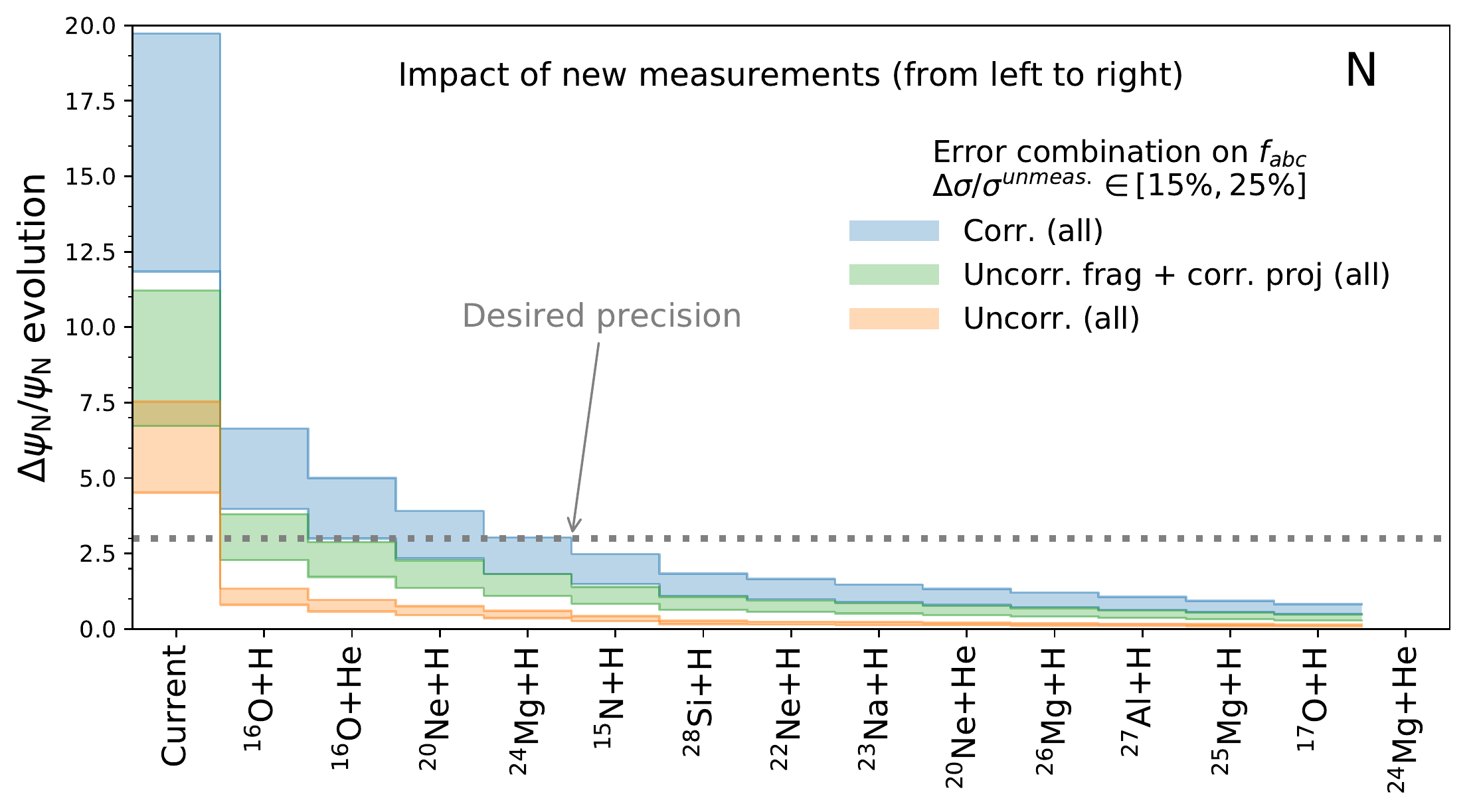}
\caption{Evolution of error on the calculated C and N fluxes as if new reactions (from left to right) are measured with a perfect accuracy. The plot is read from left to right, with the first bin giving the currently estimated uncertainty on the flux (no new cross section measurement). The three sets of curves correspond to three different assumptions made on the cross-section errors, namely correlated Eq.~(\ref{eq:uncertainty_sumCorr}), uncorrelated Eq.~(\ref{eq:uncertainty_sumUncorr}), or a mixture of these two Eq.~(\ref{eq:uncertainty_sumMix}). The shaded areas is obtained by varying the assumption made on the current uncertainty on all cross section, $\Delta\sigma_r^{\rm current}$, between $15\%$ and $25\%$.\label{fig:err_evol_CN}}
\end{figure*}

\onecolumngrid
\section{Plots for inelastic cross sections of Li, Be, B, and N isotopes in reactions with protons \label{app:plots_xs_inel}}

The total cross section is usually divided into elastic and inelastic parts. Elastic scattering keeps both projectile and target nuclei intact and thus does not influence the CR propagation. The inelastic cross section can be subdivided into the quasi-elastic and the production cross sections. The former leads only to the breakup of the projectile nucleus, whereas in the latter case also new particles are produced (pions, kaons, etc).

\begin{figure*}[!ht]
\includegraphics[width=0.495\textwidth]{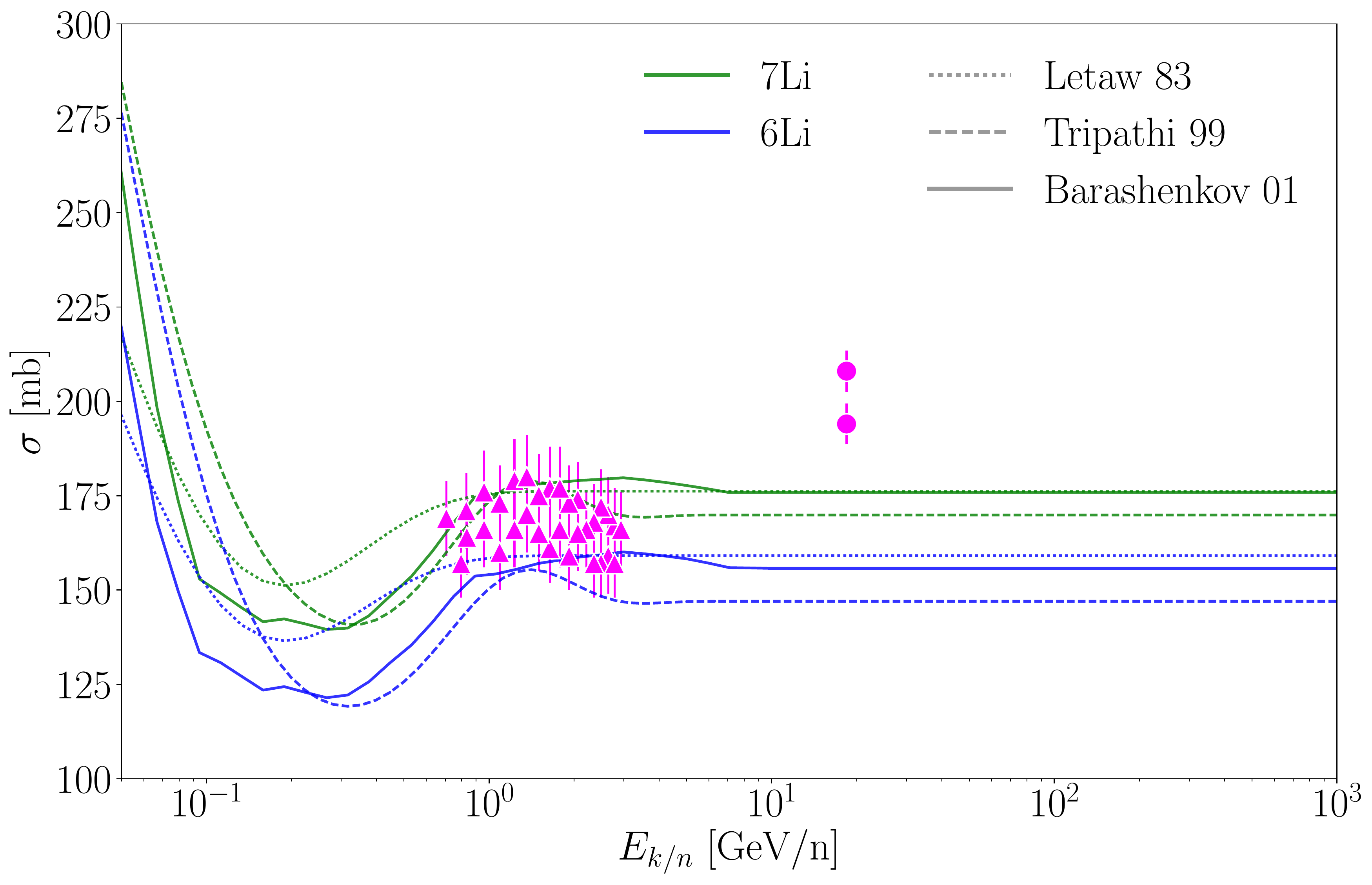}
\includegraphics[width=0.495\textwidth]{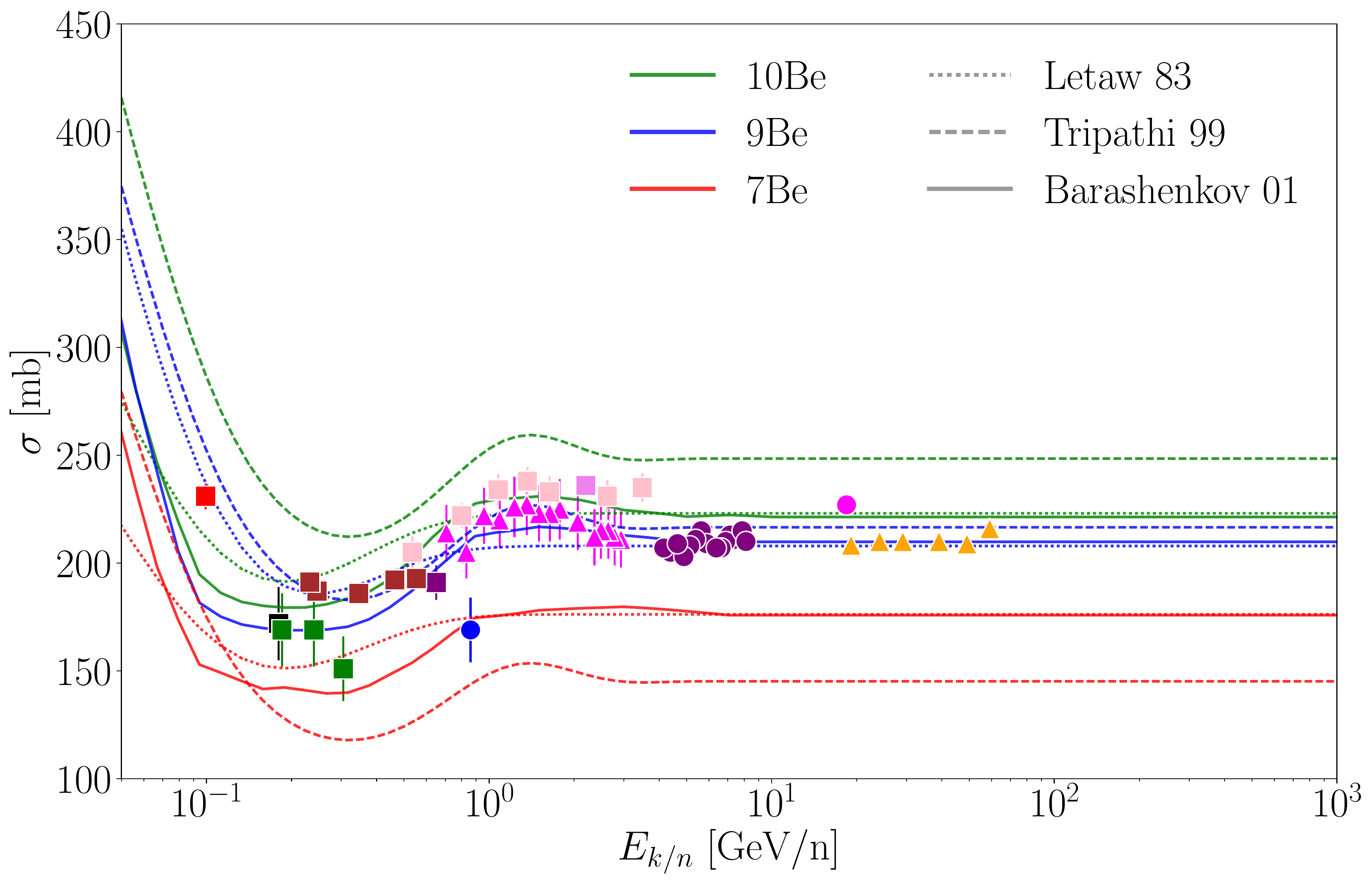}
\includegraphics[width=0.495\textwidth]{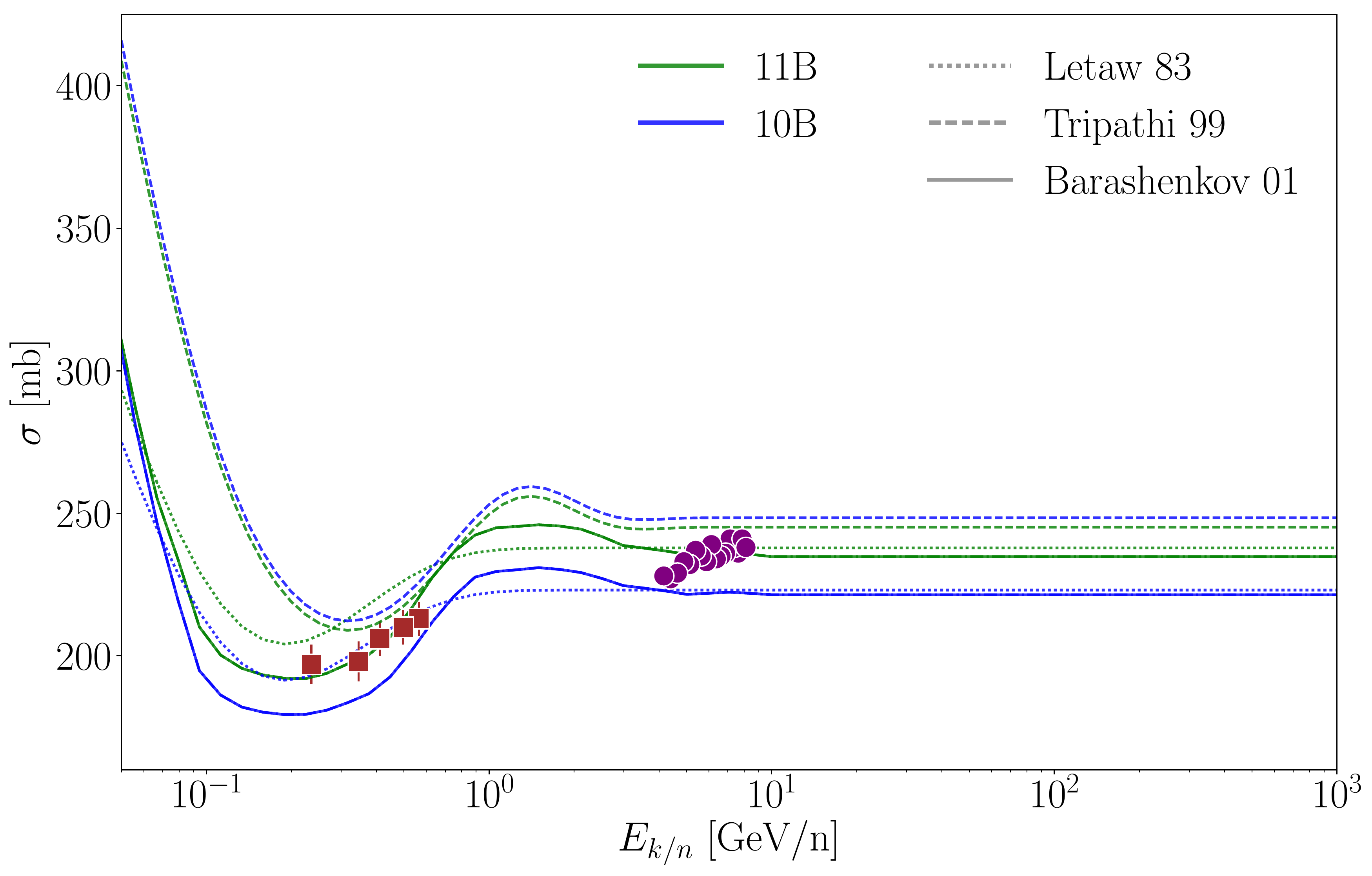}
\includegraphics[width=0.495\textwidth]{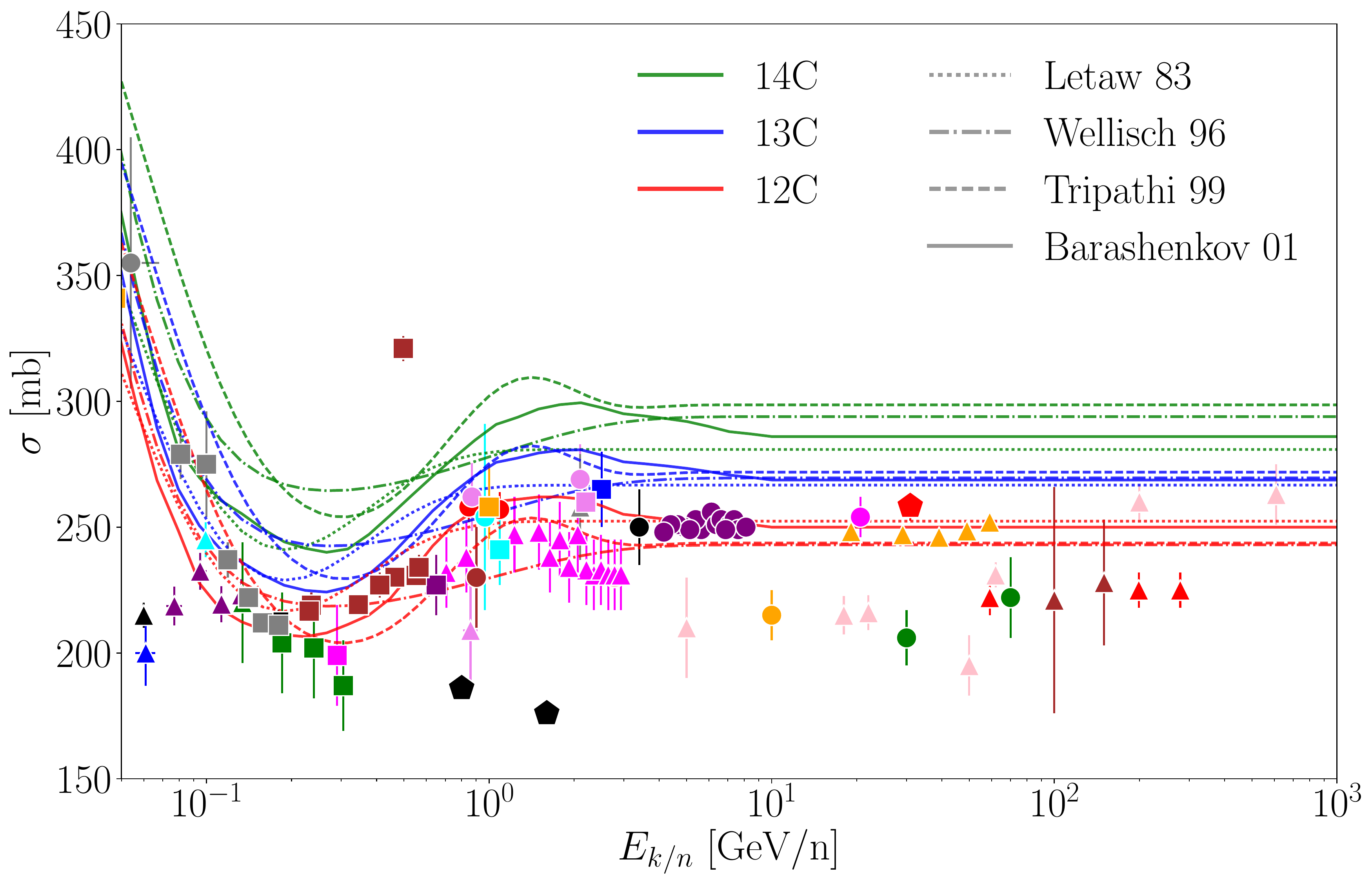}
\includegraphics[width=0.495\textwidth]{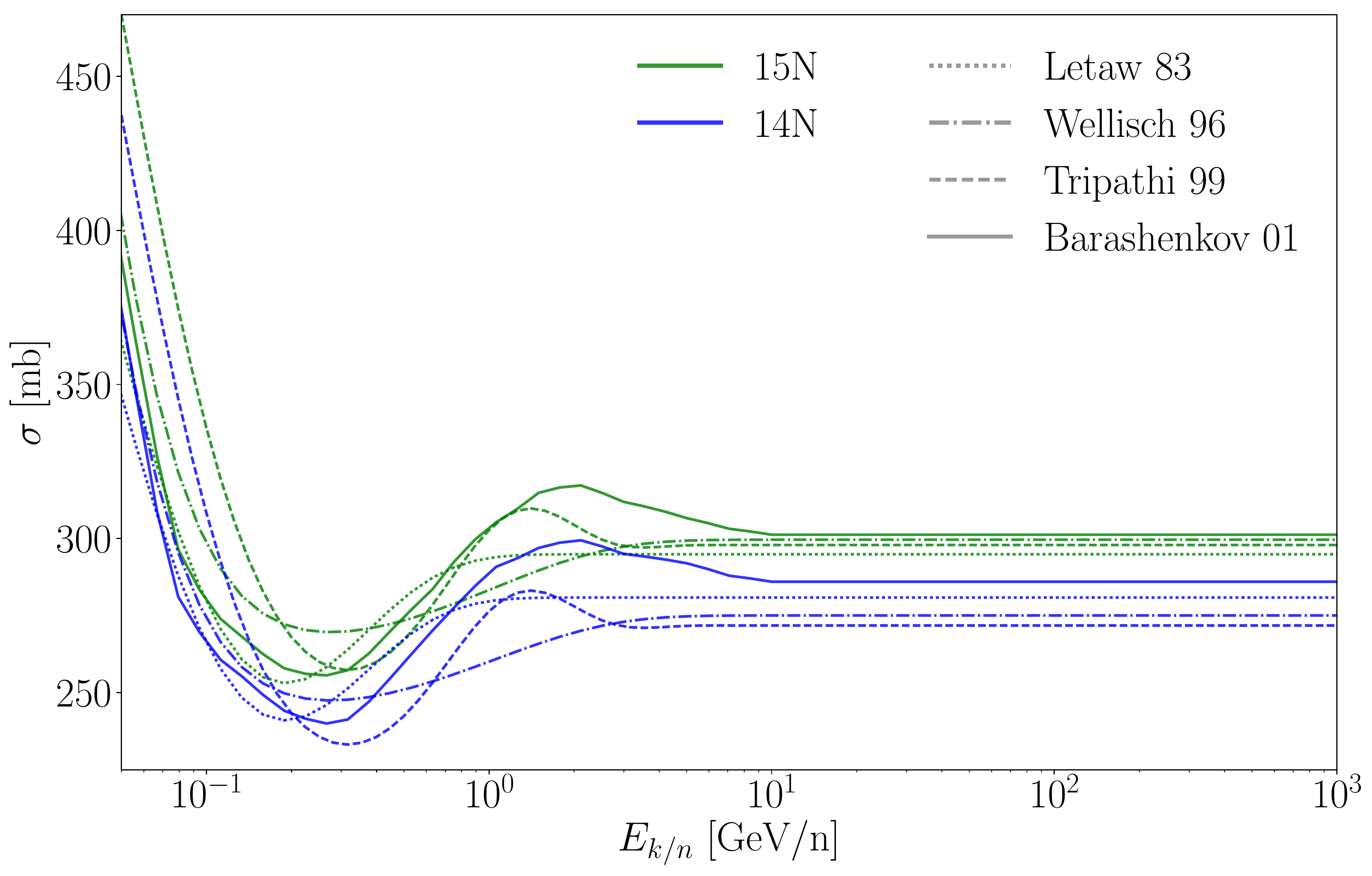}
\includegraphics[width=0.495\textwidth]{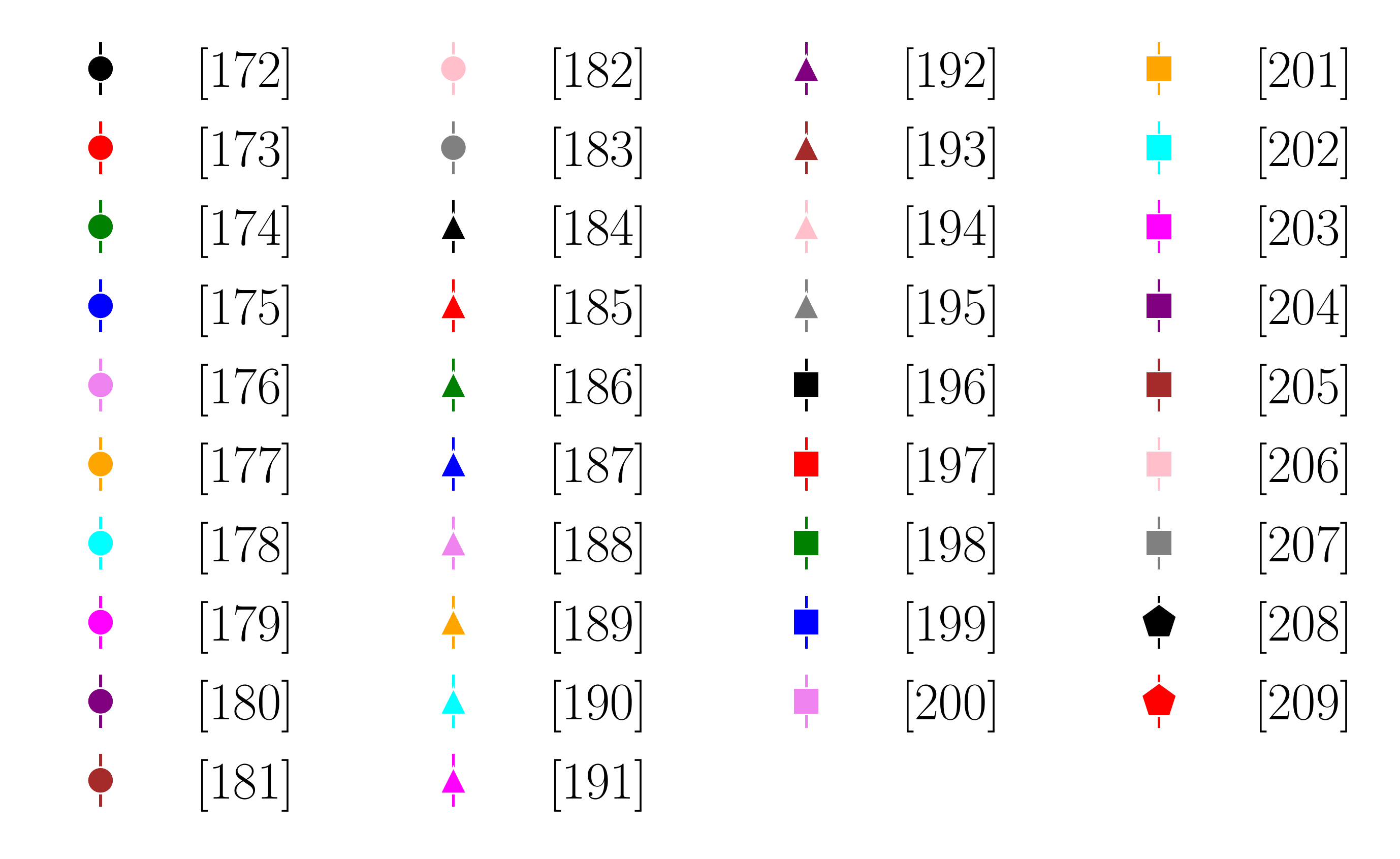}
\caption{Inelastic cross-sections of Li, Be, B, C, N isotopes in reactions with protons. For the lithium cross-section, both $^{6}$Li and $^{7}$Li are shown. For Be, B and C cases only data for isotopes $^{9}$Be, $^{12}$C and natural samples are available. These data are extracted from the information presented in \cite{bobchenko1979measurement}, which was stacked into tables by S.~Mashnik. Further data are from the EXFOR database \cite{OTUKA2014272}. Full tables including lower energy data points are available on request.
\label{fig:xs_inel_N}}
\end{figure*}

The inelastic cross sections of CR nuclei are not frequently discussed in papers dedicated to the accuracy of the nuclear data. CR transport calculations are not very sensitive to their exact value, because: first, a typical fragmentation rate is smaller than other rates, such as escape (diffusion) rate and ionization energy losses; second, errors in the fragmentation rate can be compensated by the adjustments in the source abundances of the corresponding species. If the source abundances are fixed, then the uncertainties in the total inelastic cross sections from the models showed in Fig.~\ref{fig:xs_inel_N}, translate into a maximal uncertainty of $\sim$10\% for the B/C ratio at 1~GeV/n (not shown). Since the accuracy of astrophysical measurements nowadays exceeds the accuracy of the cross section calculations, we feel it is necessary to access the accuracy of all relevant cross sections including the total inelastic cross sections.

Figure \ref{fig:xs_inel_N} shows total inelastic cross sections of isotopes of Li, Be, B, C, and N in reactions with protons. We use a collection of data from \cite{bobchenko1979measurement}, which was assembled into tables by S.~Mashnik\footnote{\url{http://www.oecd-nea.org/dbdata/bara.html}}, and the EXFOR\footnote{\url{https://www-nds.iaea.org/exfor/exfor.htm}} database \cite{2018arXiv180205714Z, OTUKA2014272}. The data are plotted together with parametrizations proposed in \cite{1983ApJS...51..271L,1999NIMPB.155..349T,BarPol1994,1996PhRvC..54.1329W}. Note that typos in the published formulas \cite{1996PhRvC..54.1329W} were corrected (Wellisch and Axen, private communication).

Parametrizations proposed in \cite{1983ApJS...51..271L,BarPol1994} are consistent above 2--3 GeV/n dependently on the species. Parametrization by \cite{1999NIMPB.155..349T} is clearly off from \cite{1983ApJS...51..271L,BarPol1994} by $\sim$10\% (Li) or $\sim$10--25\% ($^{10}$Be, $^7$Be) in a non-systematic way, while agrees well with \cite{1983ApJS...51..271L,BarPol1994} for $^9$Be, and C and N isotopes. Parametrization proposed by \cite{1996PhRvC..54.1329W} is valid for $Z>5$, so we can compare it only with predictions for C and N, where it agrees well with other parametrizations above a few GeV/n. Below 2--3 GeV/n all parametrizations demonstrate different behaviour with significant scattering between their predictions that can be as large as 50-100\% at $\sim$100 MeV/n, see, e.g., predictions by \cite{BarPol1994} and \cite{1999NIMPB.155..349T} for N. Especially disappointing is the absence of data in the energy range below $\sim$1 GeV/n, where the measurements of isotopic abundances of CR species are only available. Tuning propagated CR abundances to the data in this energy range could lead to significant errors at higher energies if incorrect cross sections are used.

\section{Plots for production cross sections of the most important reactions with protons \label{app:plots_xs_prod}}

In this Appendix we show the plots of the most important reaction channels discussed in the paper. For each shown reaction, all existing data above $\sim$100 MeV/n are plotted together with available parametrizations. Only reactions with hydrogen target are presented. Measurements with He beam are significantly more complicated and the data are scarce, so we do not include them.

We display, sorted by growing projectile atomic number $Z$, the cross-sections for which measurements are available and whose flux impact is $>0.01\%$ in either Li, Be, B, or C (flux impact $>1\%$ are highlighted in boldface)\footnote{The list of all reactions regardless of their flux impact value is provided as Supplementary Material.}. This means that plots of some potentially important channels (i.e.\ with a large flux impact) are not shown if such data points do not exist or are missing in our database: the list of important channels with missing data is indicated in the next-to-last column of Tables \ref{tab:xs_Li} to \ref{tab:xs_N} in appendix \ref{app:table_reactions}. This review is based on the GALPROP cross-section data base assembled in the file {\tt isotope\_cs.dat} supplemented by some other references pointed out in \cite{2018JCAP...01..055R}. The references and some details of this data base are provided in Table \ref{tab:ref}. See also \cite{2012RadM...47..315N} for an independent attempt at listing all reactions ever measured in the MeV/n-GeV/n energy range.

\begin{center}
\begin{table}[!ht]
\caption{References of the legend of the fragmentation cross-sections plots (GALPROP and \cite{2018JCAP...01..055R}).\label{tab:ref}}
\begin{tabular}[c]{c c p{8cm}}
\hline
\hline
Ref. in plot & Reference & Comments\\
\hline
 {[Ep69]}  &    \cite{epherre1969comparison} & Most of the data also in {[RV84]} except some ghost nuclei \\ 
 {[LM69]}  &    \cite{1969PhRv..177.1548L}   & \\
 {[Ni72]}  &    \cite{1972NuPhA.181..329N}   & \\
 {[Fo77]}  &    \cite{fontes1977b}           & \\ 
 {[Ra79]}  &    \cite{1979PhRvC..20..787R}   & \\
 {[Ol83]}  &    \cite{1983PhRvC..28.1602O}   & \\
 {[Gla93]} &    \cite{1993ZPhyC..60..421G}   & \\
 {[Ab94]}  &    \cite{1994NuPhA.569..753A}   &  \\
 {[ST98]}  &    \cite{1998ApJ...501..911S}   & \\
 {[Ko99]}  &    \cite{1999ICRC....4..267K}   & \\
 {[Bli01]} &    \cite{2001PAN....64..907B}   & \\
 {[Ko02]}  &    \cite{korejwo2002isotopic}   & \\[0.3cm]
&\multicolumn{2}{l}{\em Compilations and data bases}\\
 {[RV84]}   &   \cite{1984ADNDT..31..359R}   & Certain cross sections in this compilation do not have associated error bars. In these cases it is assumed that the relative error is 10\% for those cross sections whose value $>$10 mb, 20\% for cross sections $<$10 mb, and 30\% for cross sections $<$1 mb. \\
 {TOBV}   &            -     & Target is natural Si (old database, no reference) \\
 {NUCLEX} &  \cite{NUCLEX}   &   \\[0.3cm]

&\multicolumn{2}{l}{\em Michel and Leya's group}\\
 {[Mi95]}  & \cite{1995NIMPB.103..183M}  &  \\
 {[cMi95-}cem] & \cite{1995NIMPB.103..183M}  & cMi95-cem are cumulative cross section data with ghost nuclei subtracted using CEM code \cite{2004AdSpR..34.1288M} \\
 {[Sc96]}  & \cite{1996NIMPB.114...91S}    &  \\[0.3cm]

&\multicolumn{2}{l}{\em Webber and/or the Transport collaboration}\\
 {[We90]}  & \cite{doi:10.1063/1.39166} &  \\
 {[W90]r}  & \cite{1990PhRvC..41..547W} & Relative error as given in {[We98]} \\
 {[W90]i}  & \cite{1990PhRvC..41..547W} & Relative error $\rm A=5\%$, $\rm B=10\%$, $\rm C=20\%$, $\rm D=30\%$\\
 {[We96]}  & \cite{1997ApJ...488..730R} &
 These data were provided from Webber (private comm.). Relative errors assumed: 10\% for those cross sections whose value $>$10 mb, 20\% for cross sections $<$10 mb. \\
{[Ch97a]}  & \cite{1997ApJ...479..504C} &  \\
{[Kn97]}   & \cite{1997PhRvC..56..398K} &  \\
{[Ch97b]}  & \cite{1997PhRvC..56.1536C} & \\
 {[We98]}  & \cite{1998ApJ...508..949W} &
 The errors given in the paper are significantly underestimated. Relative errors adopted: $\rm B =5\%$; $\rm C =10\%$; $\rm D =20\%$; $\rm E =30\%$. \\
 {[We98prc]} &  \cite{1998PhRvC..58.3539W} &
Tables 7,12-15. For Table 7, adopted relative errors are: 10\% for those cross sections whose value $>$10 mb, 20\% for cross sections $<$10 mb.  In other Tables, relative errors adopted: $\rm A =3\%$; $\rm B =4\%$; $\rm C =7\%$; $\rm D =10\%$; $\rm E =18\%$; $\rm F =26\%$.\\
\hline
\end{tabular}
\end{table}
\end{center}

Together with the data we also draw the benchmark parametrizations which were used in the computation of Tables \ref{tab:channels_Li}--\ref{tab:xs_N}, and are described in more details in the paper. Note that the widely used parametrization of W03 \cite{2003ApJS..144..153W} includes the contributions of ghost nuclei: in order to compare this parametrization with the data, we subtract them, based on the ghost contributions calculated from W98 parametrization \cite{1998PhRvC..58.3539W}.
Both the original (with ghosts included, denoted W03c for cumulative) and rescaled (without ghosts, denoted W03*) values are shown. The proportion of the ghost nuclei contribution to the cumulative cross section for a particular reaction is an important quantity, but to our knowledge, it has never been reported explicitly. The last column in Tables \ref{tab:xs_Li} to \ref{tab:xs_N} in Appendix \ref{app:table_reactions} shows the ratio of the cumulative (see Eq.~\ref{eq:ghosts}) to the direct production cross sections, for reactions for which $\sigma^{\rm c}/\sigma>1.05$. This number can be $\gtrsim$1 for some important reactions, emphasizing even further the need for their precise measurements.

\clearpage
\begin{longtable}{@{}ccc@{}}
 \\ 
 \\ [3pt] 
\multicolumn{3}{c}{\bf Z=3{ \bf projectiles: $^{x}$Li + H $\rightarrow$ $^{A}_ZX$}}\\ [3pt]
\multicolumn{3}{c}{\noindent\makebox[\linewidth]{\rule{\textwidth}{0.4pt}}}\\ [3pt]
\includegraphics[width=0.32\textwidth]{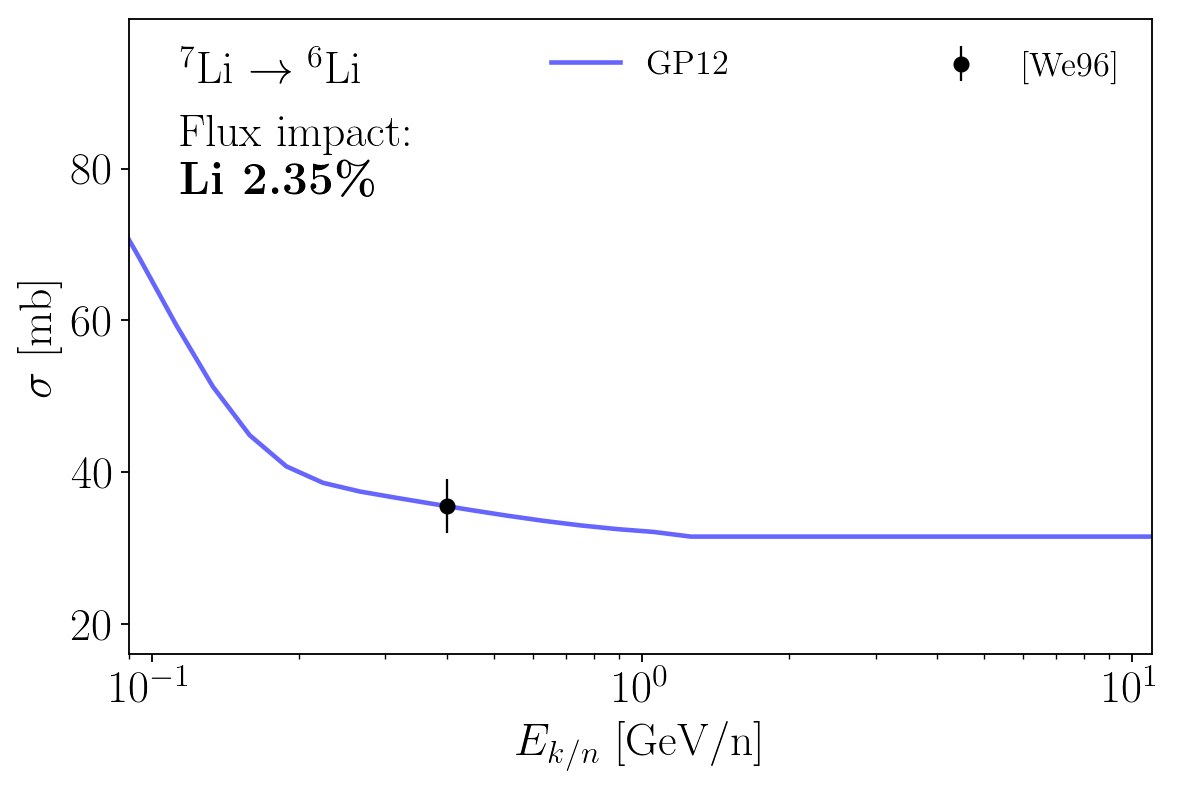}  &  &  \\ [3pt] 
\multicolumn{3}{c}{\bf Z=4{ \bf projectiles: $^{x}$Be + H $\rightarrow$ $^{A}_ZX$}}\\ [3pt]
\multicolumn{3}{c}{\noindent\makebox[\linewidth]{\rule{\textwidth}{0.4pt}}}\\ [3pt]
\includegraphics[width=0.32\textwidth]{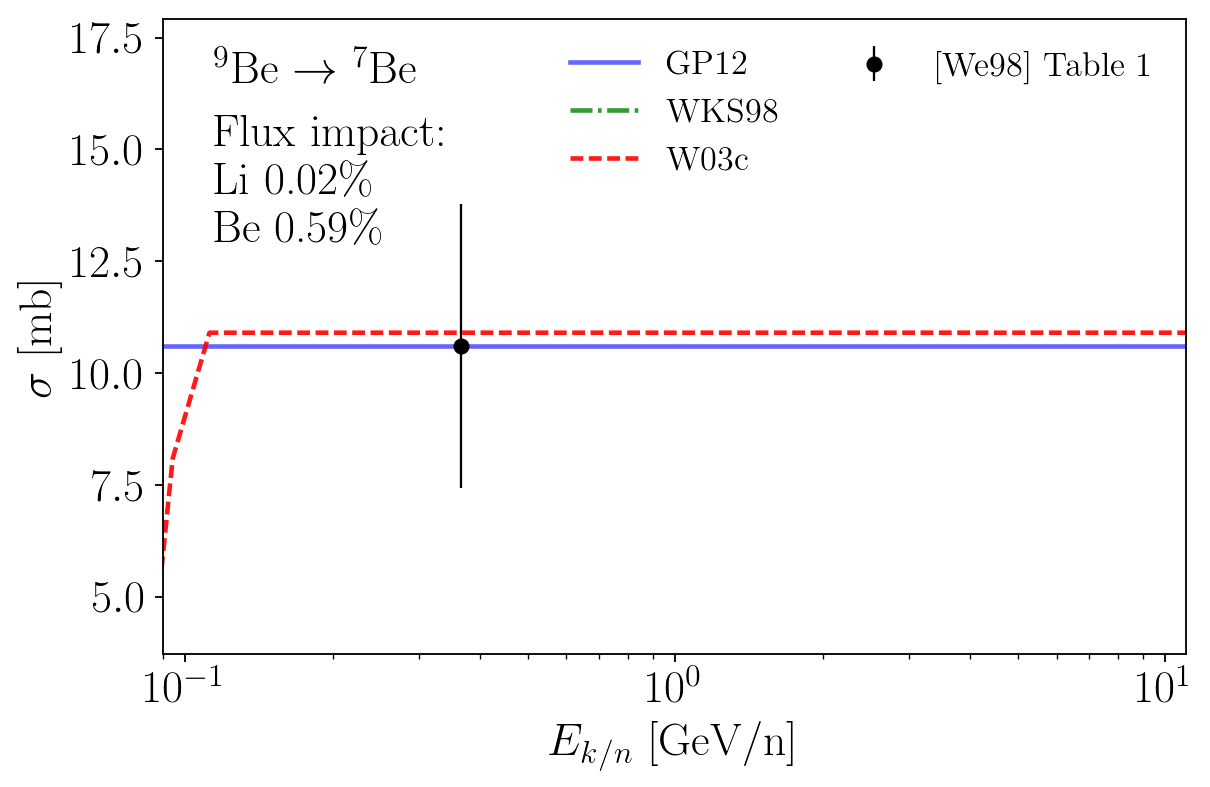} & & \\ 
 \\ [3pt] 
\multicolumn{3}{c}{\bf Z=5{ \bf projectiles: $^{x}$B + H $\rightarrow$ $^{A}_ZX$}}\\ [3pt]
\multicolumn{3}{c}{\noindent\makebox[\linewidth]{\rule{\textwidth}{0.4pt}}}\\ [3pt]
\includegraphics[width=0.32\textwidth]{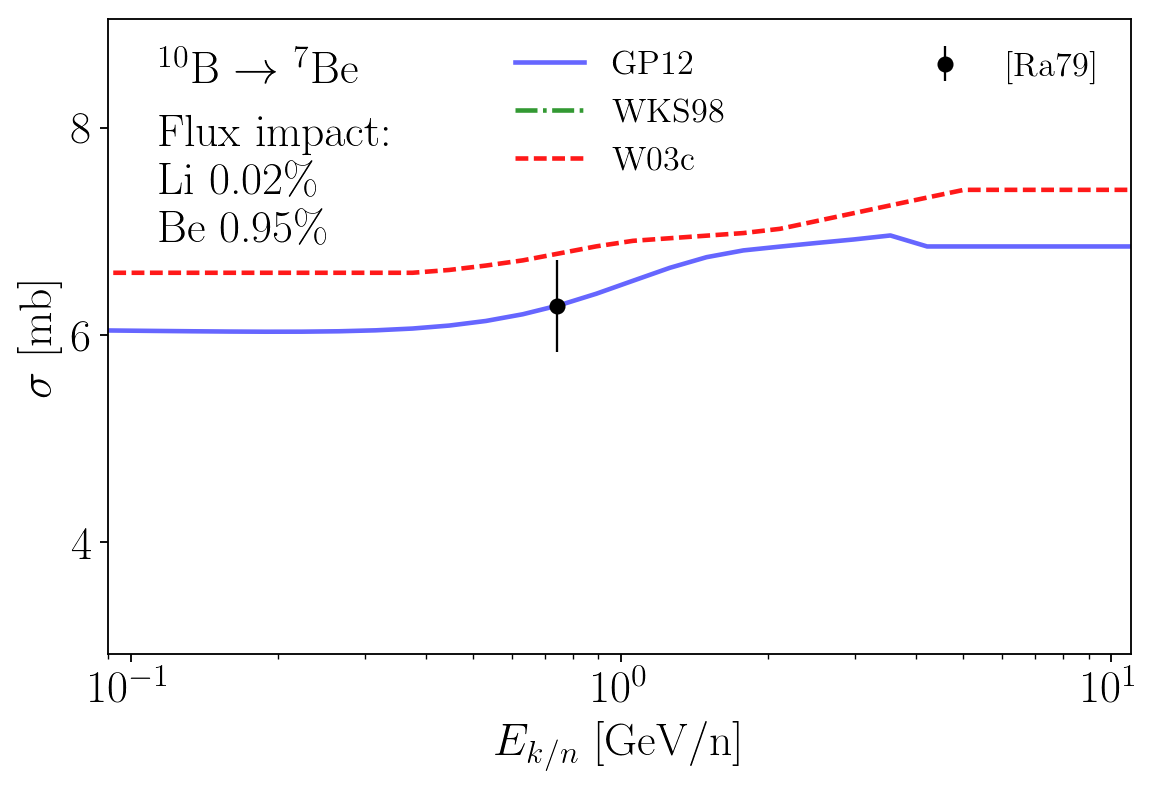}  &  
\includegraphics[width=0.32\textwidth]{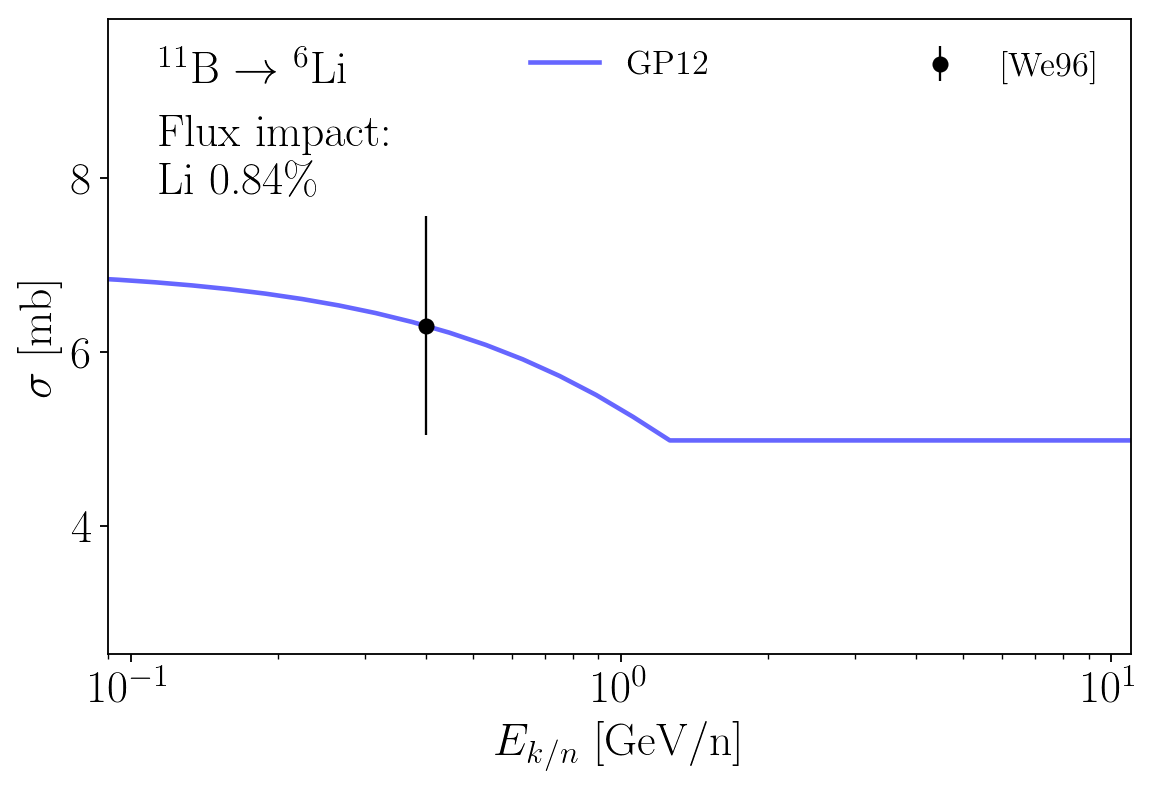}  &  
\includegraphics[width=0.32\textwidth]{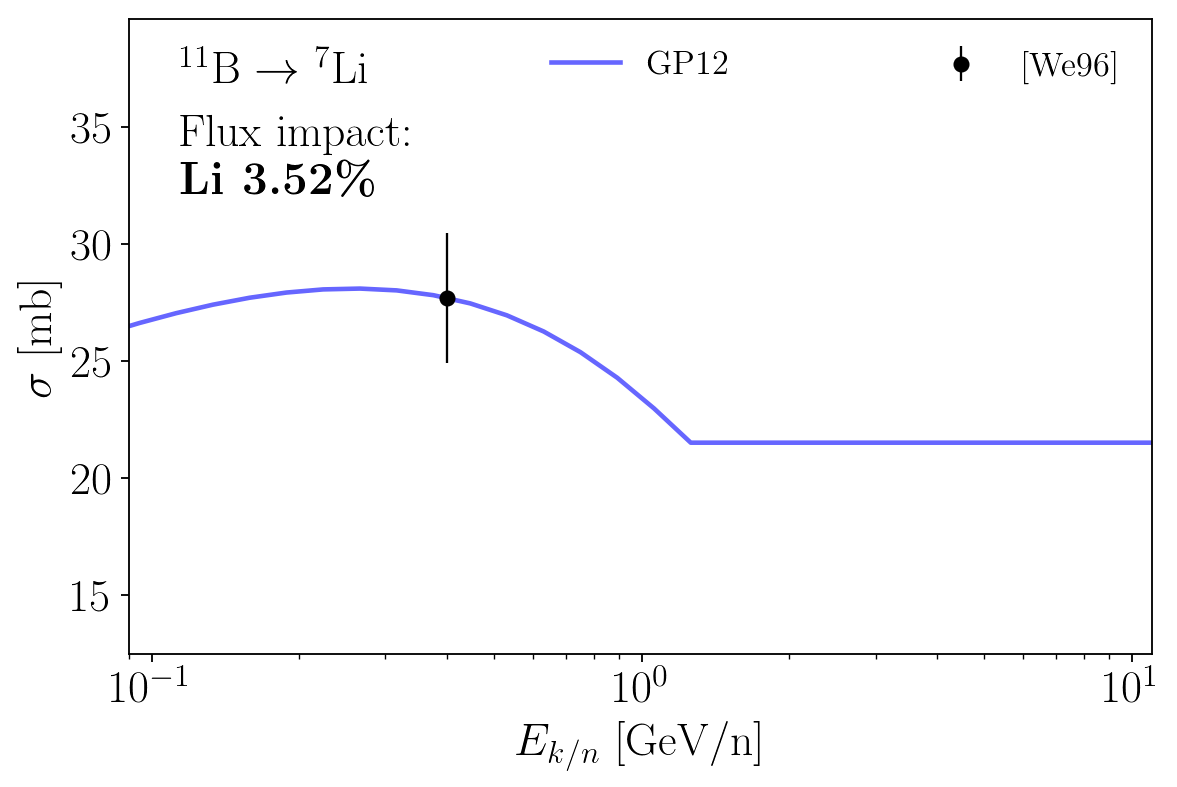}  \\ 
\includegraphics[width=0.32\textwidth]{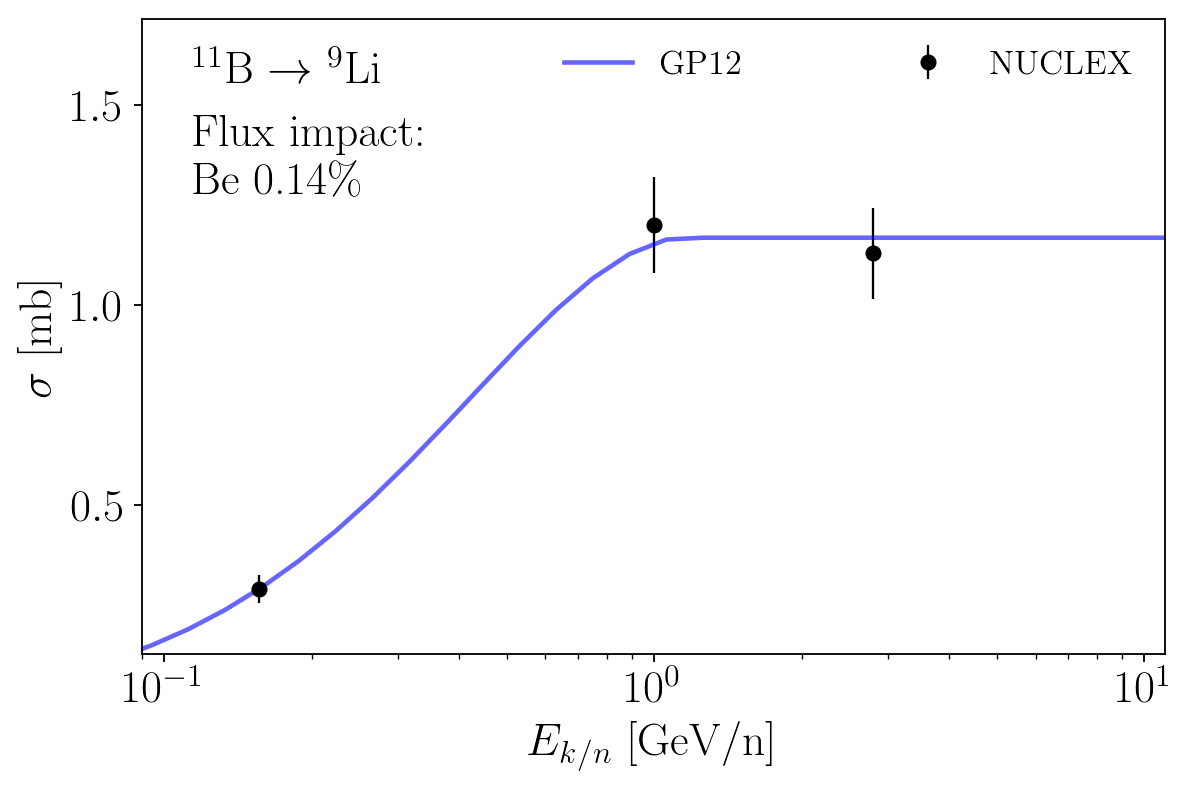}  &  
\includegraphics[width=0.32\textwidth]{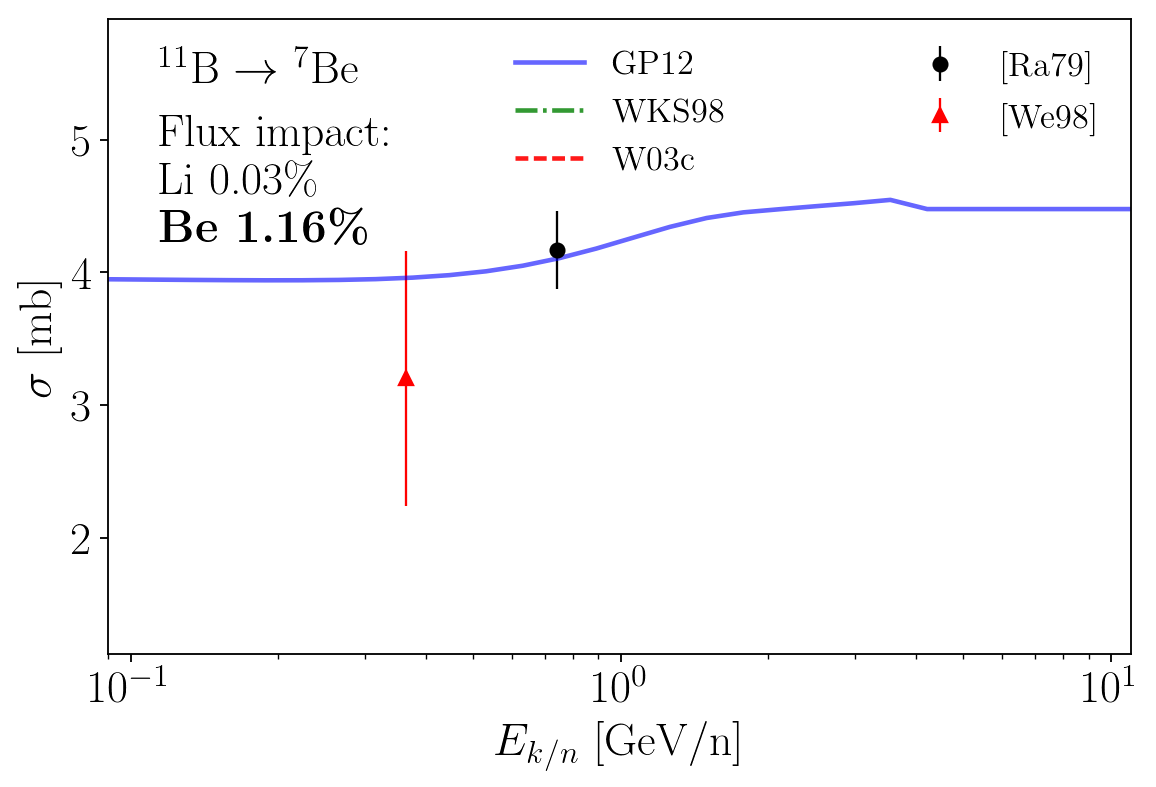}  &  
\includegraphics[width=0.32\textwidth]{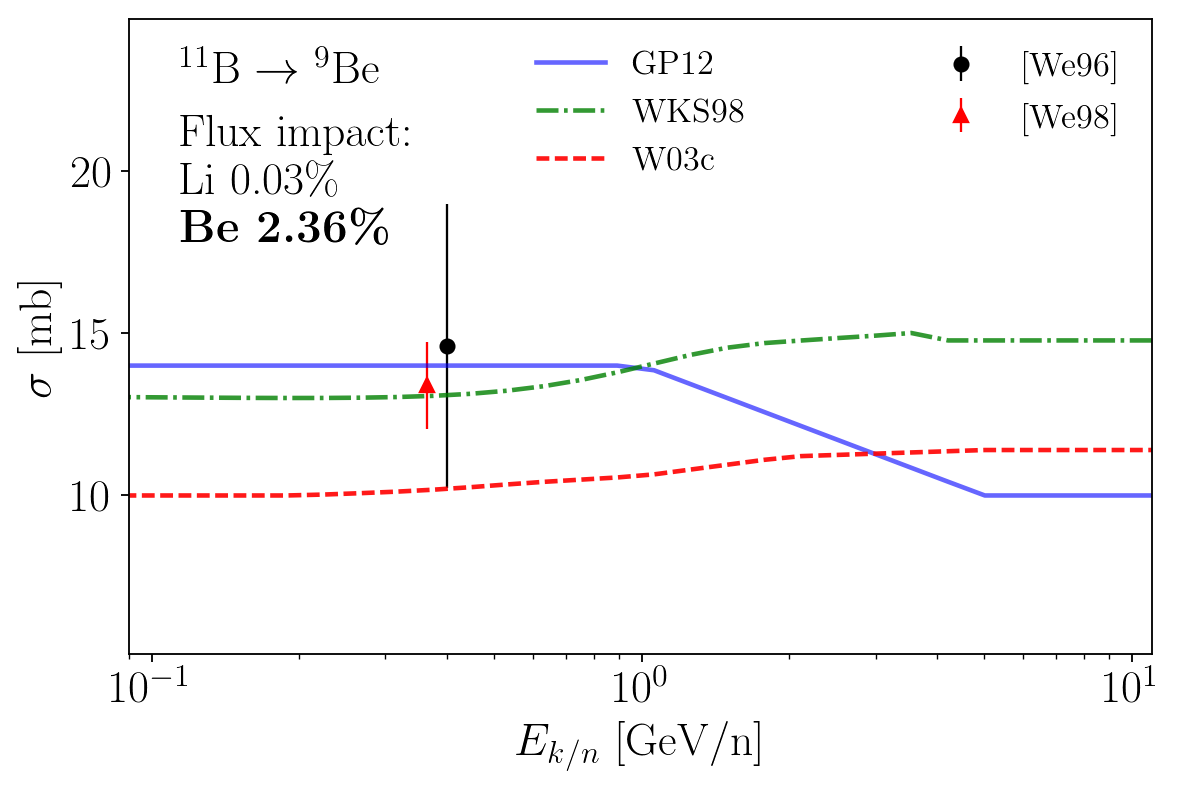}  \\ 
\includegraphics[width=0.32\textwidth]{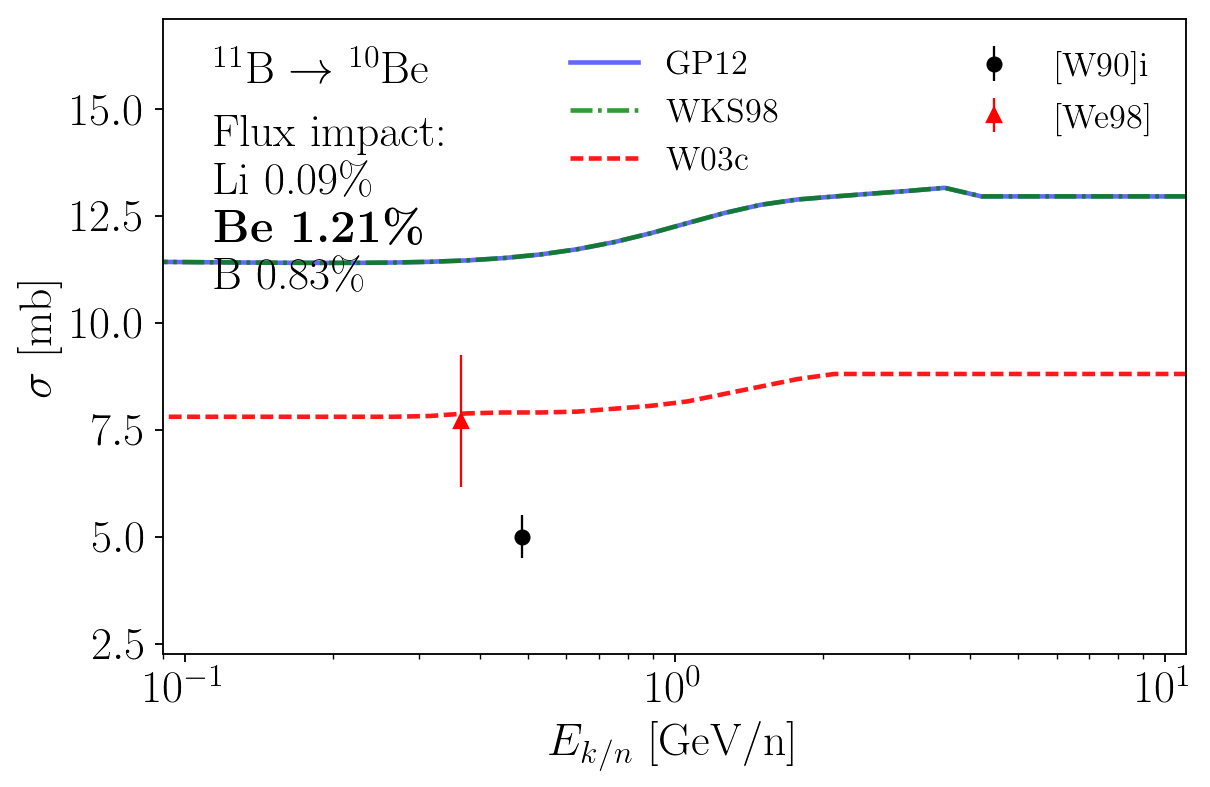}  &  
\includegraphics[width=0.32\textwidth]{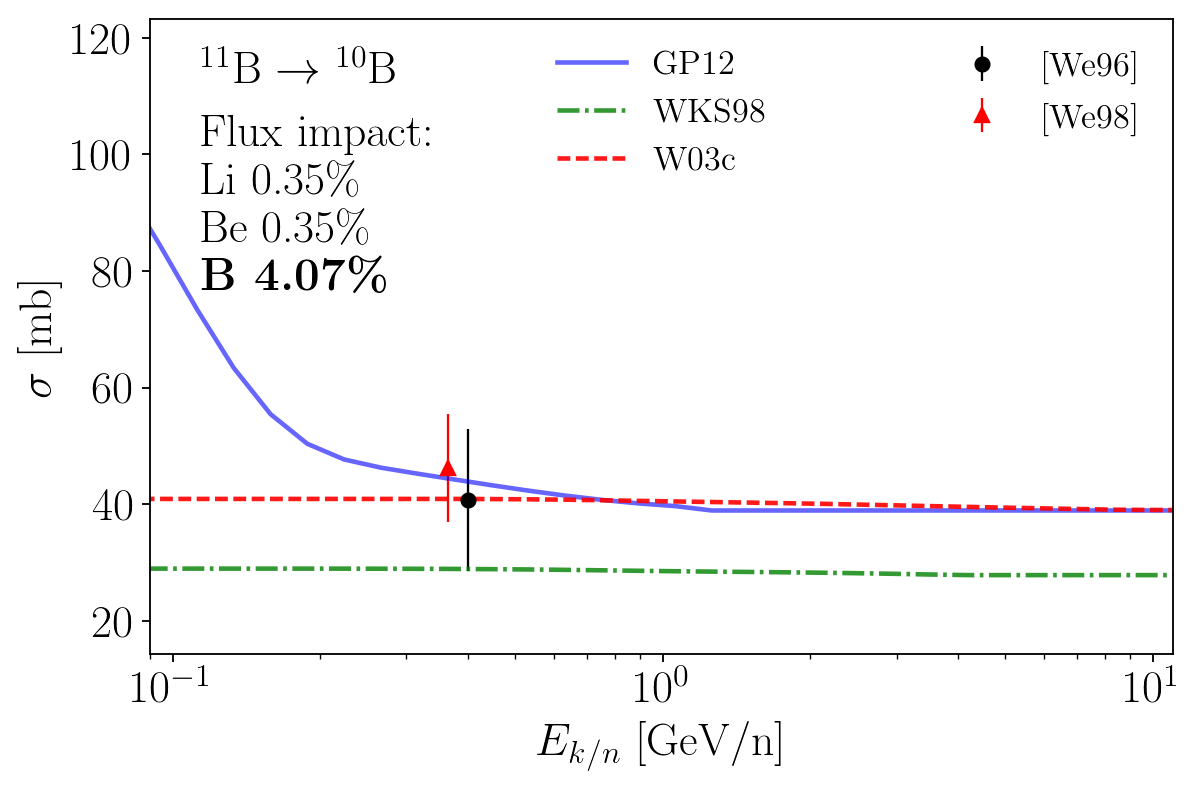}  &  \\ [3pt] 
\multicolumn{3}{c}{\bf Z=6{ \bf projectiles: $^{x}$C + H $\rightarrow$ $^{A}_ZX$}}\\ [3pt]
\multicolumn{3}{c}{\noindent\makebox[\linewidth]{\rule{\textwidth}{0.4pt}}}\\ [3pt]
\includegraphics[width=0.32\textwidth]{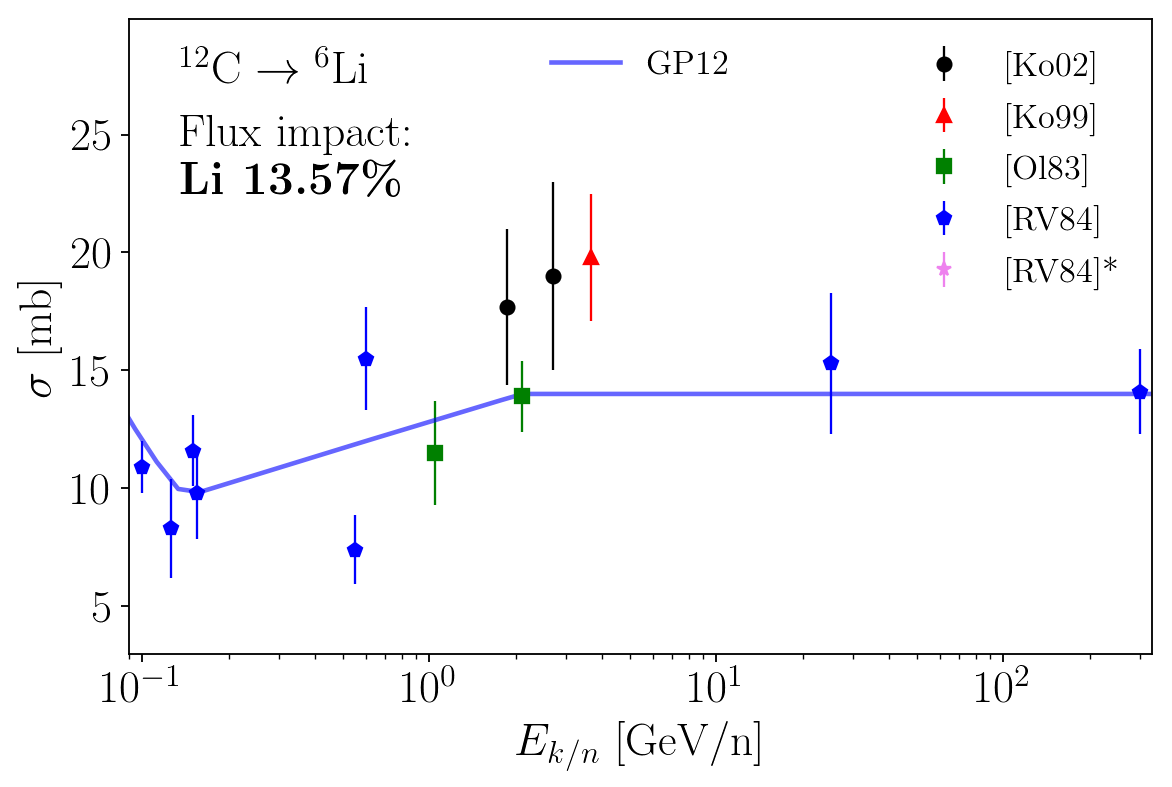}  &  
\includegraphics[width=0.32\textwidth]{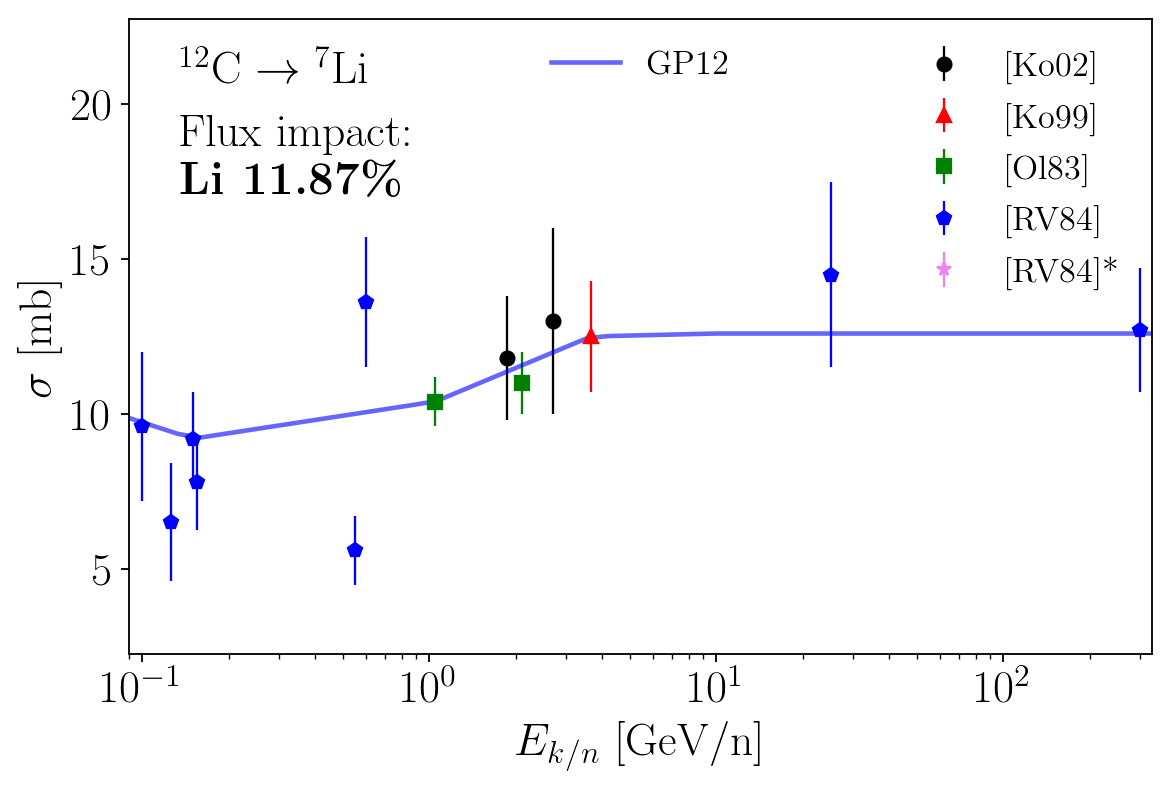}  &  
\includegraphics[width=0.32\textwidth]{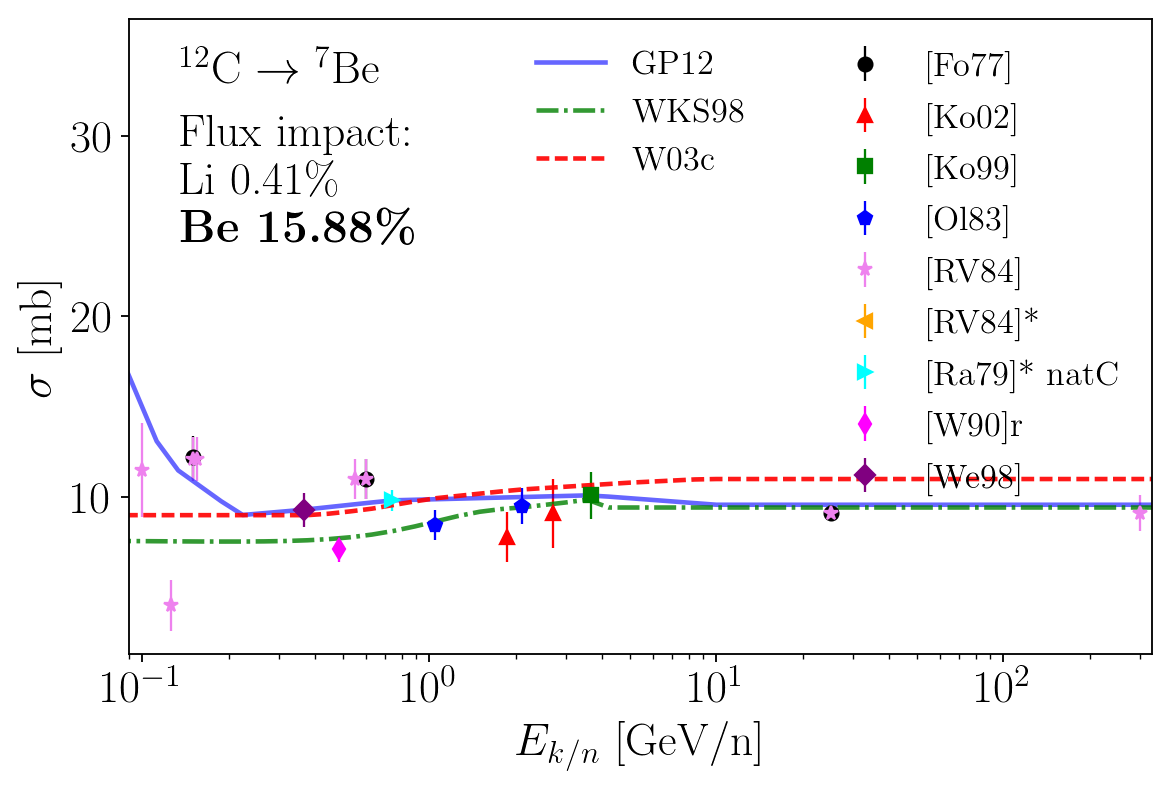}  \\ 
\includegraphics[width=0.32\textwidth]{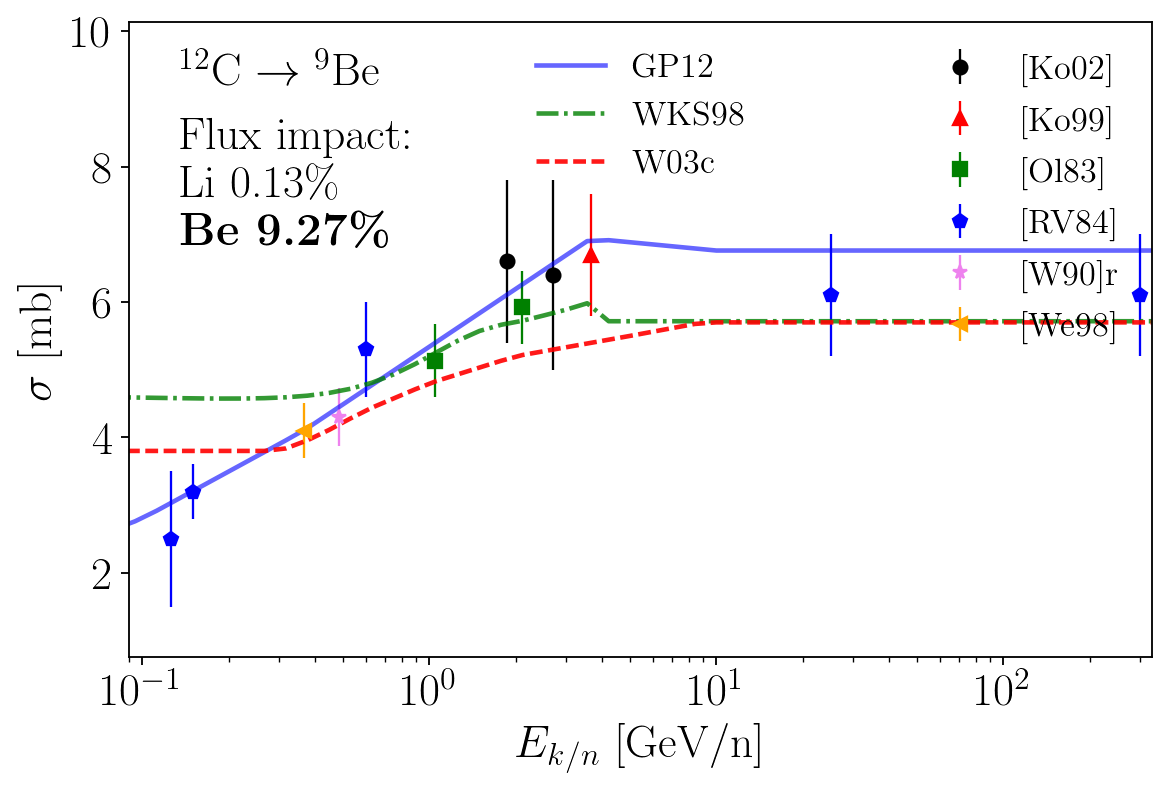}  &  
\includegraphics[width=0.32\textwidth]{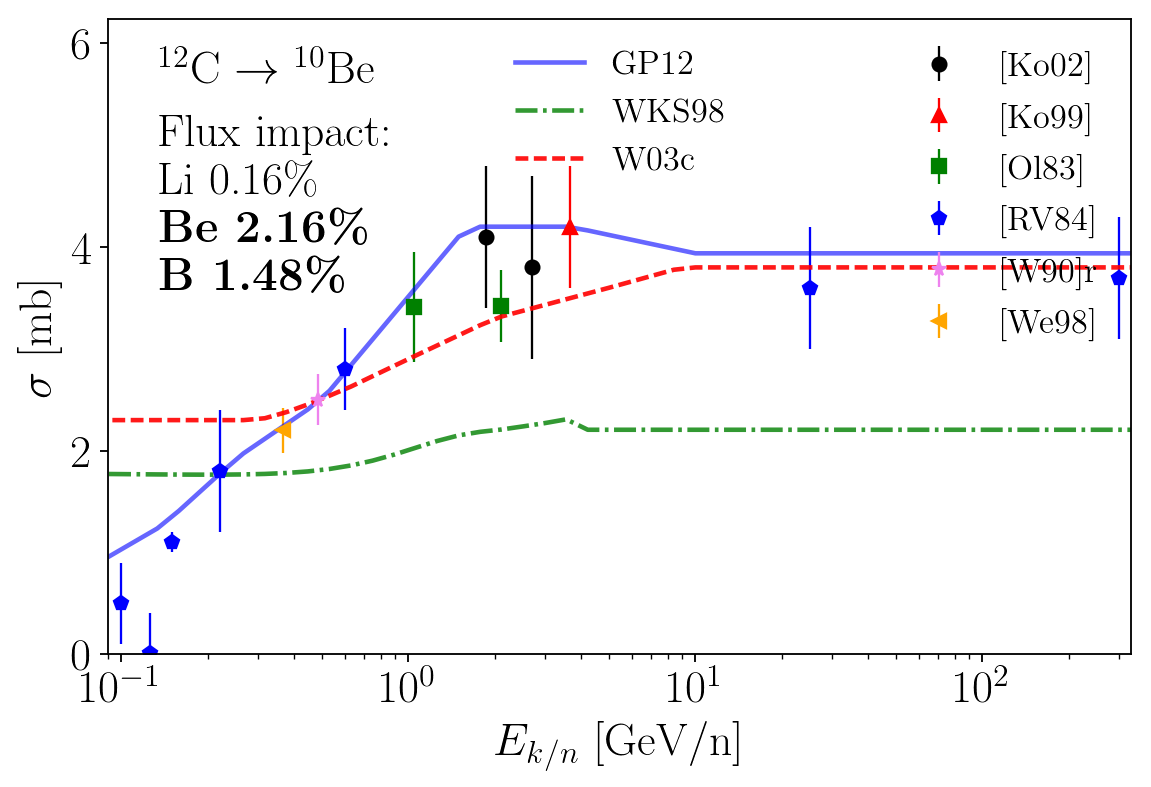}  &  
\includegraphics[width=0.32\textwidth]{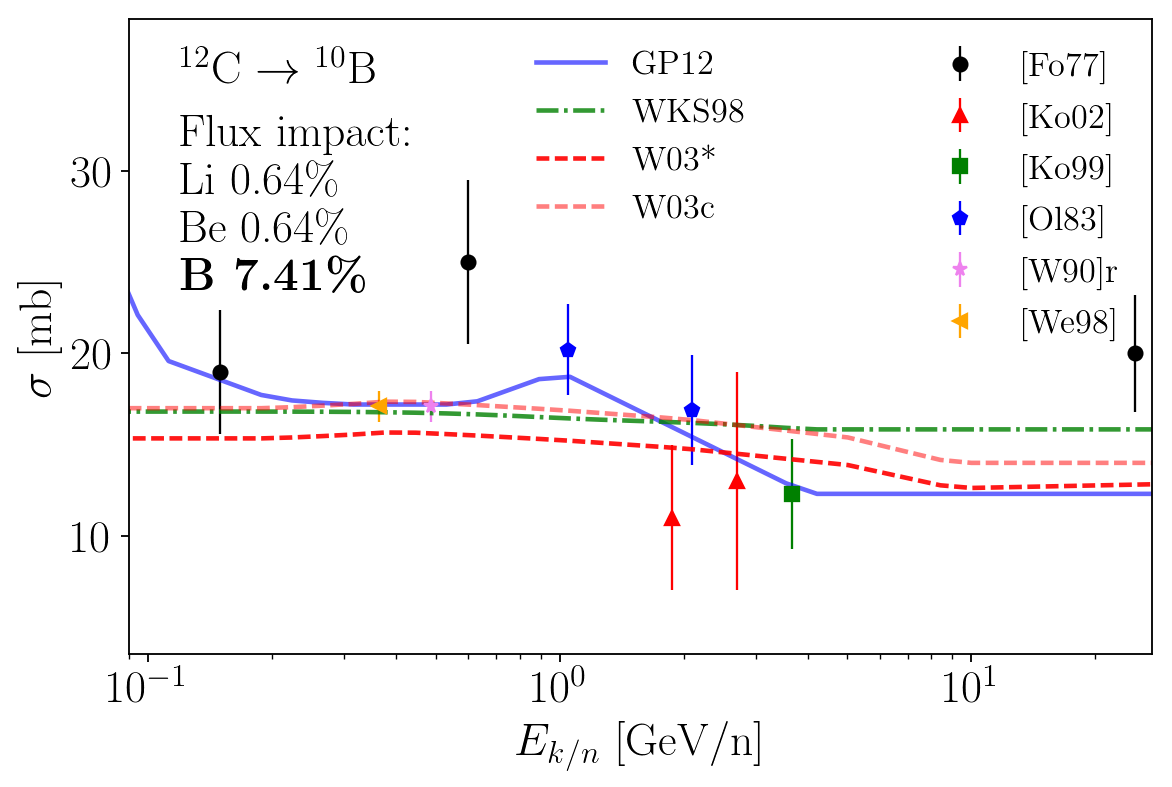}  \\ 
\includegraphics[width=0.32\textwidth]{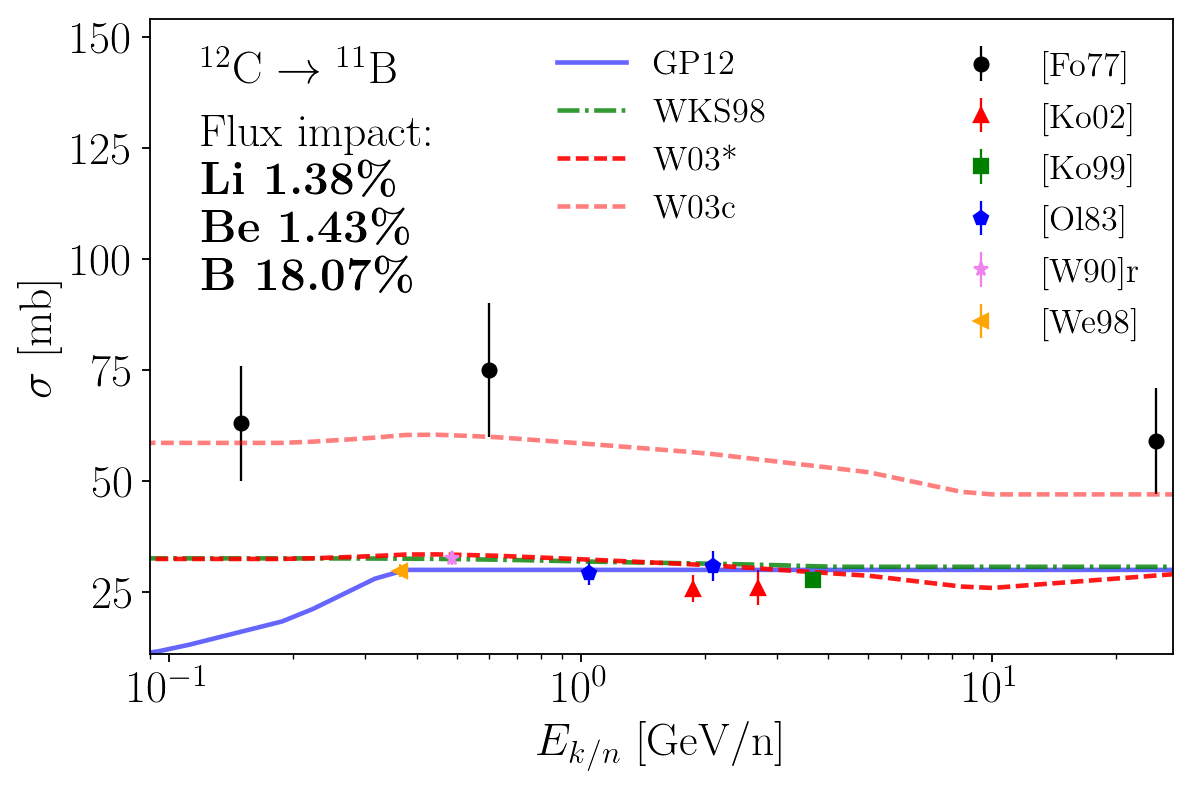}  &  
\includegraphics[width=0.32\textwidth]{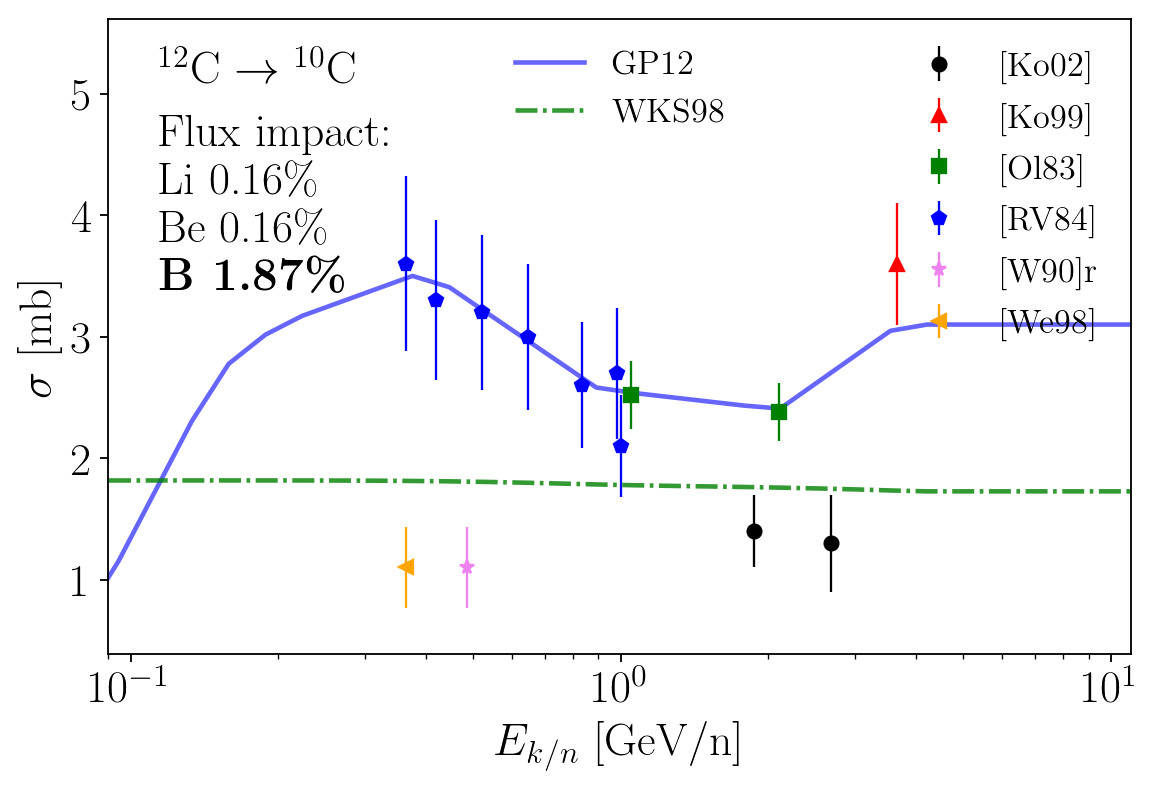}  &  
\includegraphics[width=0.32\textwidth]{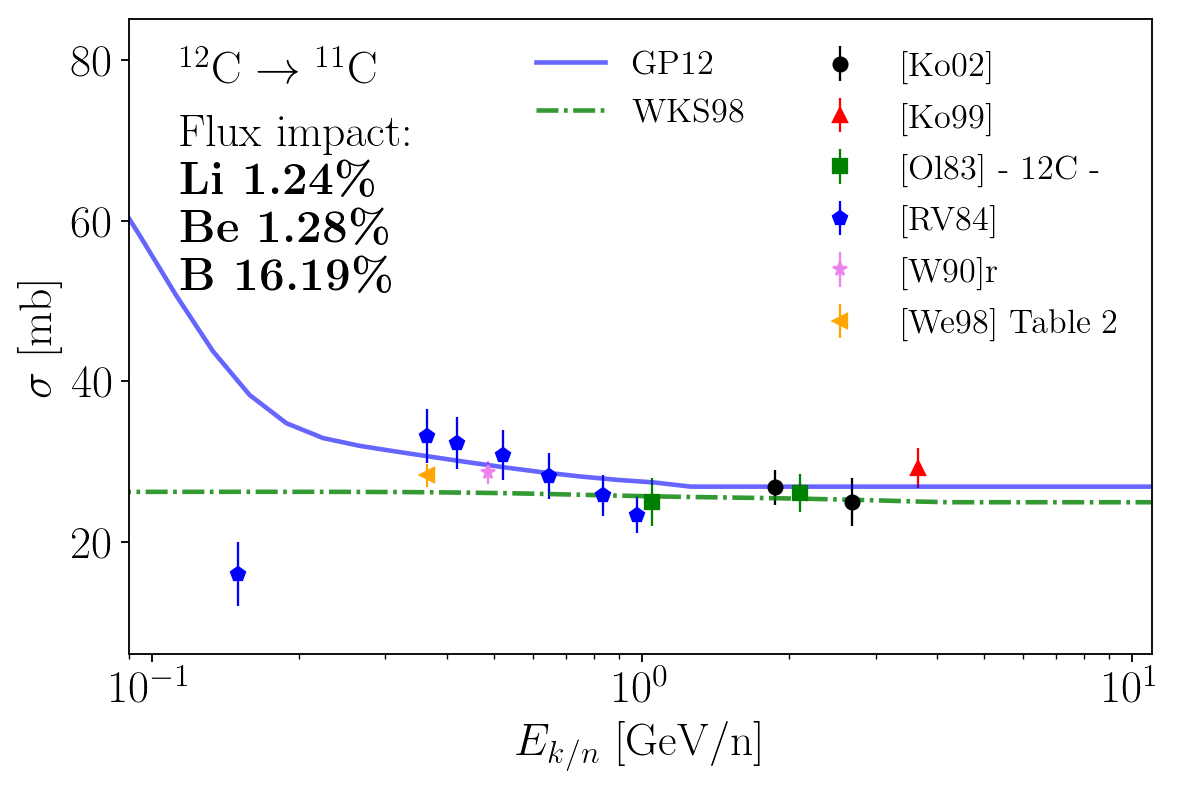}  \\ 
\includegraphics[width=0.32\textwidth]{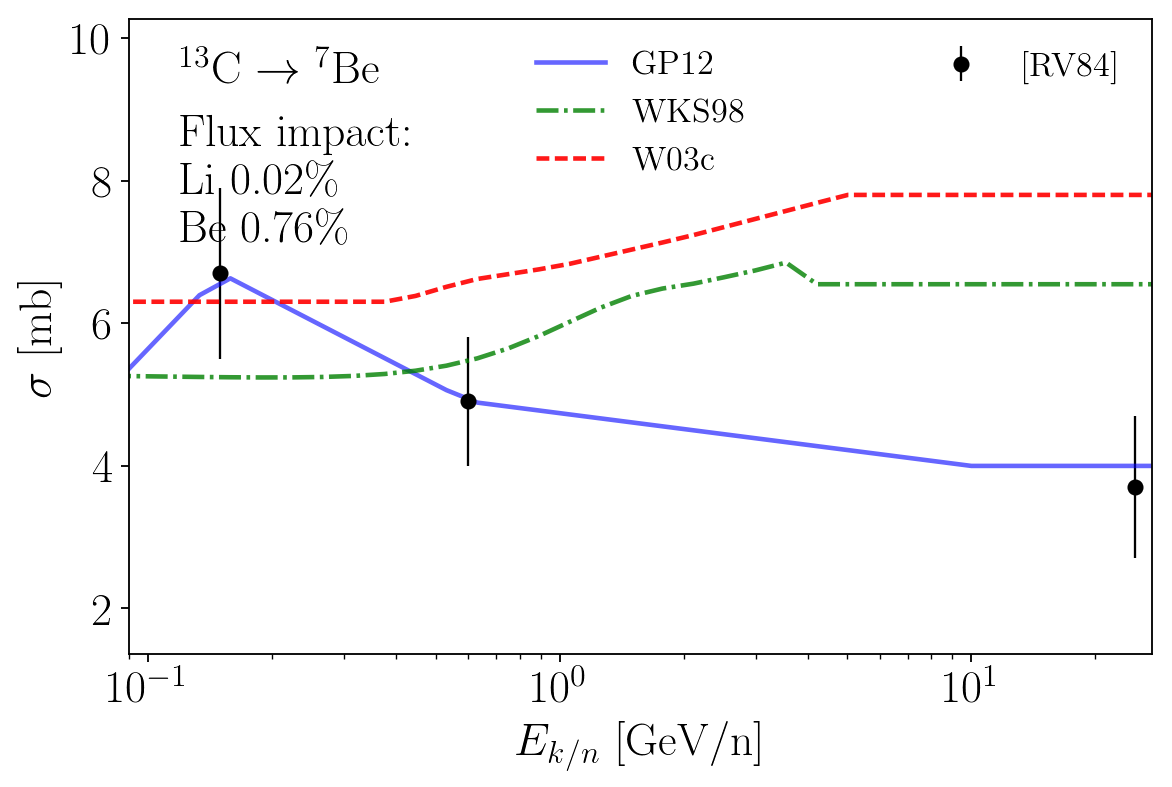}  &  
\includegraphics[width=0.32\textwidth]{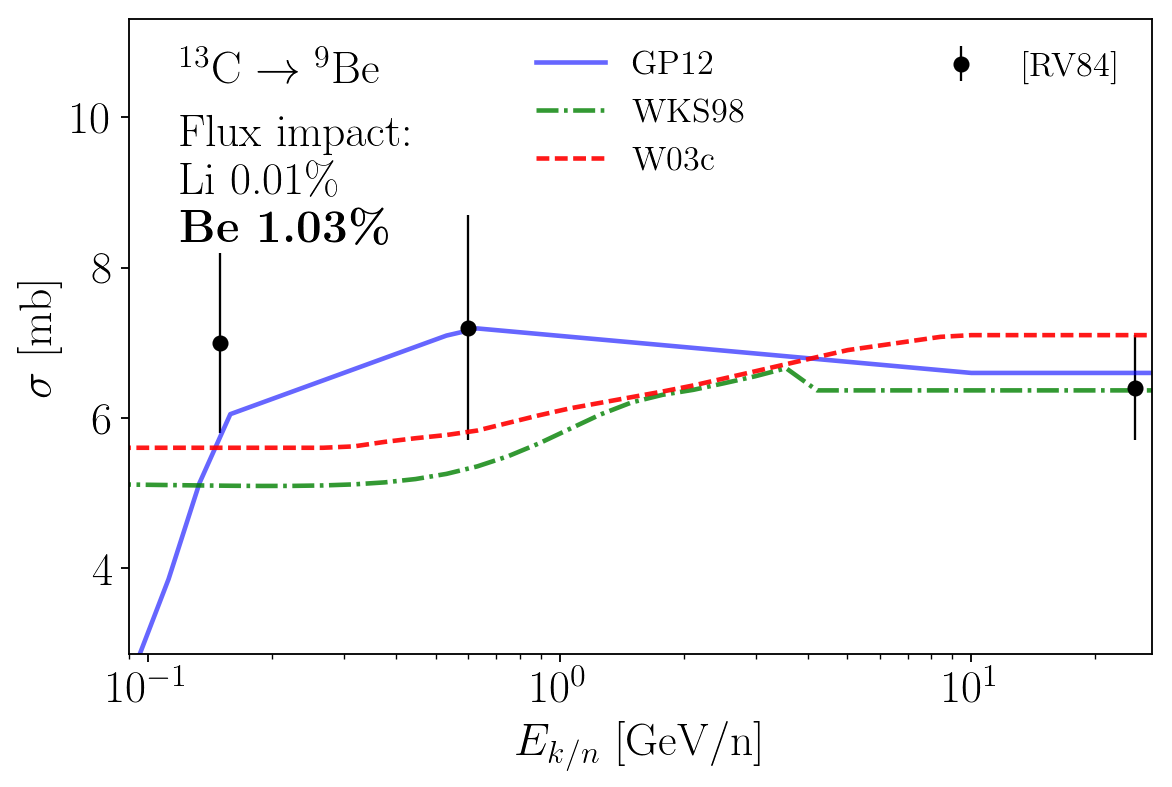}  &  
\includegraphics[width=0.32\textwidth]{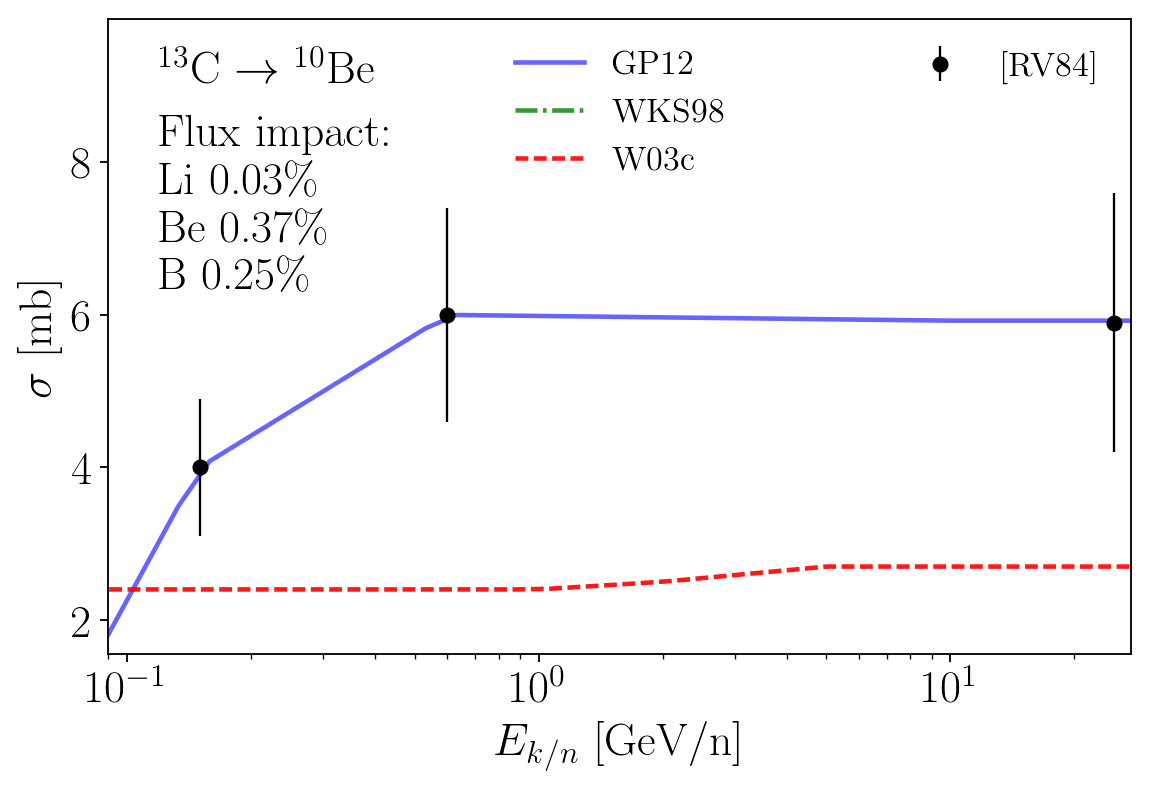}  \\ 
 \\ [3pt] 
\multicolumn{3}{c}{\bf Z=7{ \bf projectiles: $^{x}$N + H $\rightarrow$ $^{A}_ZX$}}\\ [3pt]
\multicolumn{3}{c}{\noindent\makebox[\linewidth]{\rule{\textwidth}{0.4pt}}}\\ [3pt]
\includegraphics[width=0.32\textwidth]{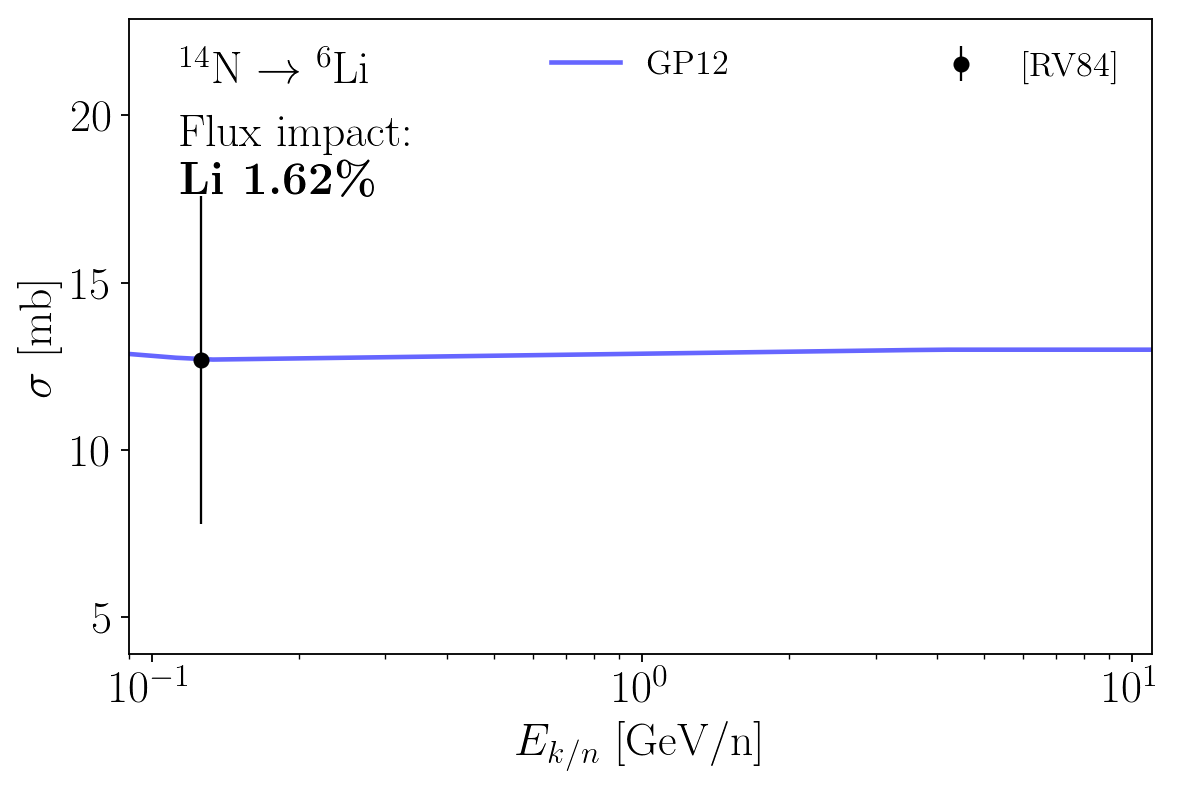}  &  
\includegraphics[width=0.32\textwidth]{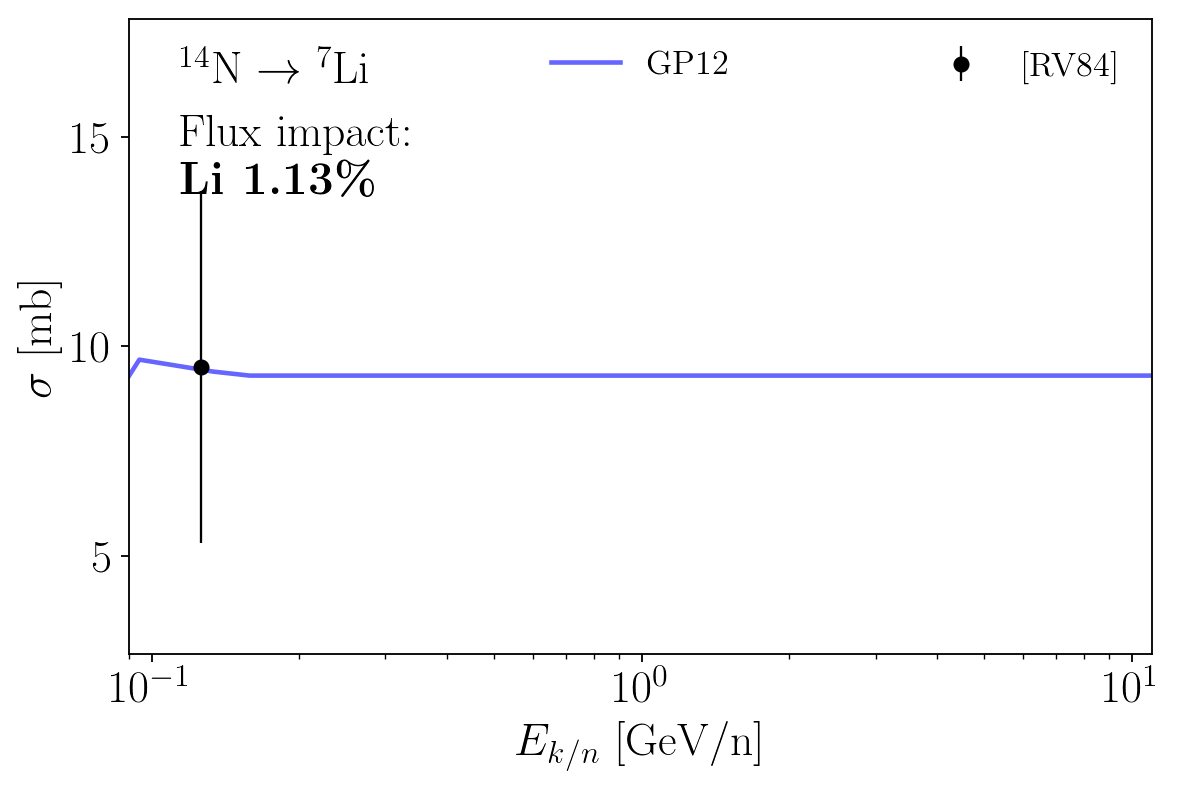}  &  
\includegraphics[width=0.32\textwidth]{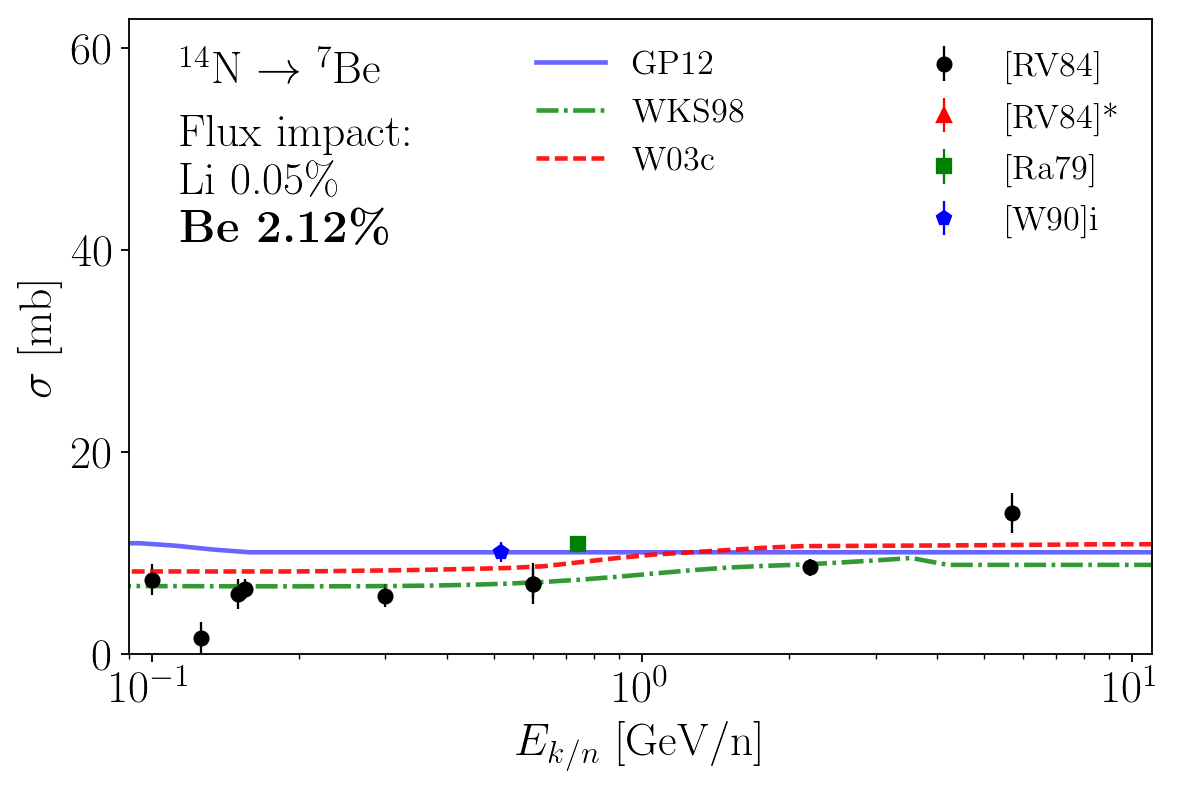}  \\ 
\includegraphics[width=0.32\textwidth]{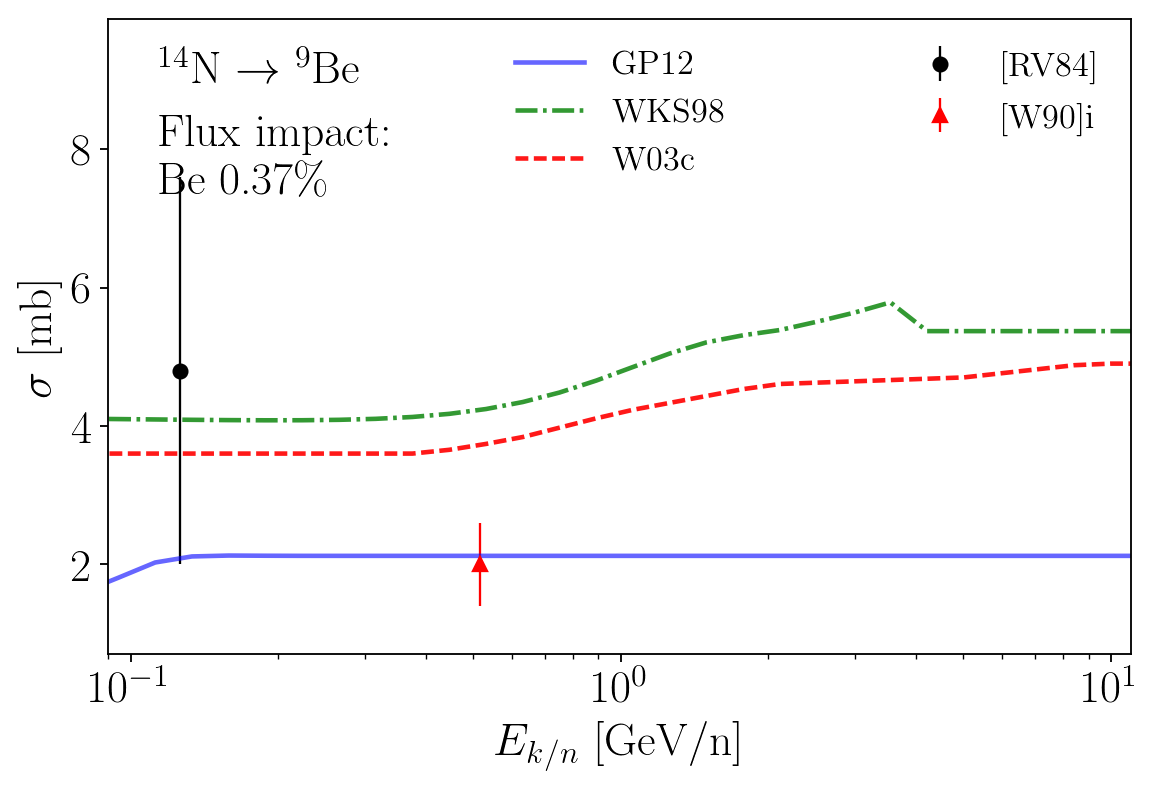}  &  
\includegraphics[width=0.32\textwidth]{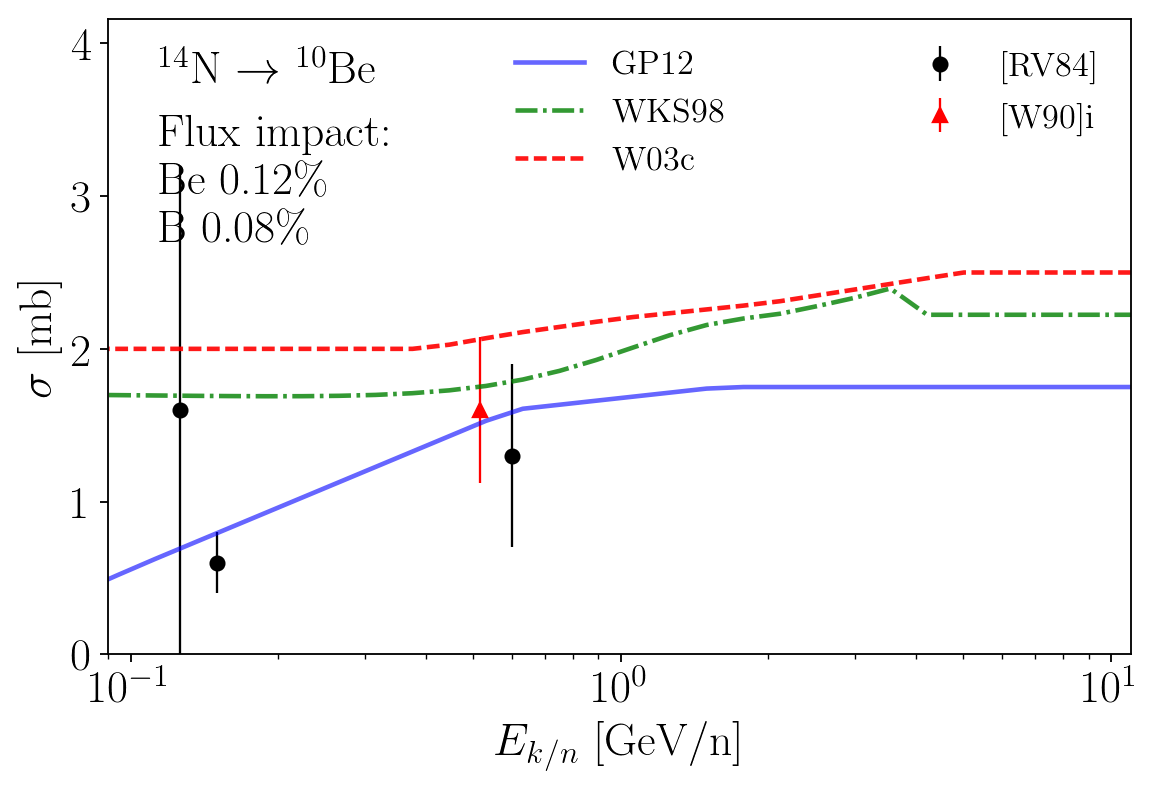}  &  
\includegraphics[width=0.32\textwidth]{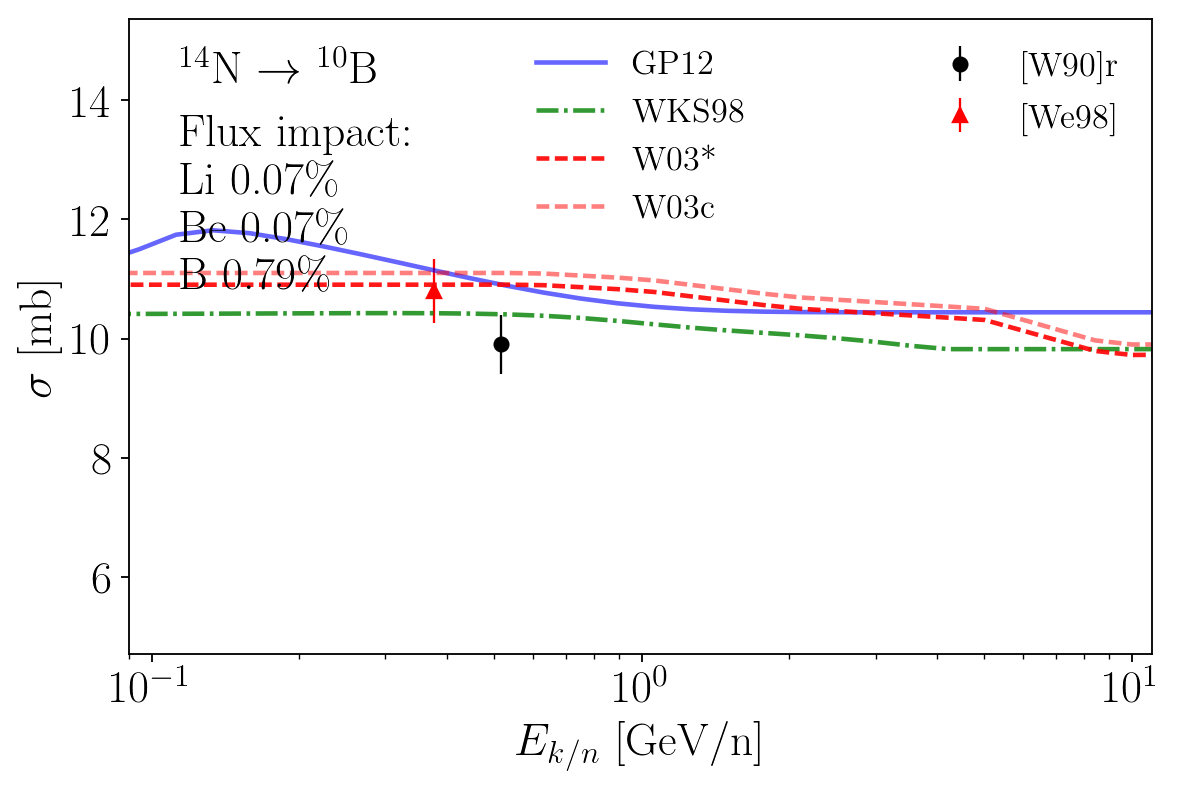}  \\ 
\includegraphics[width=0.32\textwidth]{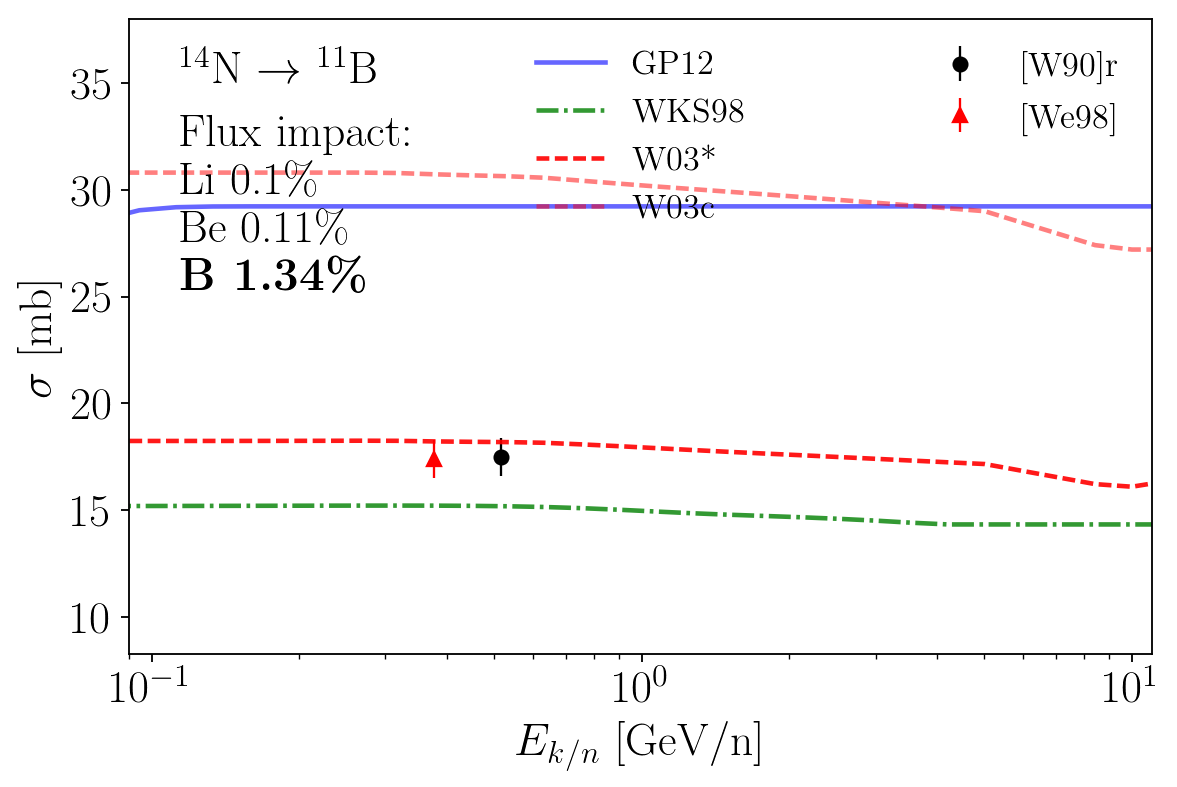}  &  
\includegraphics[width=0.32\textwidth]{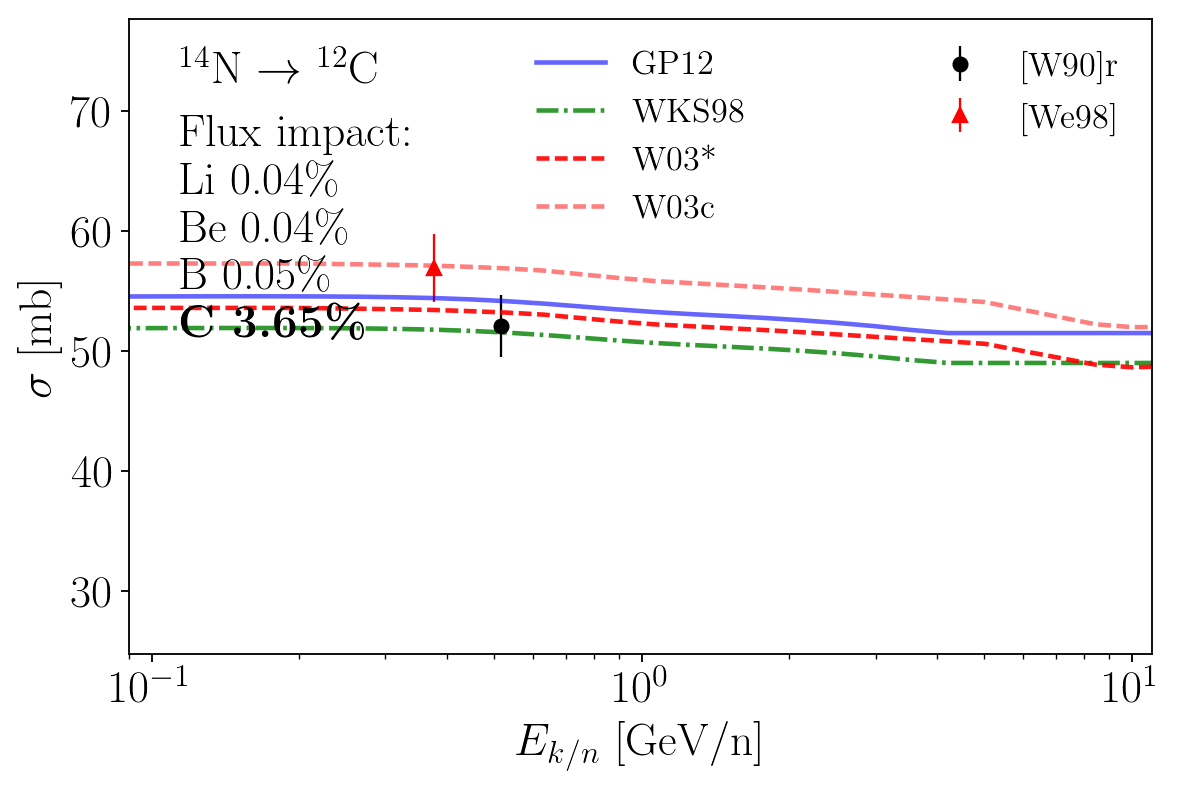}  &  
\includegraphics[width=0.32\textwidth]{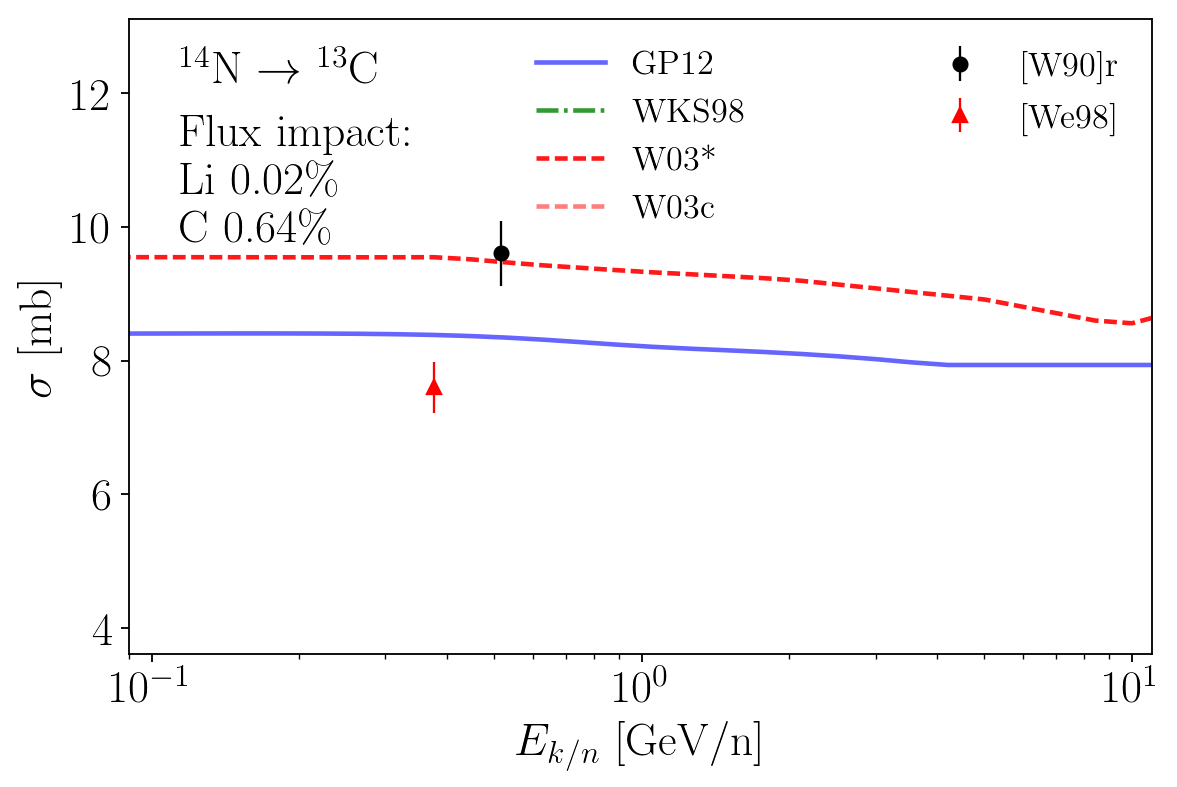}  \\ 
\includegraphics[width=0.32\textwidth]{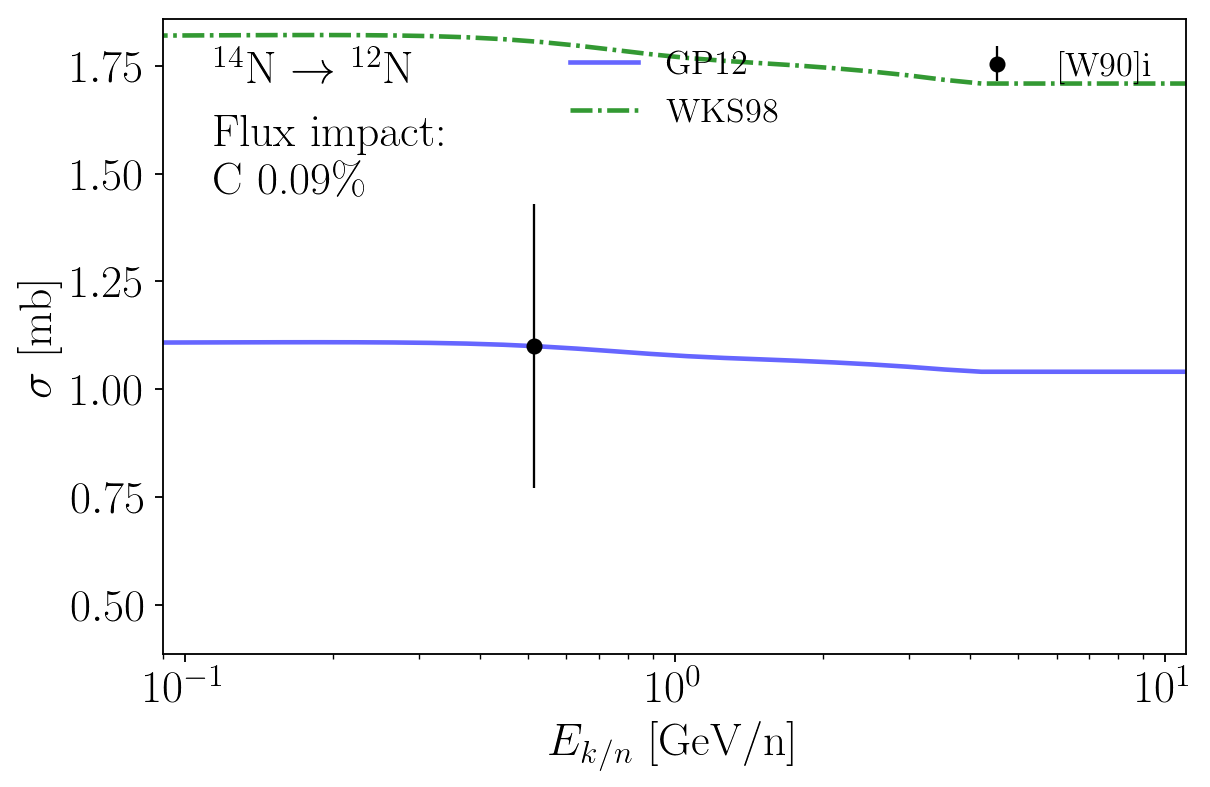}  &  
\includegraphics[width=0.32\textwidth]{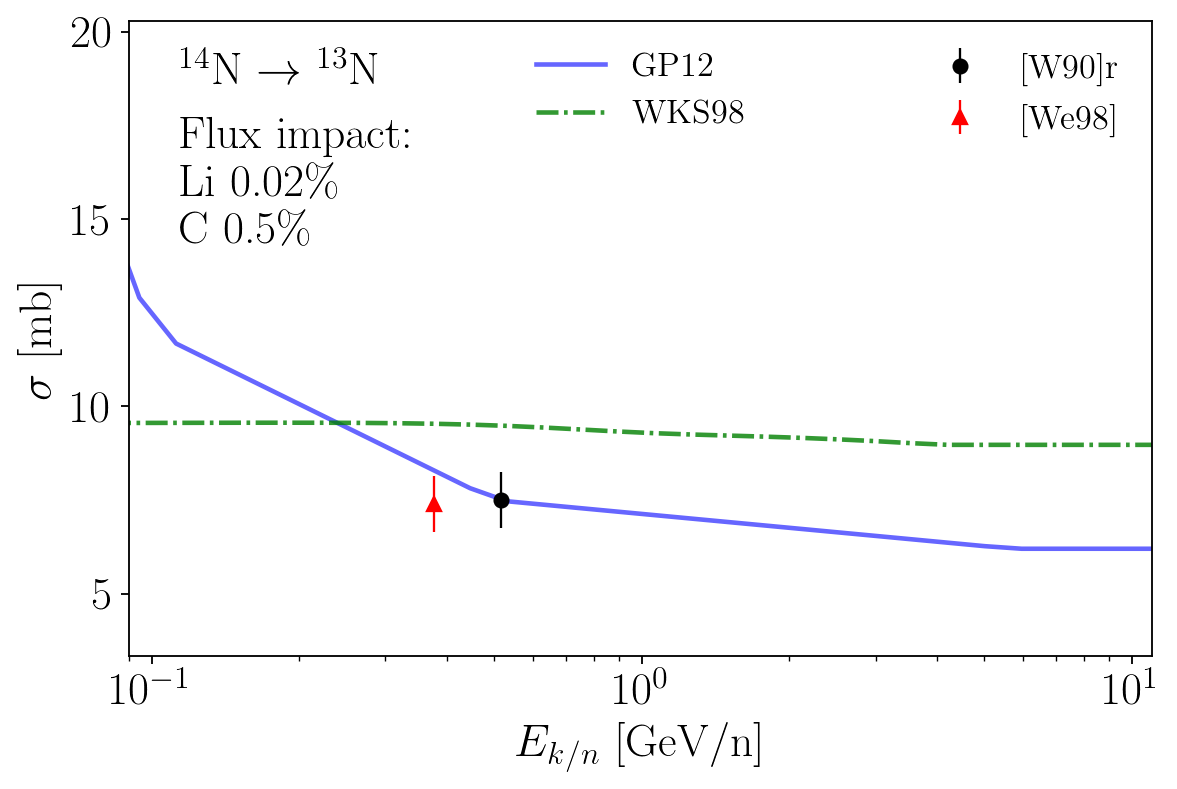}  &  
\includegraphics[width=0.32\textwidth]{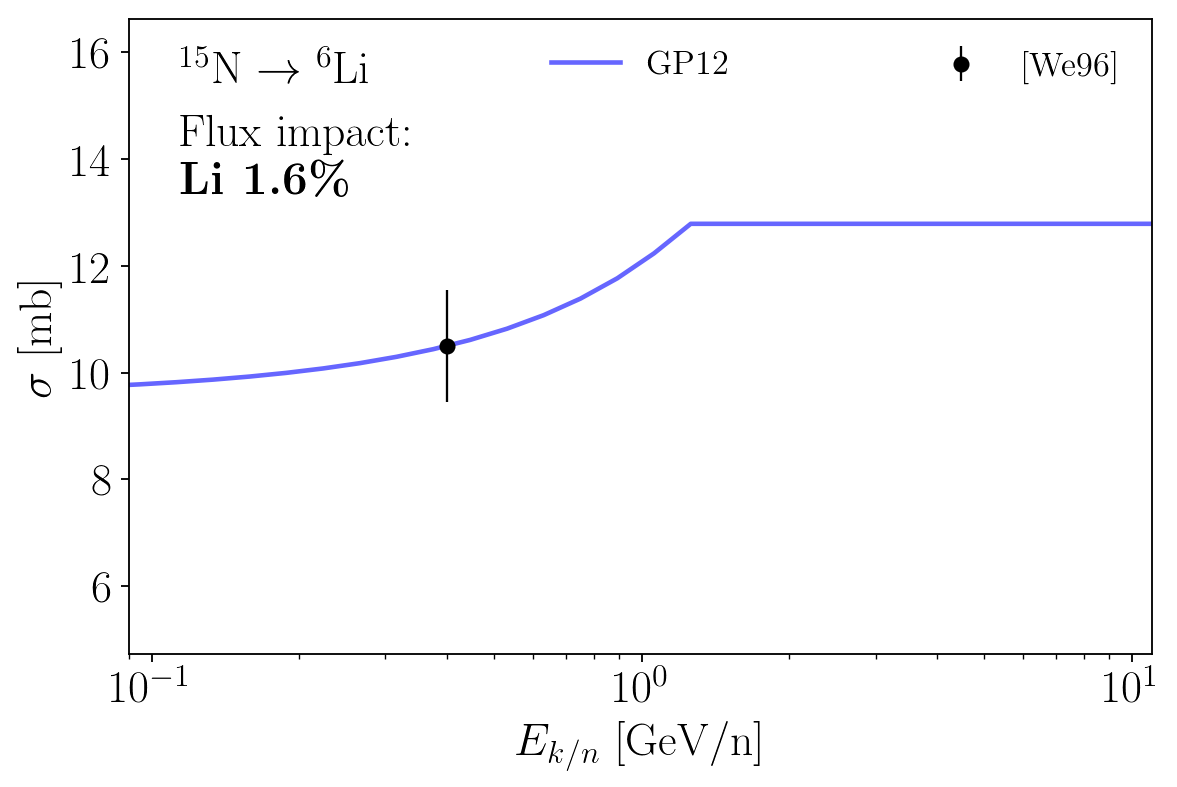}  \\ 
\includegraphics[width=0.32\textwidth]{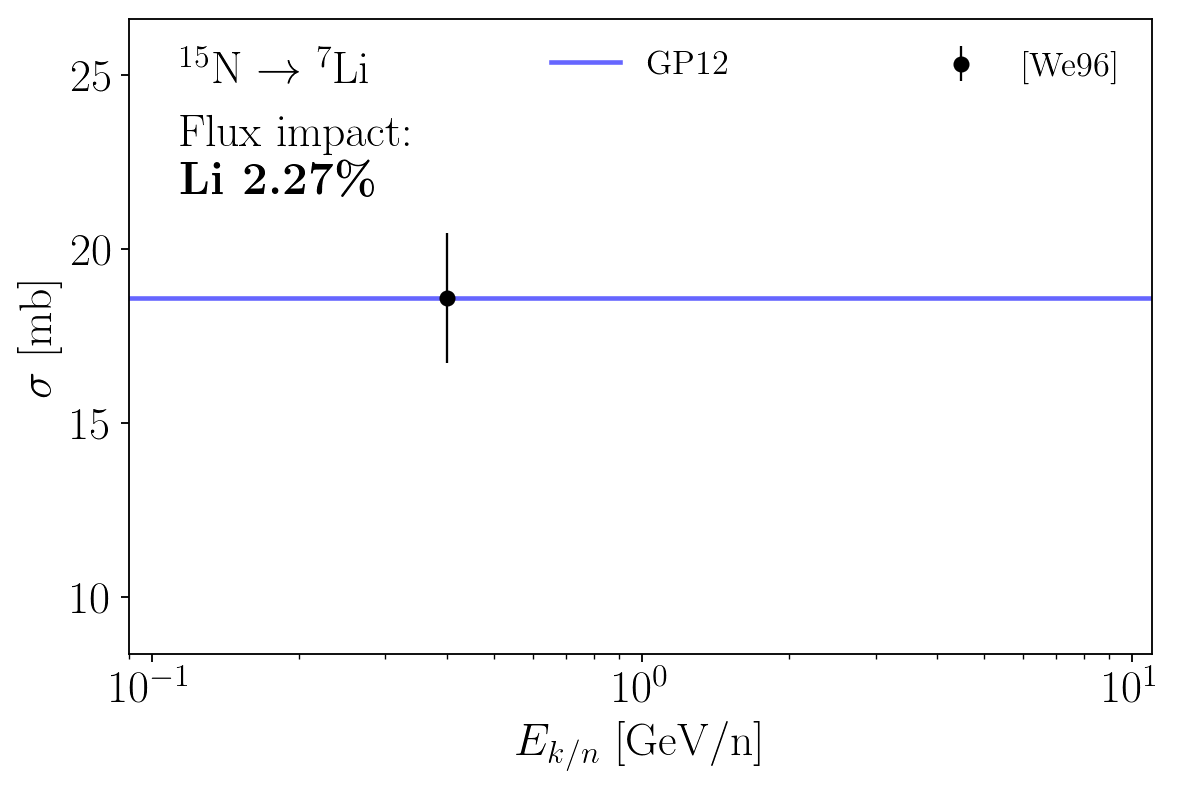}  &  
\includegraphics[width=0.32\textwidth]{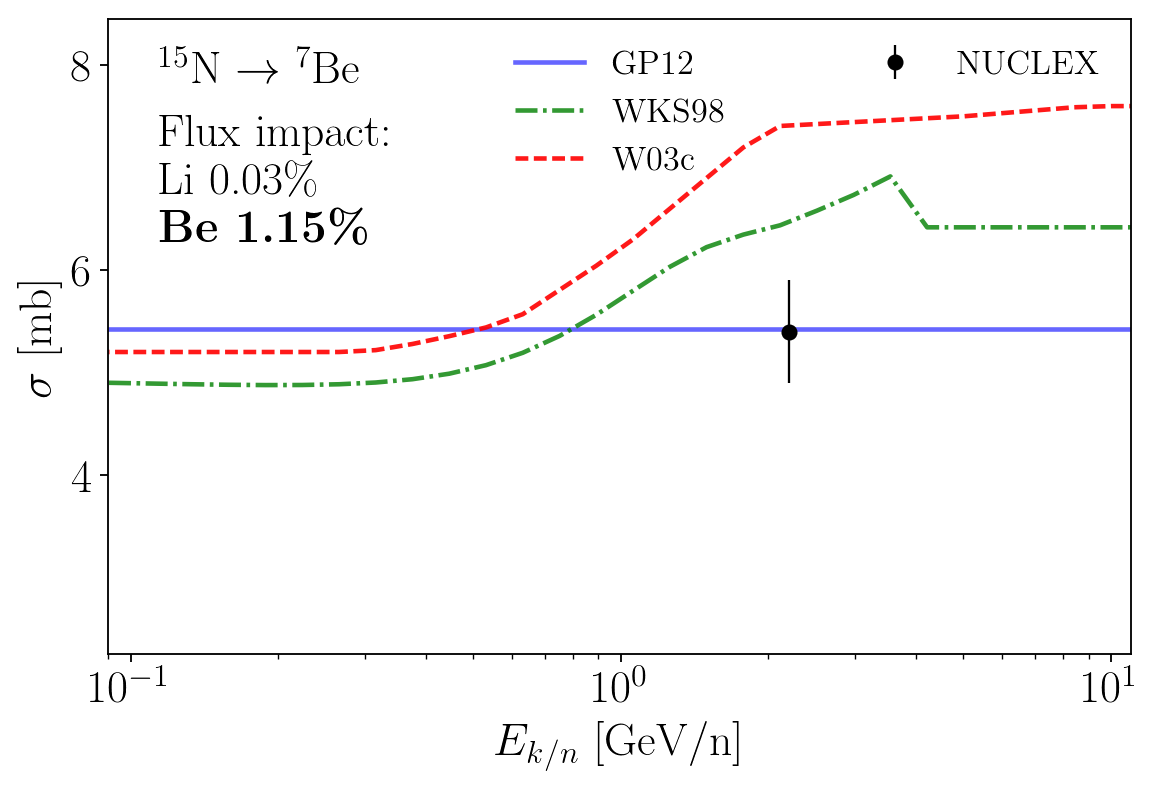}  &  
\includegraphics[width=0.32\textwidth]{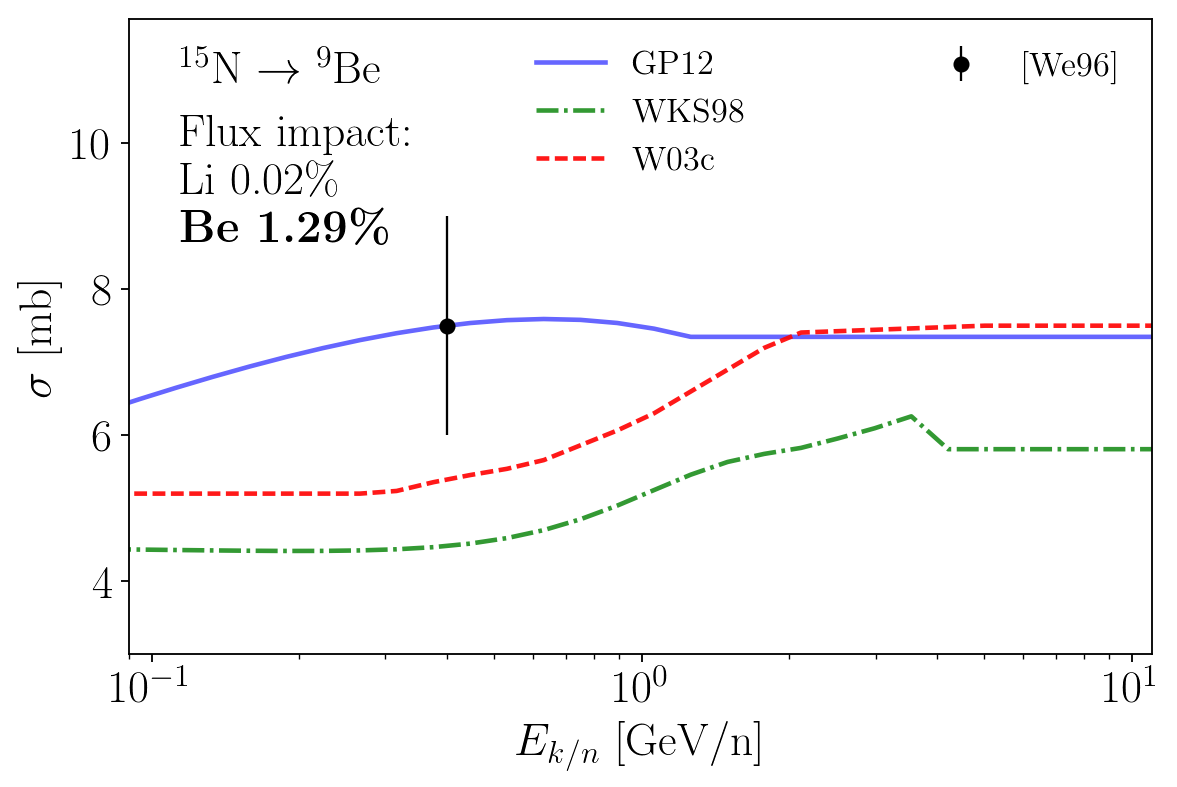}  \\   
\includegraphics[width=0.32\textwidth]{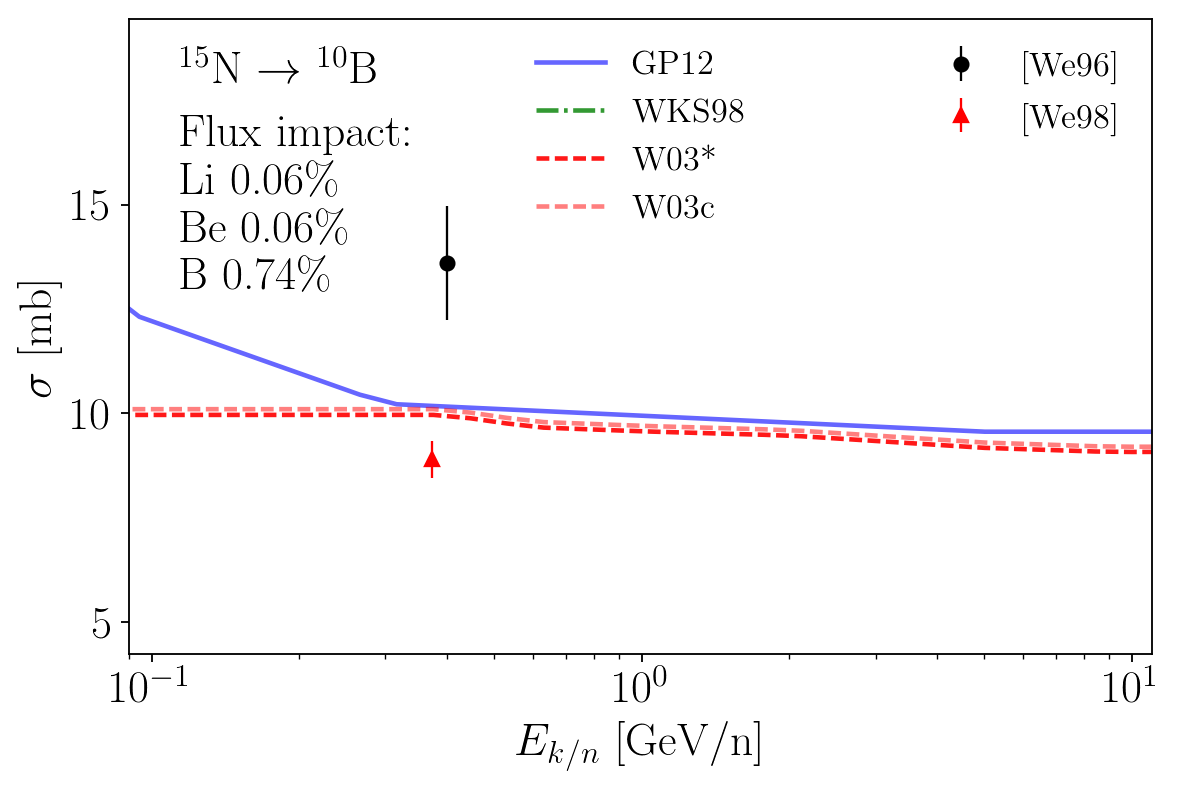}  &  
\includegraphics[width=0.32\textwidth]{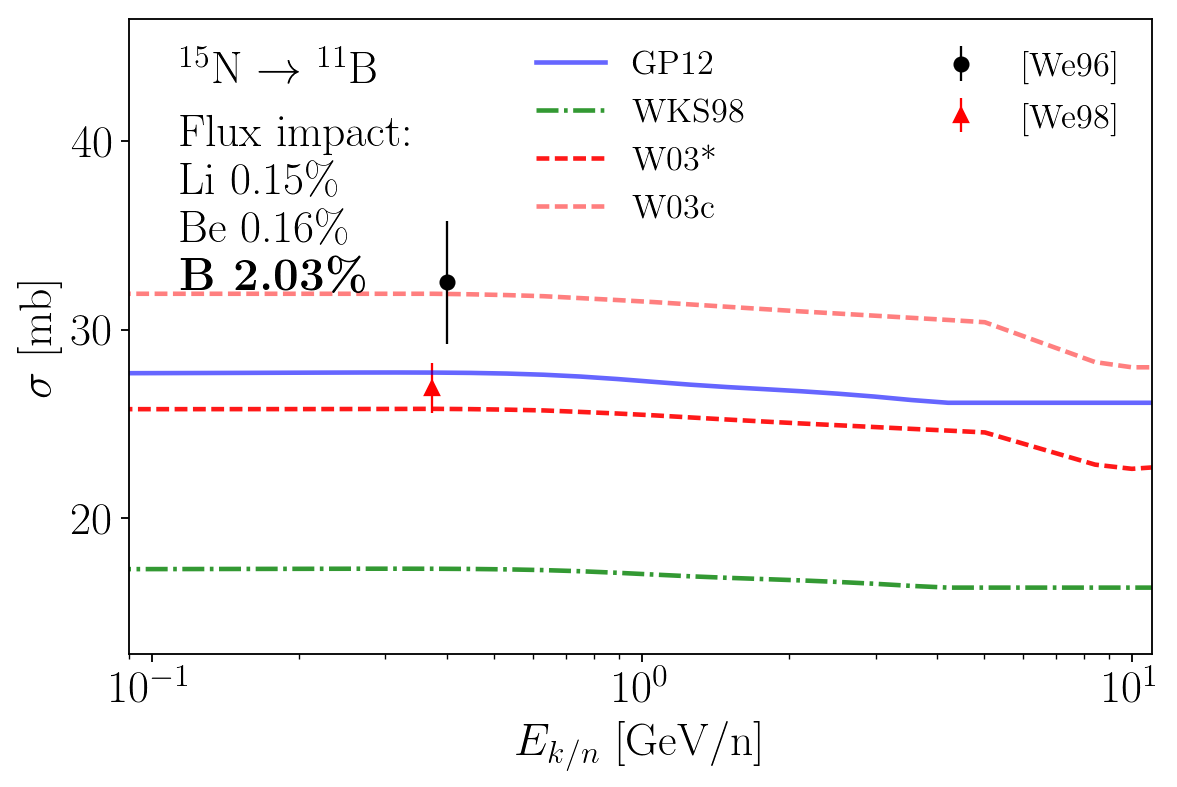}  & 
\includegraphics[width=0.32\textwidth]{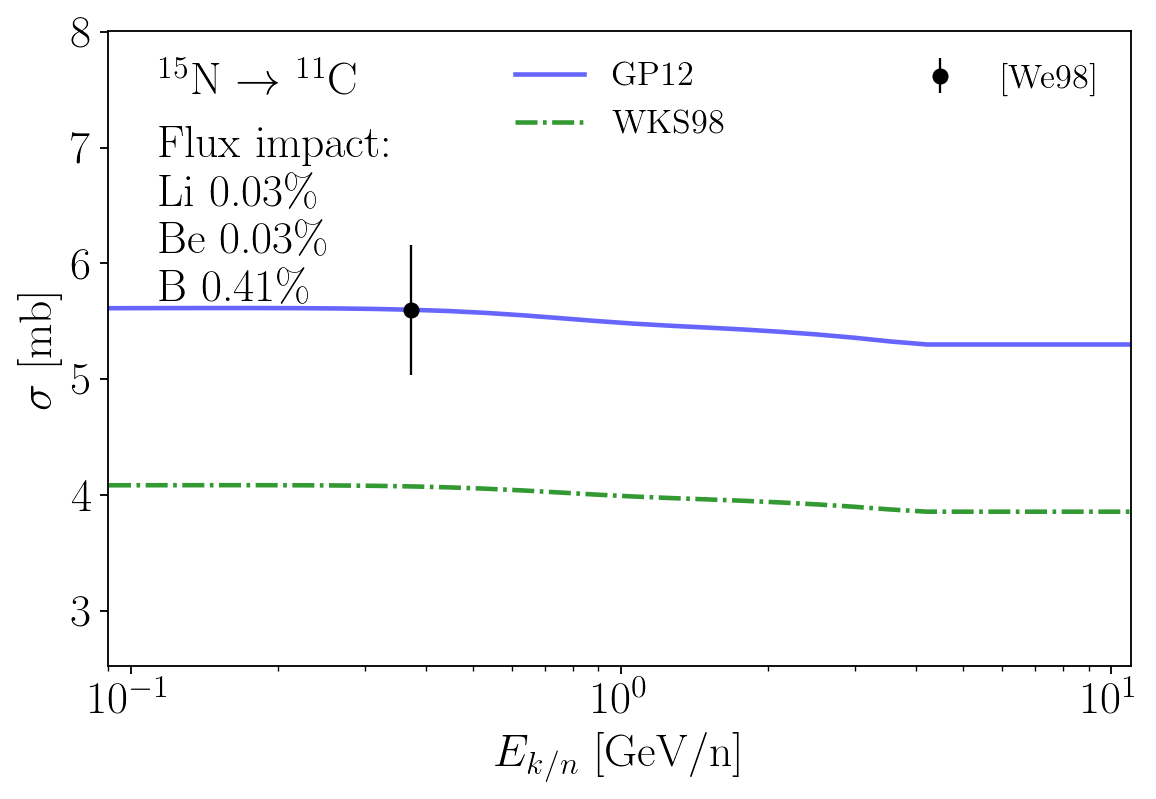}  \\  
\includegraphics[width=0.32\textwidth]{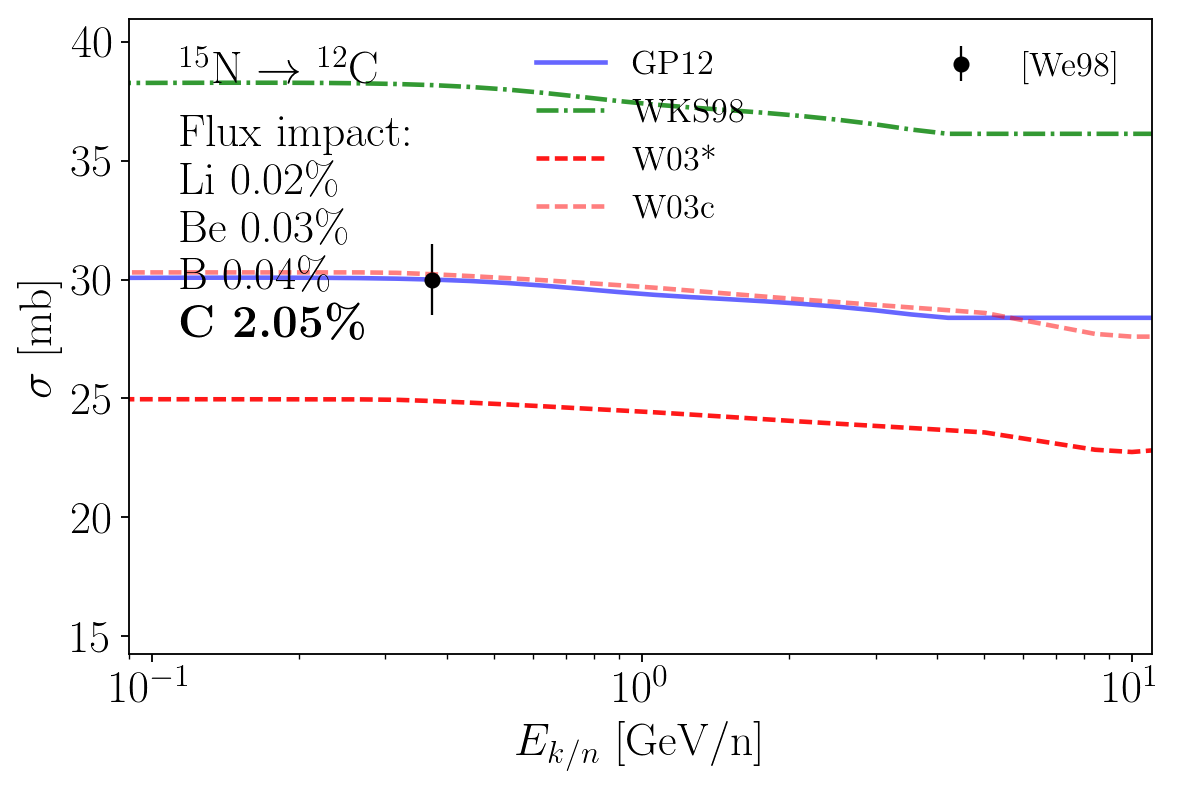}  &  
\includegraphics[width=0.32\textwidth]{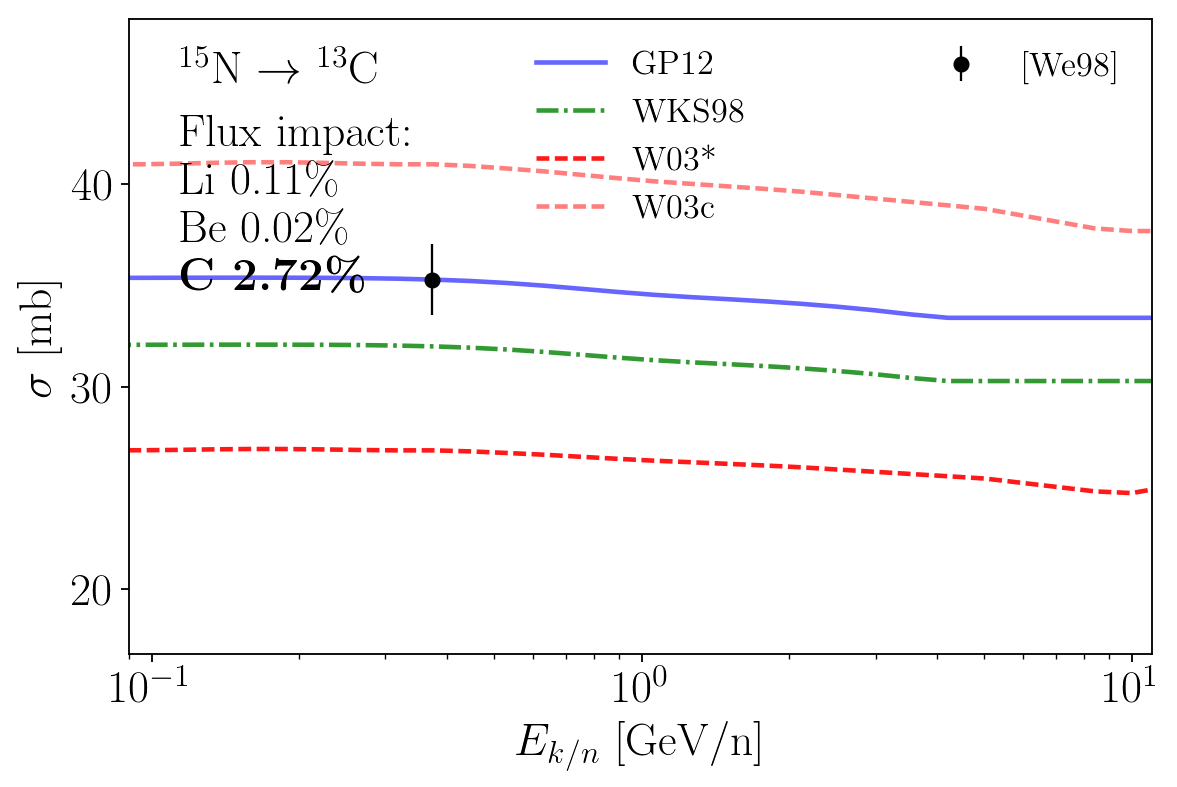}  & 
\includegraphics[width=0.32\textwidth]{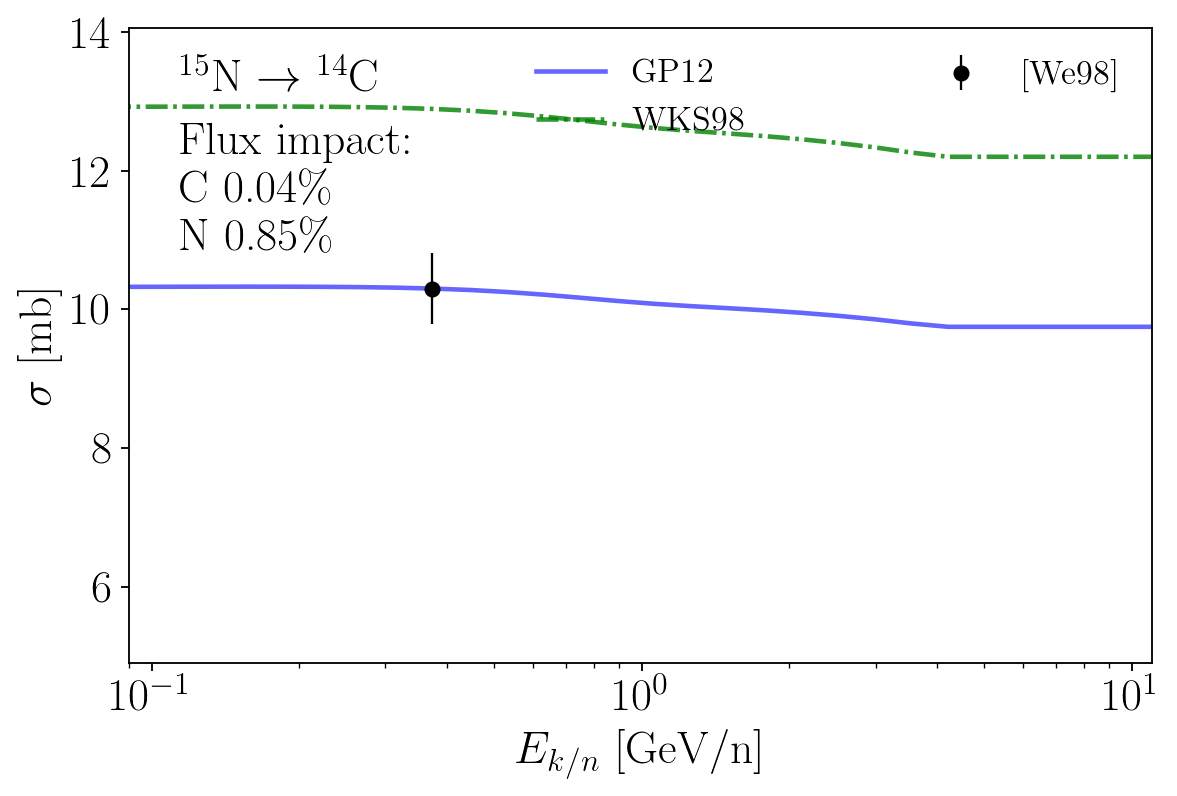}  \\  
\includegraphics[width=0.32\textwidth]{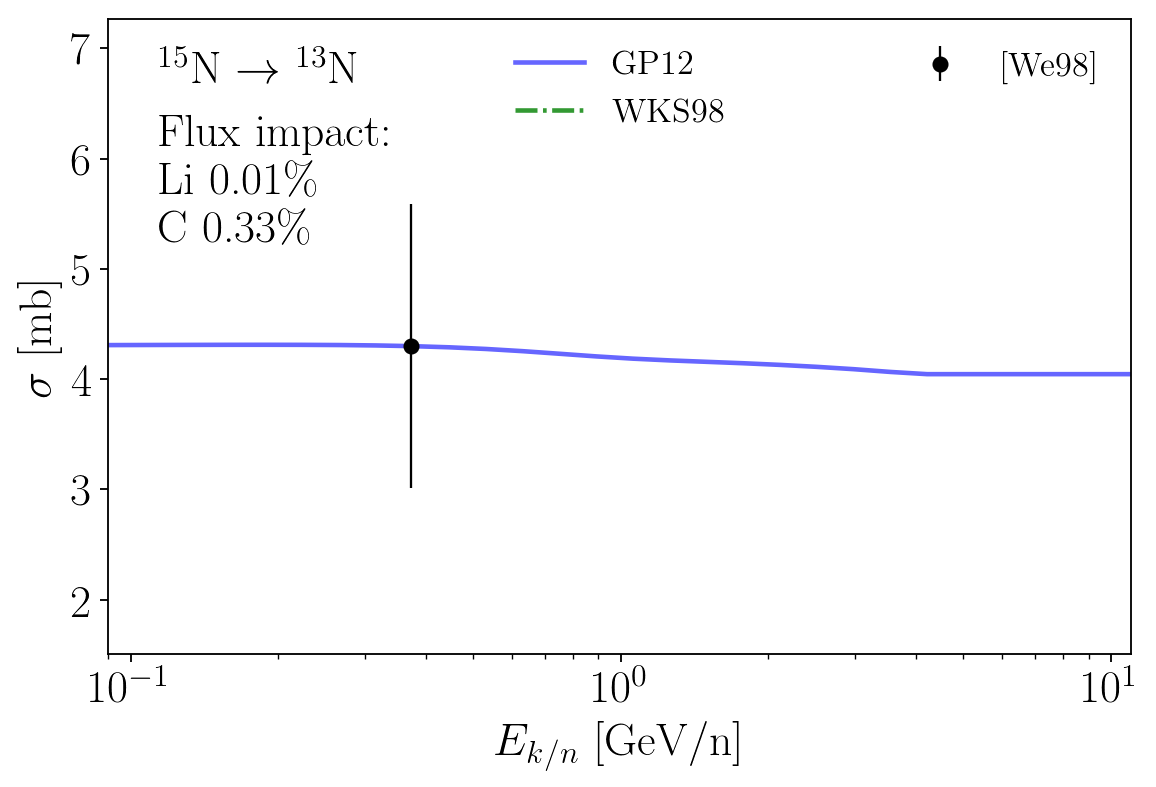}  &  
\includegraphics[width=0.32\textwidth]{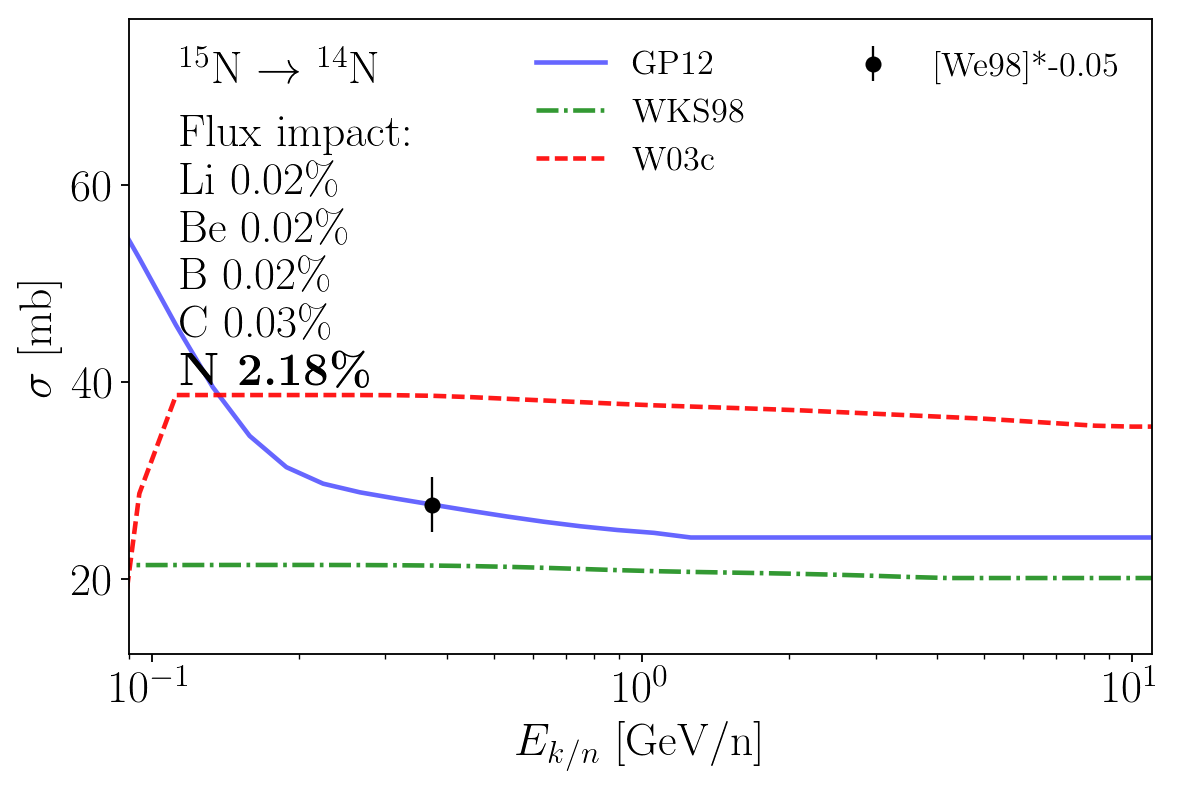}  & 
 \\ [3pt] 
\multicolumn{3}{c}{\bf Z=8{ \bf projectiles: $^{x}$O + H $\rightarrow$ $^{A}_ZX$}}\\ [3pt]
\multicolumn{3}{c}{\noindent\makebox[\linewidth]{\rule{\textwidth}{0.4pt}}}\\ [3pt]
\includegraphics[width=0.32\textwidth]{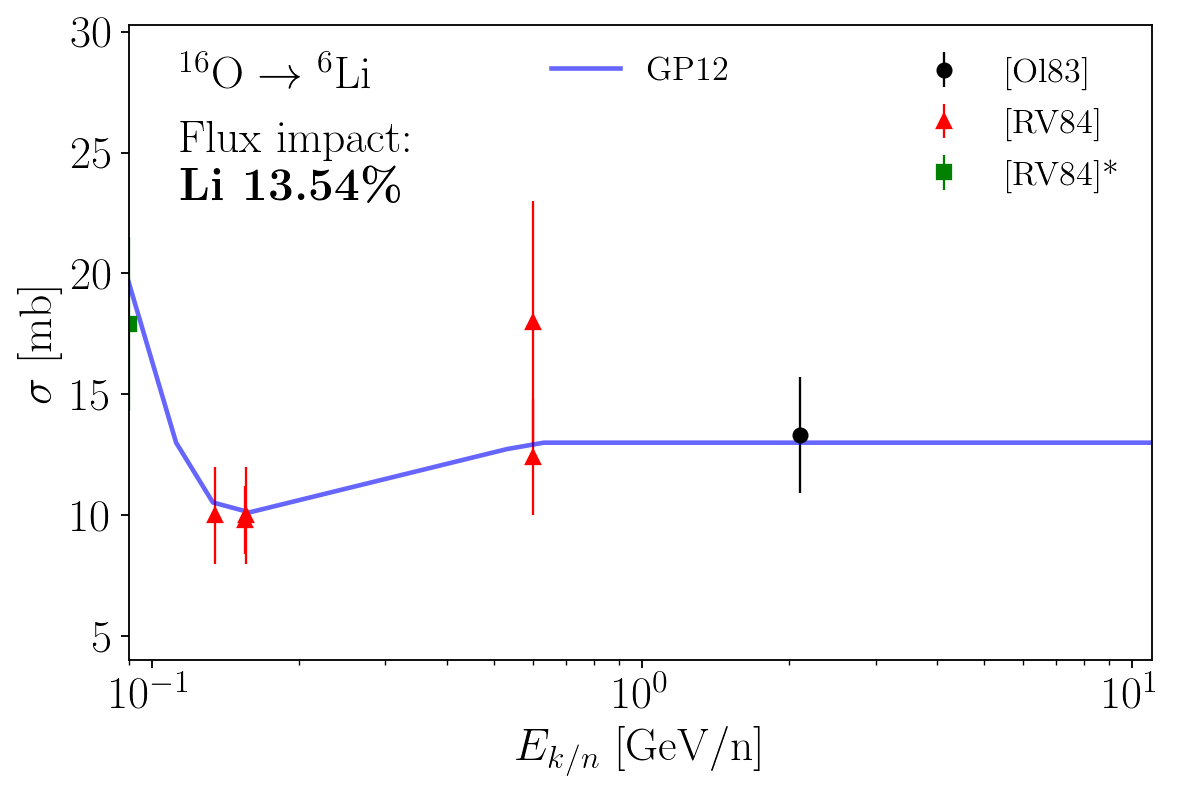}  &  
\includegraphics[width=0.32\textwidth]{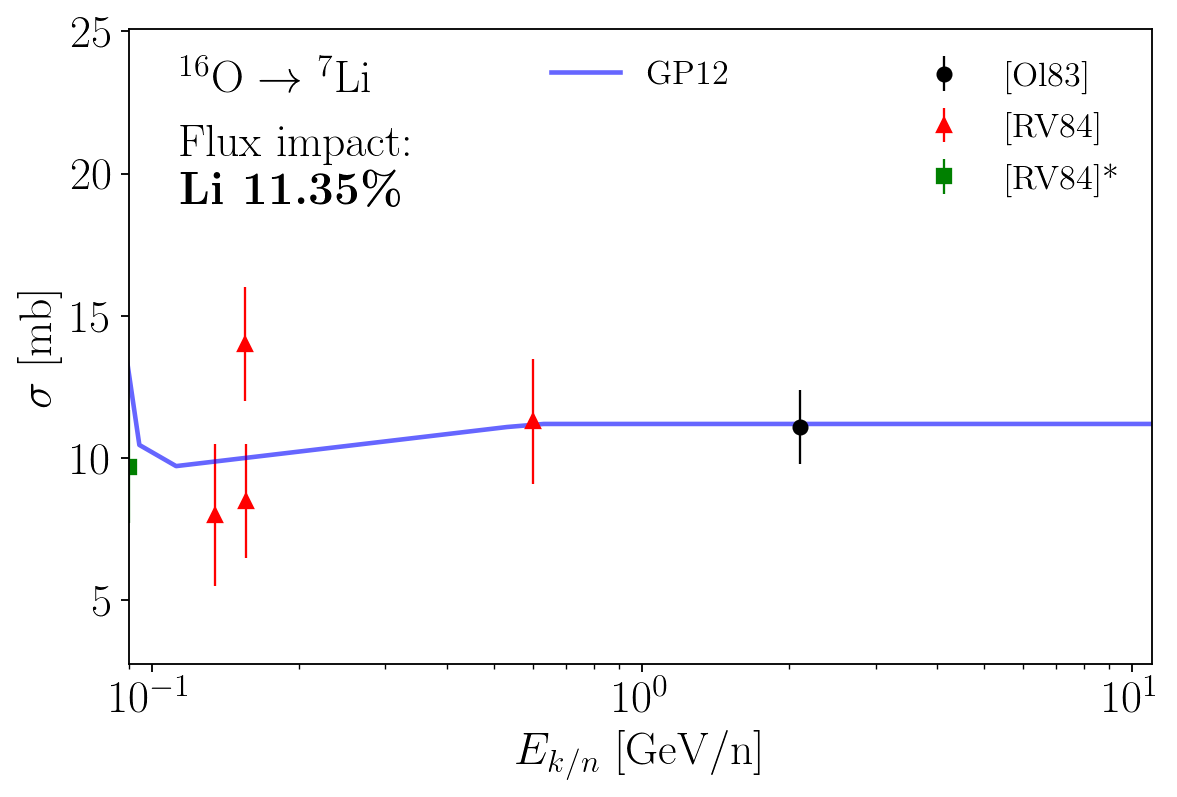}  &  
\includegraphics[width=0.32\textwidth]{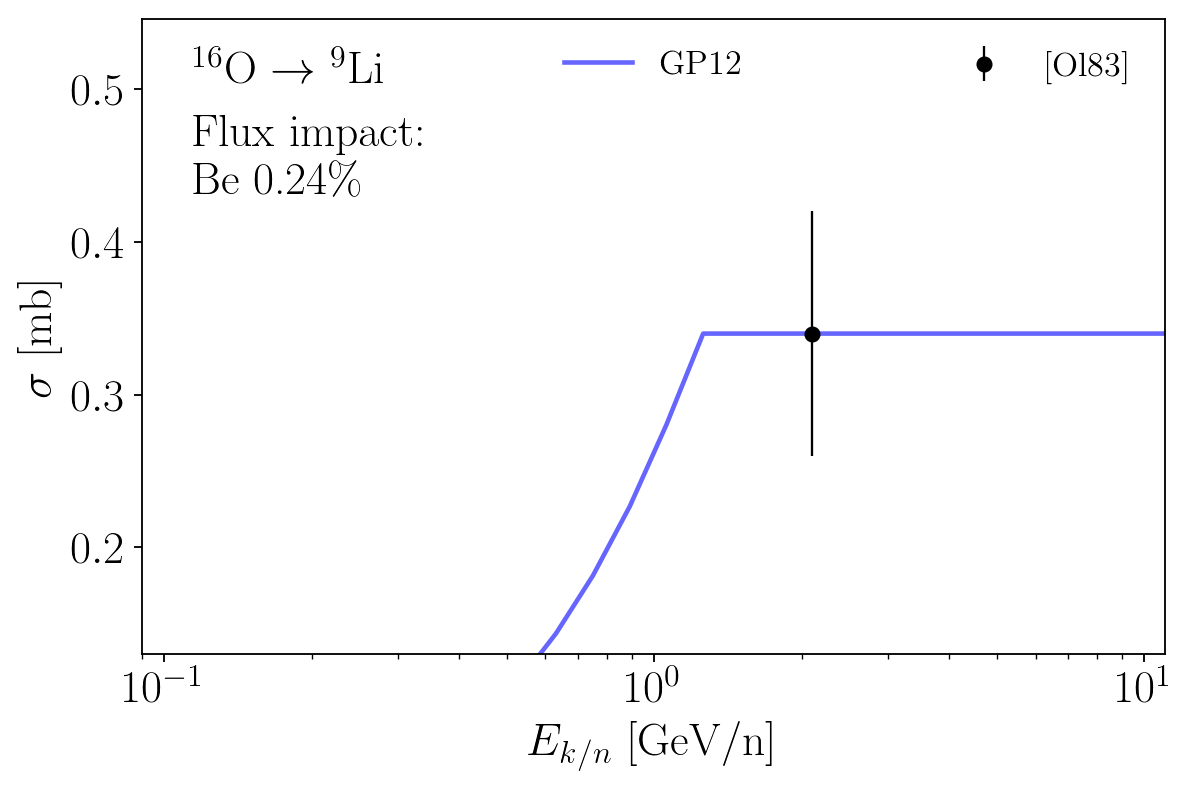}  \\ 
\includegraphics[width=0.32\textwidth]{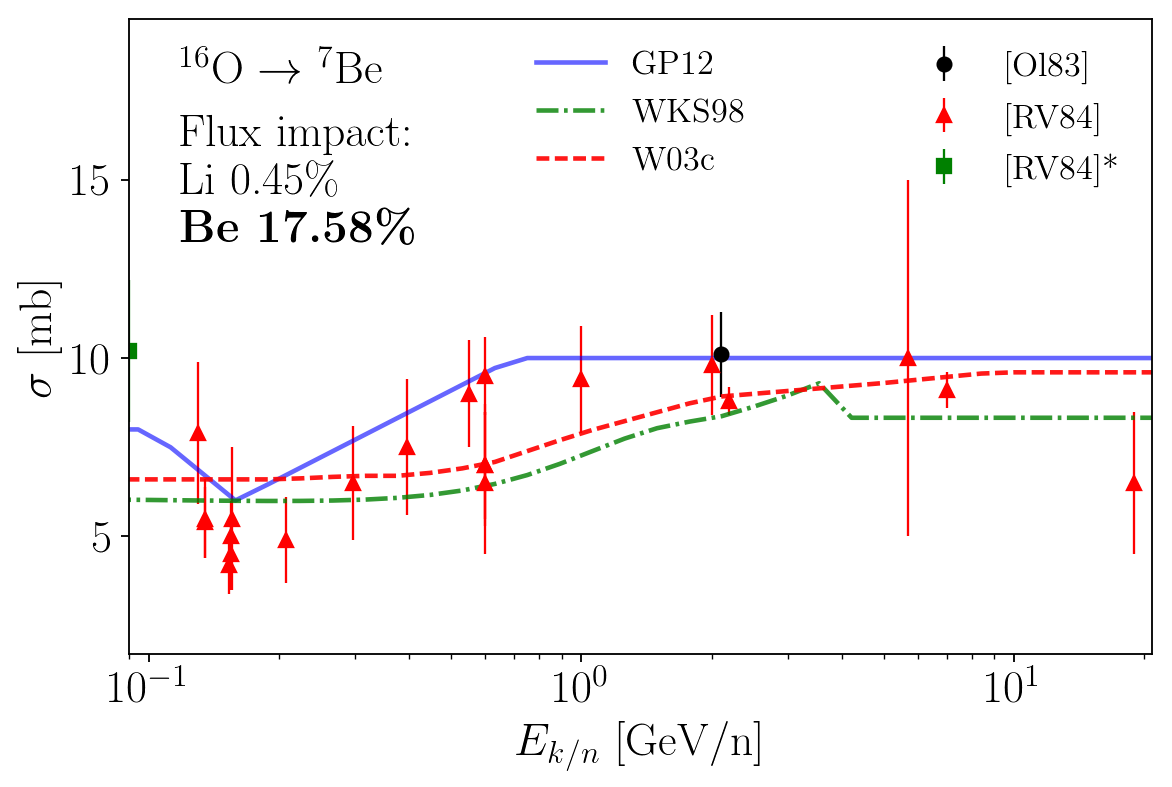}  &  
\includegraphics[width=0.32\textwidth]{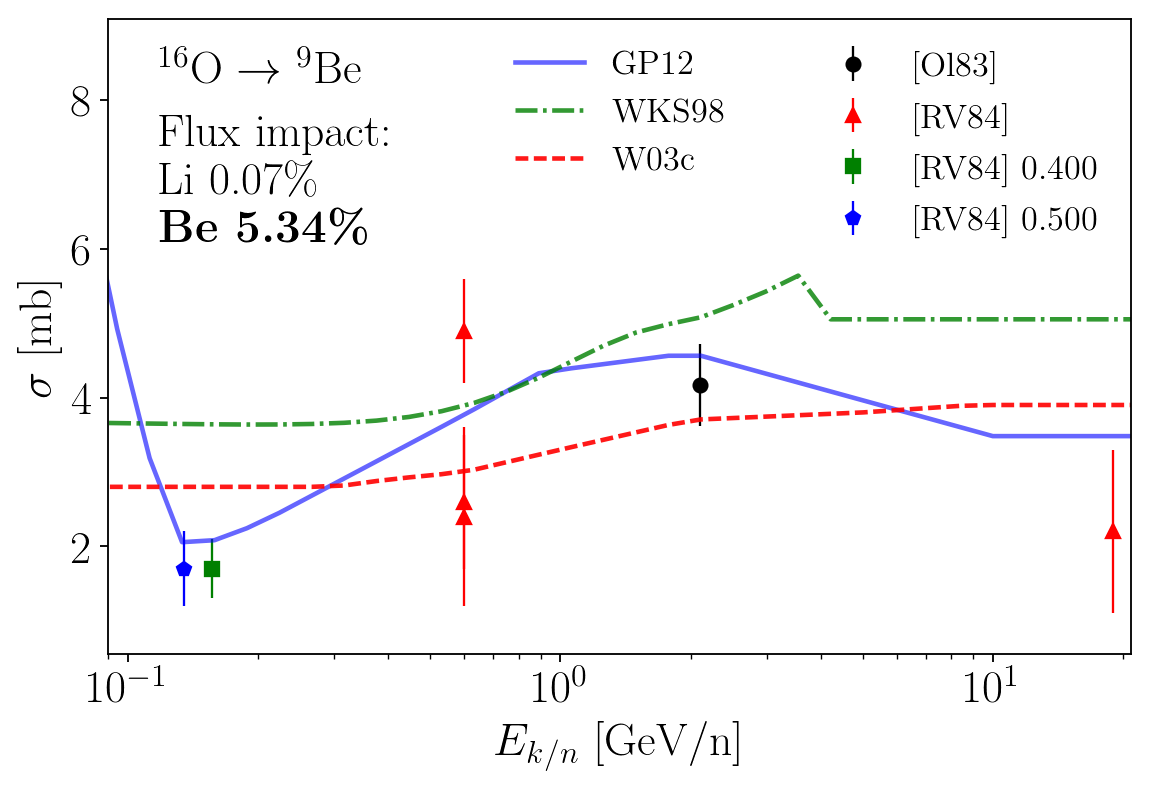}  &  
\includegraphics[width=0.32\textwidth]{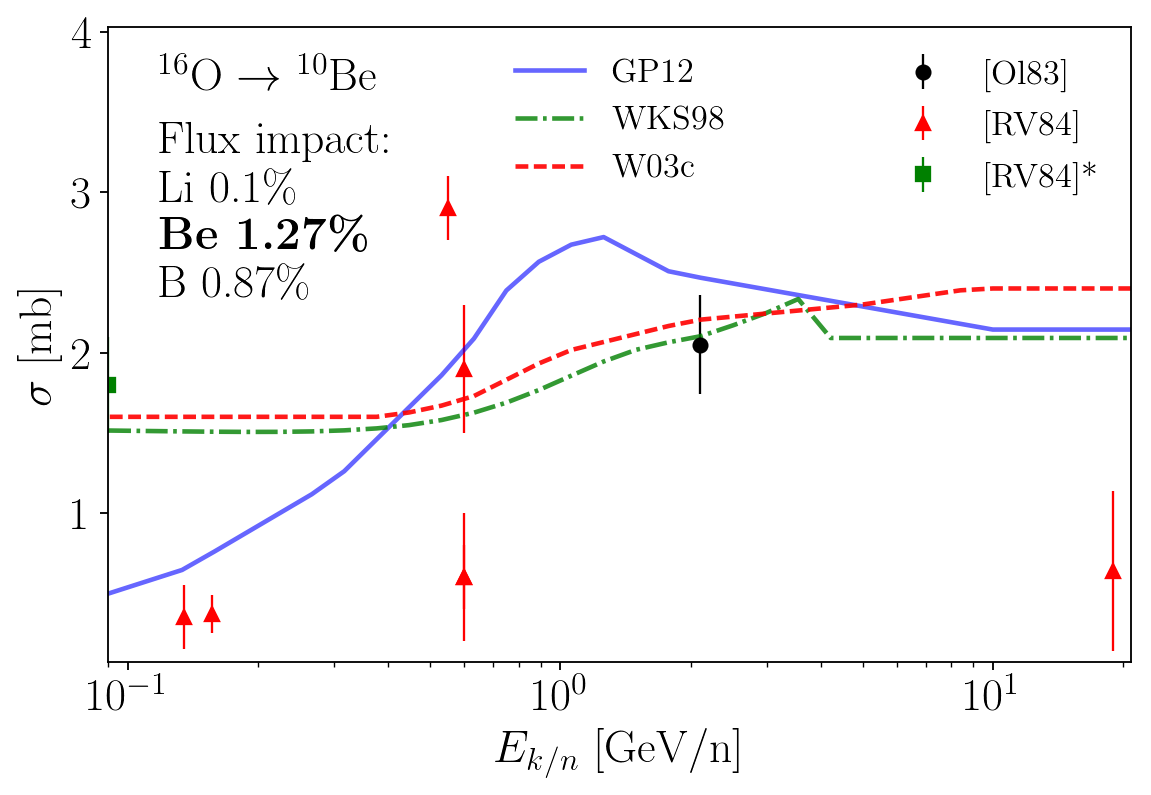}  \\ 
\includegraphics[width=0.32\textwidth]{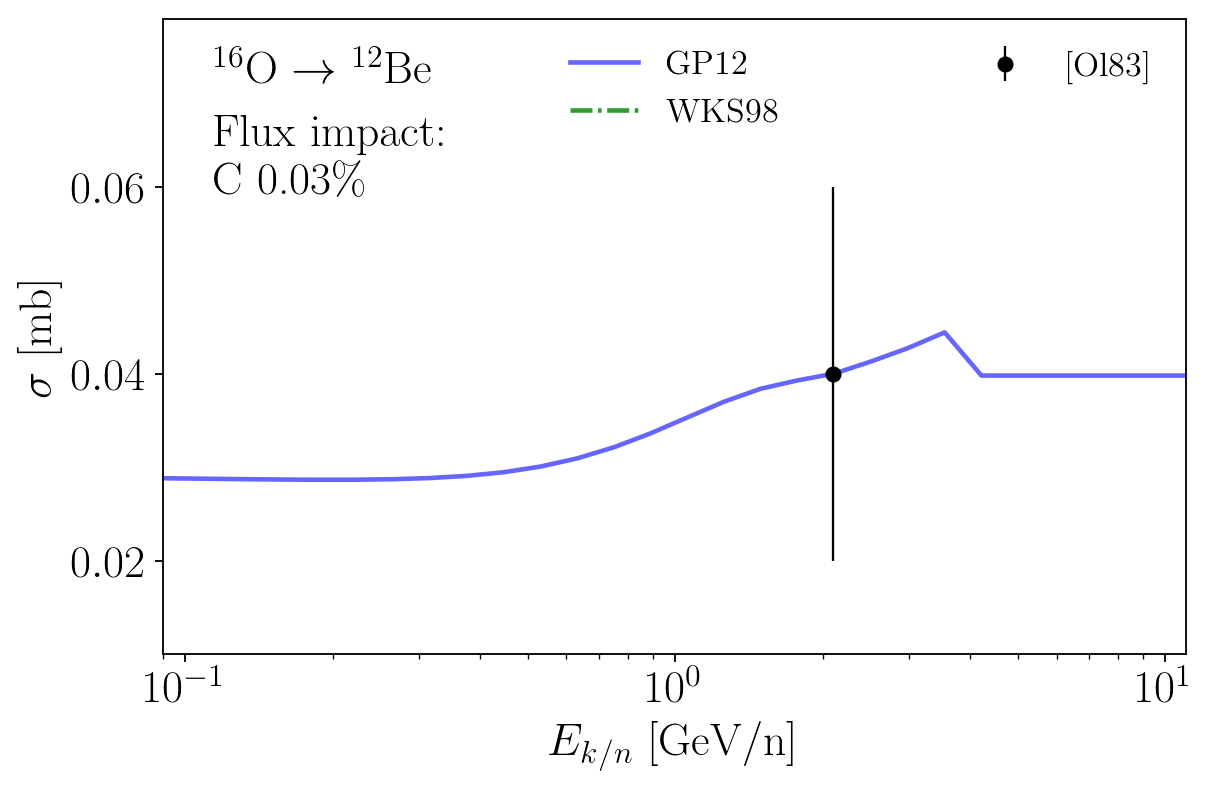}  &  
\includegraphics[width=0.32\textwidth]{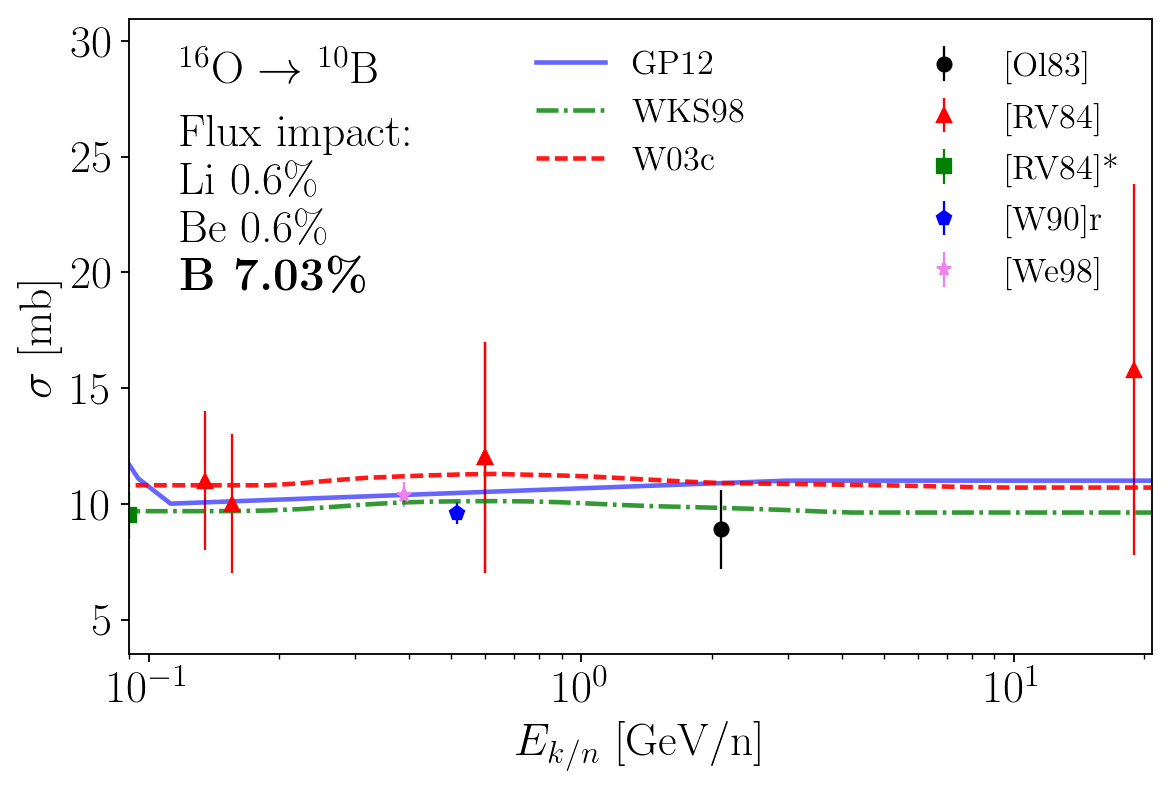}  &  
\includegraphics[width=0.32\textwidth]{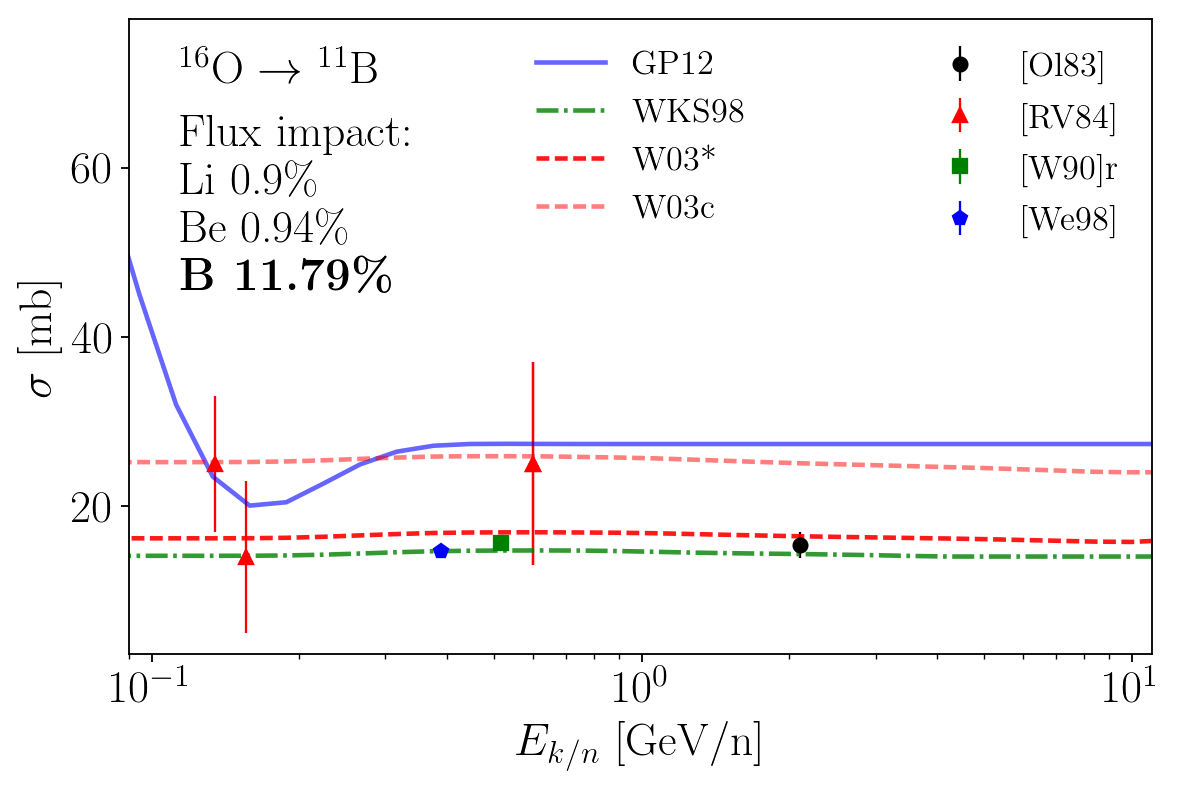}  \\ 
\includegraphics[width=0.32\textwidth]{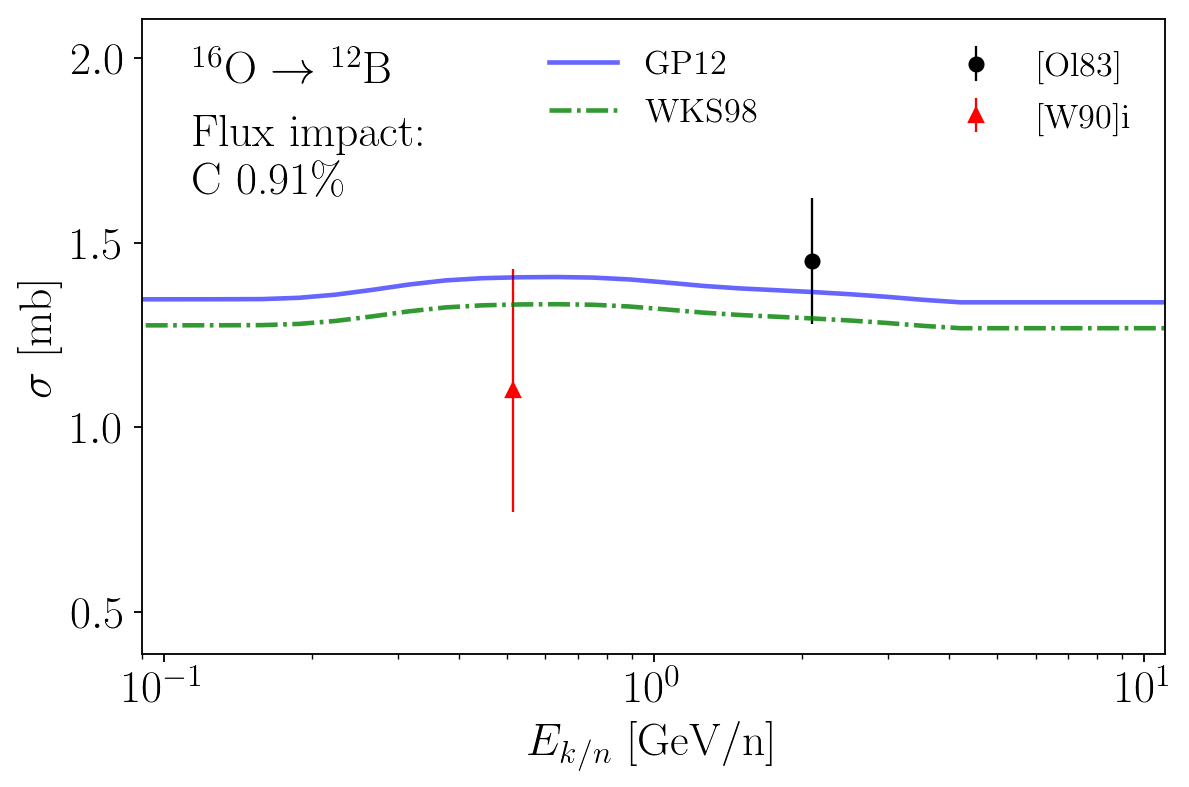}  &  
\includegraphics[width=0.32\textwidth]{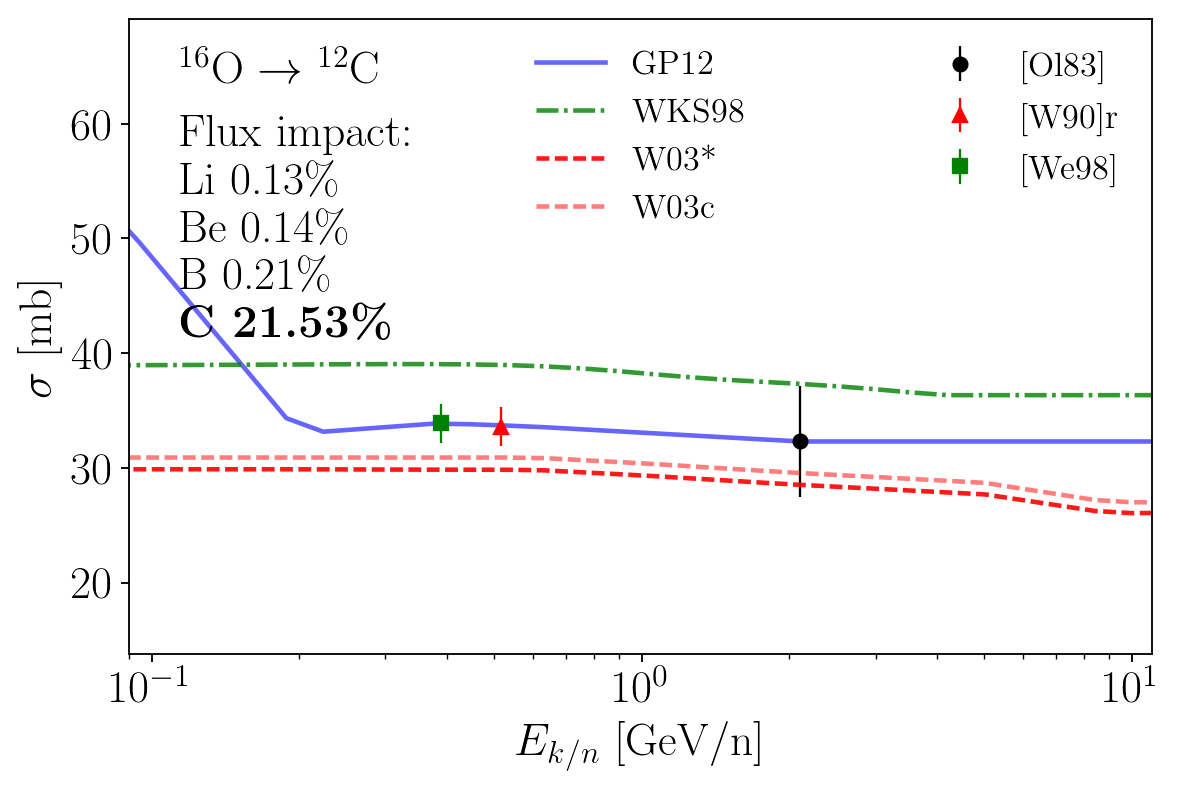}  &  
\includegraphics[width=0.32\textwidth]{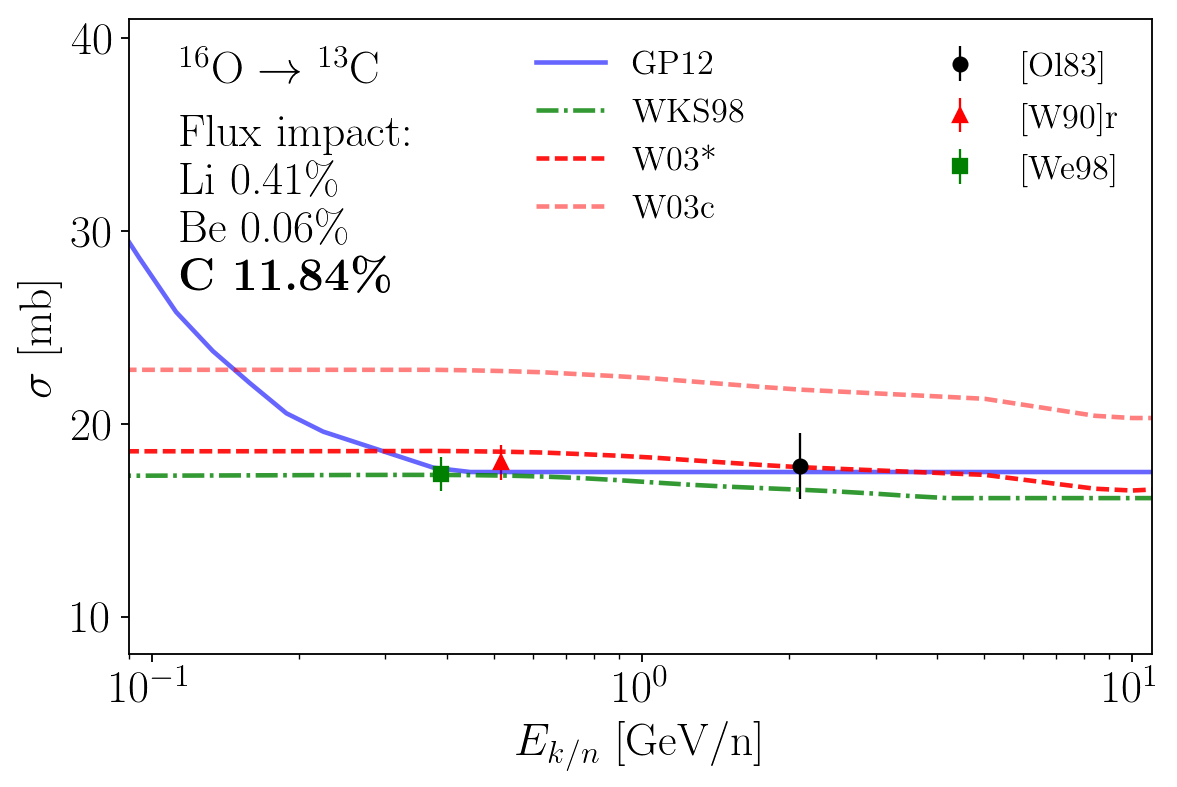}  \\ 
\includegraphics[width=0.32\textwidth]{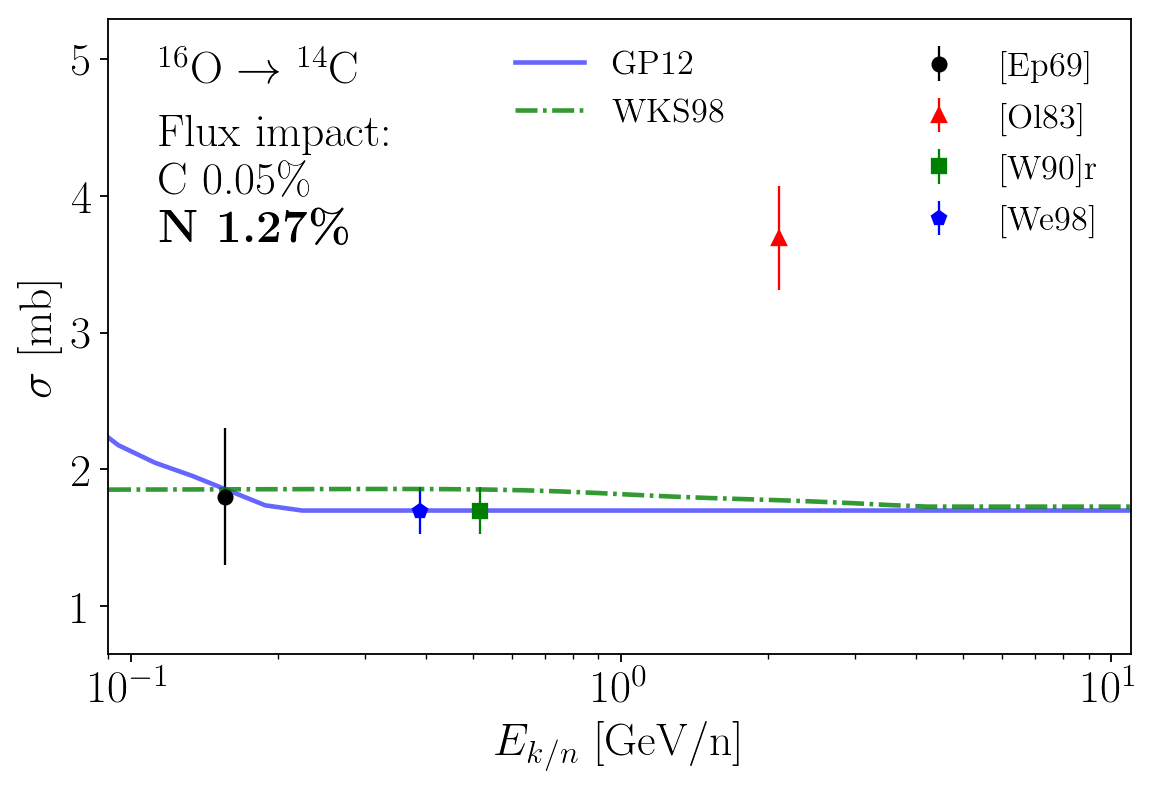}  &  
\includegraphics[width=0.32\textwidth]{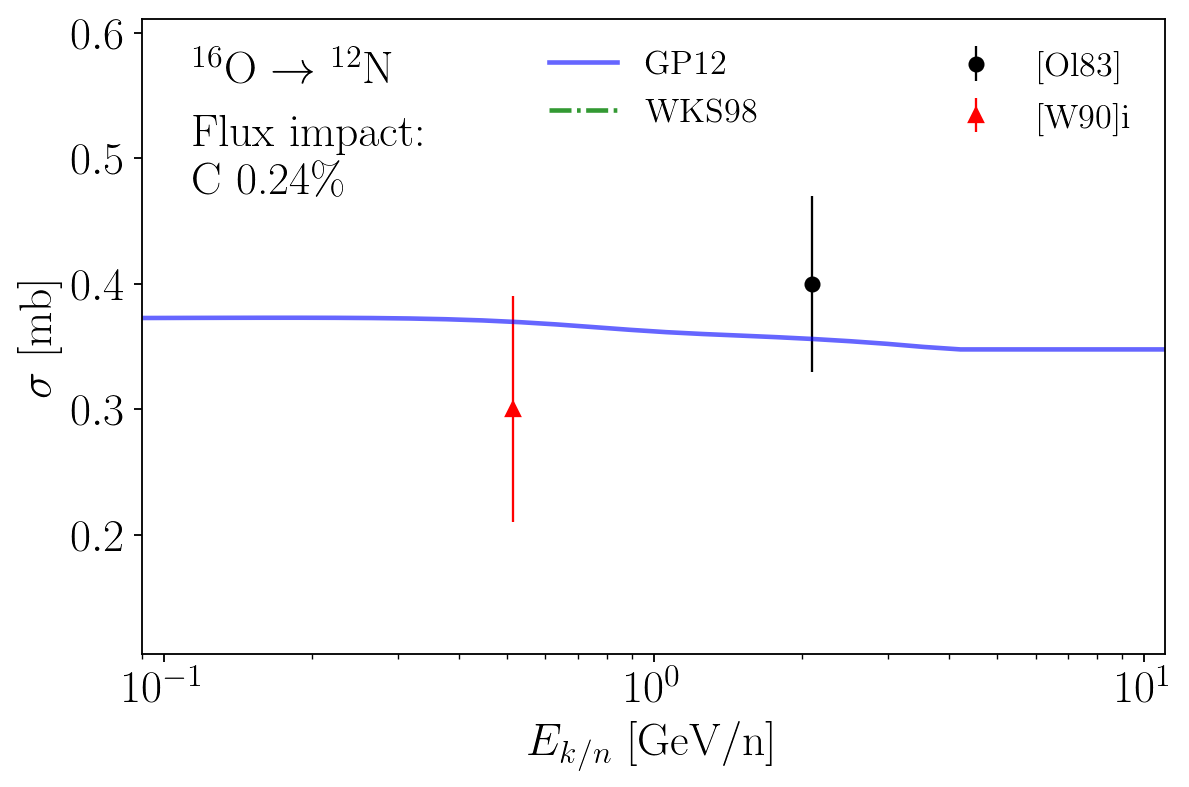}  &  
\includegraphics[width=0.32\textwidth]{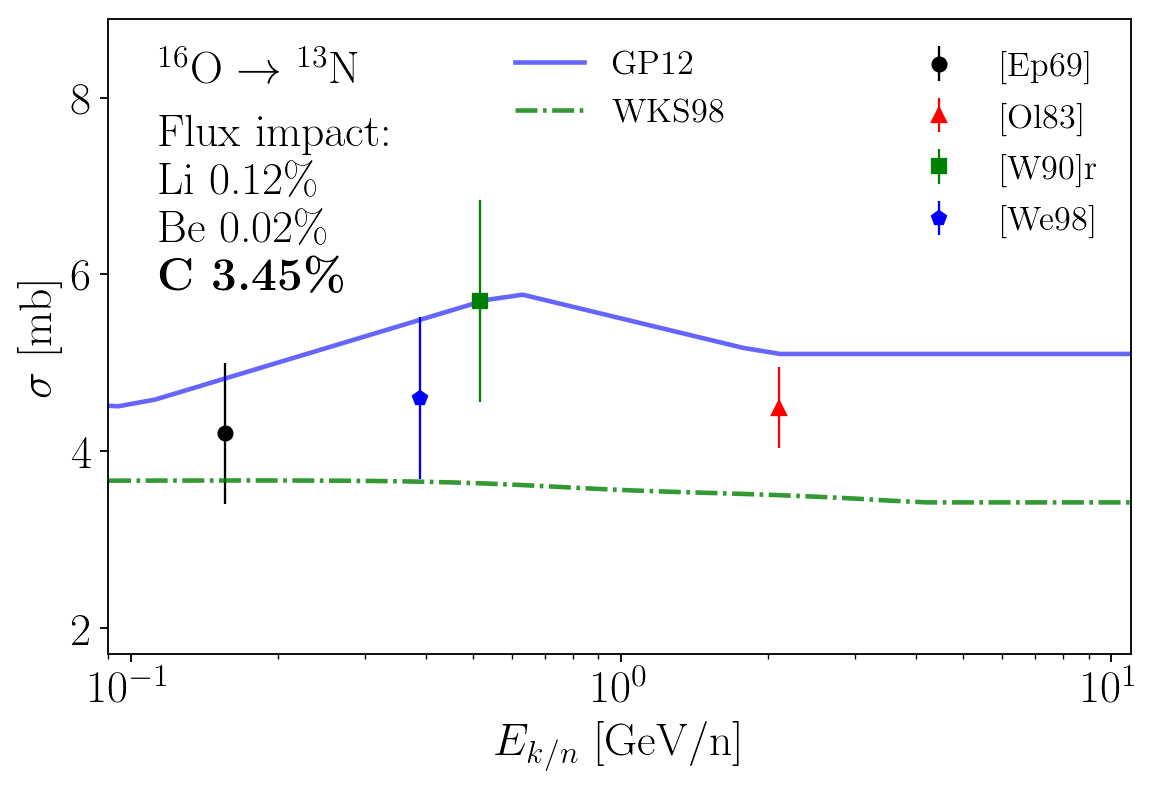}  \\ 
\includegraphics[width=0.32\textwidth]{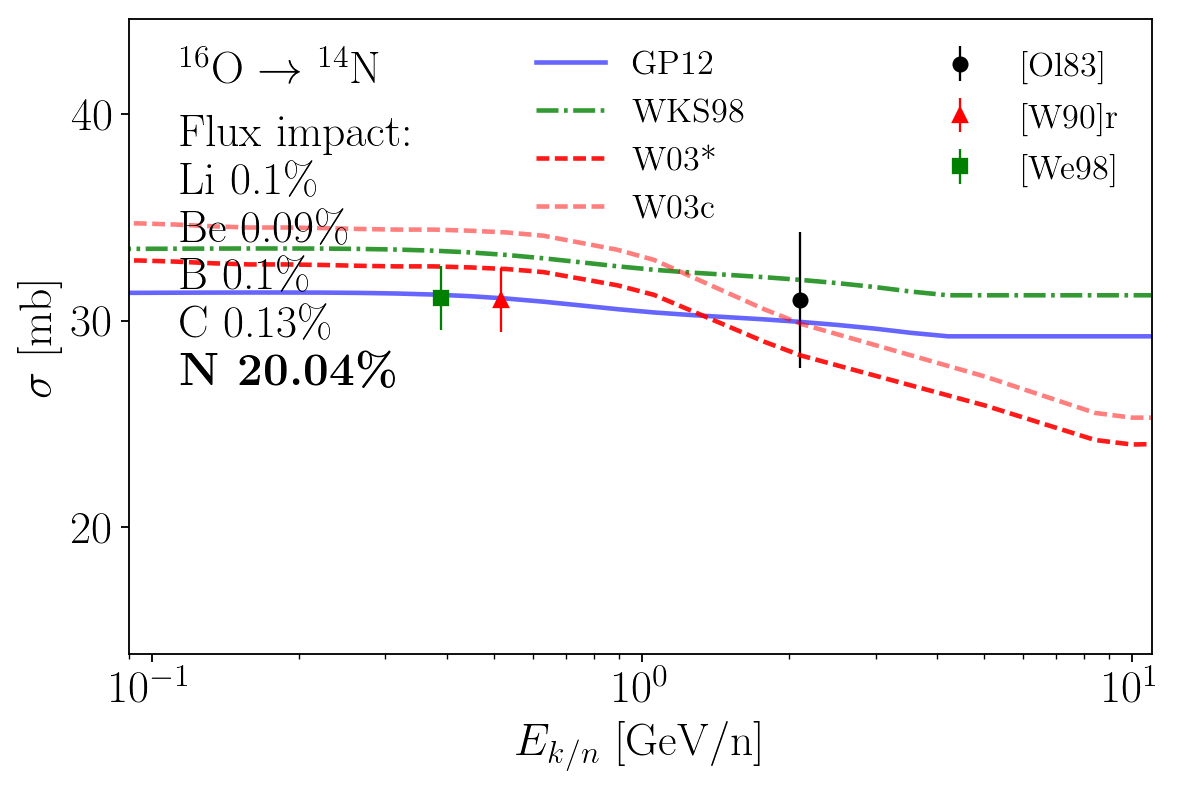}  &  
\includegraphics[width=0.32\textwidth]{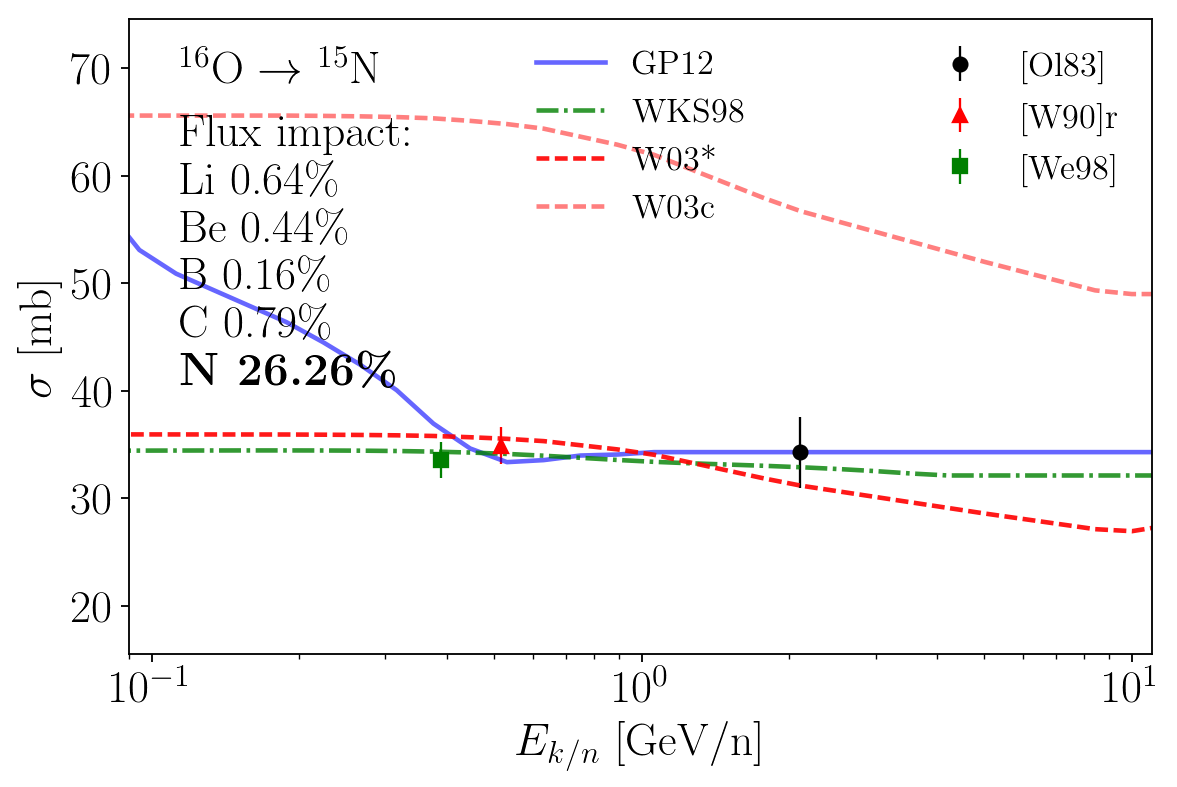}  &  
\includegraphics[width=0.32\textwidth]{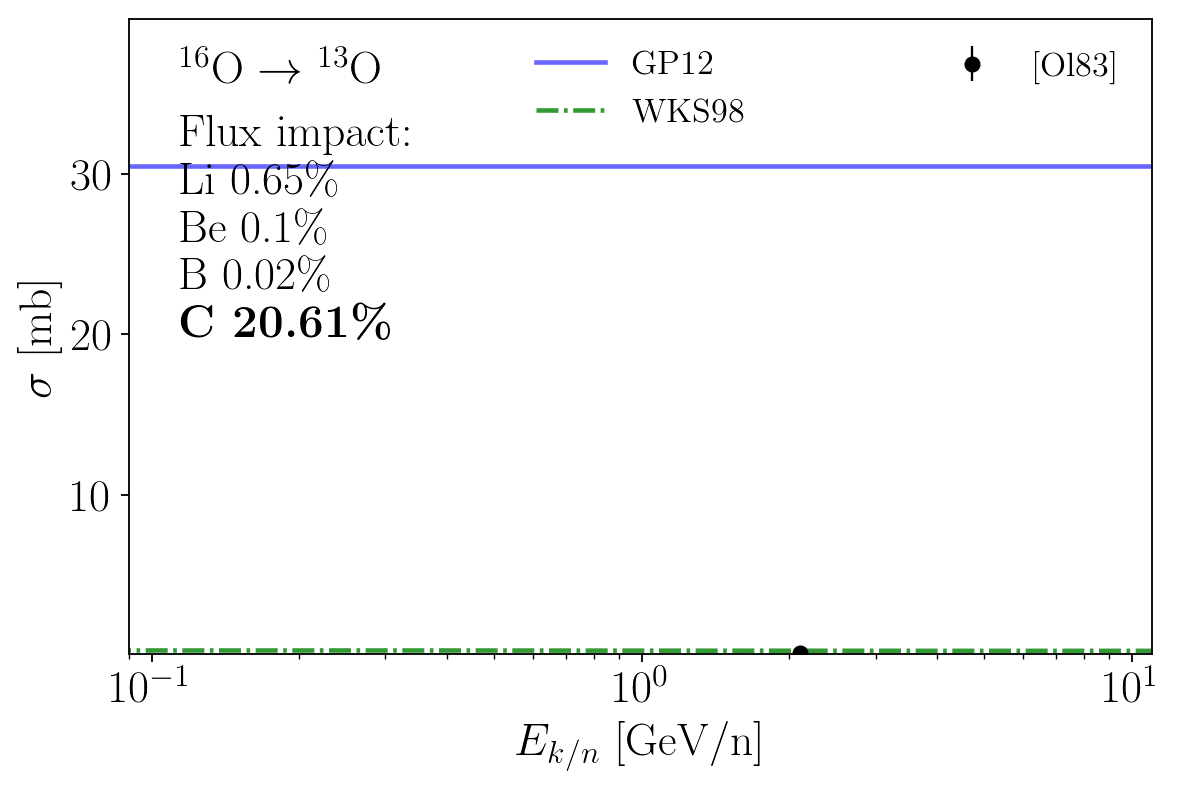}  \\ 
\includegraphics[width=0.32\textwidth]{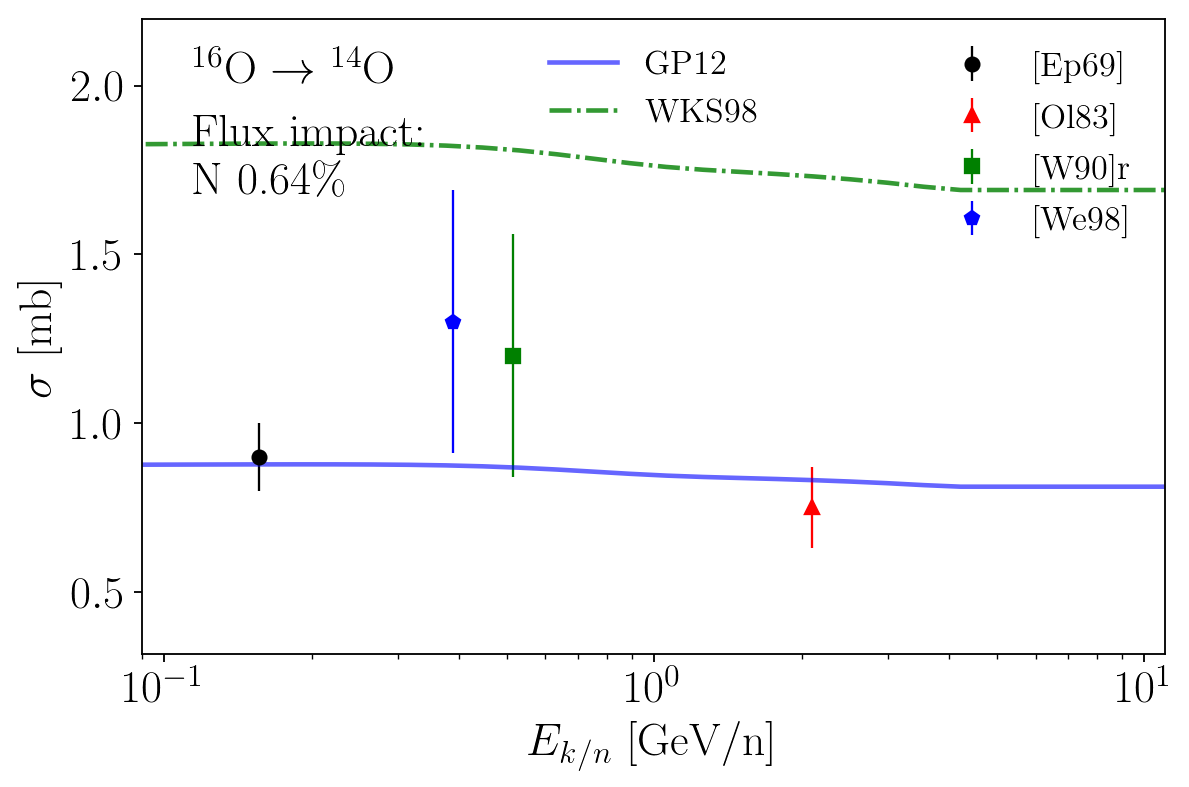}  &  
\includegraphics[width=0.32\textwidth]{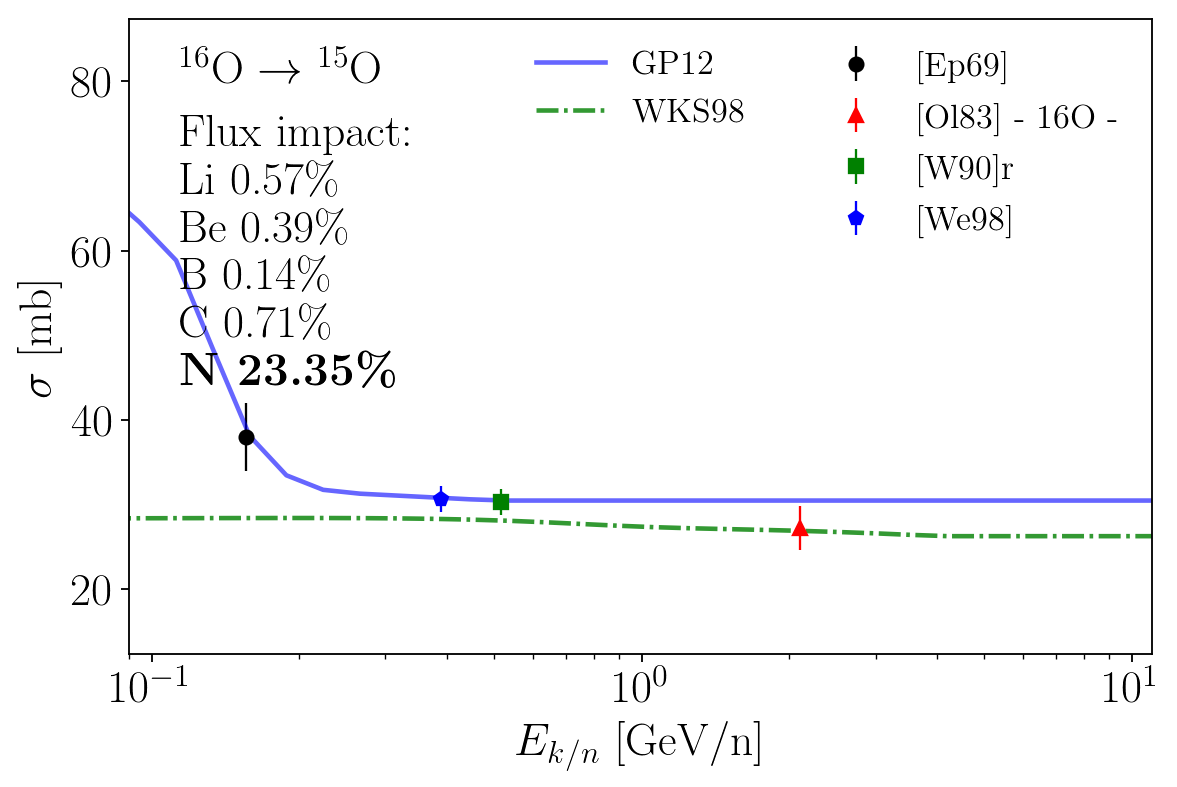}  &  \\ [3pt] 
\multicolumn{3}{c}{\bf Z=10{ \bf projectiles: $^{x}$Ne + H $\rightarrow$ $^{A}_ZX$}}\\ [3pt]
\multicolumn{3}{c}{\noindent\makebox[\linewidth]{\rule{\textwidth}{0.4pt}}}\\ [3pt]
\includegraphics[width=0.32\textwidth]{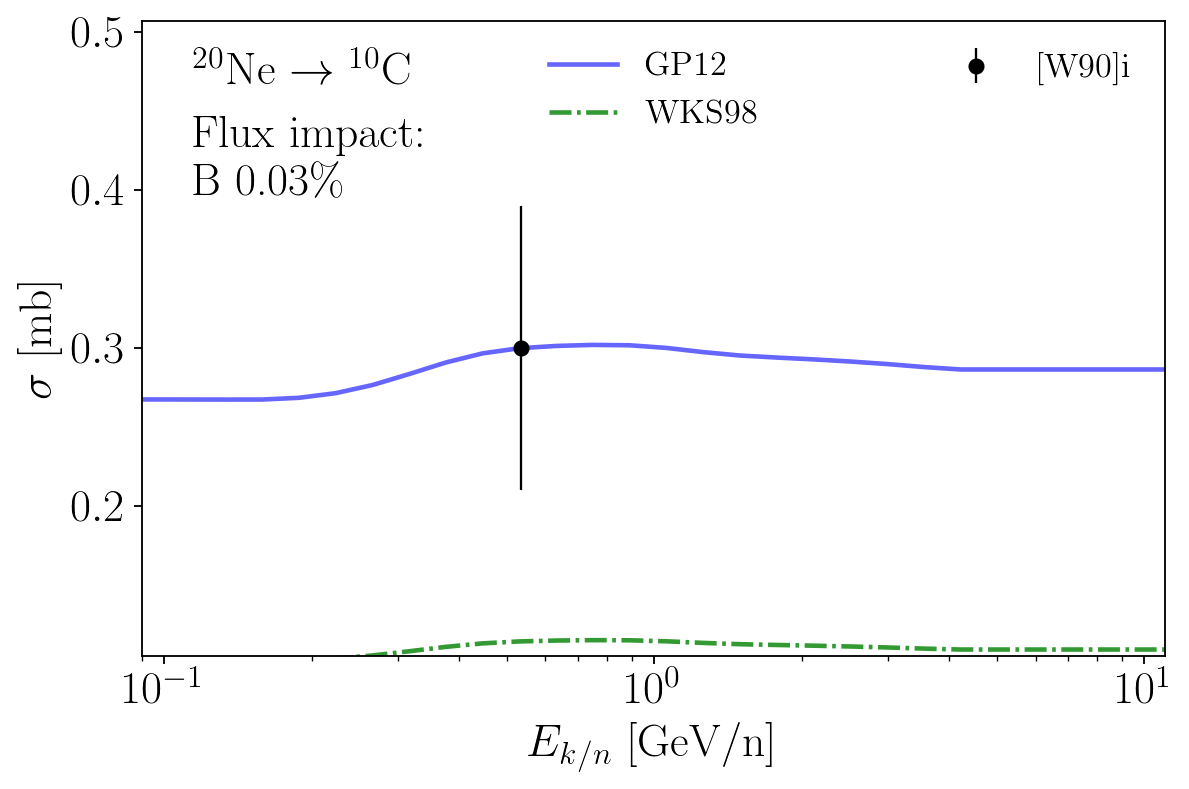}  &  
\includegraphics[width=0.32\textwidth]{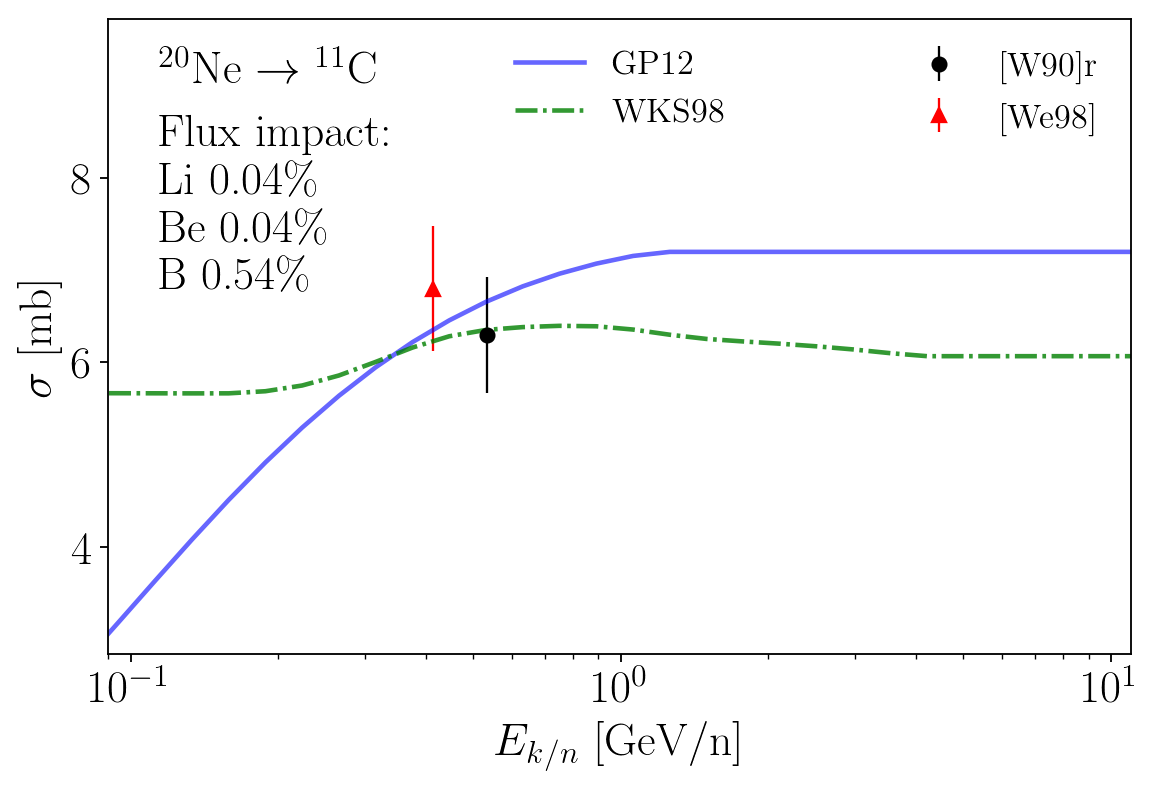}  &  
\includegraphics[width=0.32\textwidth]{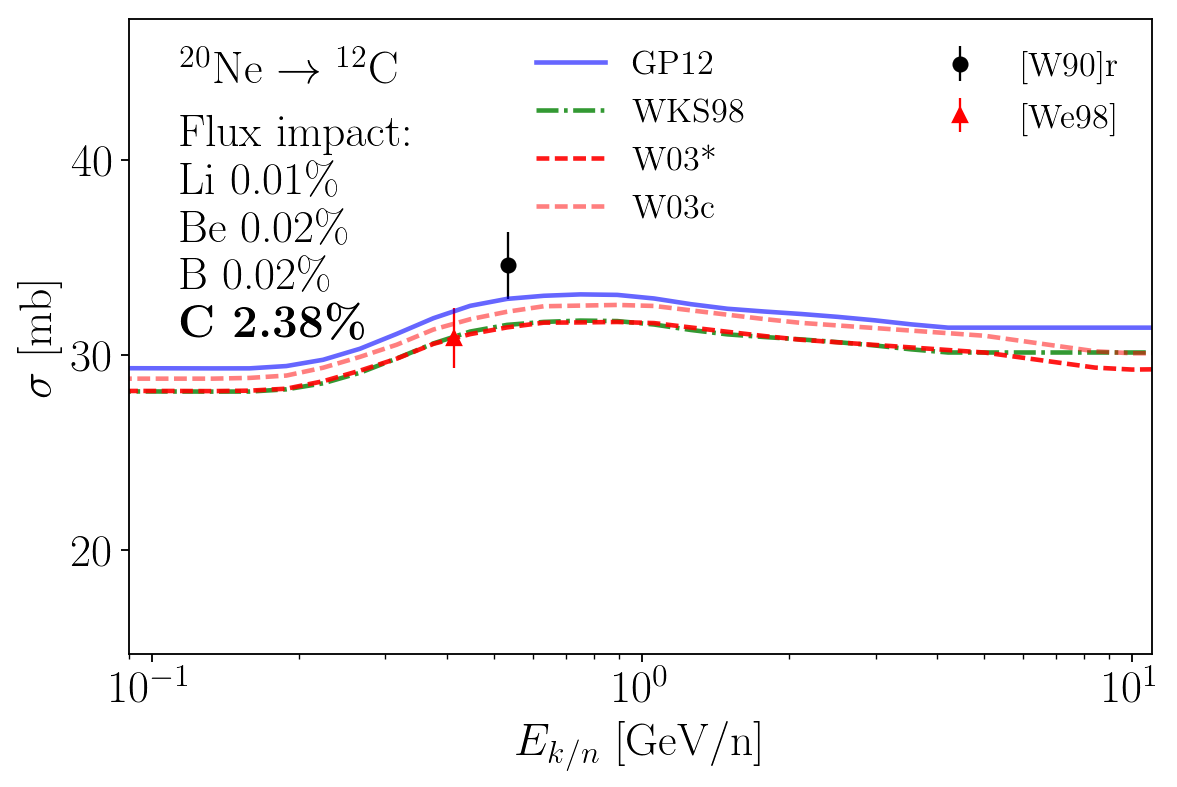}  \\ 
\includegraphics[width=0.32\textwidth]{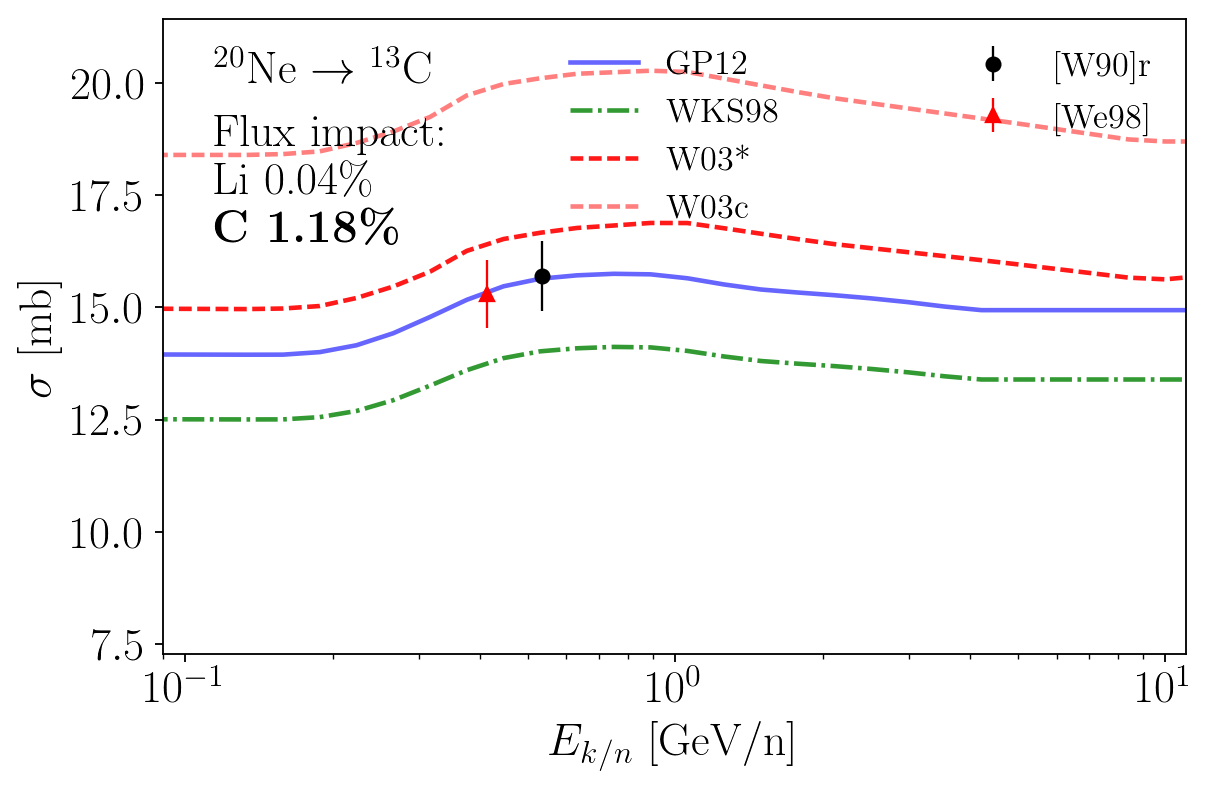}  &  
\includegraphics[width=0.32\textwidth]{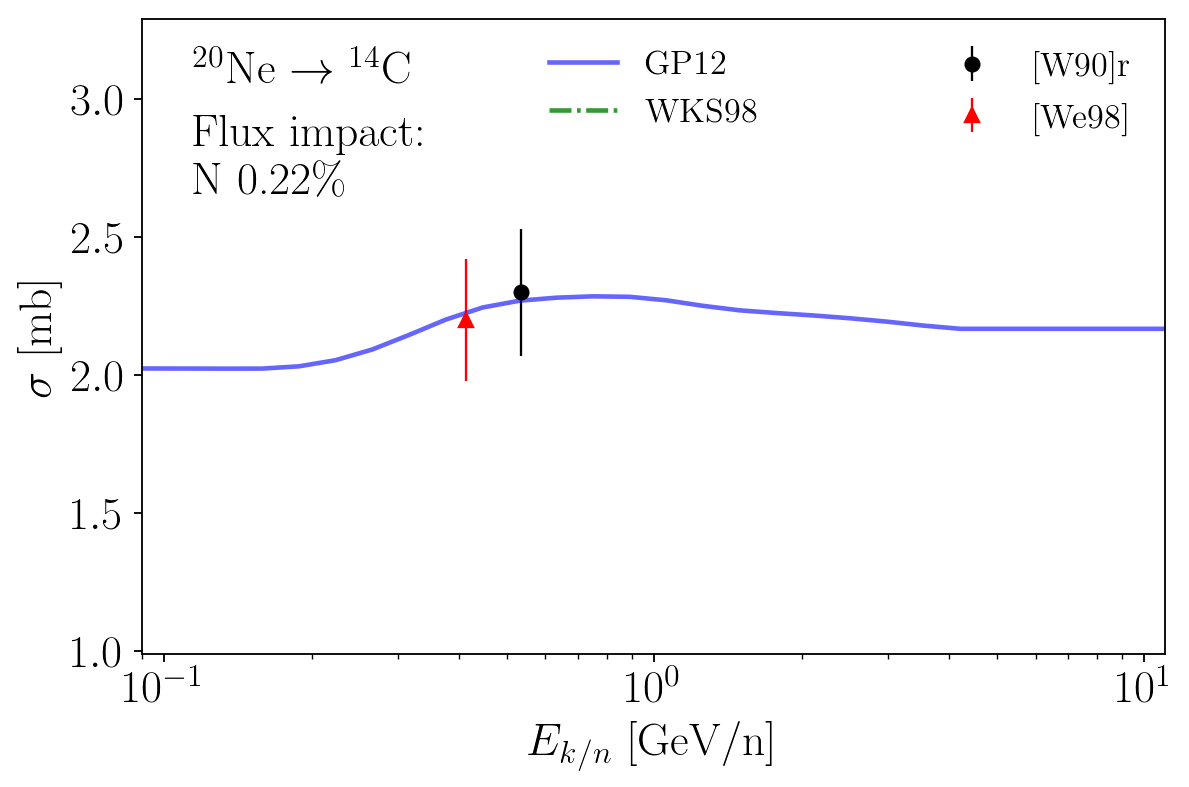}  &  
\includegraphics[width=0.32\textwidth]{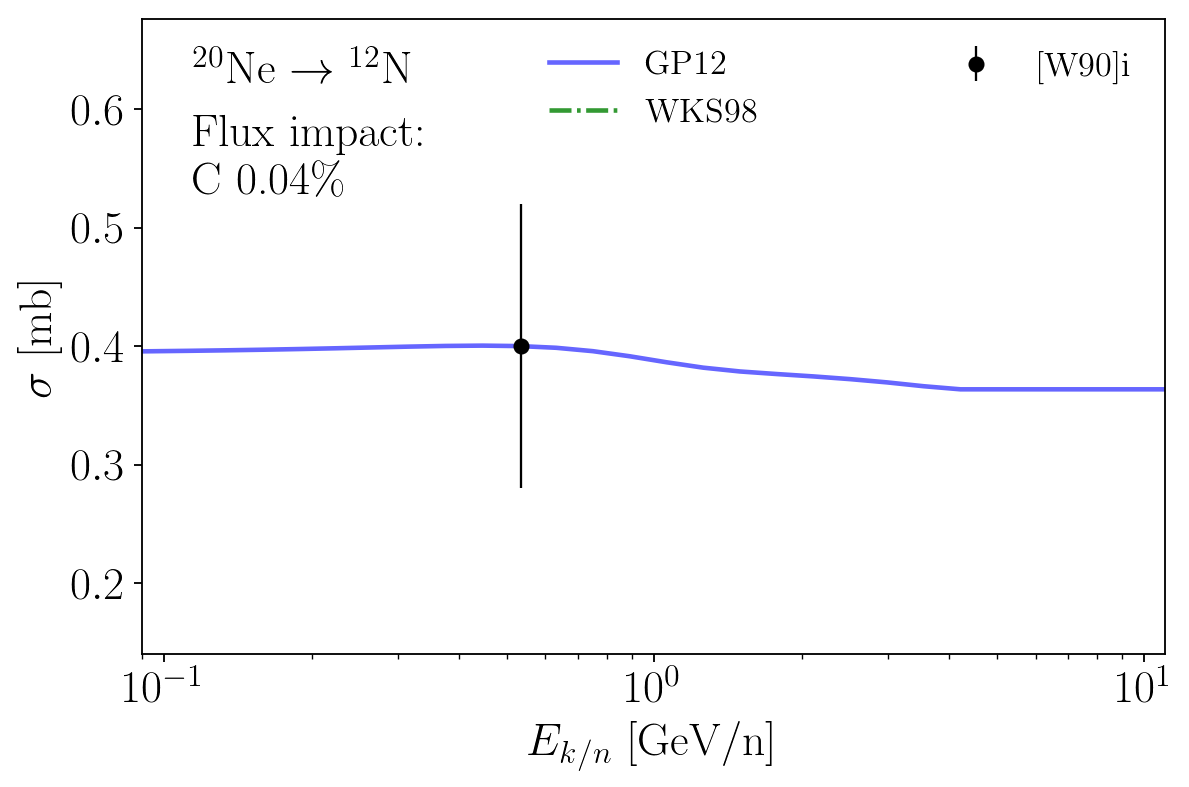}  \\ 
\includegraphics[width=0.32\textwidth]{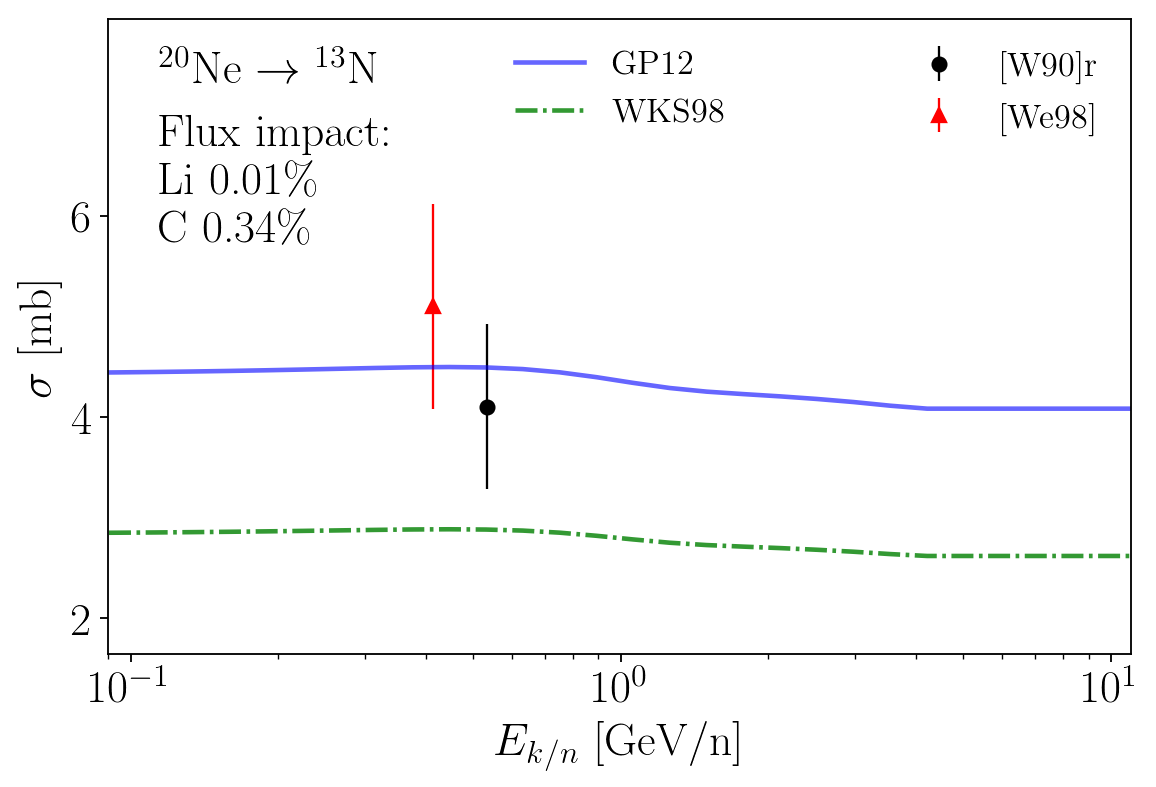}  &  
\includegraphics[width=0.32\textwidth]{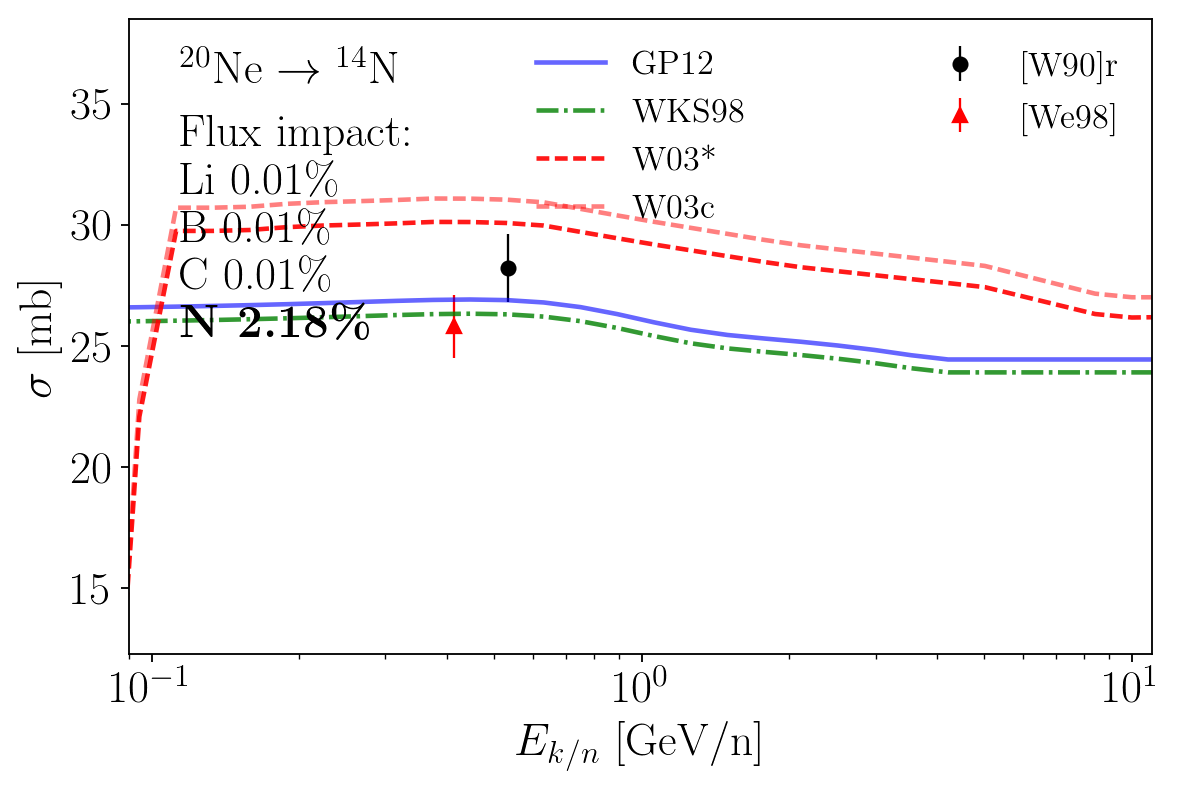}  &  
\includegraphics[width=0.32\textwidth]{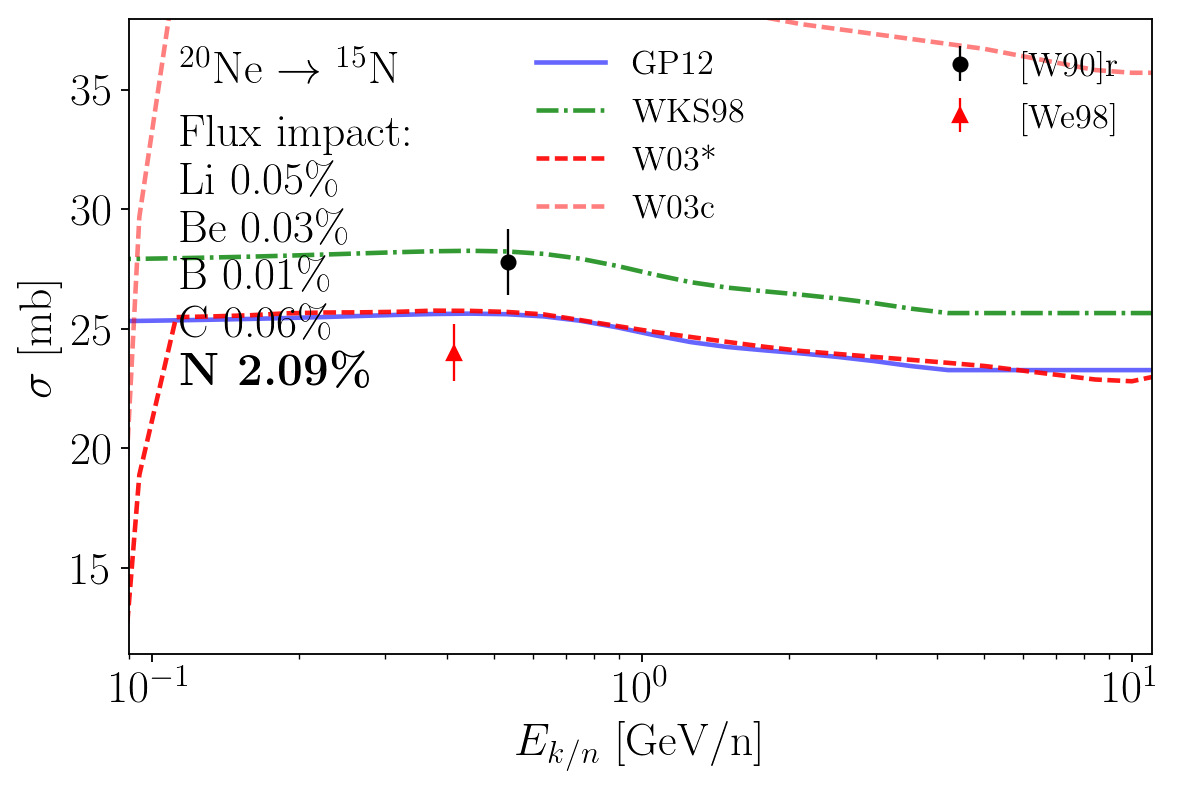}  \\ 
\includegraphics[width=0.32\textwidth]{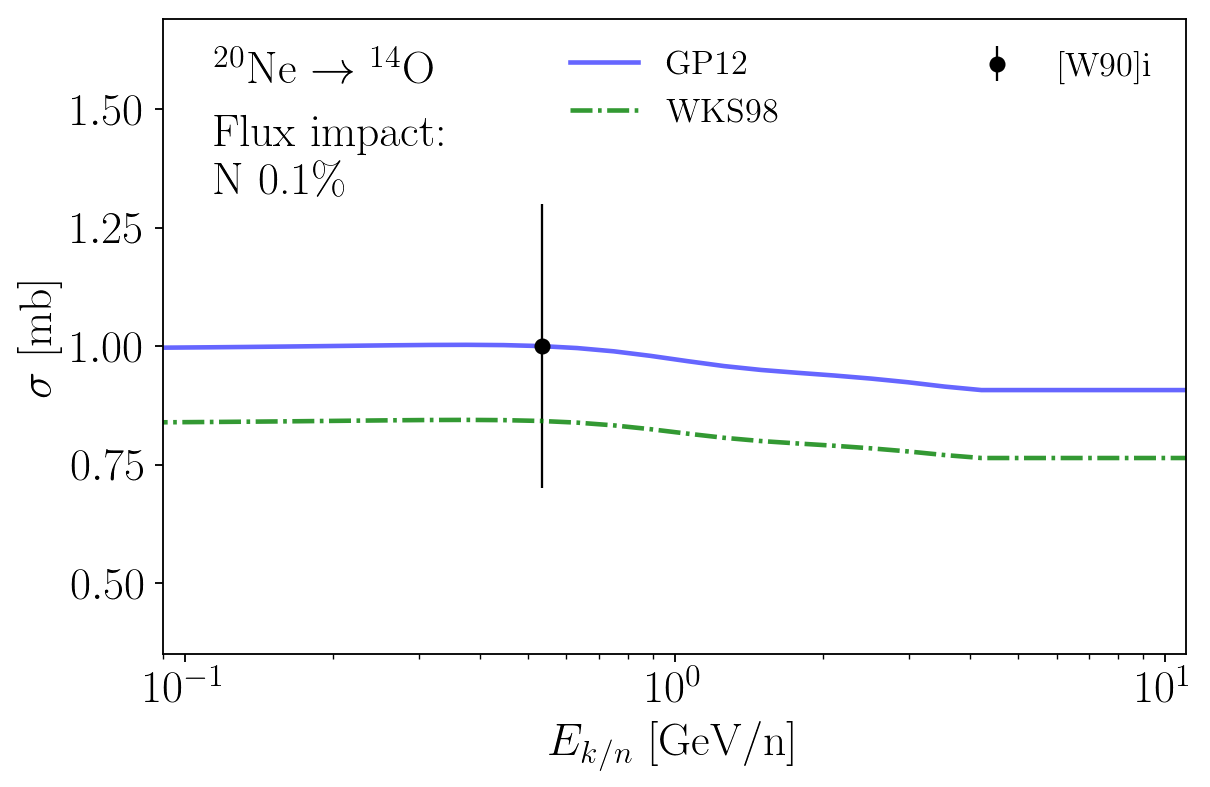}  &  
\includegraphics[width=0.32\textwidth]{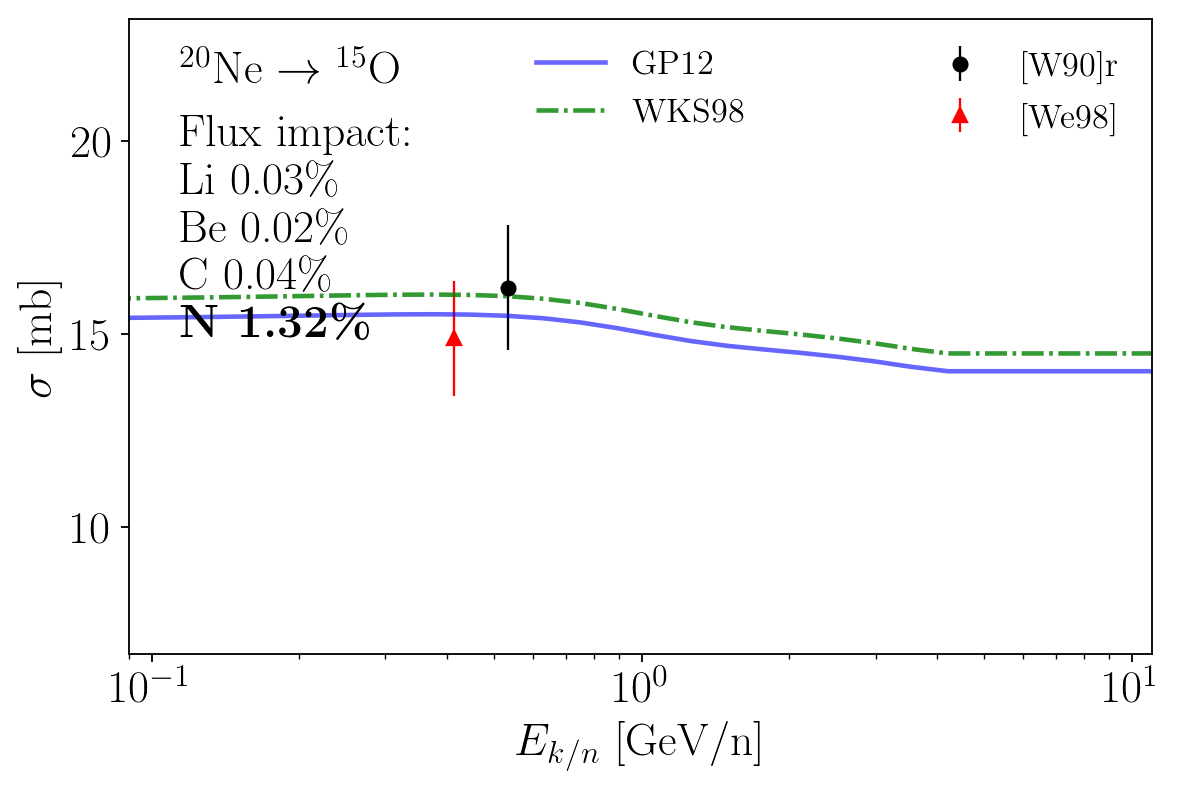}  &  
\includegraphics[width=0.32\textwidth]{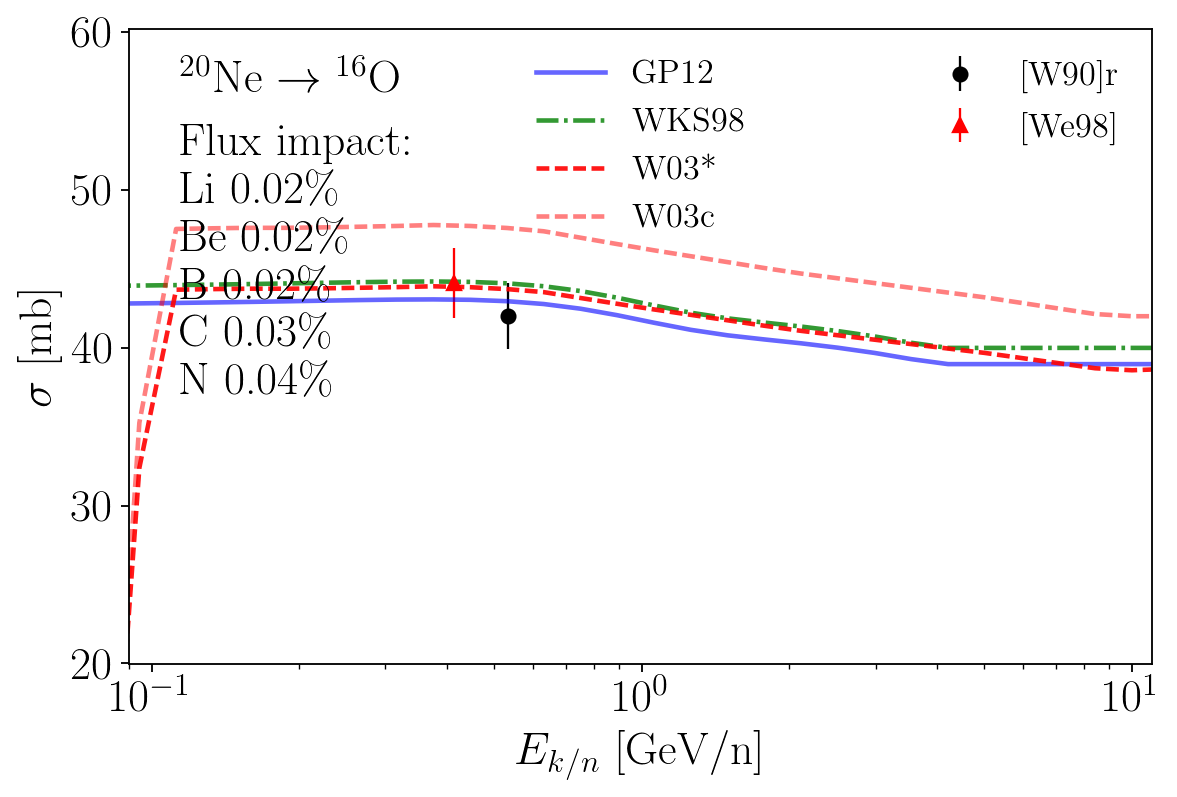}  \\ 
\includegraphics[width=0.32\textwidth]{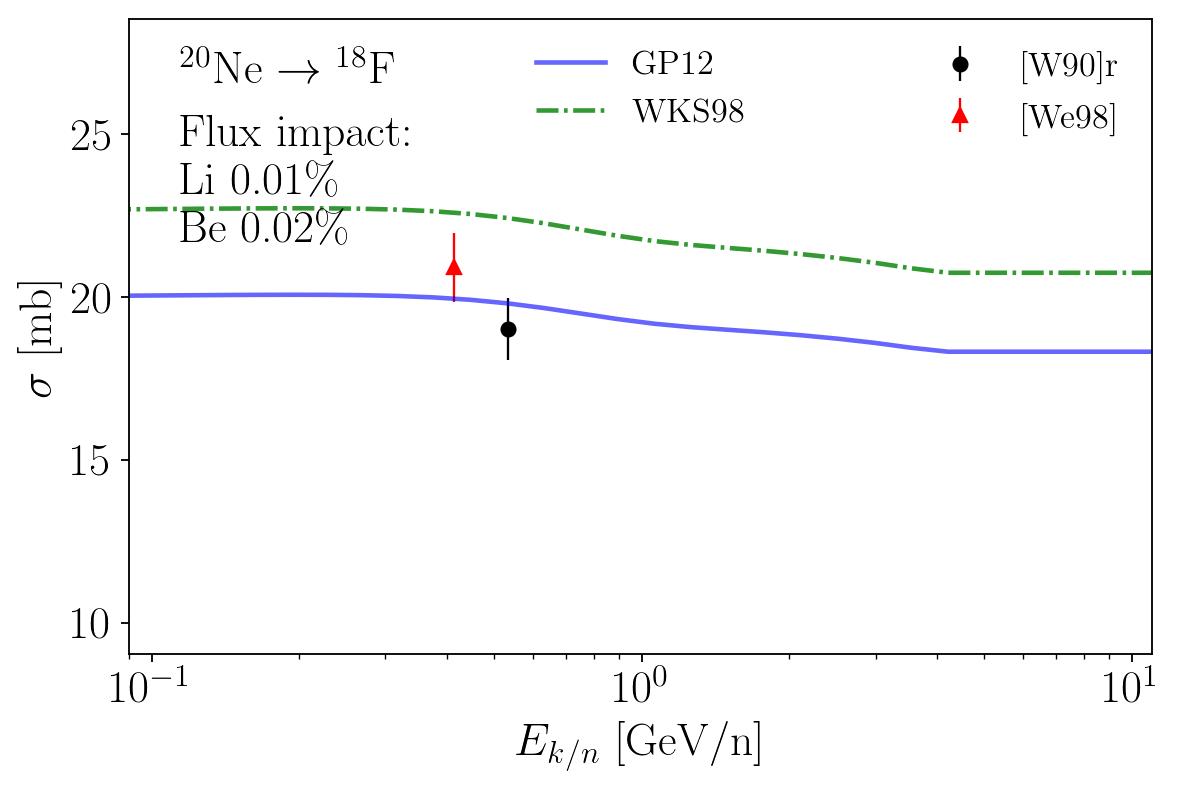}  &  
\includegraphics[width=0.32\textwidth]{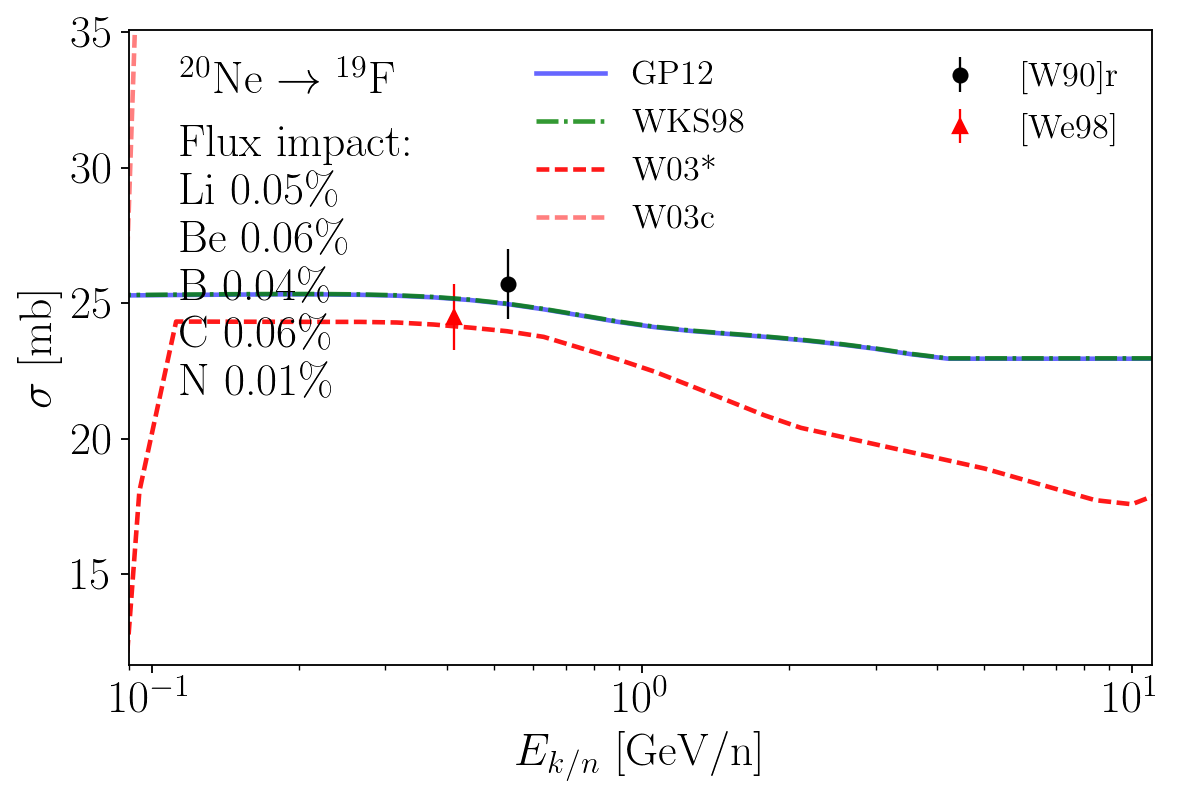}  &  
\includegraphics[width=0.32\textwidth]{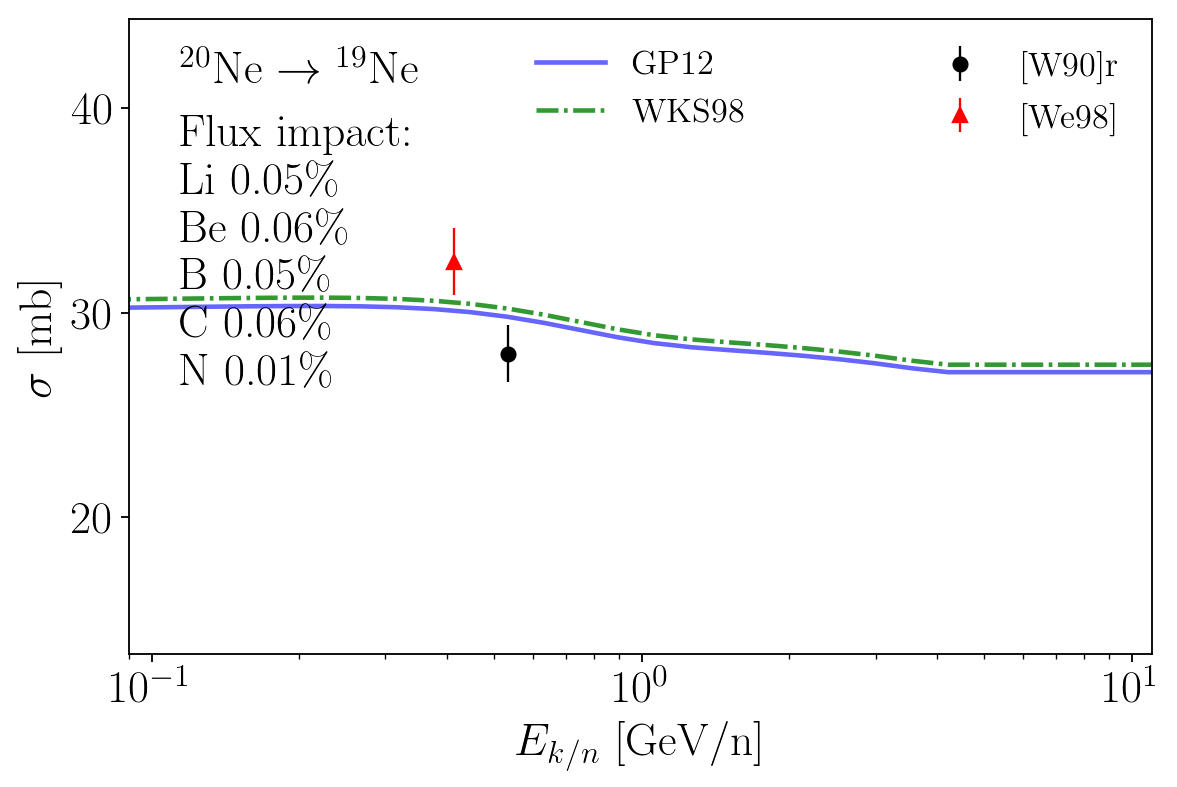}  \\ 
\includegraphics[width=0.32\textwidth]{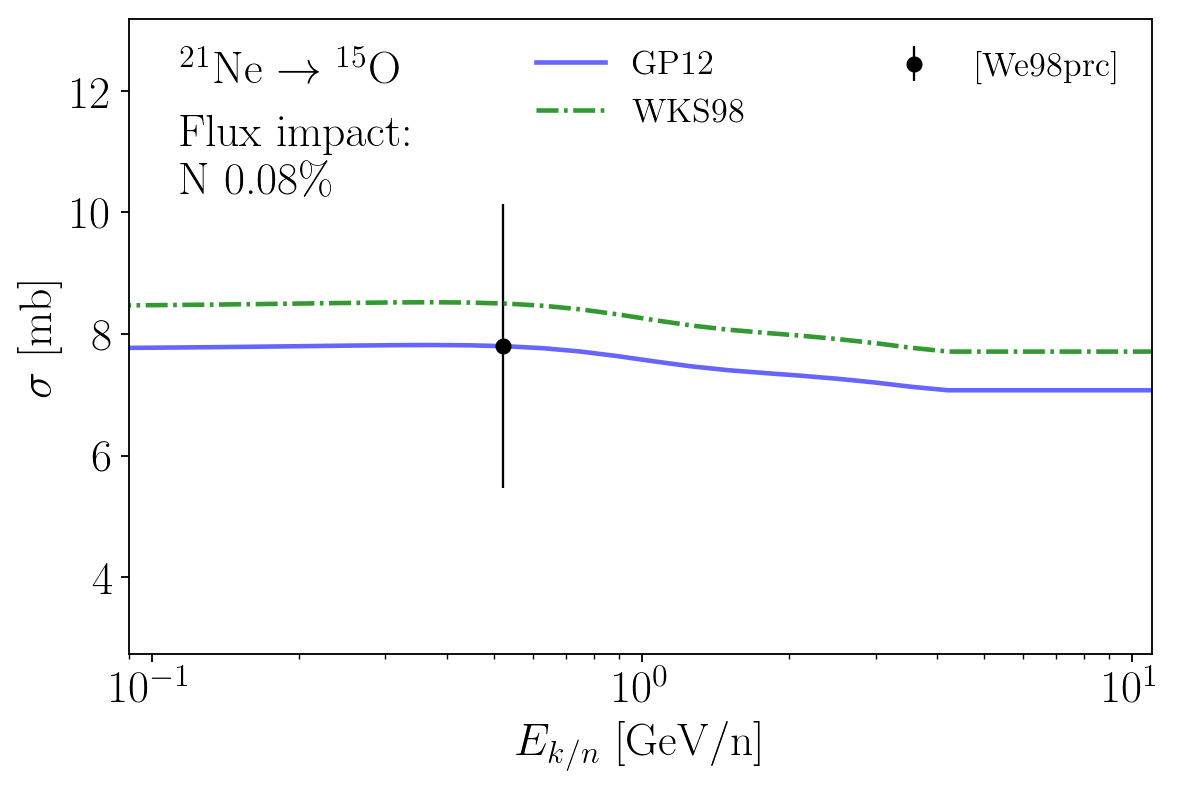}  &  
\includegraphics[width=0.32\textwidth]{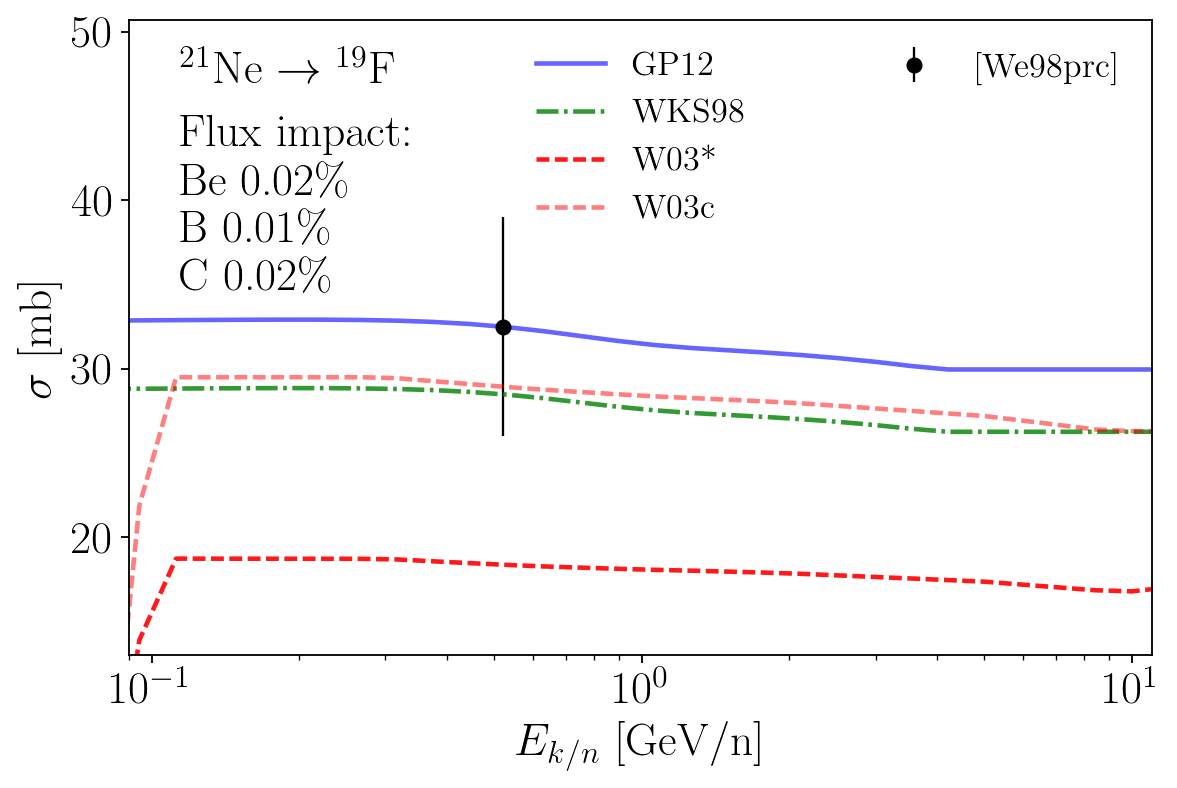}  &  
\includegraphics[width=0.32\textwidth]{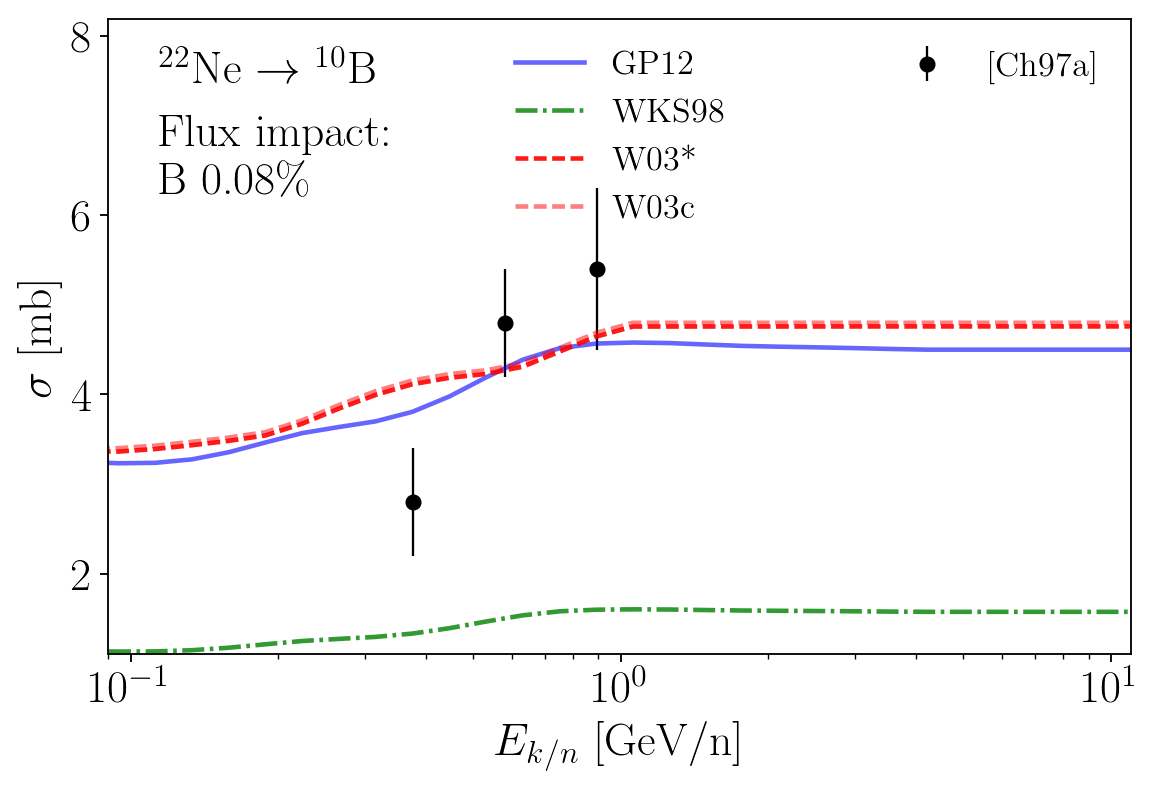}  \\ 
\includegraphics[width=0.32\textwidth]{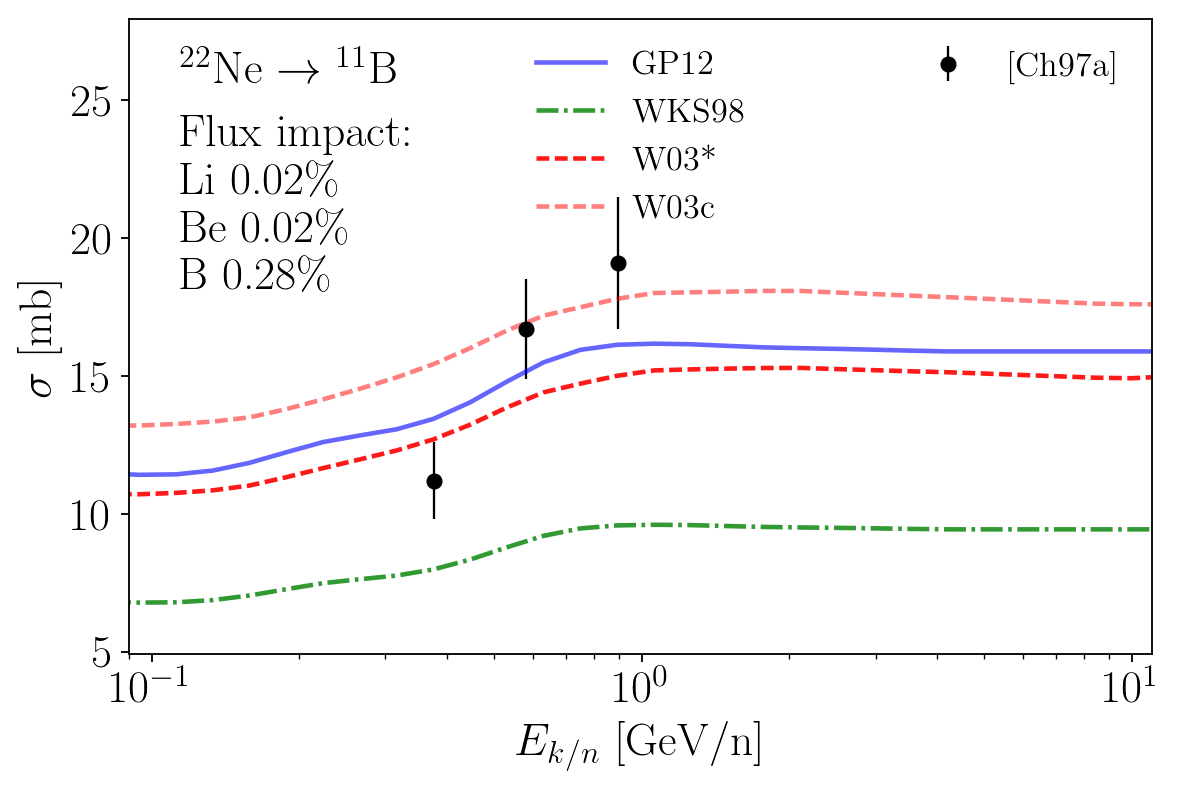}  &  
\includegraphics[width=0.32\textwidth]{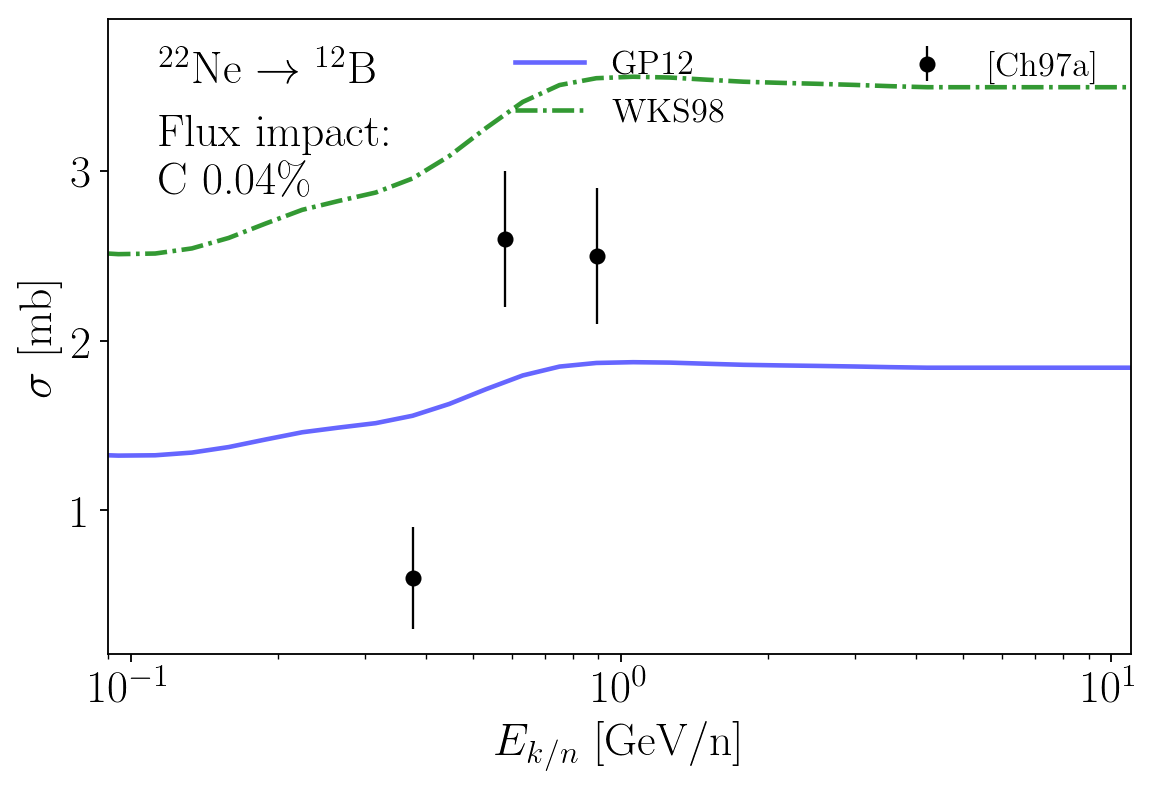}  &  
\includegraphics[width=0.32\textwidth]{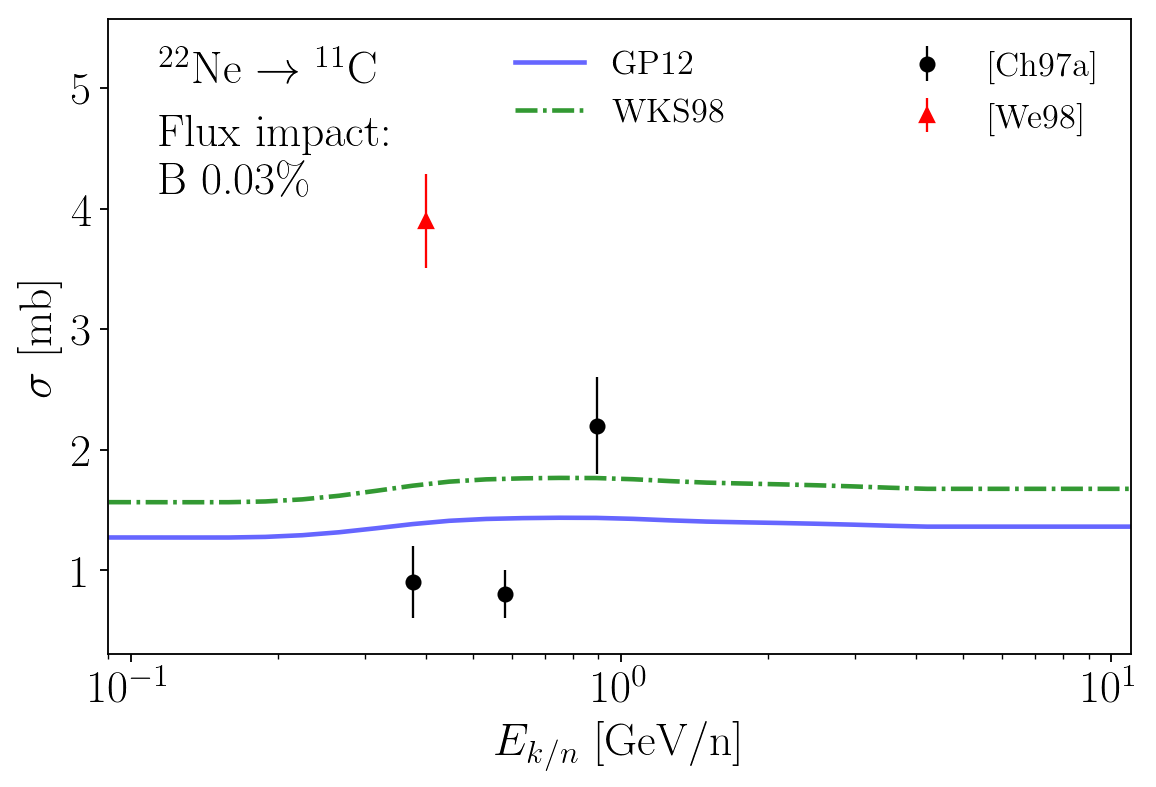}  \\ 
\includegraphics[width=0.32\textwidth]{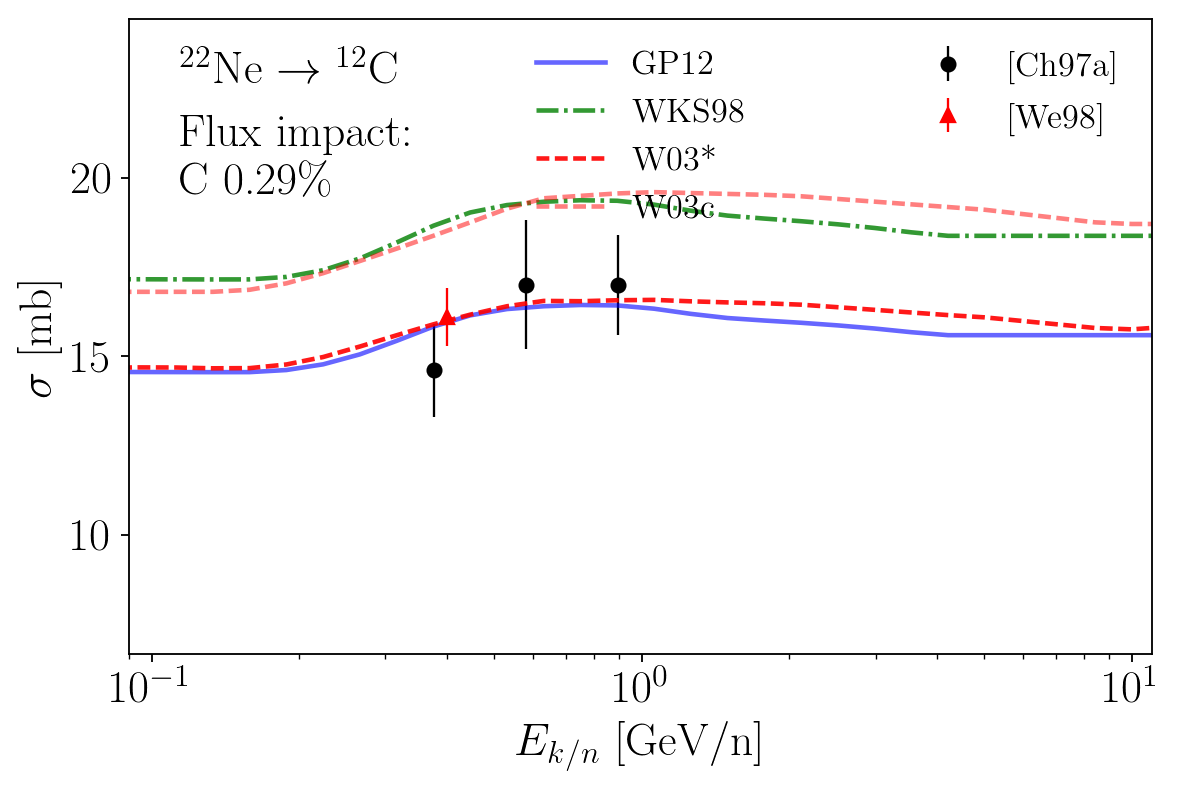}  &  
\includegraphics[width=0.32\textwidth]{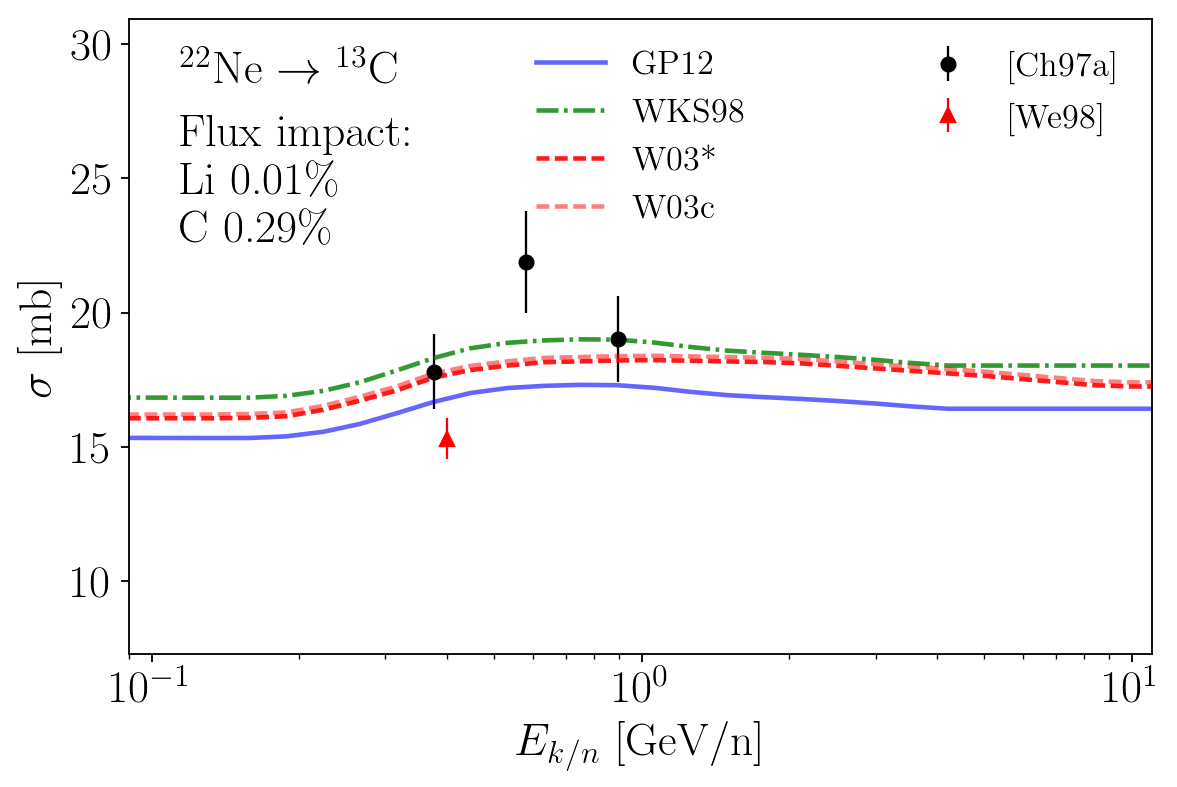}  &  
\includegraphics[width=0.32\textwidth]{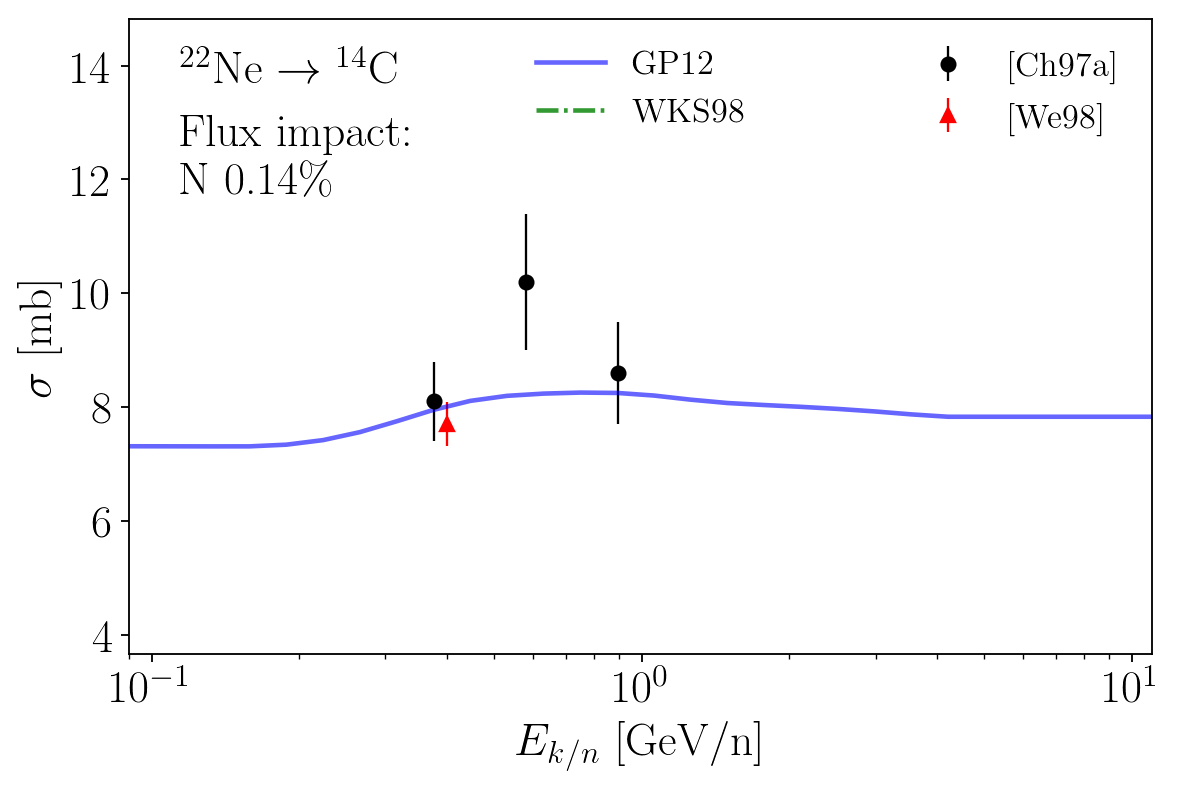}  \\ 
\includegraphics[width=0.32\textwidth]{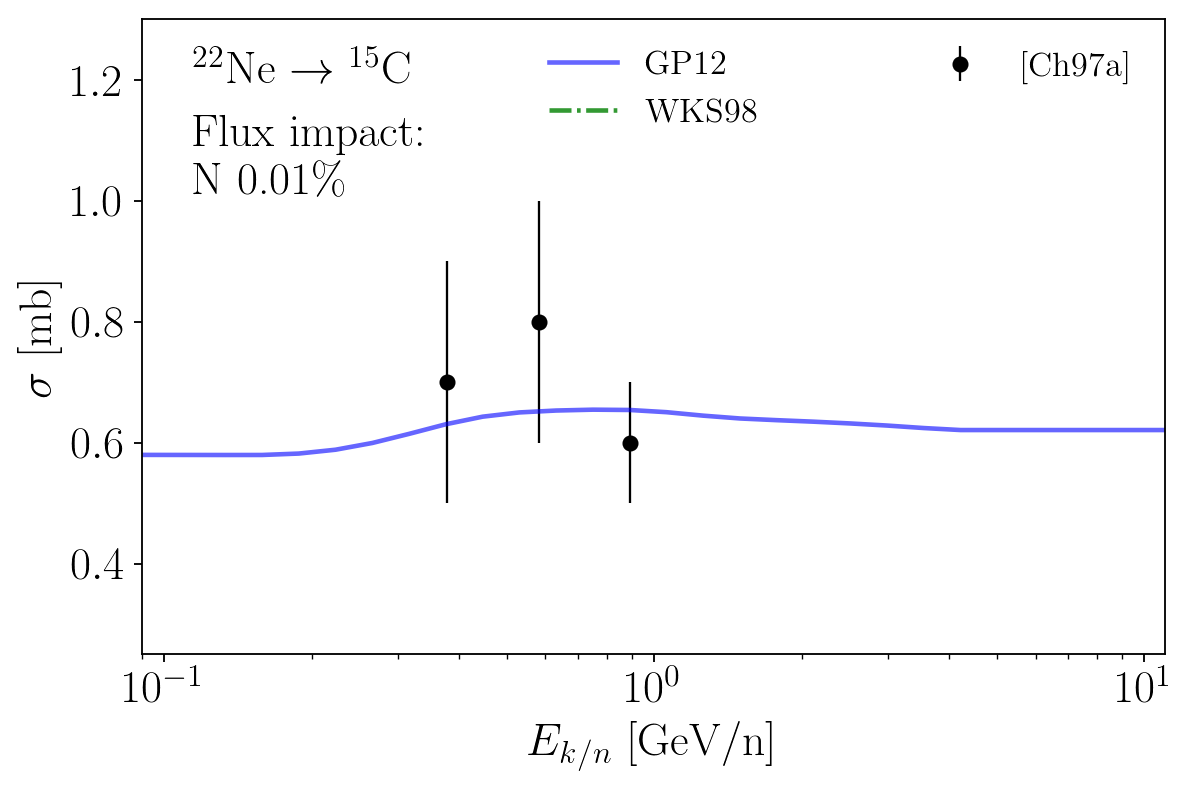}  &  
\includegraphics[width=0.32\textwidth]{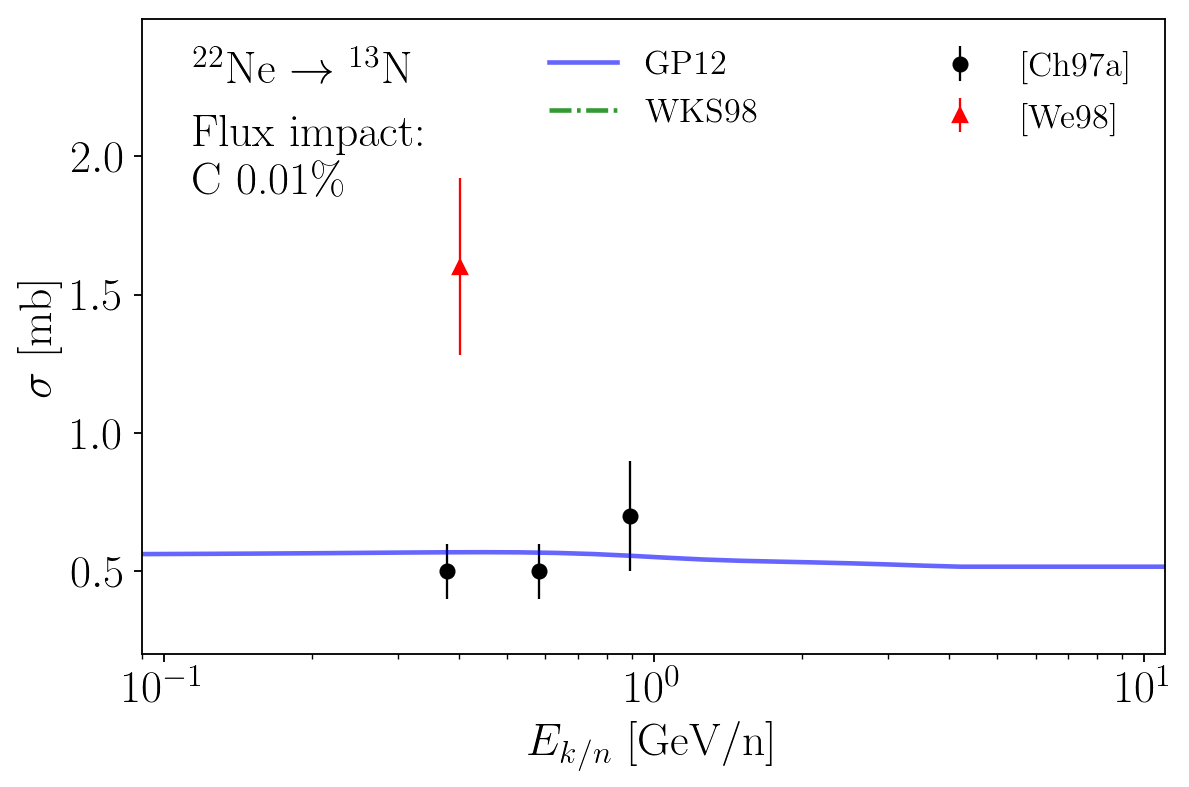}  &  
\includegraphics[width=0.32\textwidth]{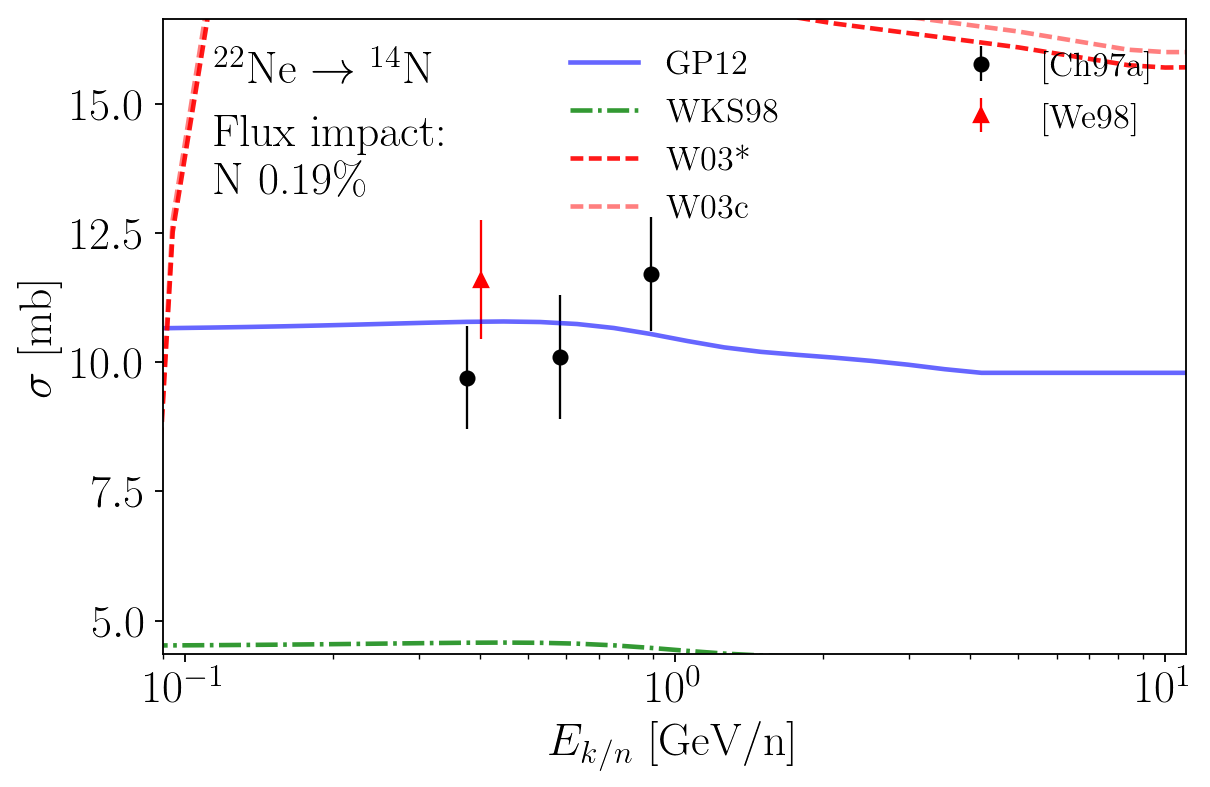}  \\ 
\includegraphics[width=0.32\textwidth]{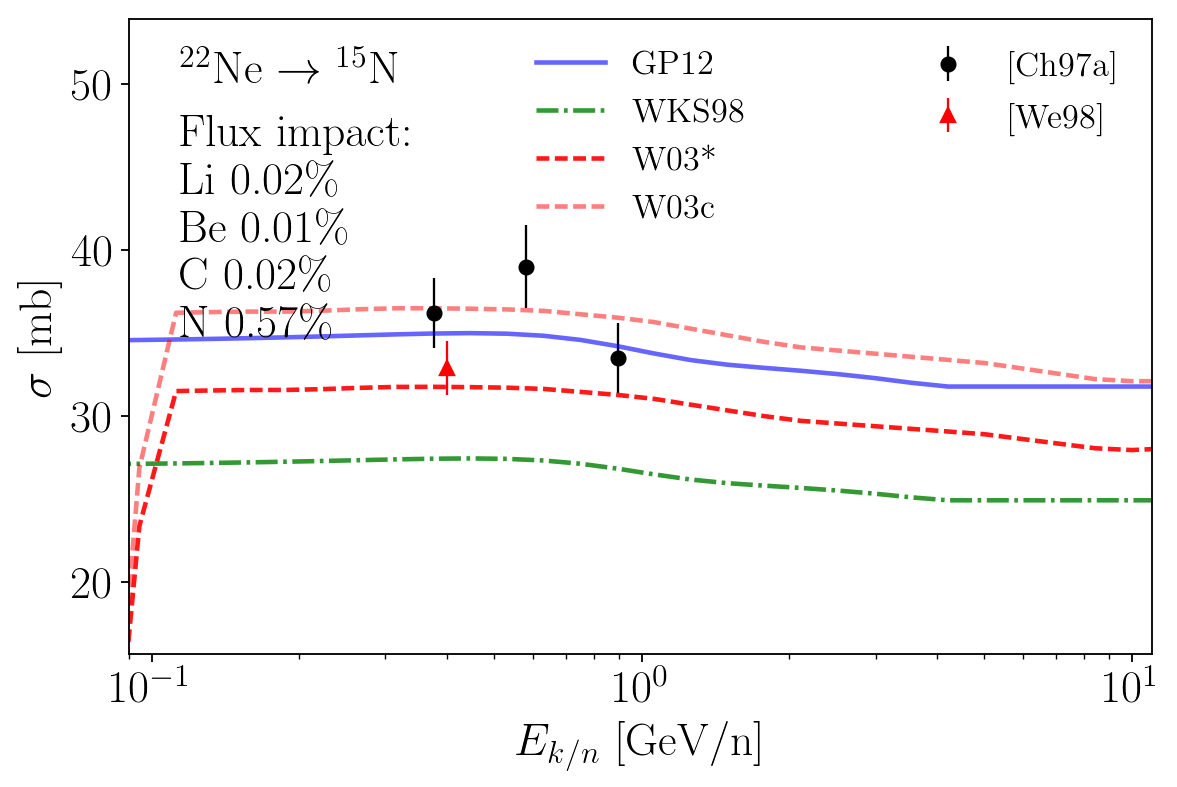}  &  
\includegraphics[width=0.32\textwidth]{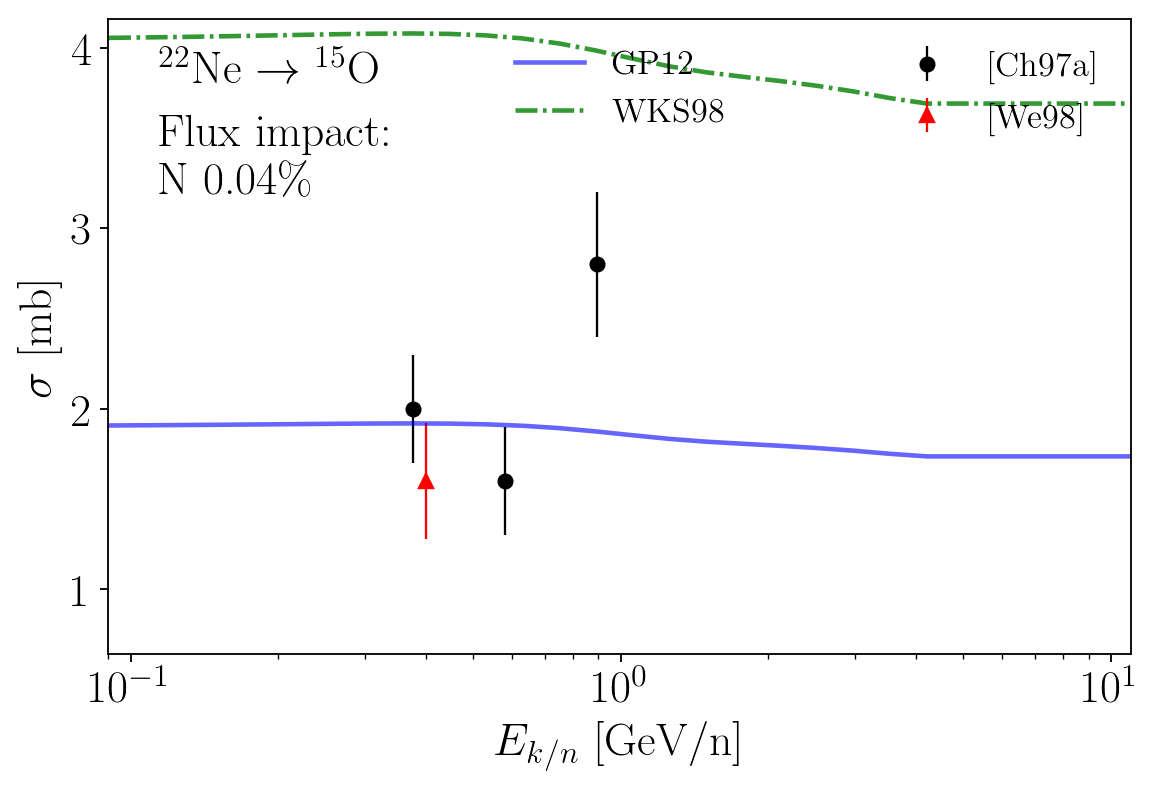}  &  
\includegraphics[width=0.32\textwidth]{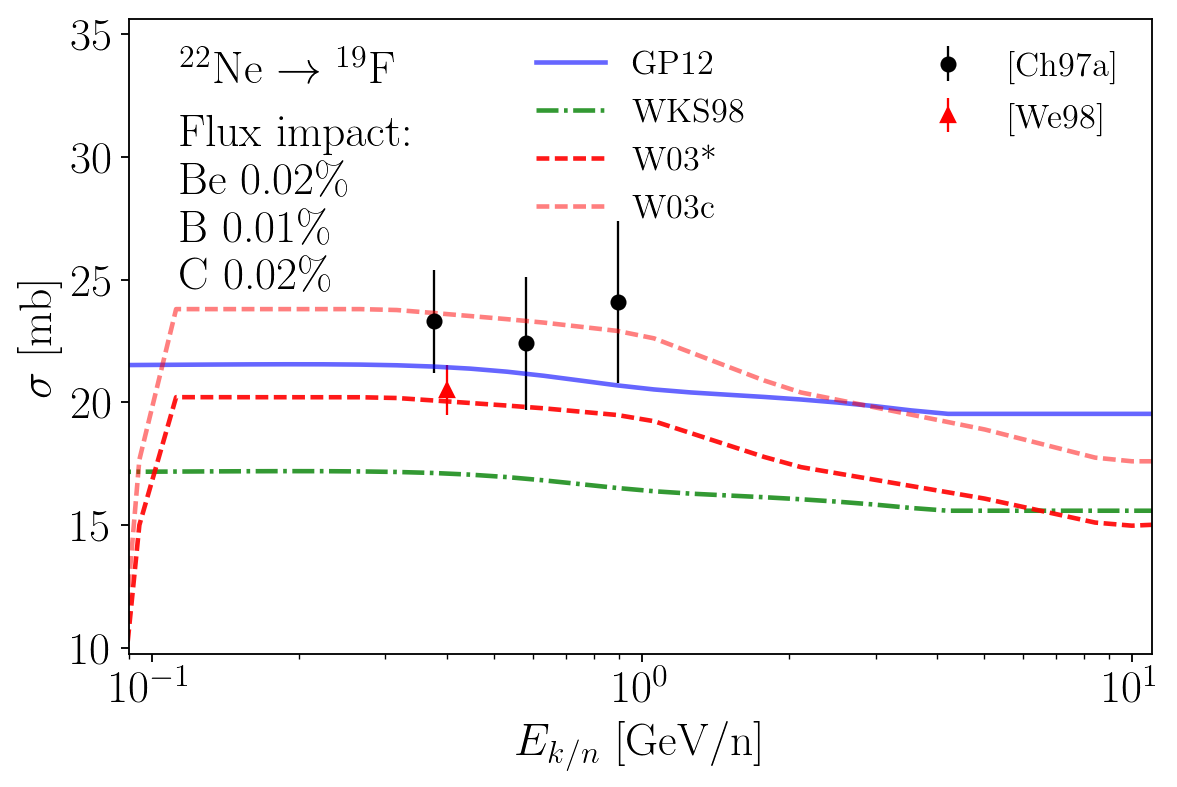}  \\ 
 \\ [3pt] 
\multicolumn{3}{c}{\bf Z=11{ \bf projectiles: $^{x}$Na + H $\rightarrow$ $^{A}_ZX$}}\\ [3pt]
\multicolumn{3}{c}{\noindent\makebox[\linewidth]{\rule{\textwidth}{0.4pt}}}\\ [3pt]
\includegraphics[width=0.32\textwidth]{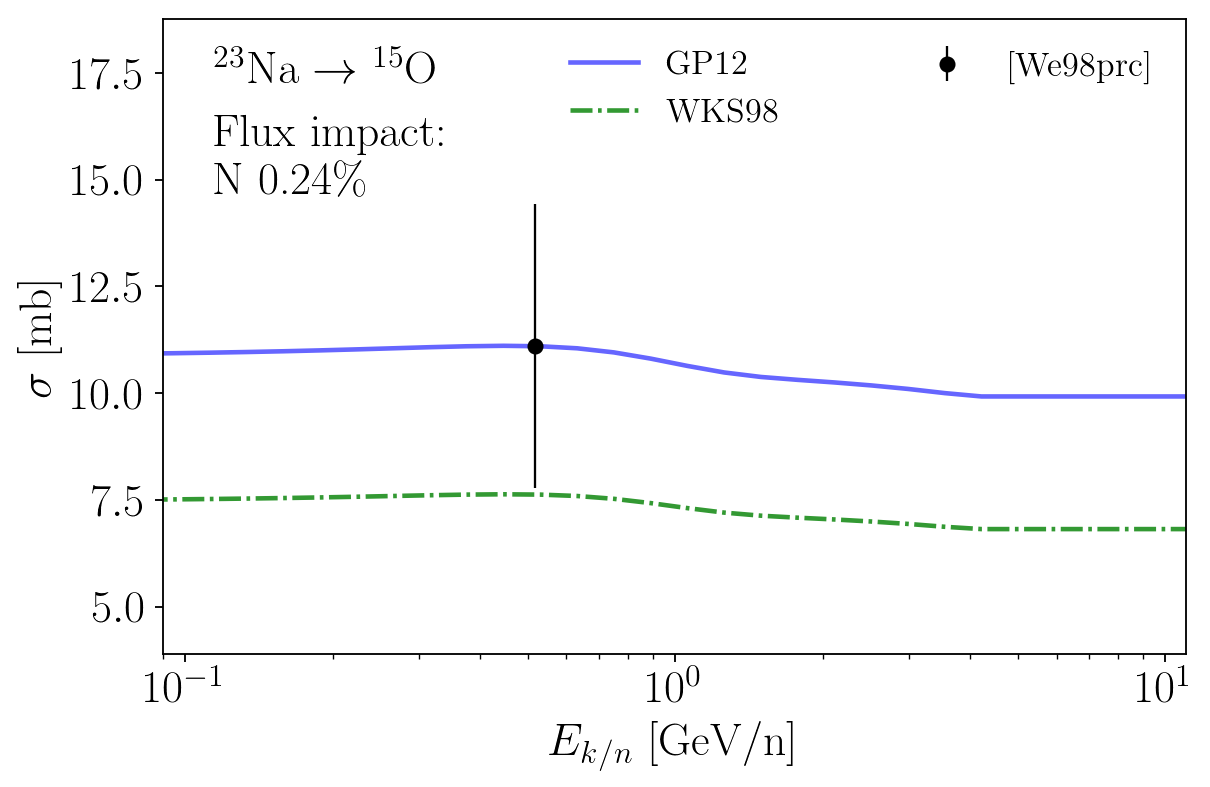}  &  
\includegraphics[width=0.32\textwidth]{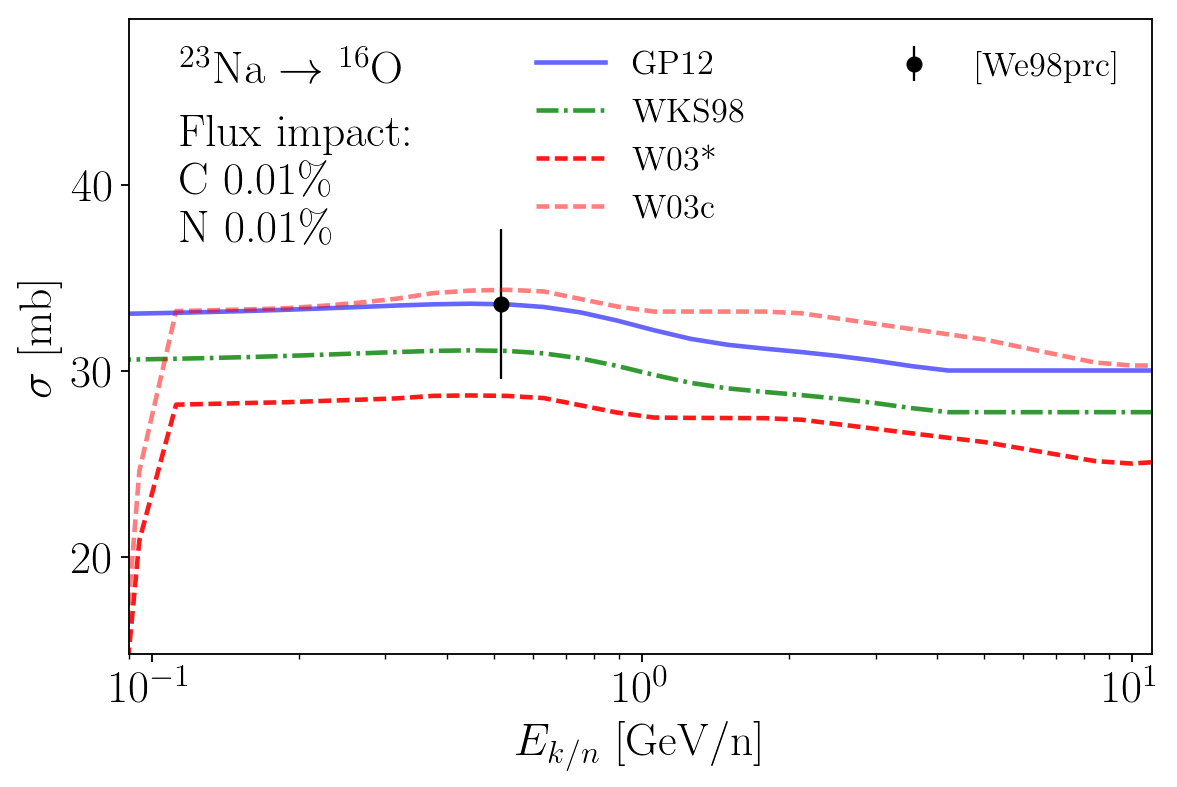}  &  
\includegraphics[width=0.32\textwidth]{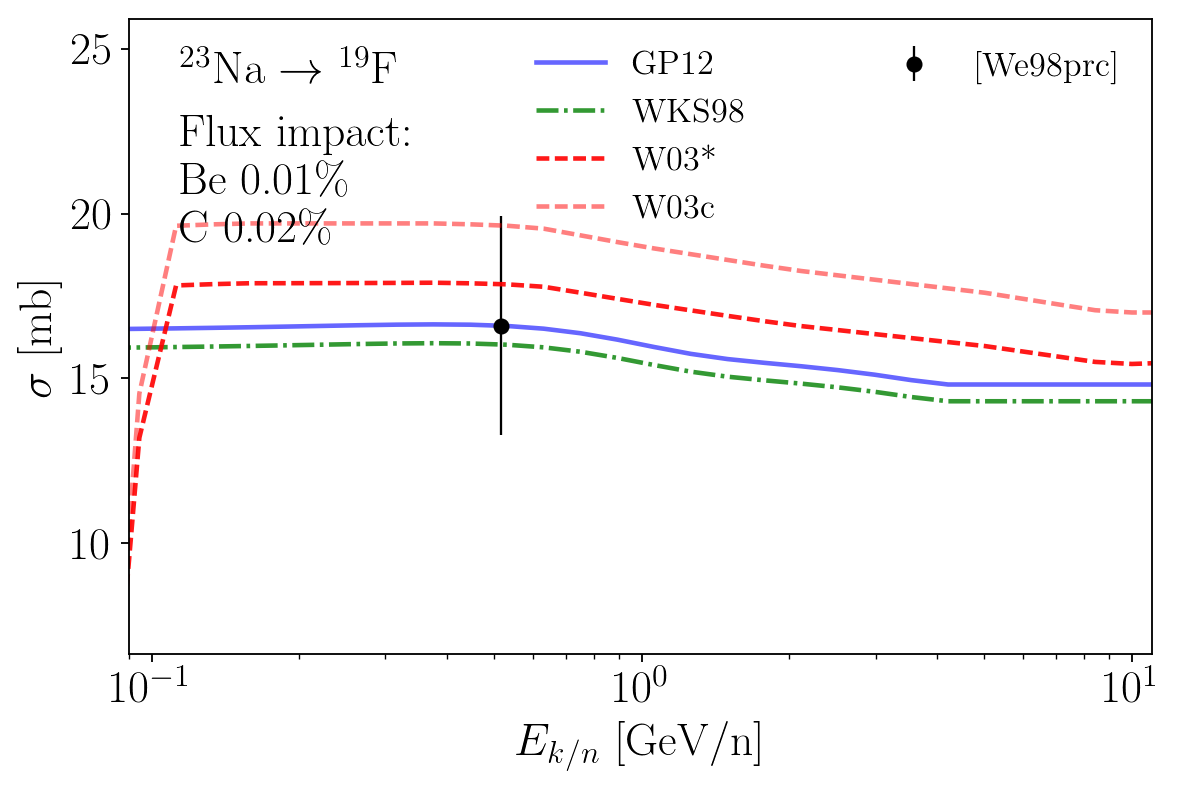}  \\ 
 \\ [3pt] 
\multicolumn{3}{c}{\bf Z=12{ \bf projectiles: $^{x}$Mg + H $\rightarrow$ $^{A}_ZX$}}\\ [3pt]
\multicolumn{3}{c}{\noindent\makebox[\linewidth]{\rule{\textwidth}{0.4pt}}}\\ [3pt]
\includegraphics[width=0.32\textwidth]{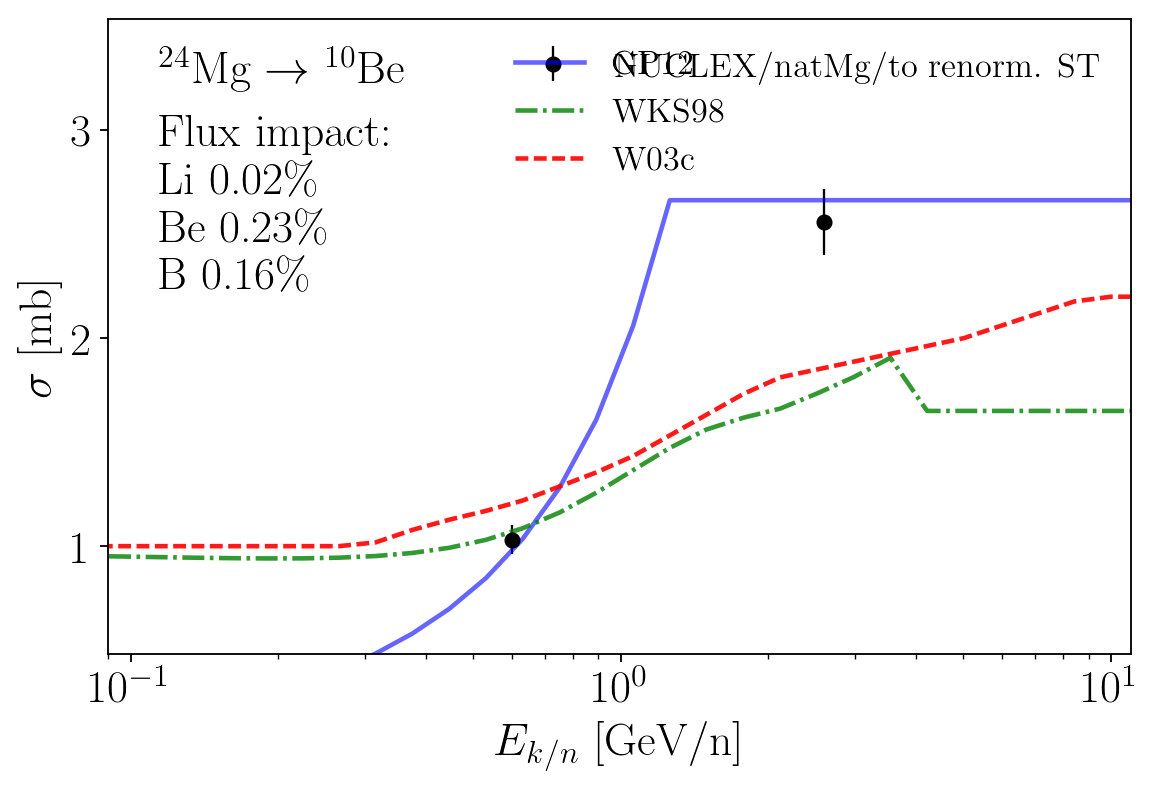}  &  
\includegraphics[width=0.32\textwidth]{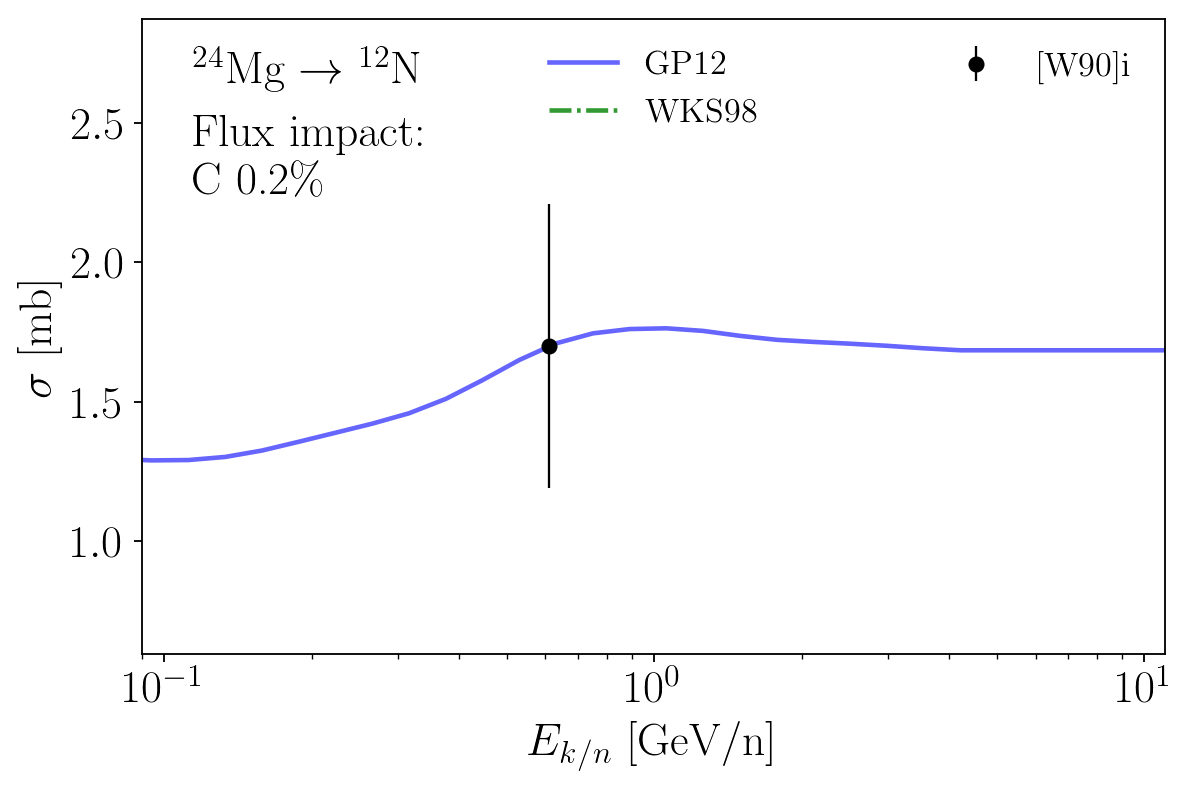}  &  
\includegraphics[width=0.32\textwidth]{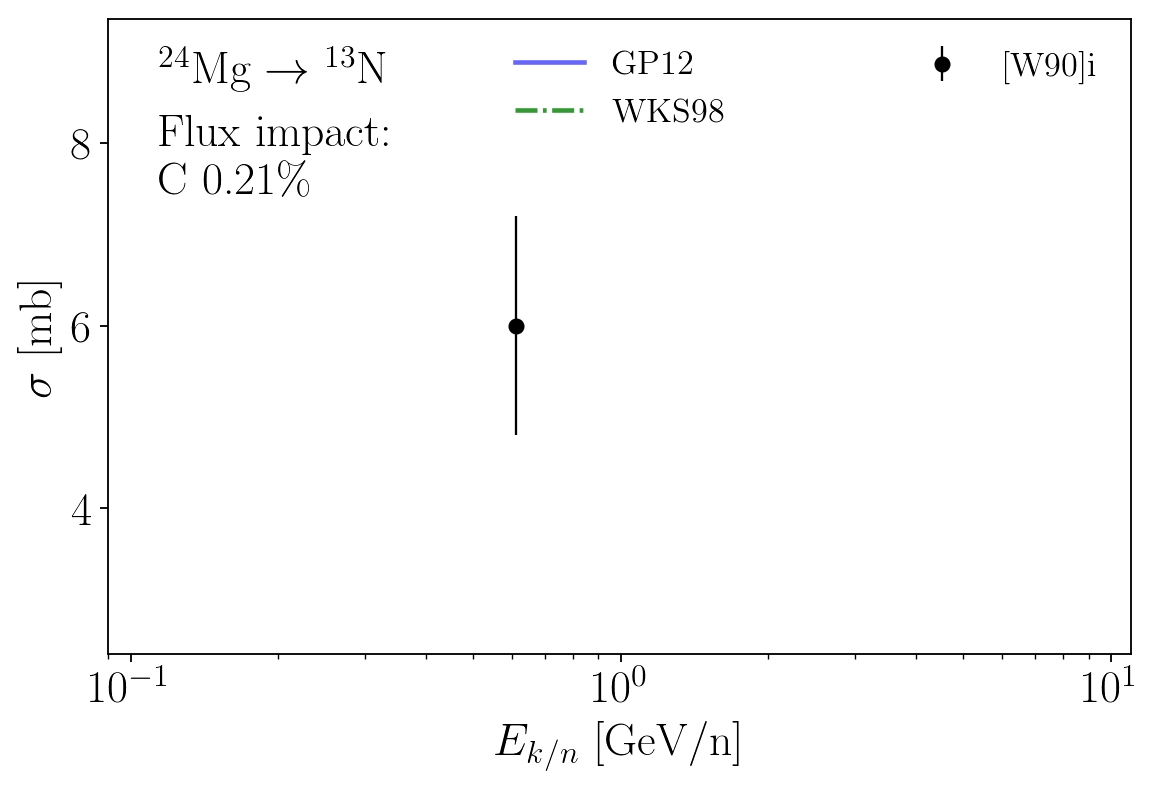}  \\ 
\includegraphics[width=0.32\textwidth]{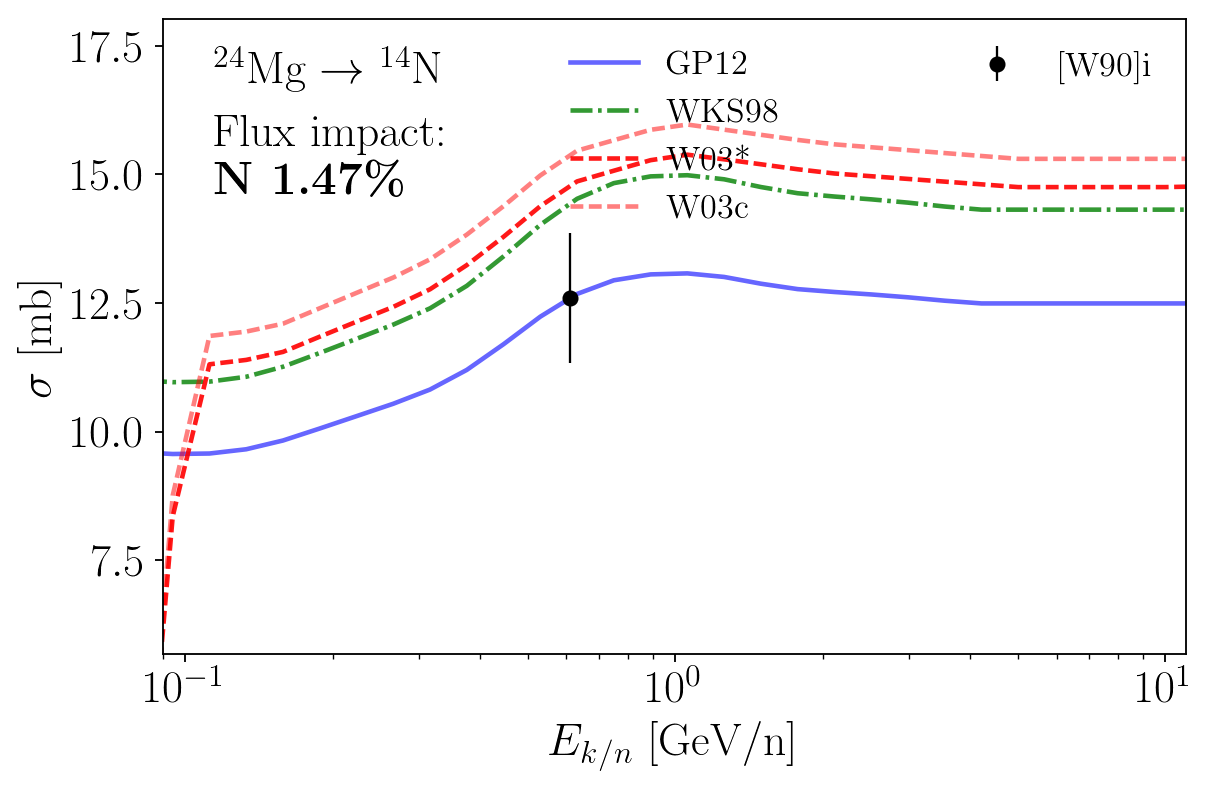}  &  
\includegraphics[width=0.32\textwidth]{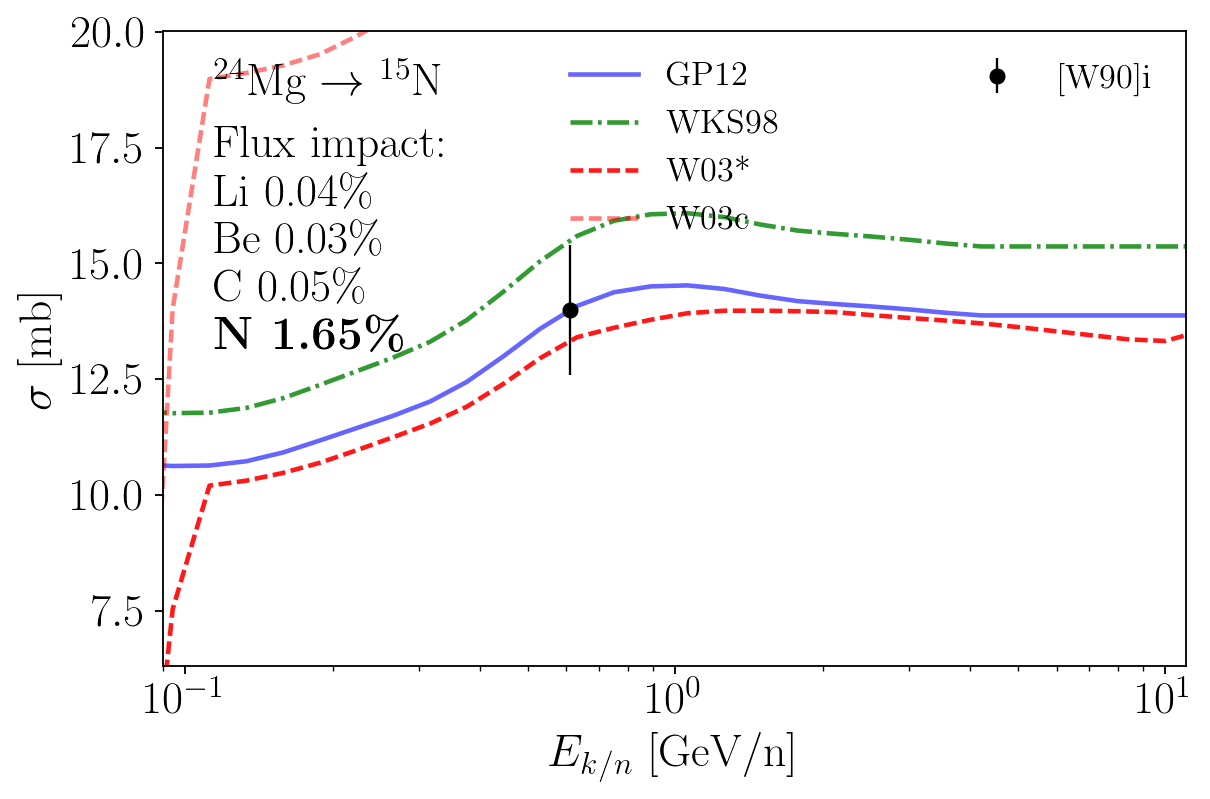}  &  
\includegraphics[width=0.32\textwidth]{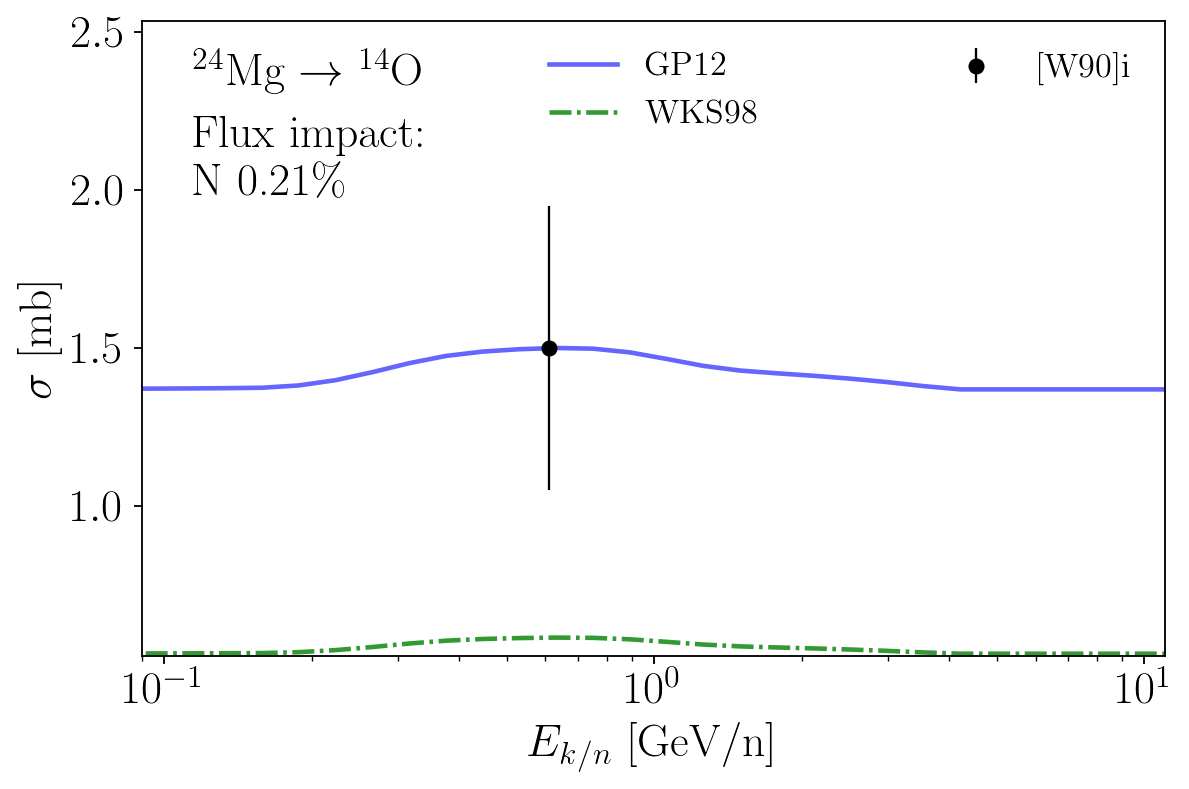}  \\ 
\includegraphics[width=0.32\textwidth]{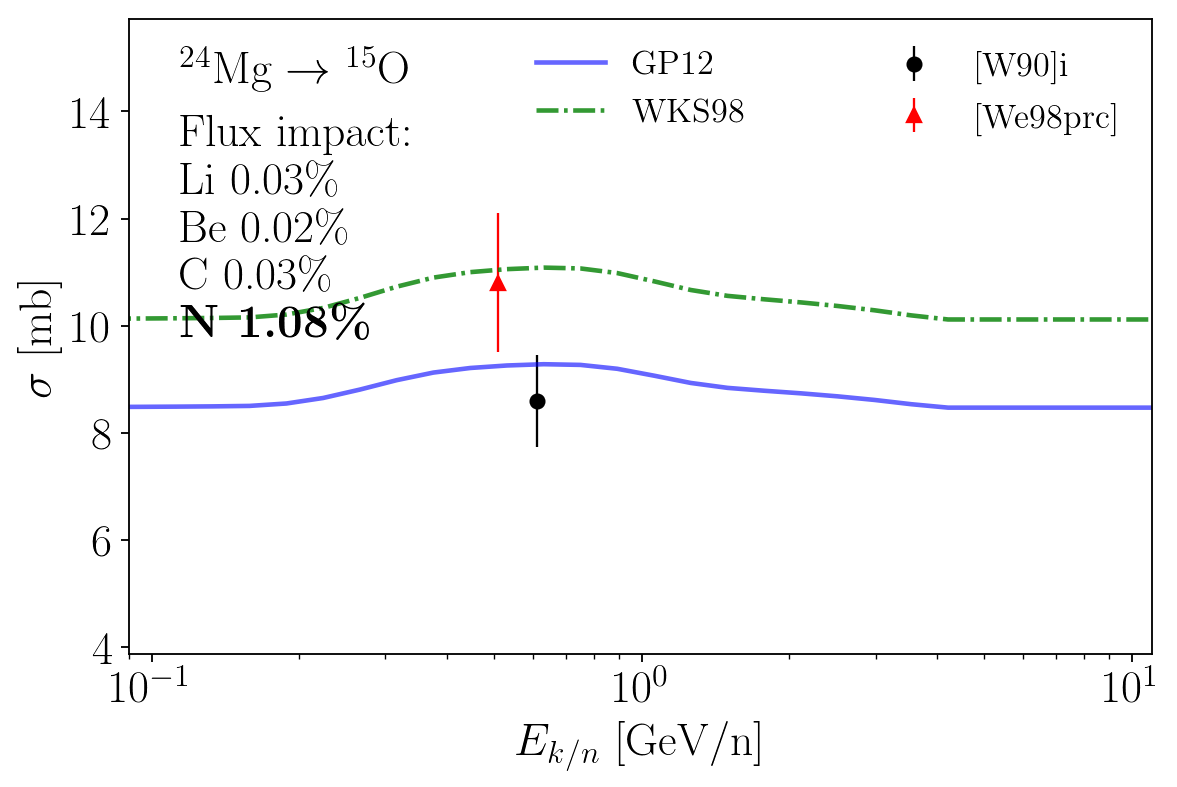}  &  
\includegraphics[width=0.32\textwidth]{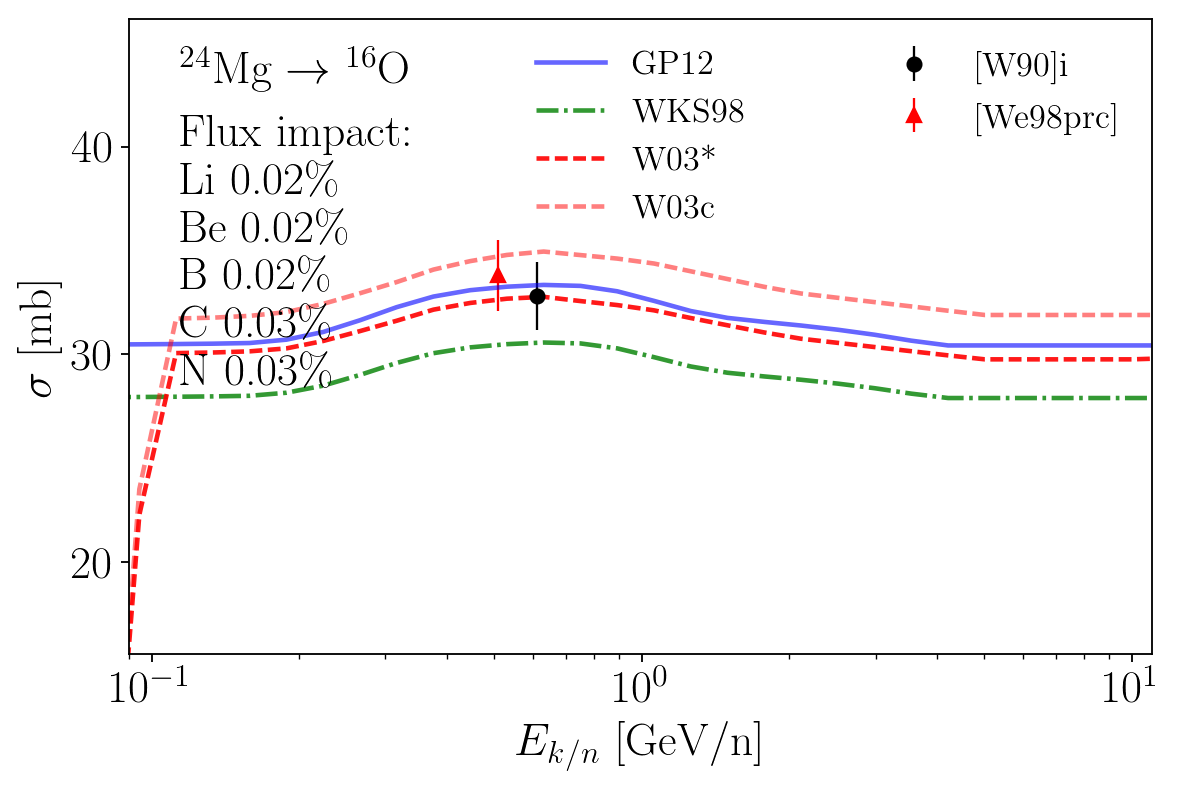}  &  
\includegraphics[width=0.32\textwidth]{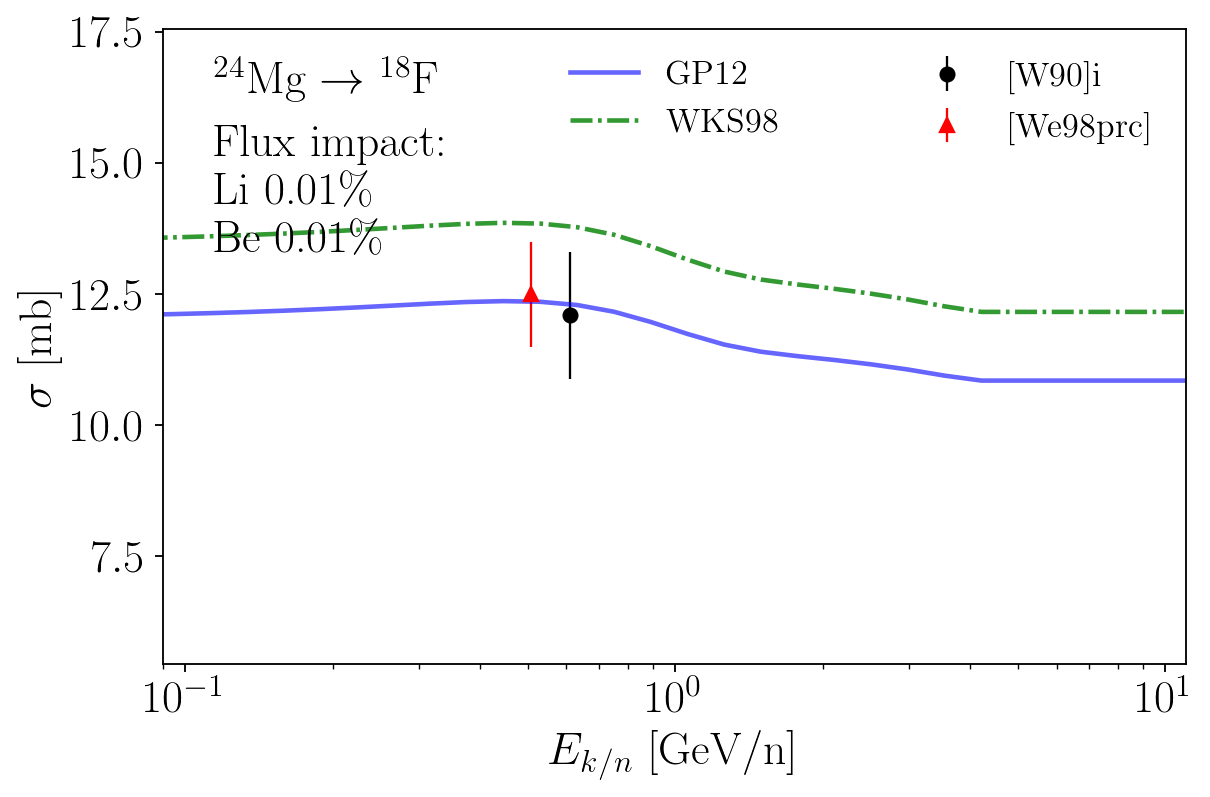}  \\ 
\includegraphics[width=0.32\textwidth]{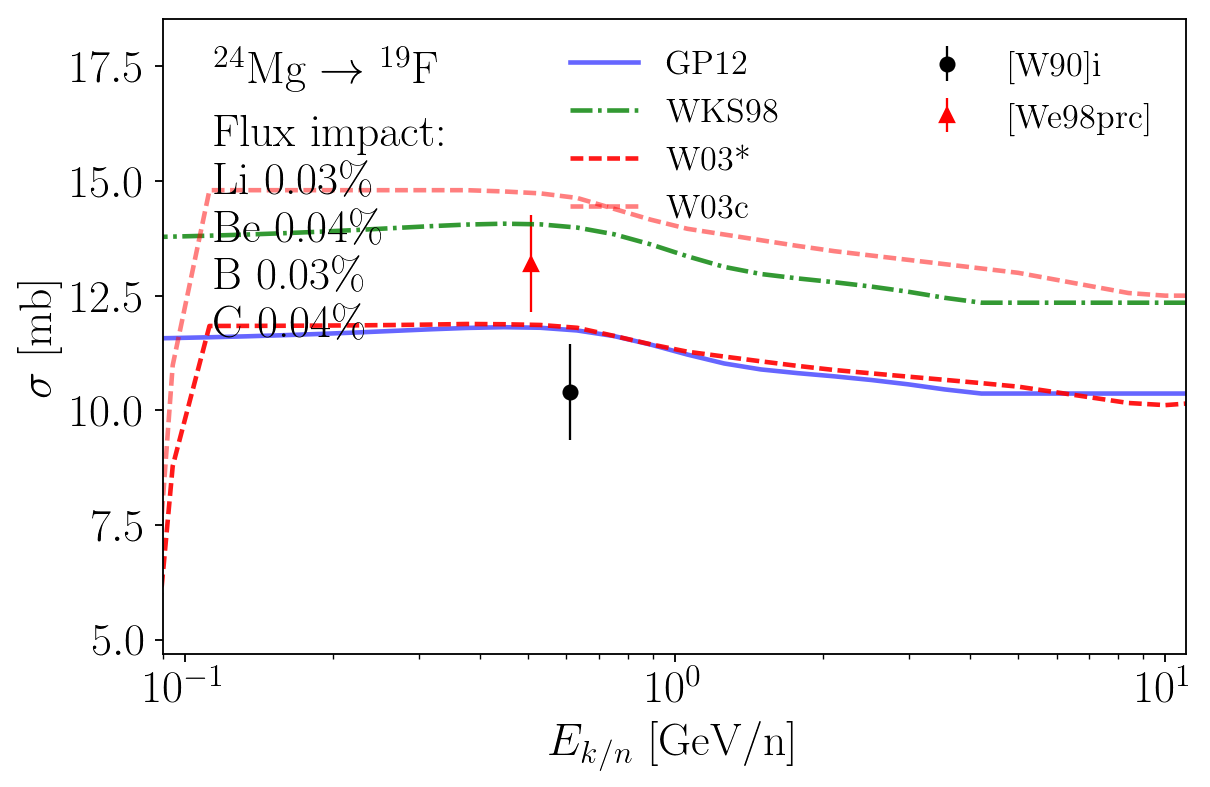}  &  
\includegraphics[width=0.32\textwidth]{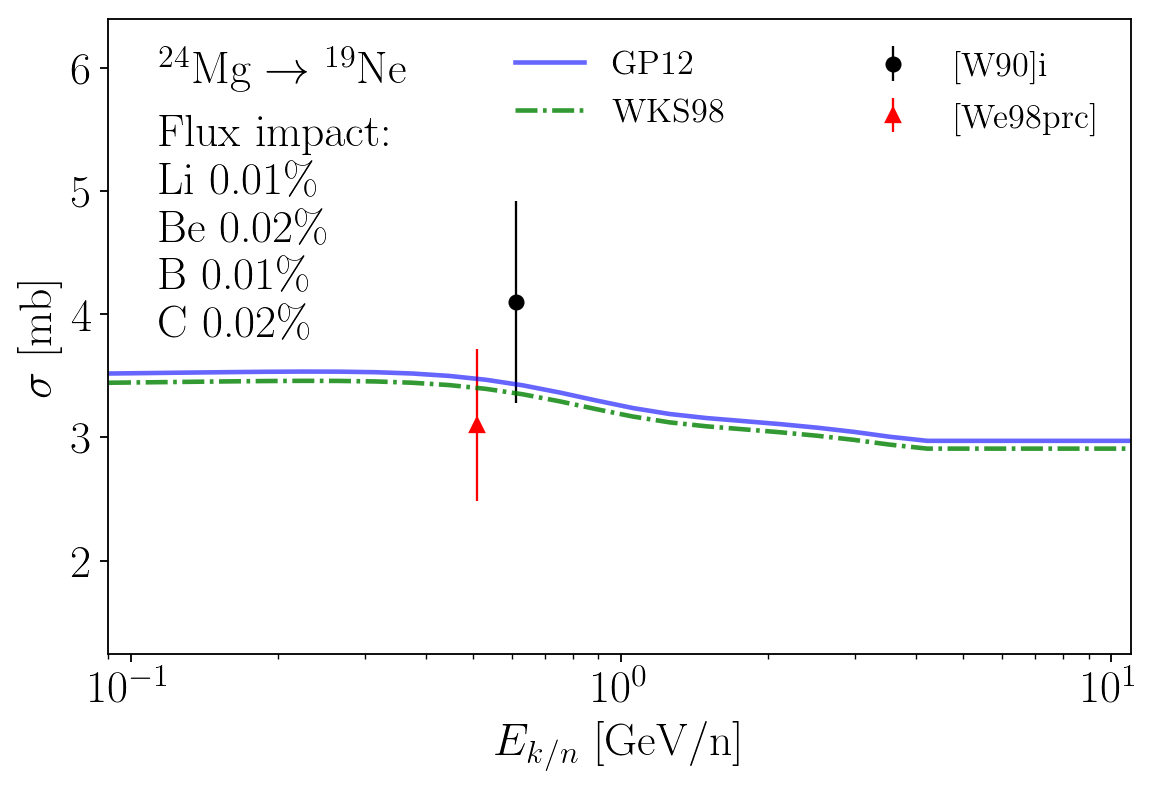}  &  
\includegraphics[width=0.32\textwidth]{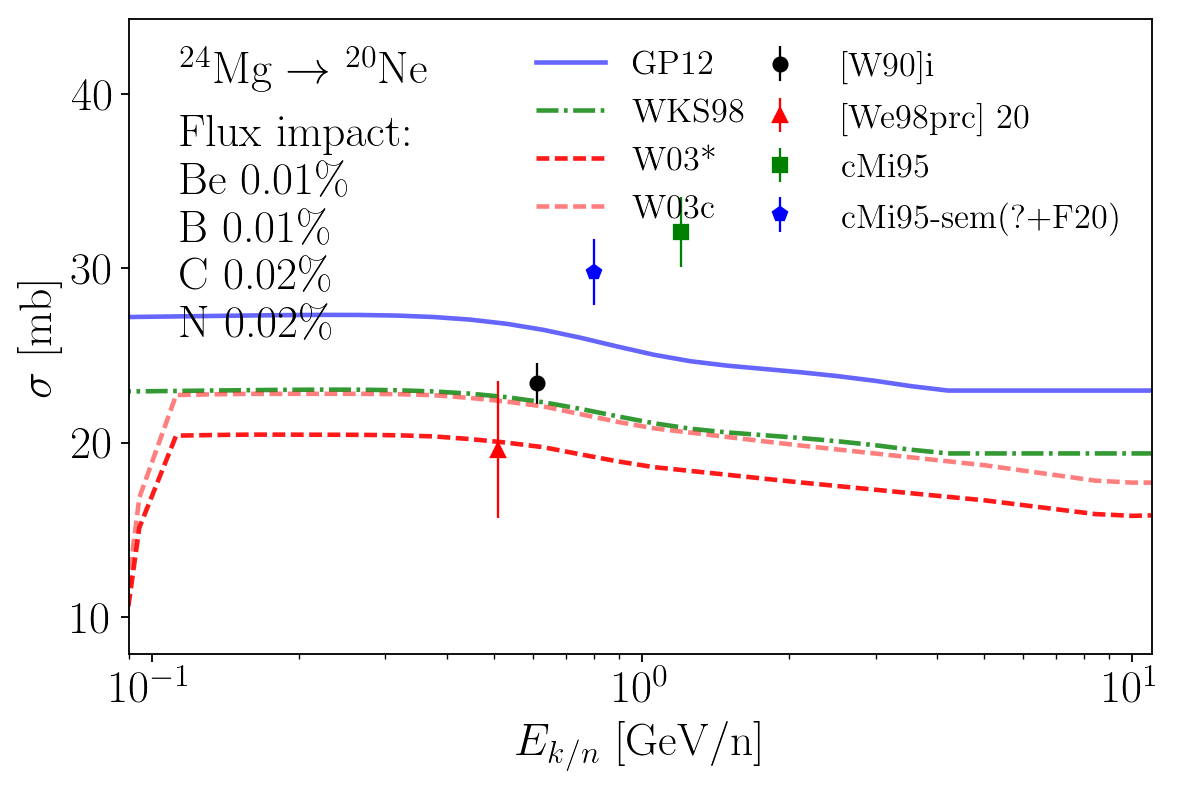}  \\ 
\includegraphics[width=0.32\textwidth]{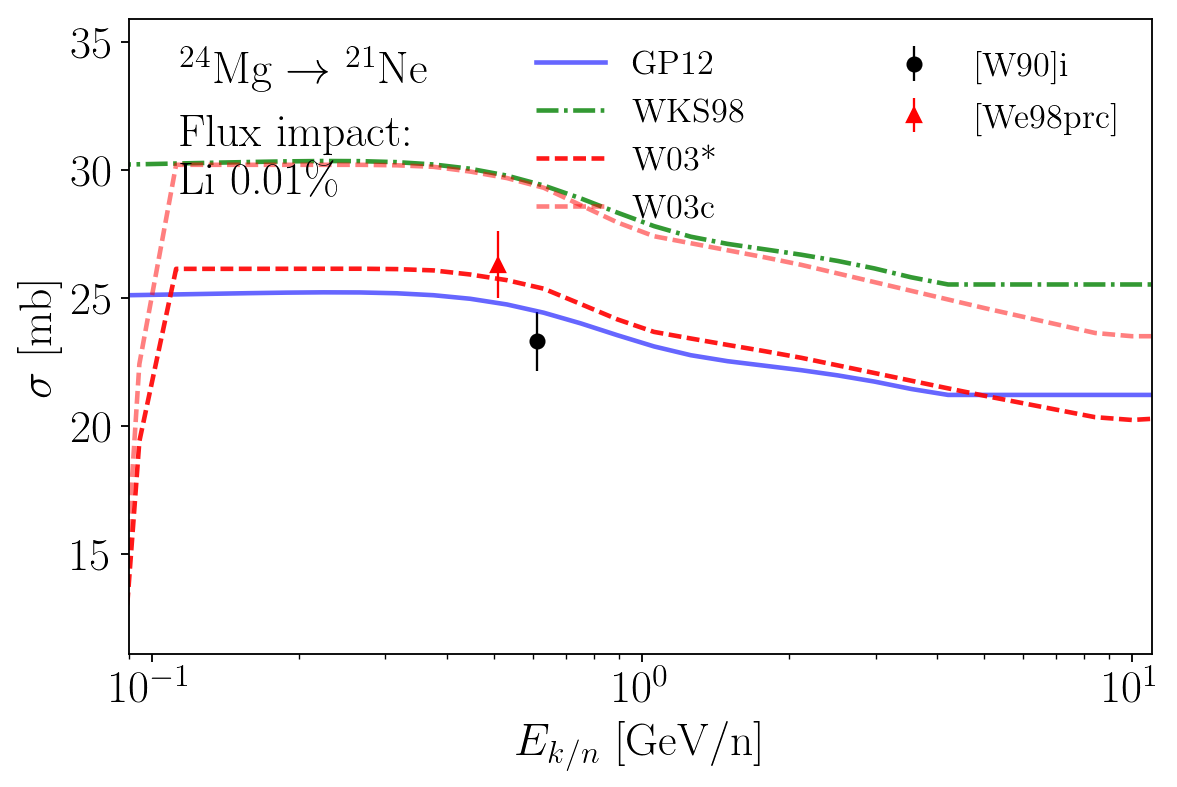}  &  
\includegraphics[width=0.32\textwidth]{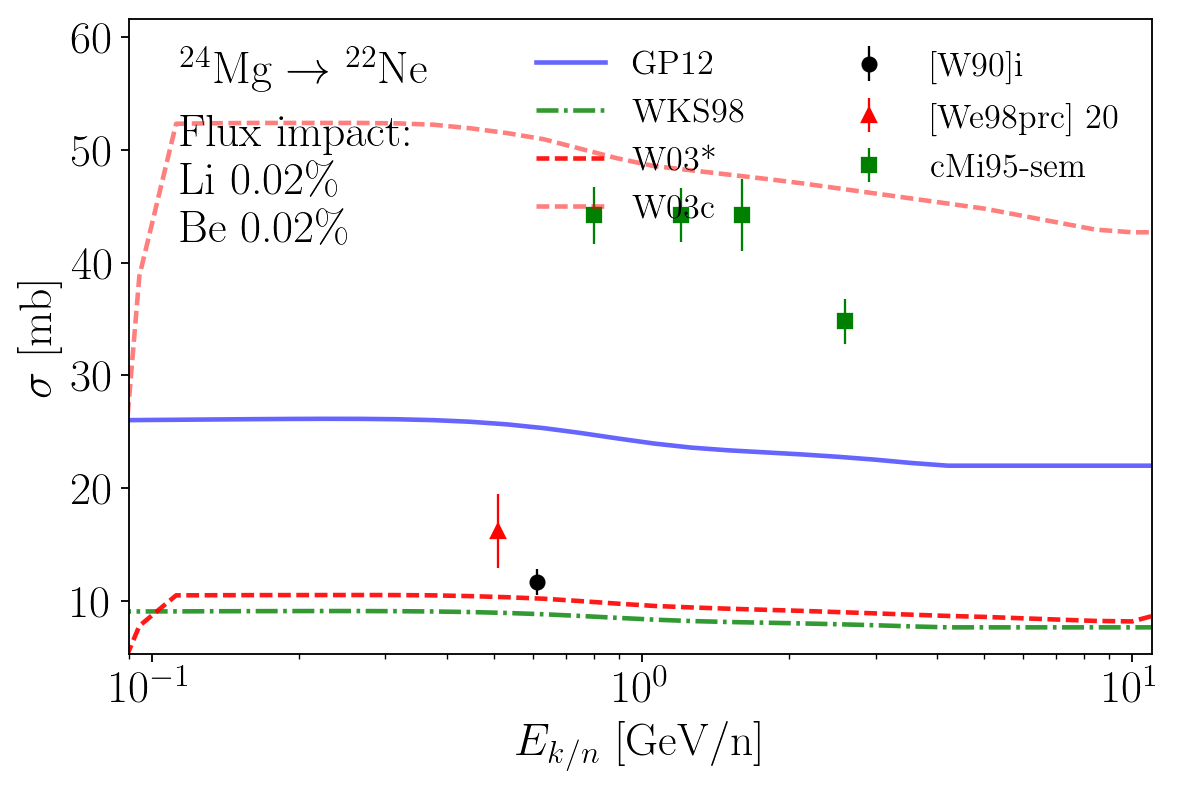}  &  
\includegraphics[width=0.32\textwidth]{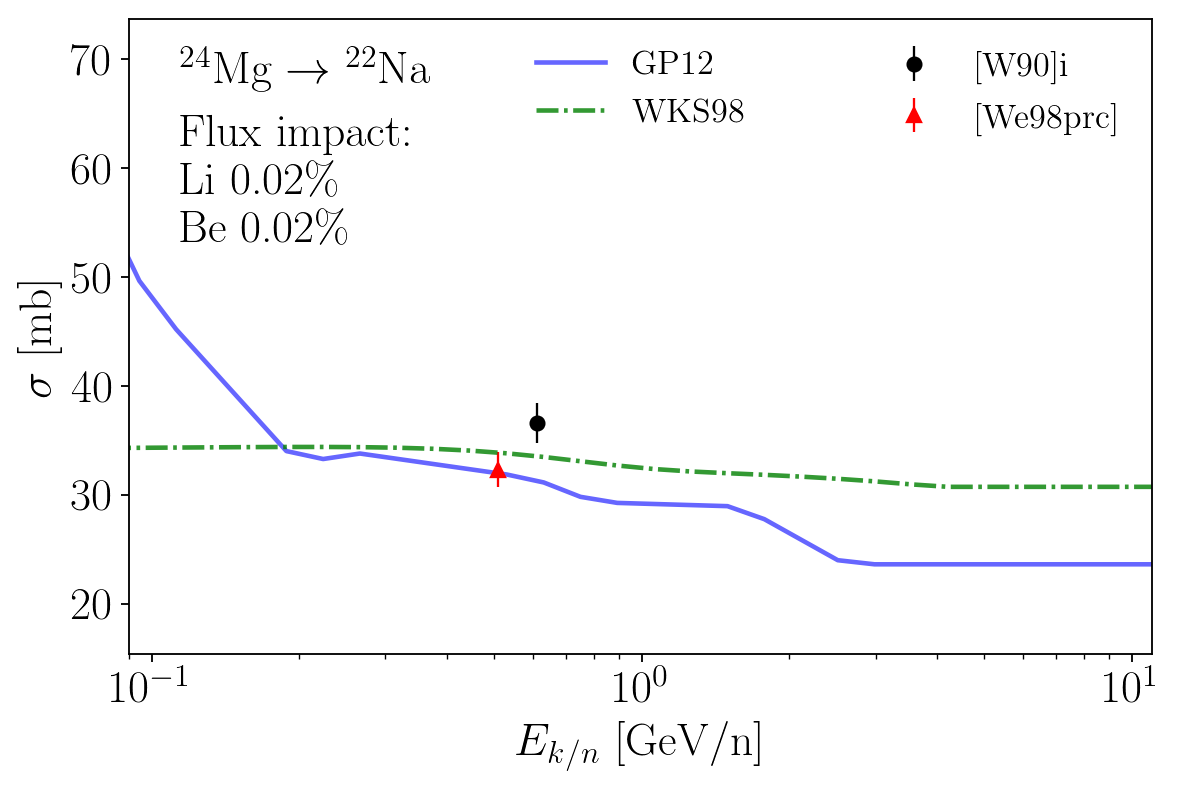}  \\ 
\includegraphics[width=0.32\textwidth]{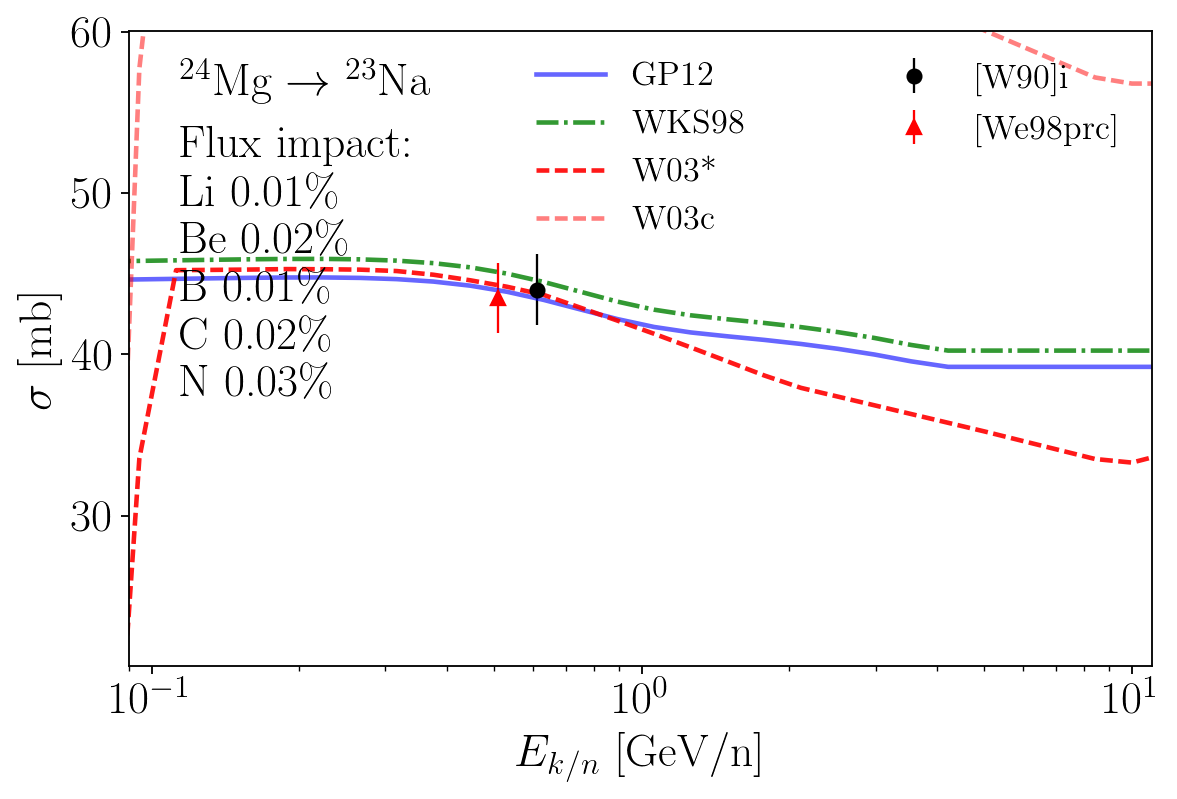}  &  
\includegraphics[width=0.32\textwidth]{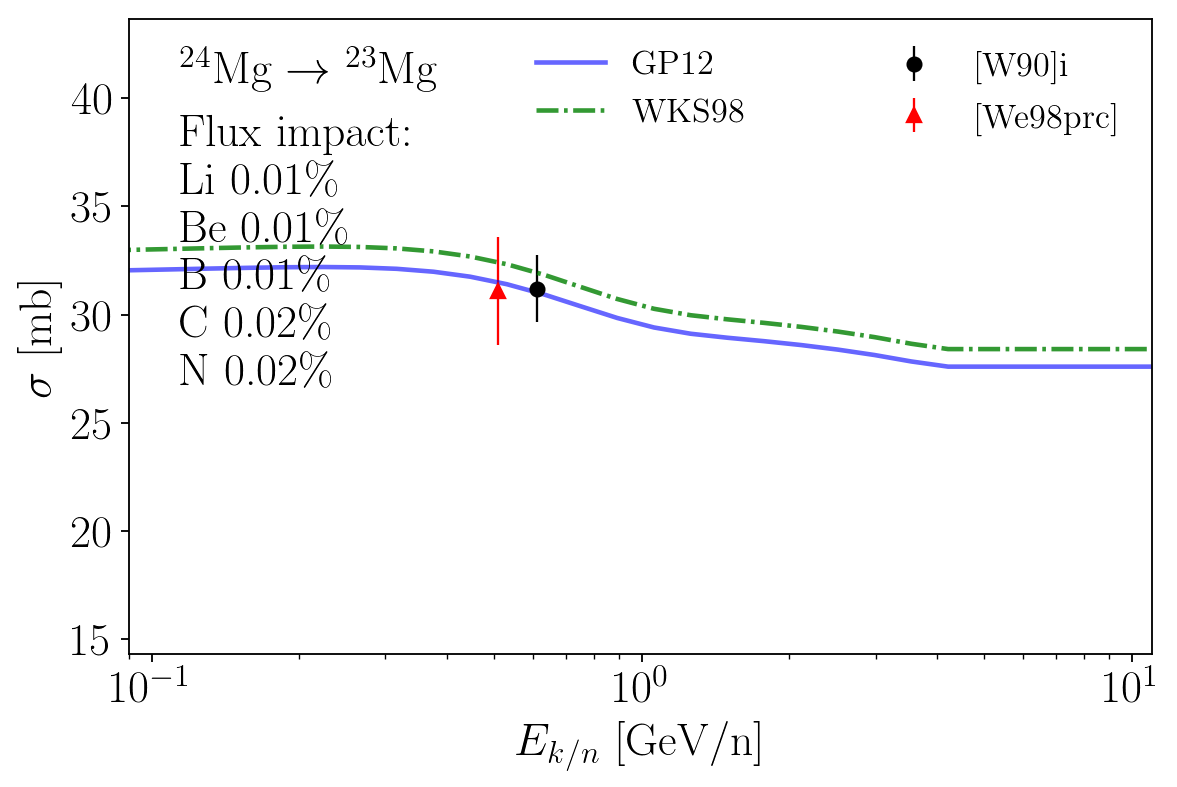}  &  
\includegraphics[width=0.32\textwidth]{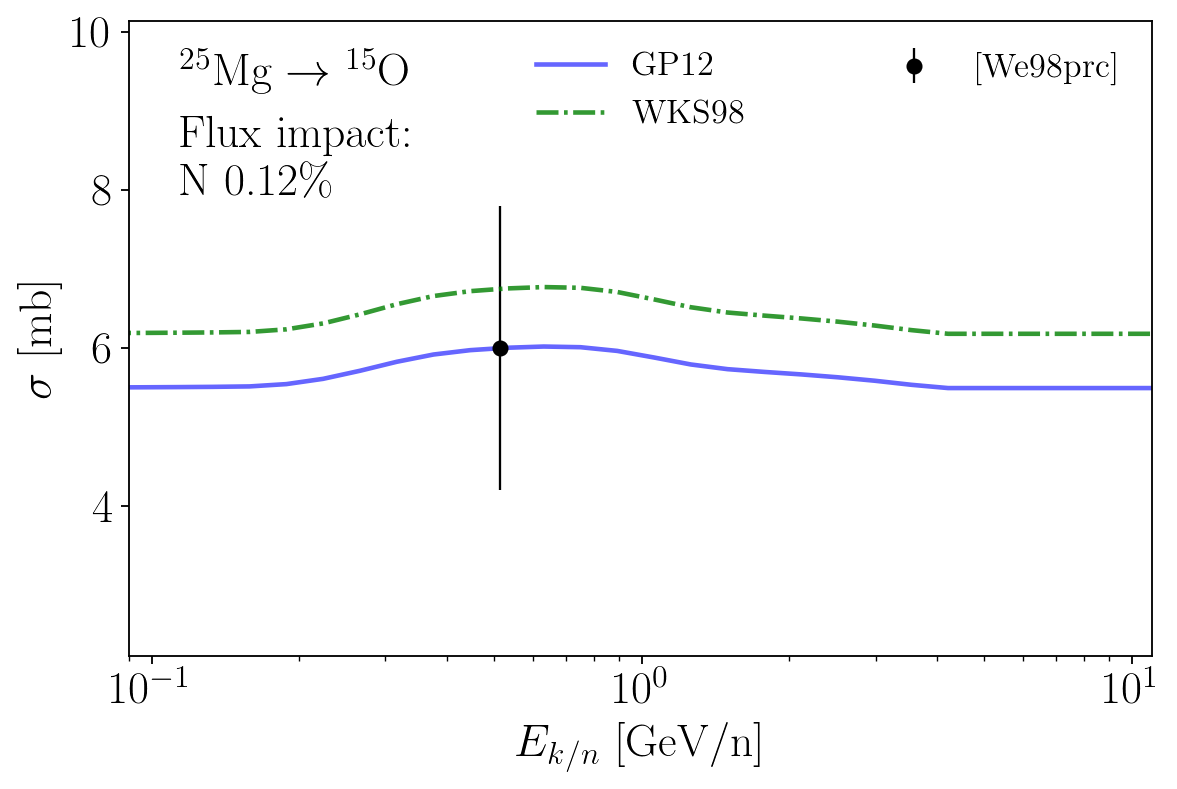}  \\ 
\includegraphics[width=0.32\textwidth]{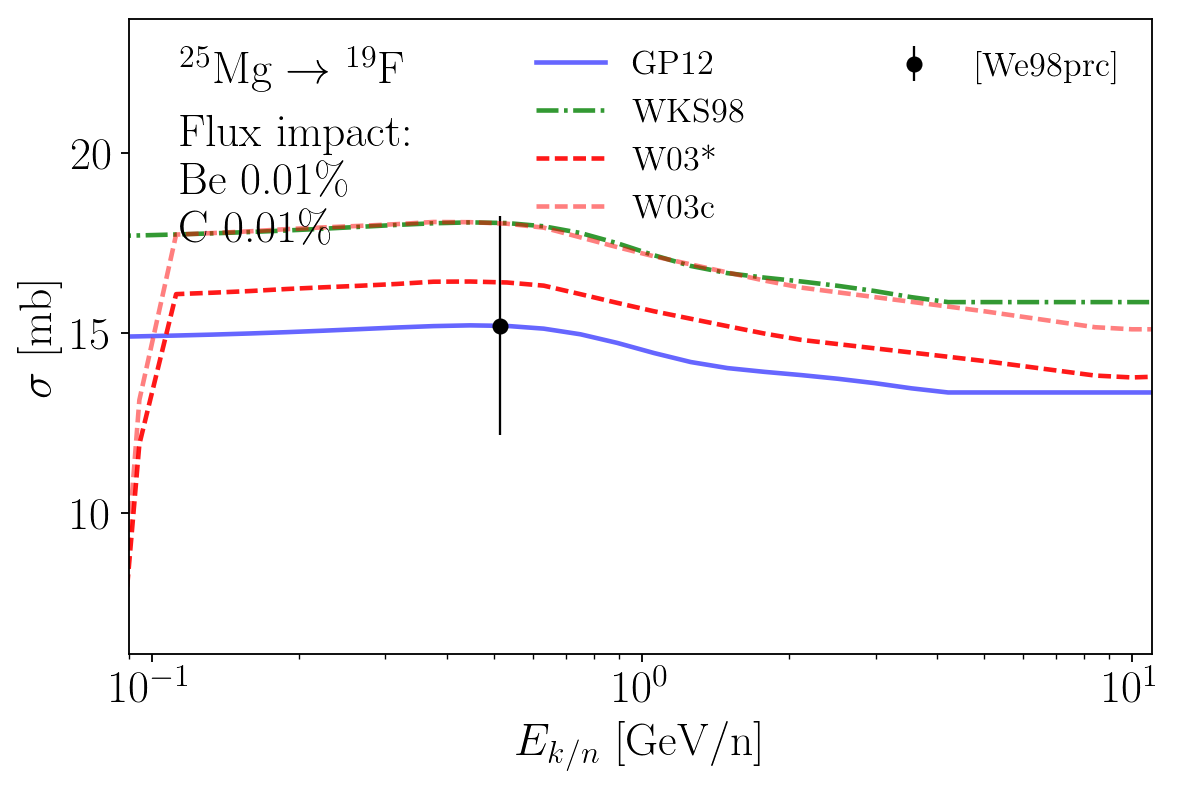}  &  
\includegraphics[width=0.32\textwidth]{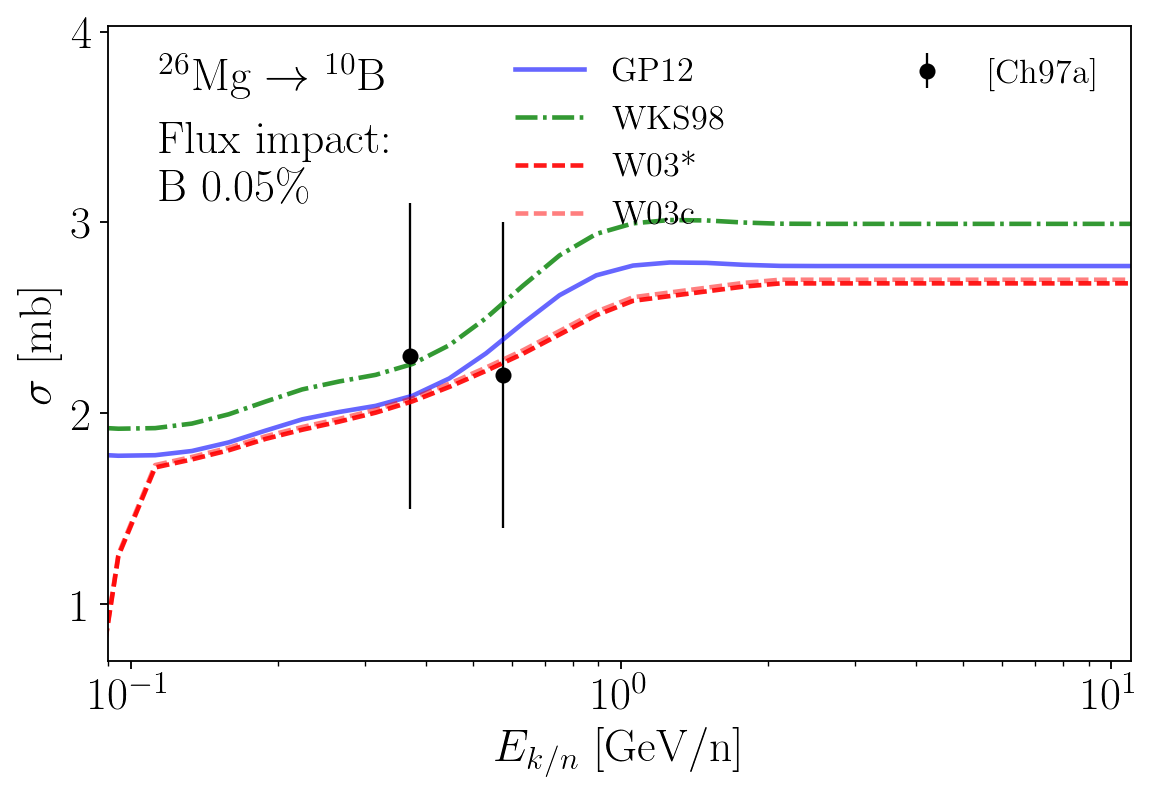}  &  
\includegraphics[width=0.32\textwidth]{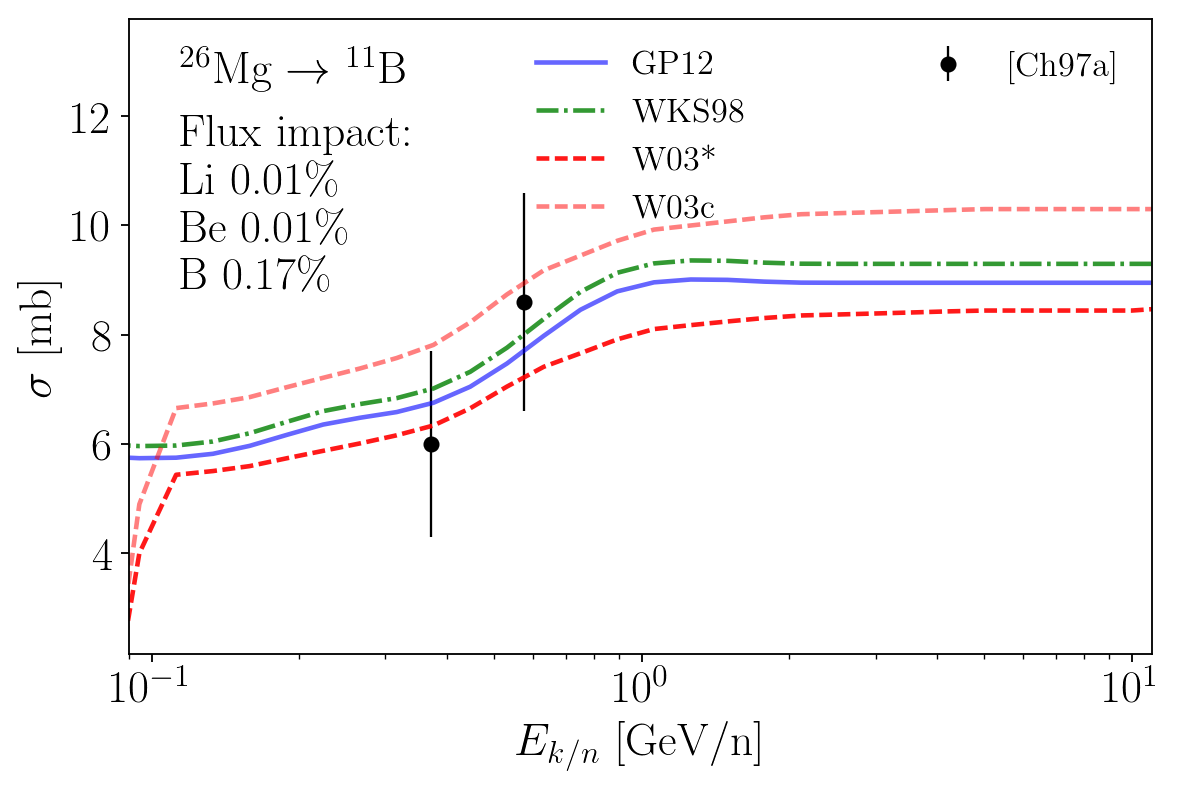}  \\ 
\includegraphics[width=0.32\textwidth]{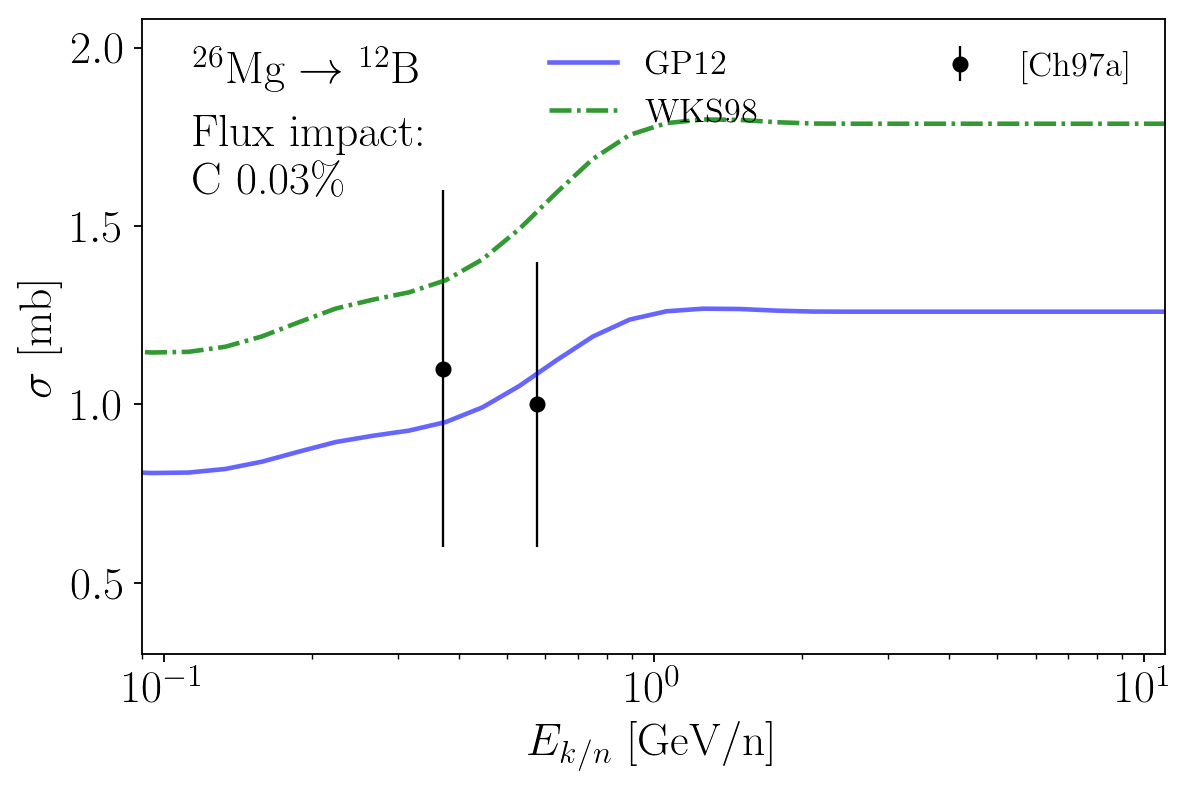}  &  
\includegraphics[width=0.32\textwidth]{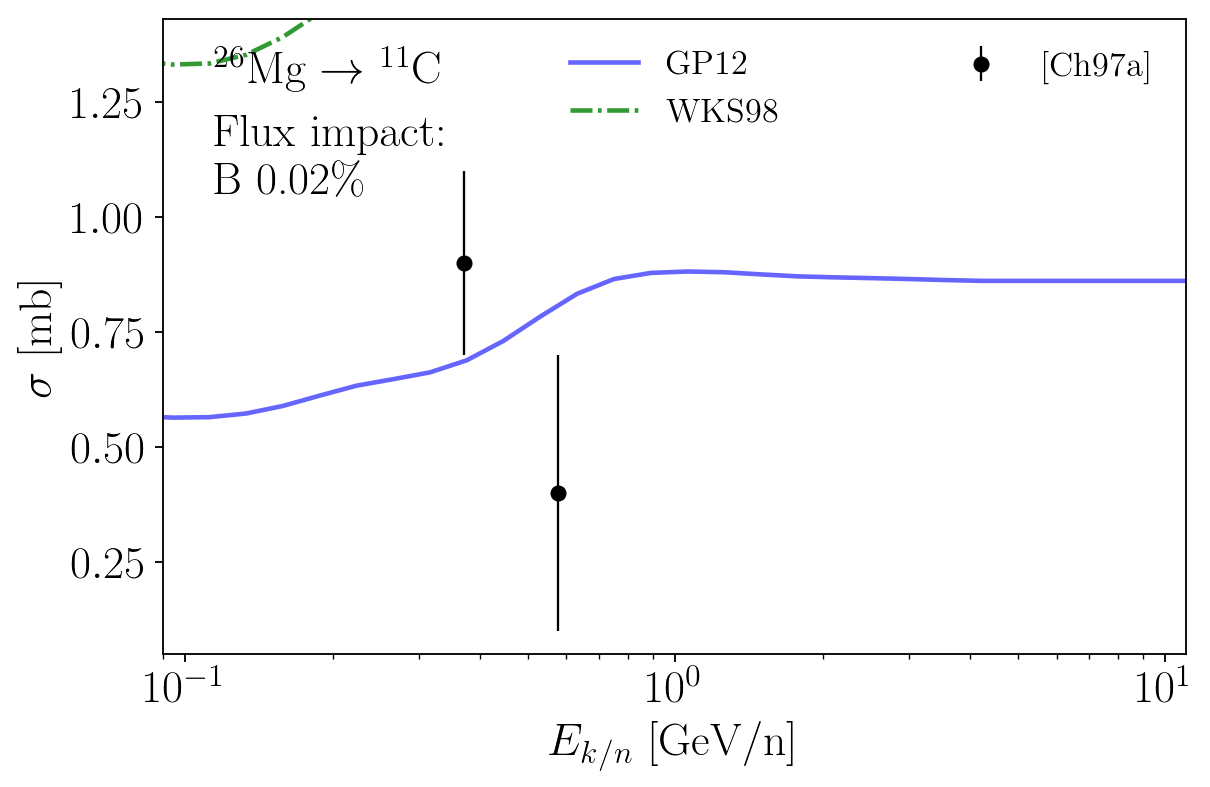}  &  
\includegraphics[width=0.32\textwidth]{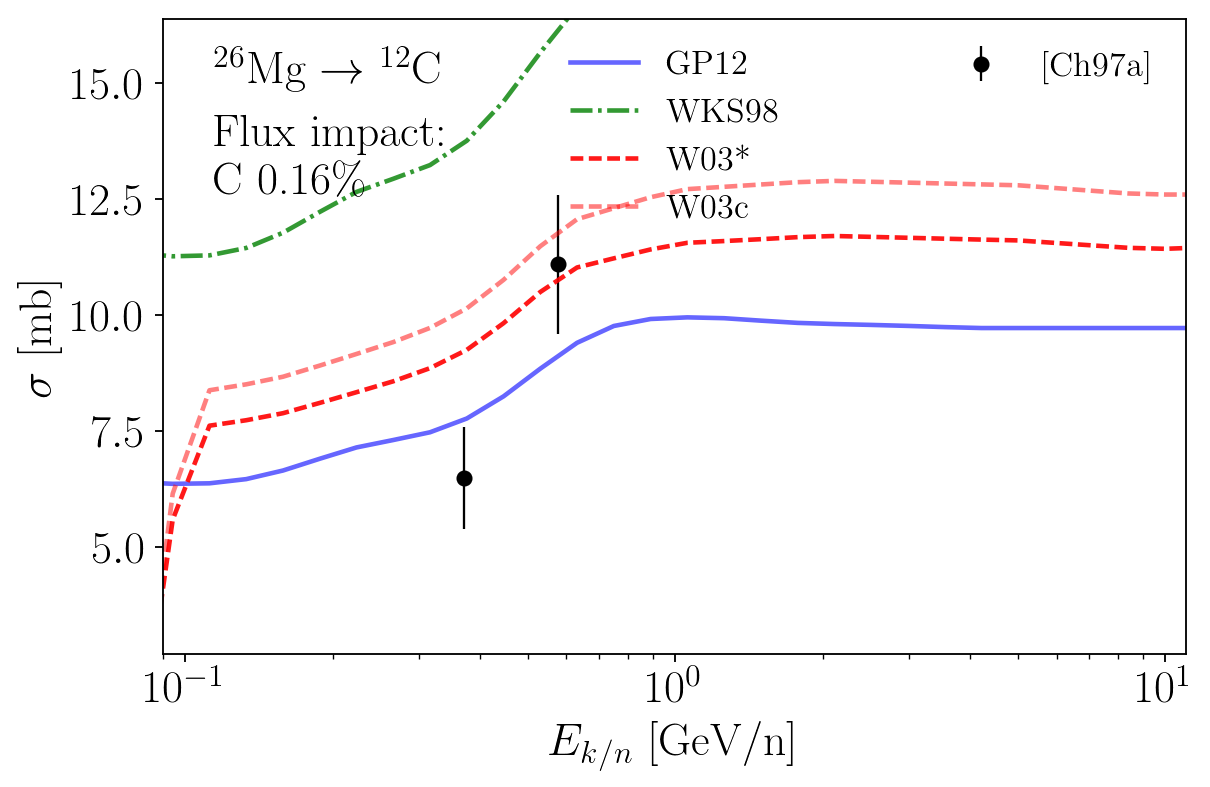}  \\ 
\includegraphics[width=0.32\textwidth]{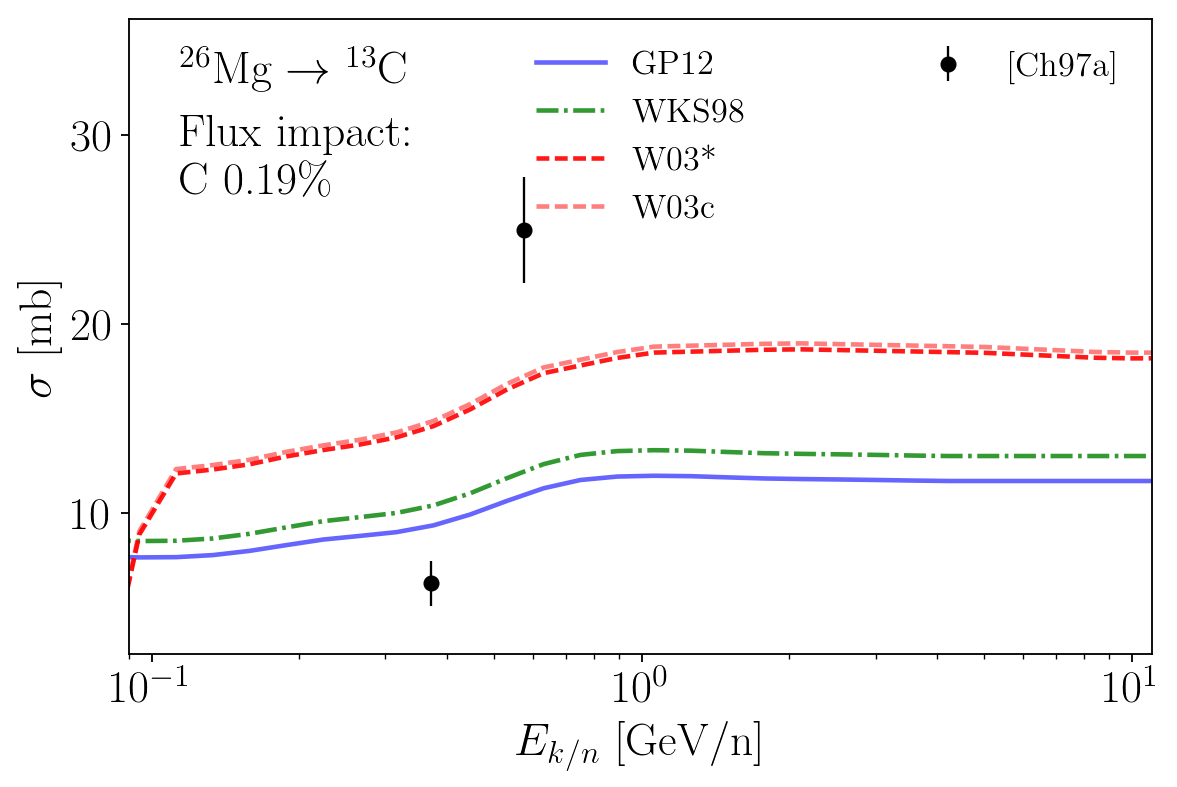}  &  
\includegraphics[width=0.32\textwidth]{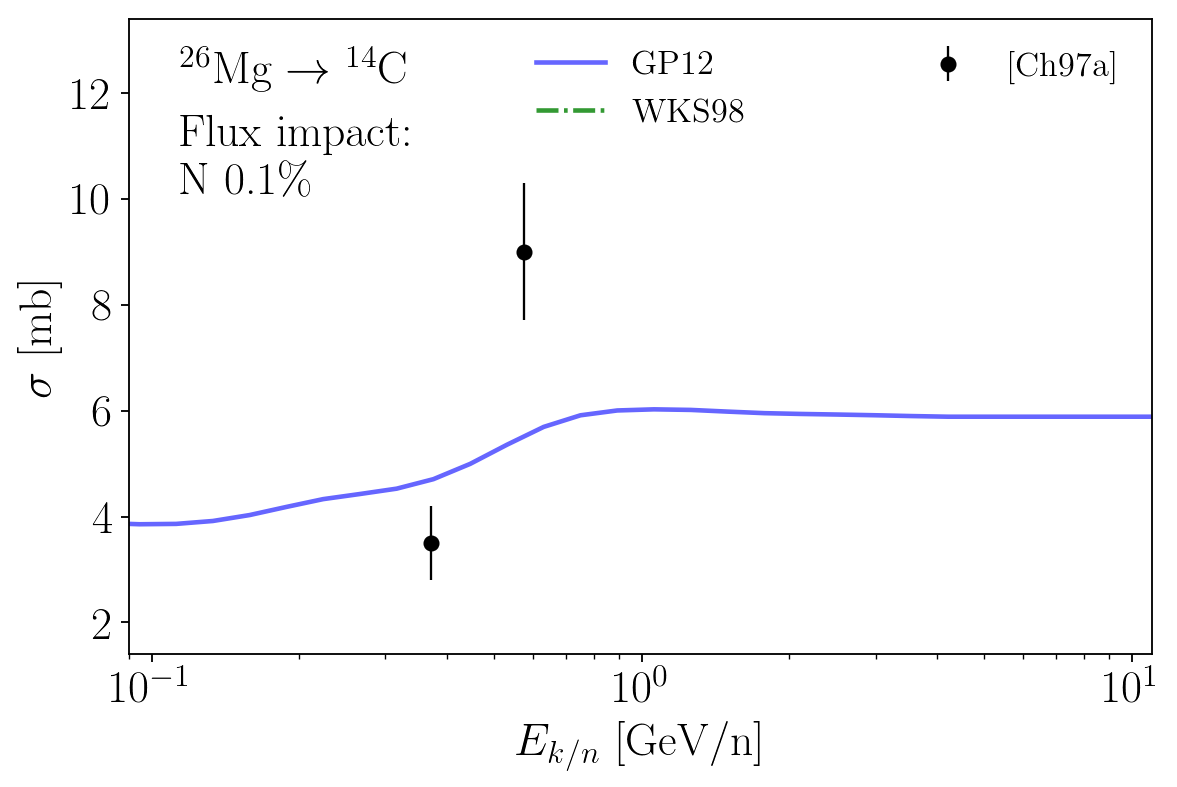}  &  
\includegraphics[width=0.32\textwidth]{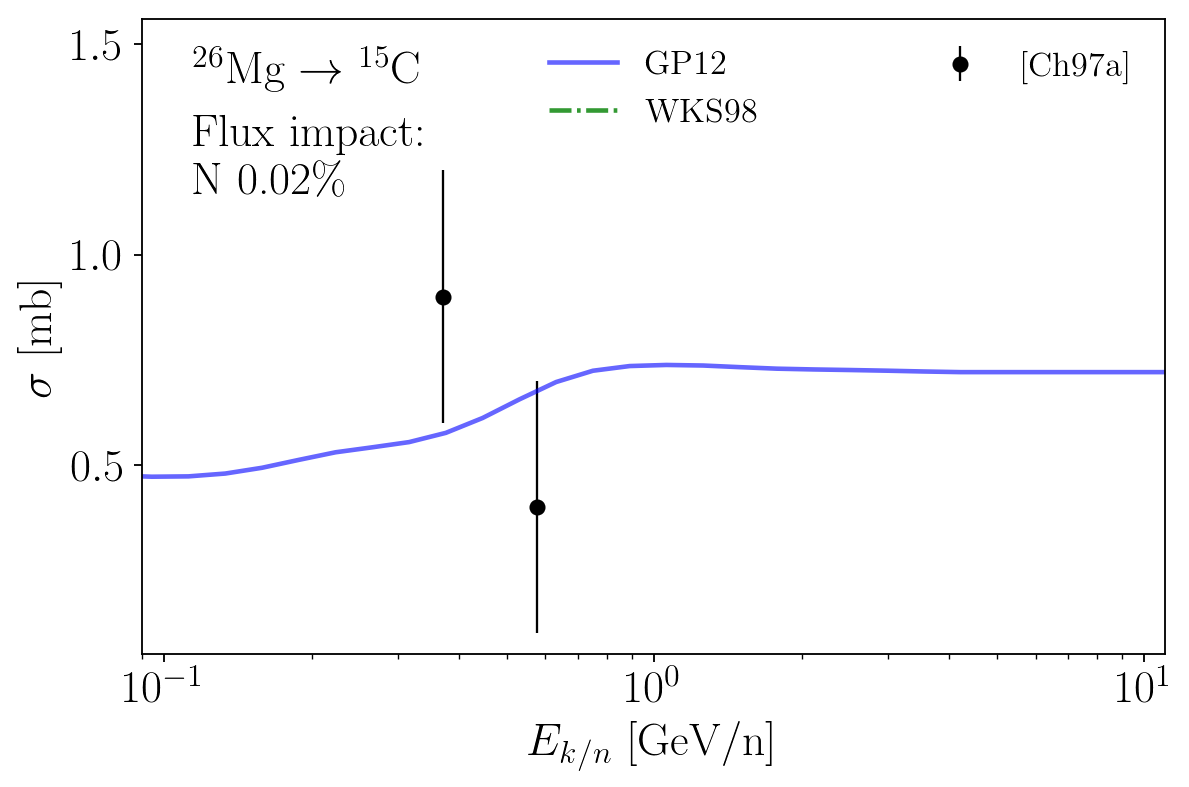}  \\ 
\includegraphics[width=0.32\textwidth]{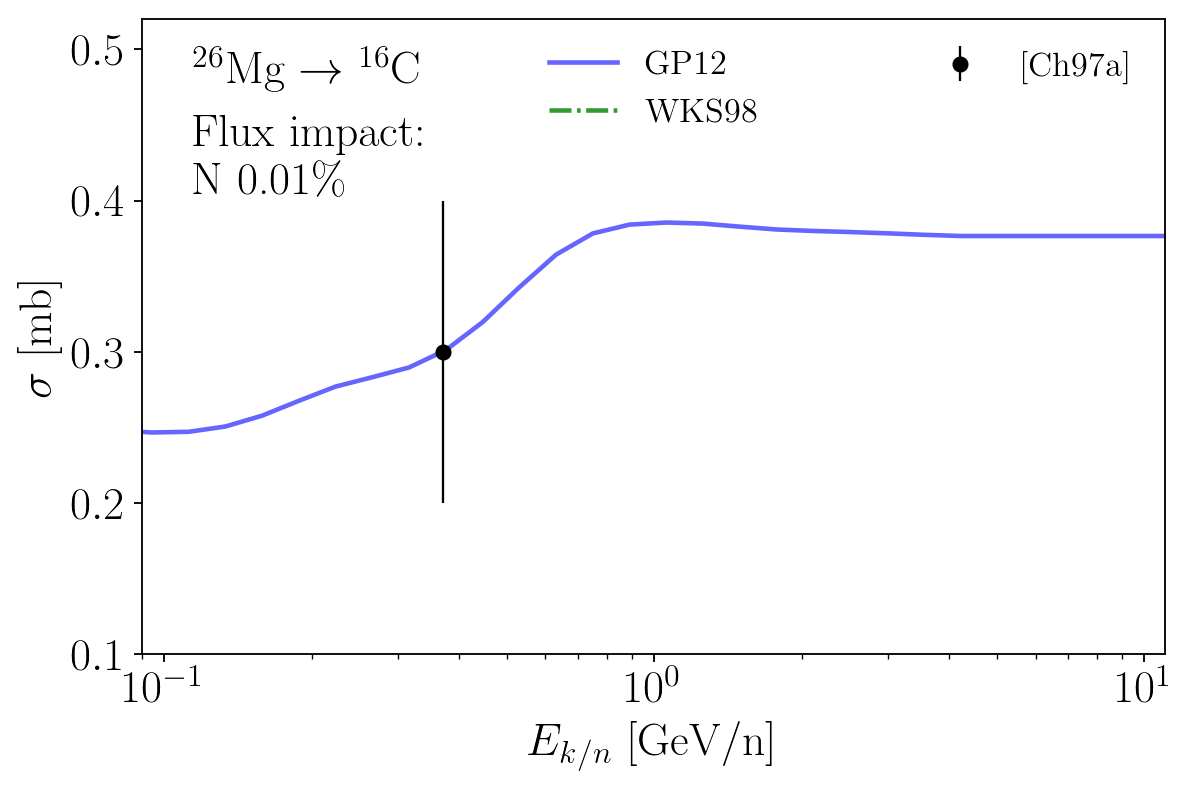}  &  
\includegraphics[width=0.32\textwidth]{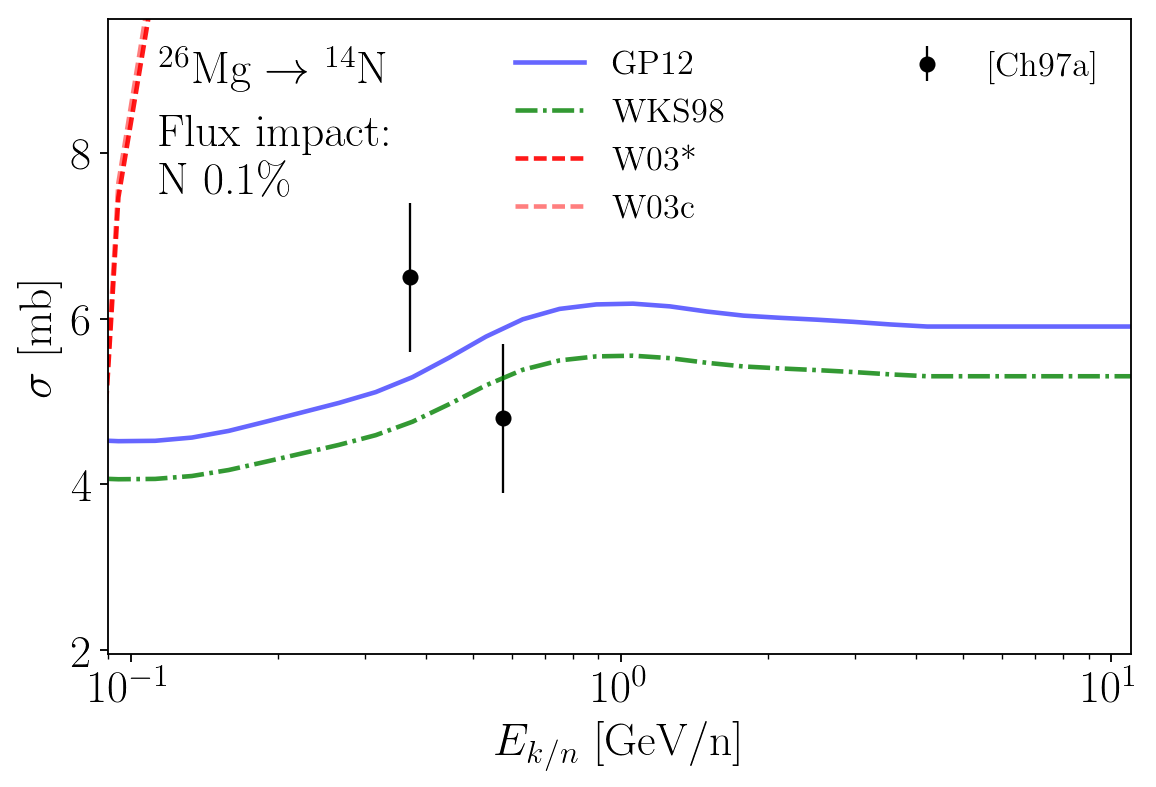}  &  
\includegraphics[width=0.32\textwidth]{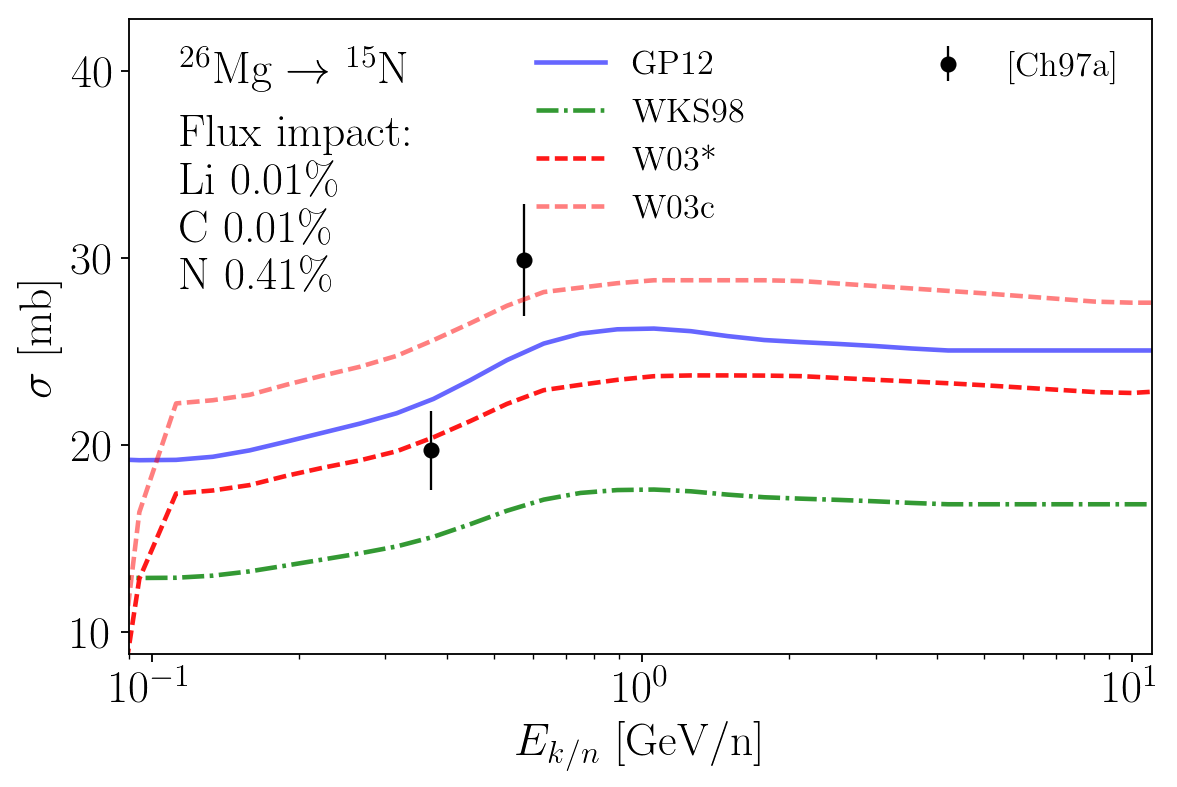}  \\ 
\includegraphics[width=0.32\textwidth]{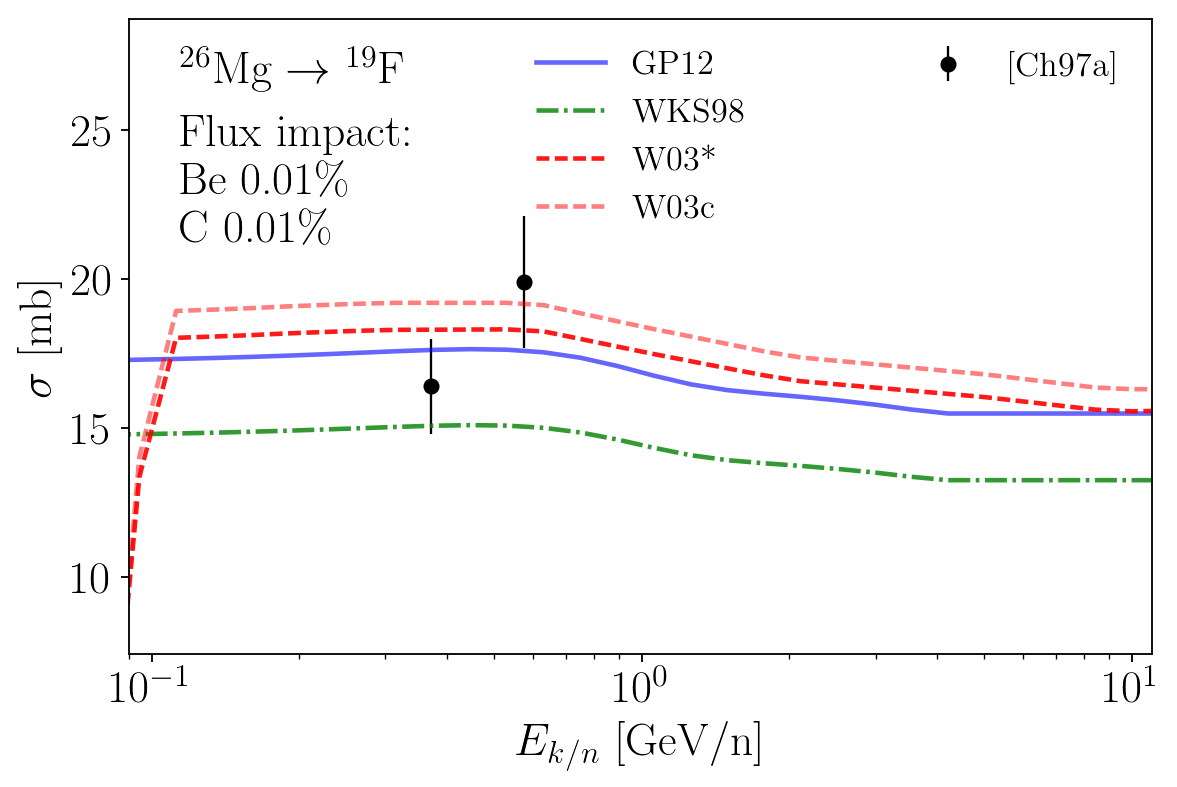}  &  &  \\ [3pt] 
\multicolumn{3}{c}{\bf Z=13{ \bf projectiles: $^{x}$Al + H $\rightarrow$ $^{A}_ZX$}}\\ [3pt]
\multicolumn{3}{c}{\noindent\makebox[\linewidth]{\rule{\textwidth}{0.4pt}}}\\ [3pt]
\includegraphics[width=0.32\textwidth]{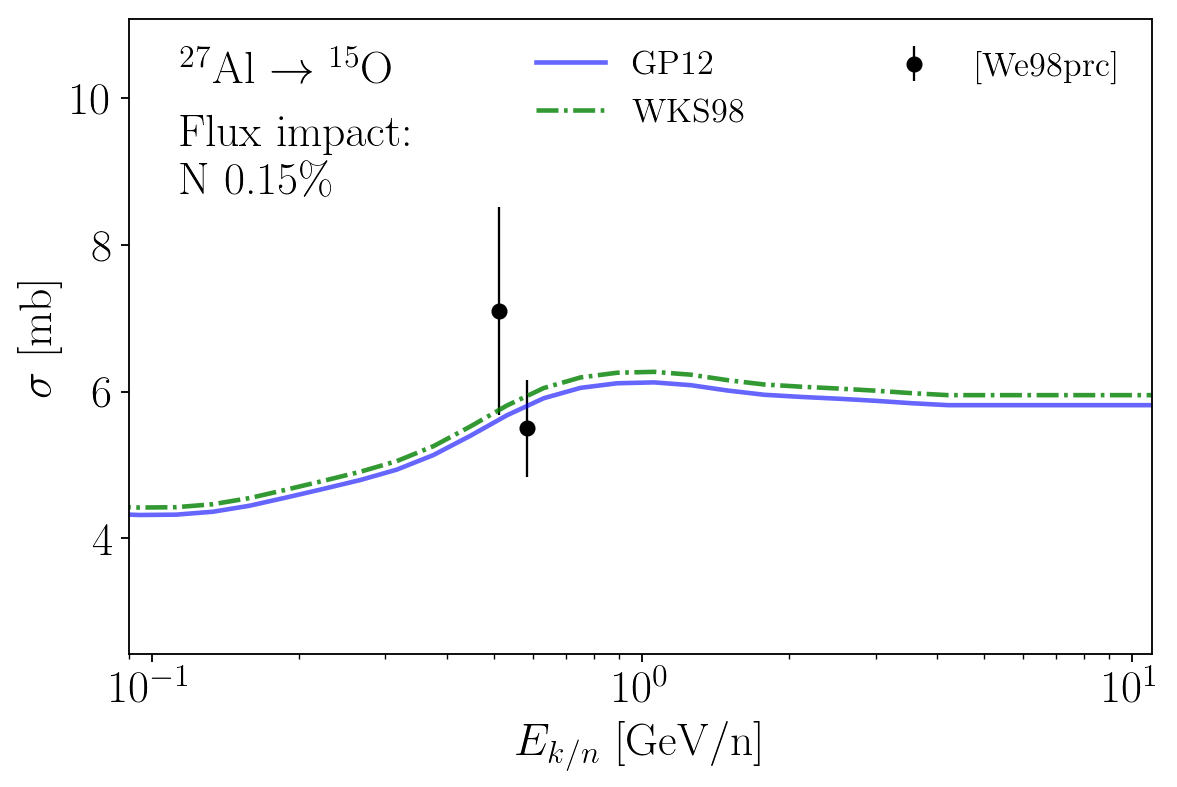}  &  
\includegraphics[width=0.32\textwidth]{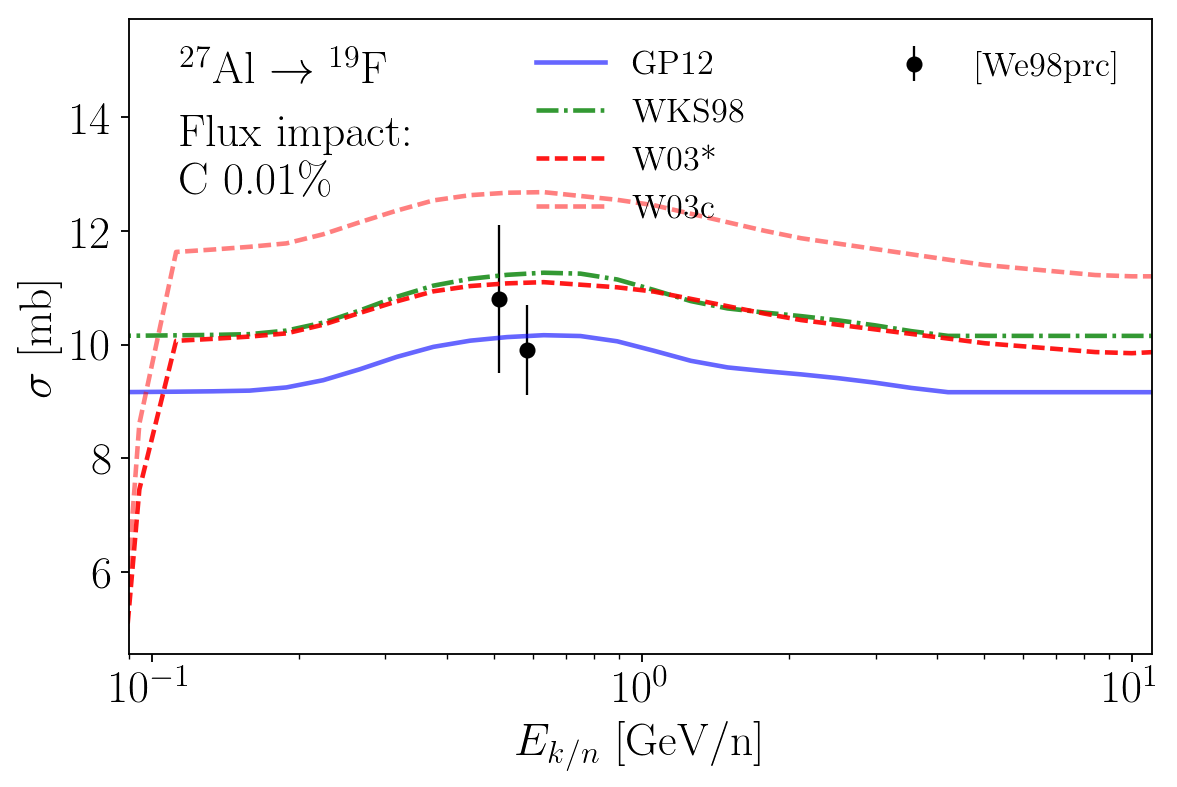}  &  \\ [3pt] 
\multicolumn{3}{c}{\bf Z=14{ \bf projectiles: $^{x}$Si + H $\rightarrow$ $^{A}_ZX$}}\\ [3pt]
\multicolumn{3}{c}{\noindent\makebox[\linewidth]{\rule{\textwidth}{0.4pt}}}\\ [3pt]
\includegraphics[width=0.32\textwidth]{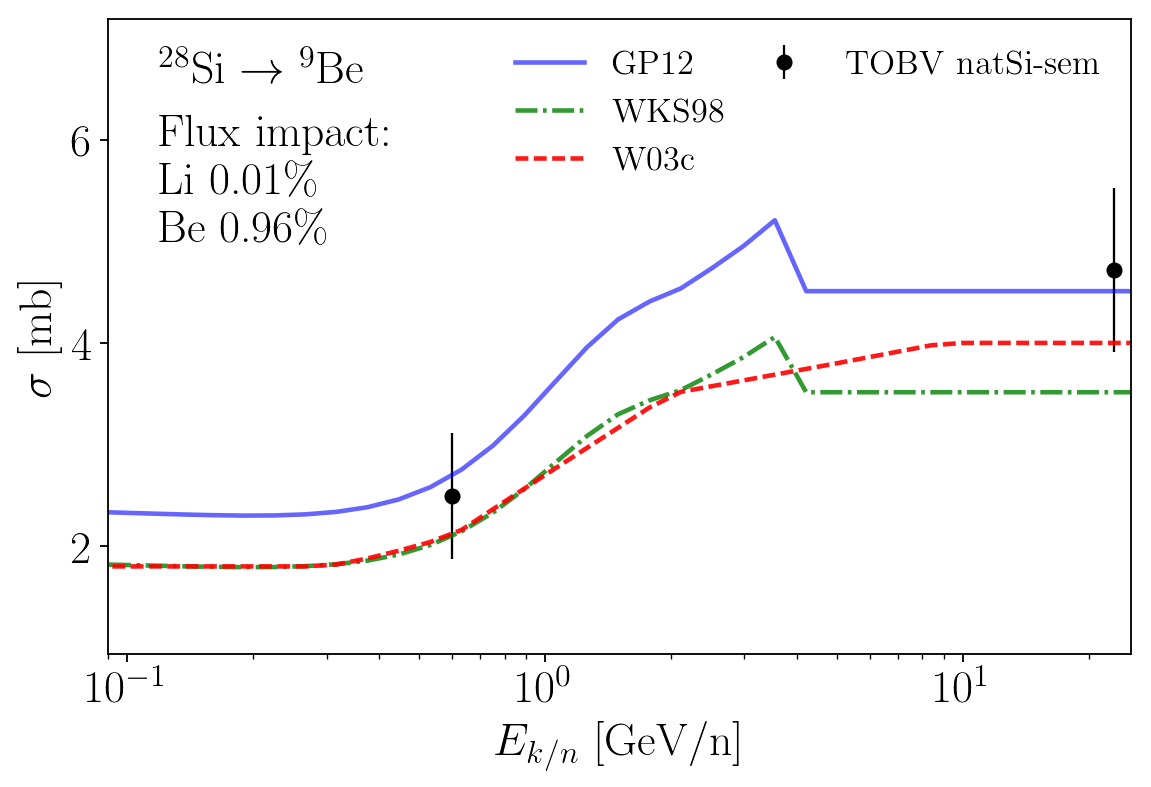}  &  
\includegraphics[width=0.32\textwidth]{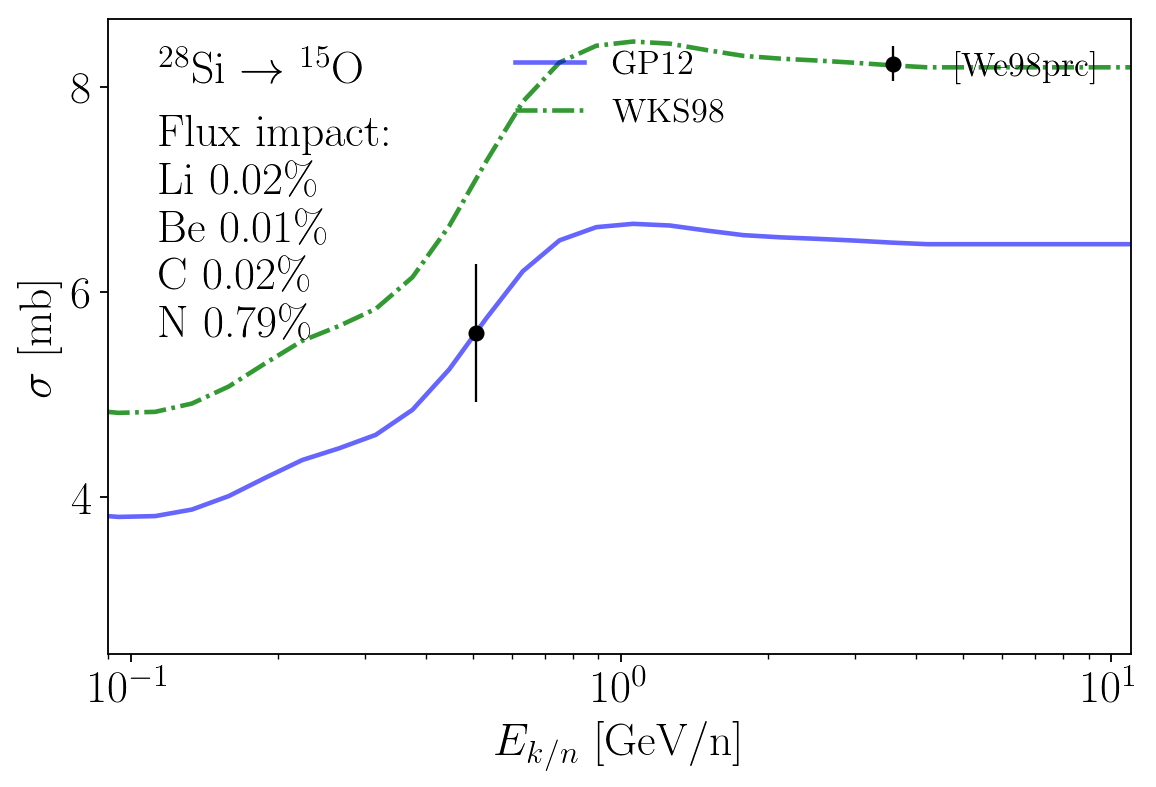}  &  
\includegraphics[width=0.32\textwidth]{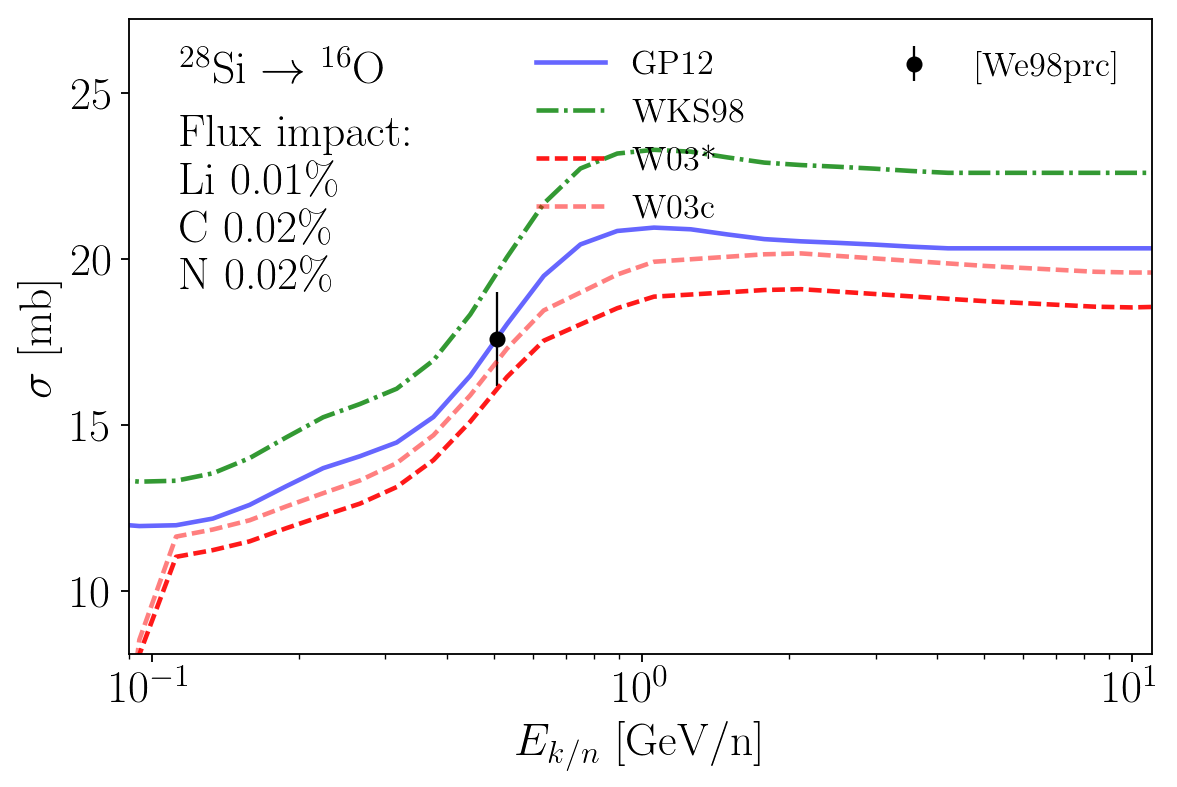}  \\ 
\includegraphics[width=0.32\textwidth]{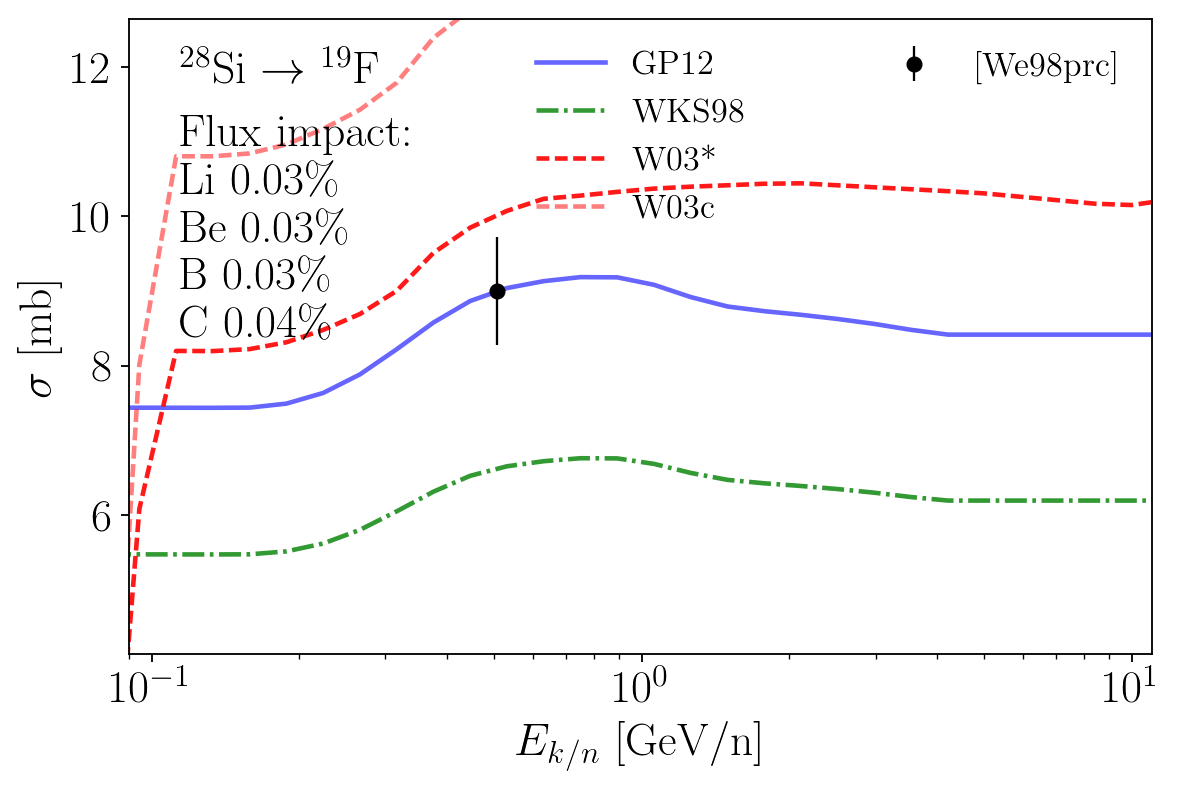}  &  
\includegraphics[width=0.32\textwidth]{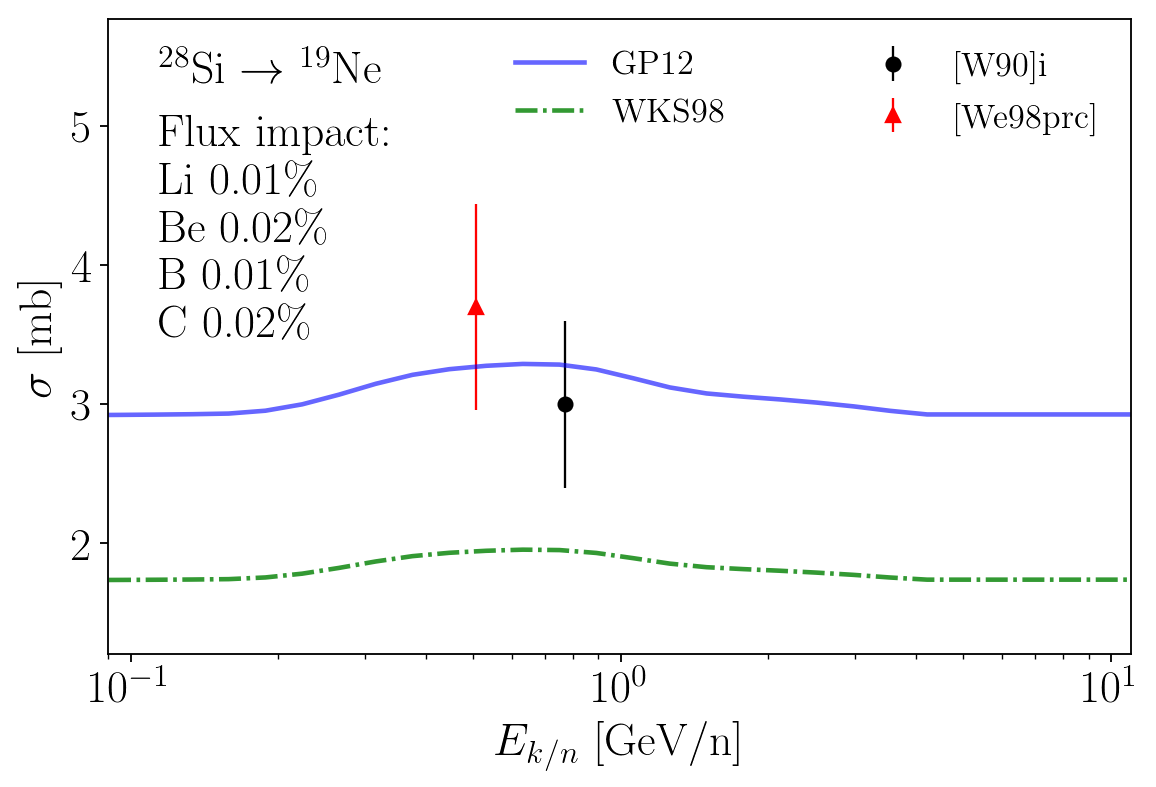}  &  
\includegraphics[width=0.32\textwidth]{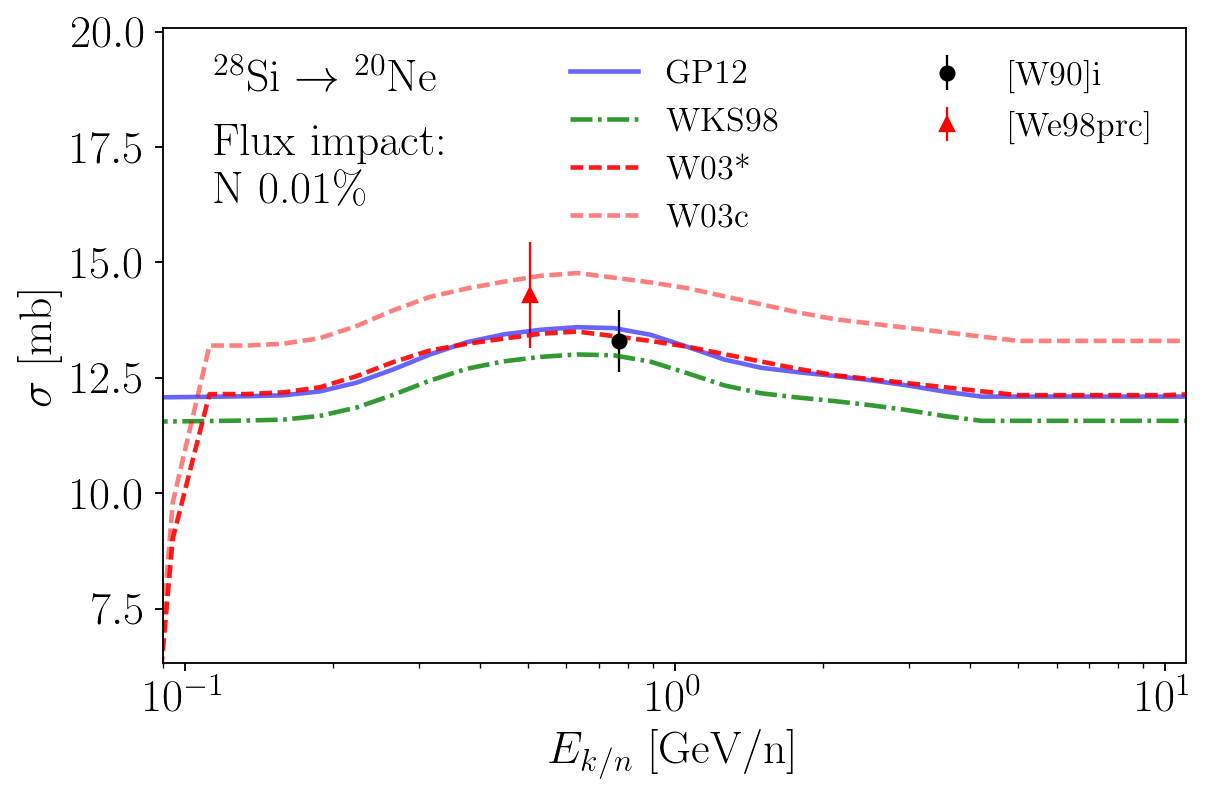}  \\ 
\includegraphics[width=0.32\textwidth]{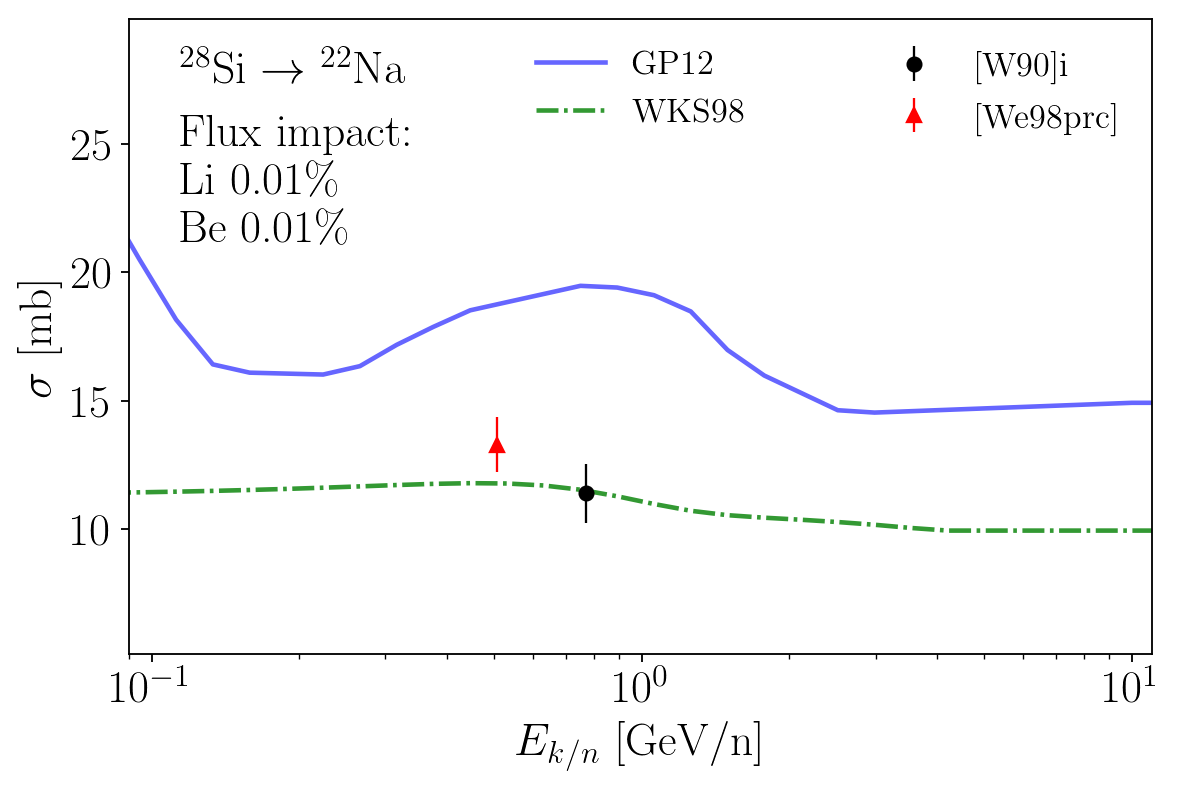}  &  
\includegraphics[width=0.32\textwidth]{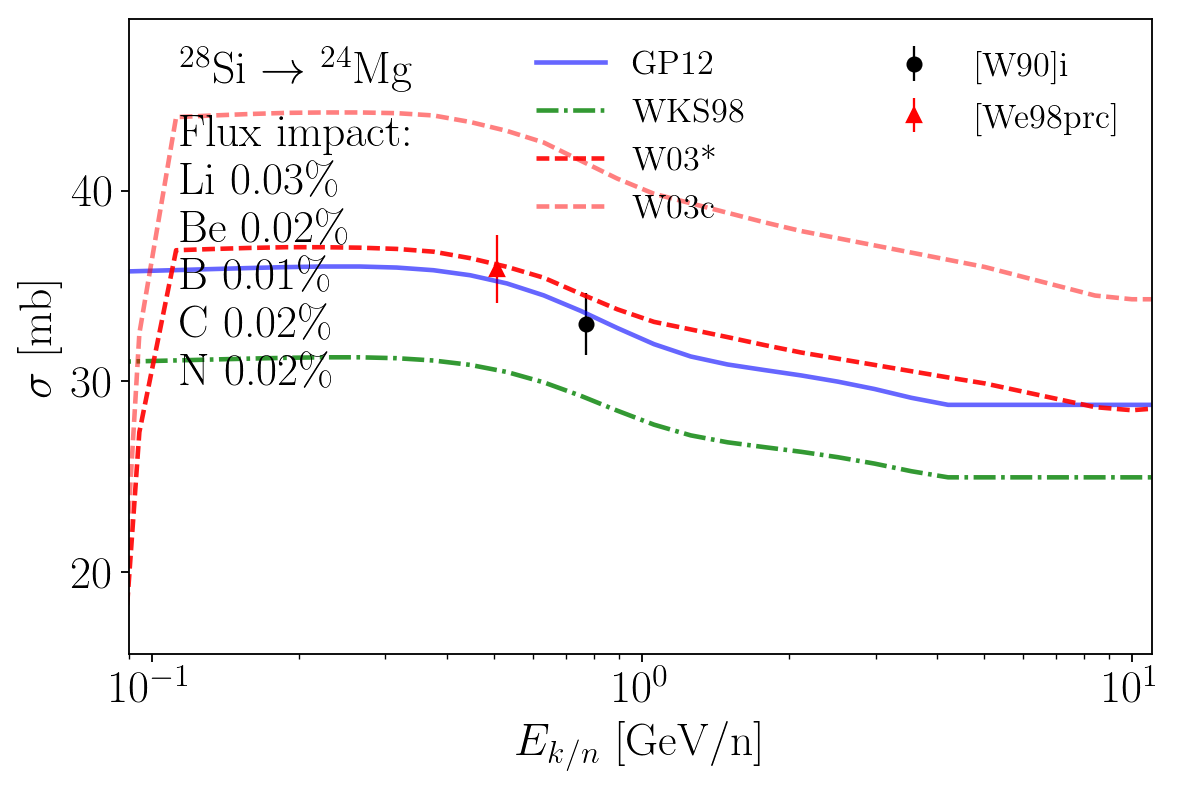}  &  
\includegraphics[width=0.32\textwidth]{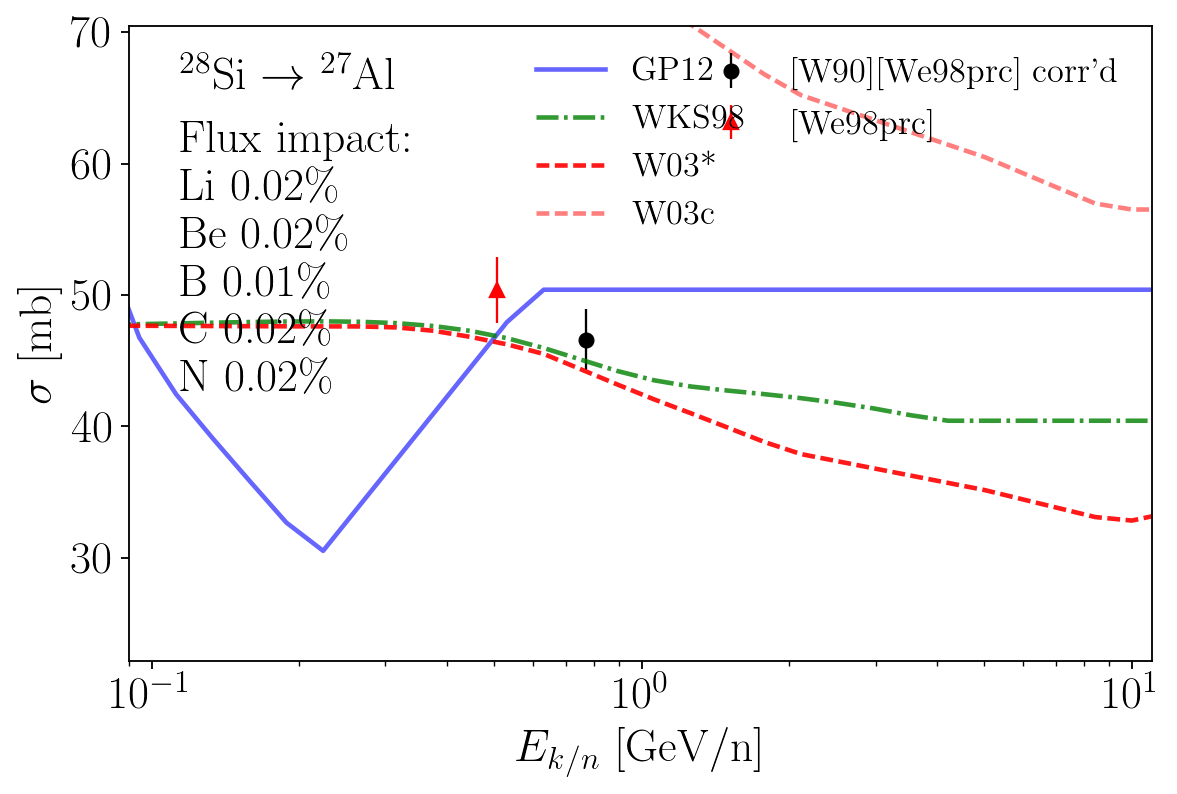}  \\ 
\includegraphics[width=0.32\textwidth]{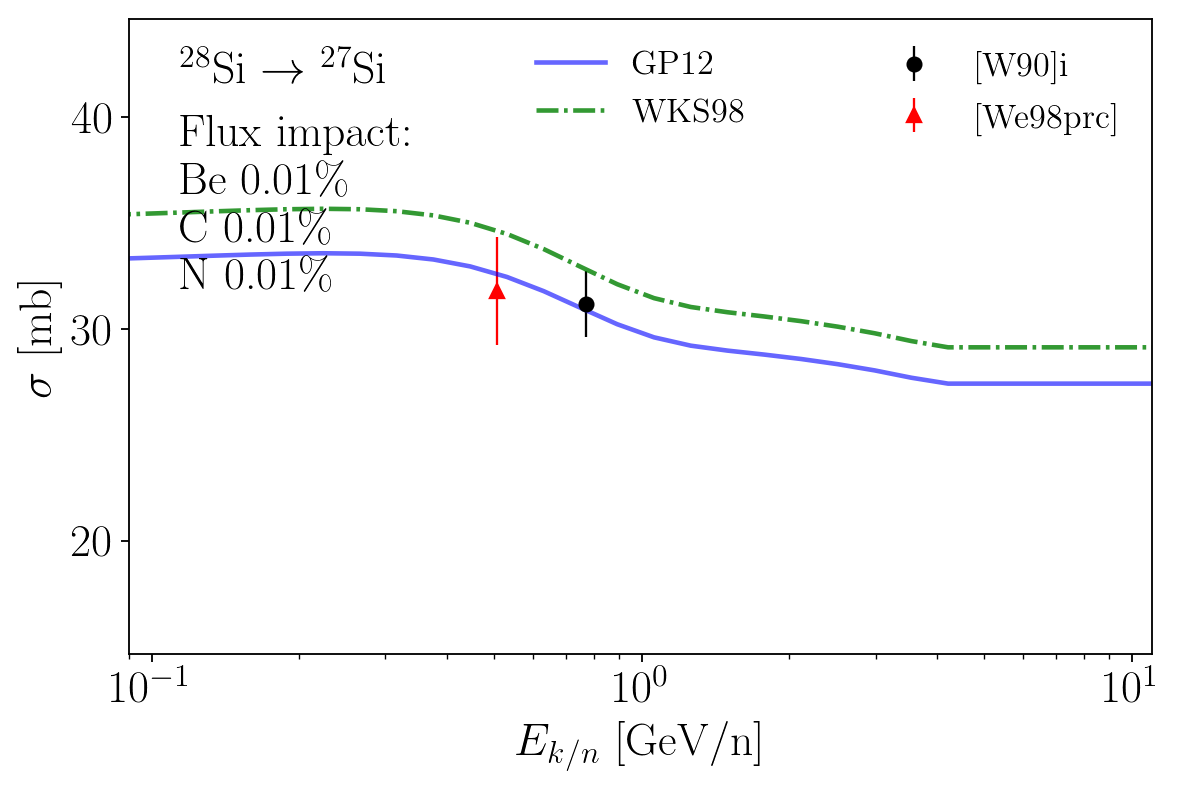}  &  
\includegraphics[width=0.32\textwidth]{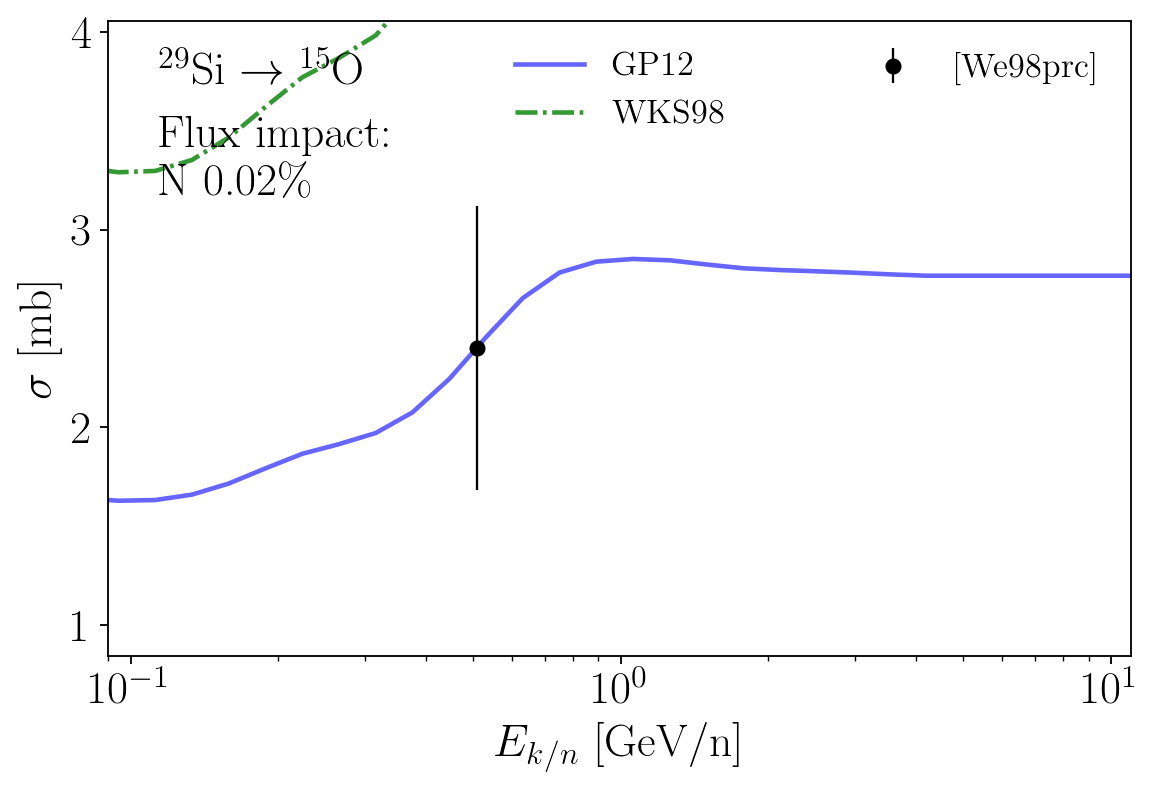}  &  \\ [3pt] 
\multicolumn{3}{c}{\bf Z=26{ \bf projectiles: $^{x}$Fe + H $\rightarrow$ $^{A}_ZX$}}\\ [3pt]
\multicolumn{3}{c}{\noindent\makebox[\linewidth]{\rule{\textwidth}{0.4pt}}}\\ [3pt]
\includegraphics[width=0.32\textwidth]{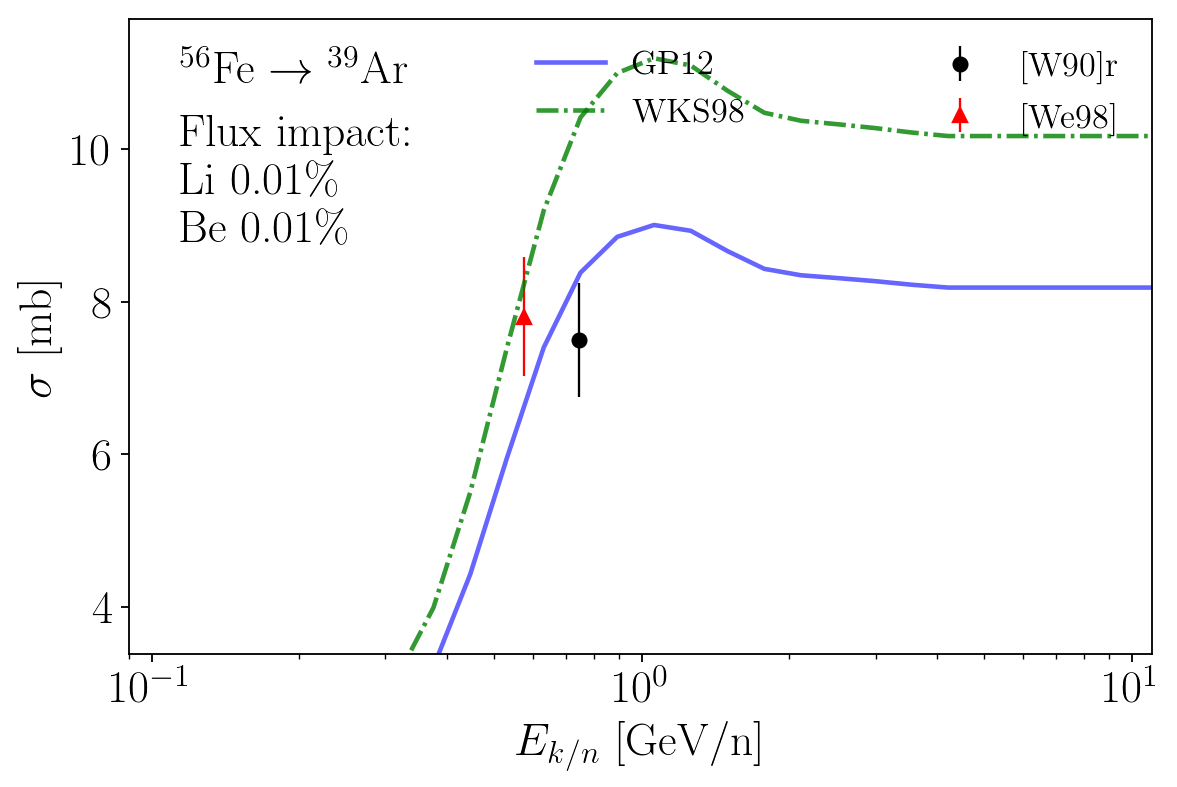}  &  
\includegraphics[width=0.32\textwidth]{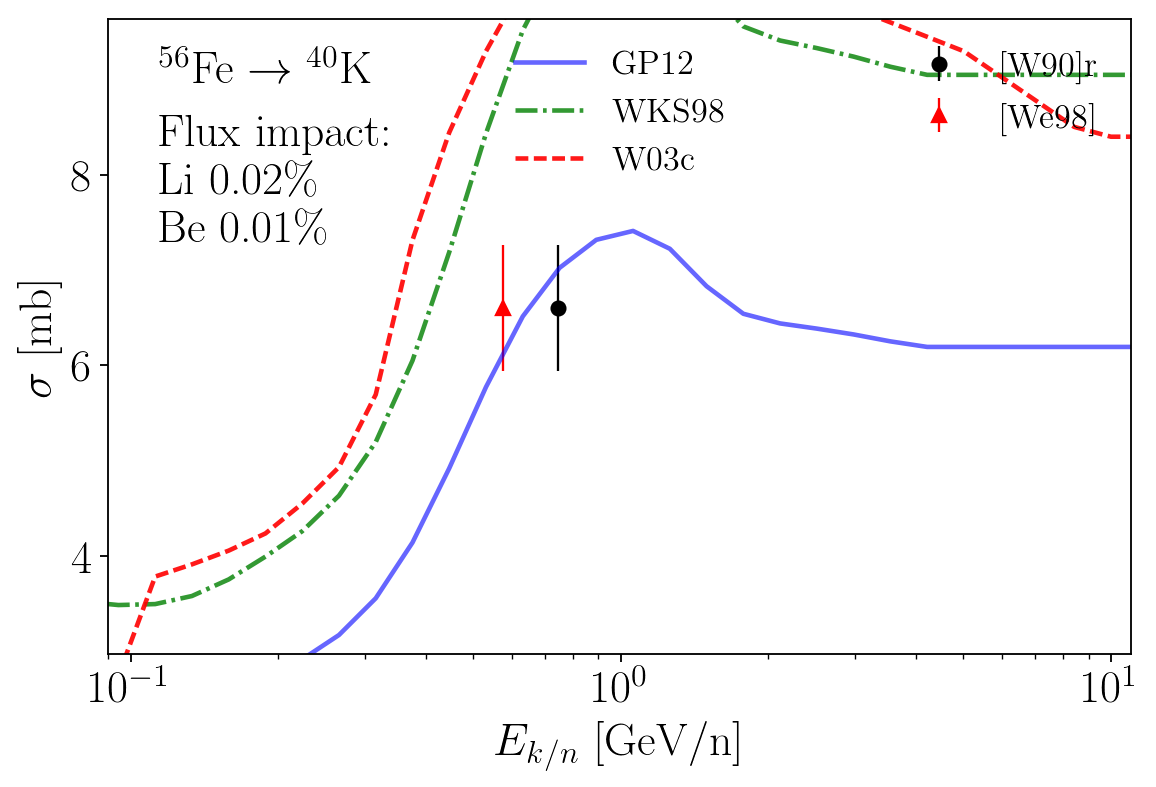}  &  
\includegraphics[width=0.32\textwidth]{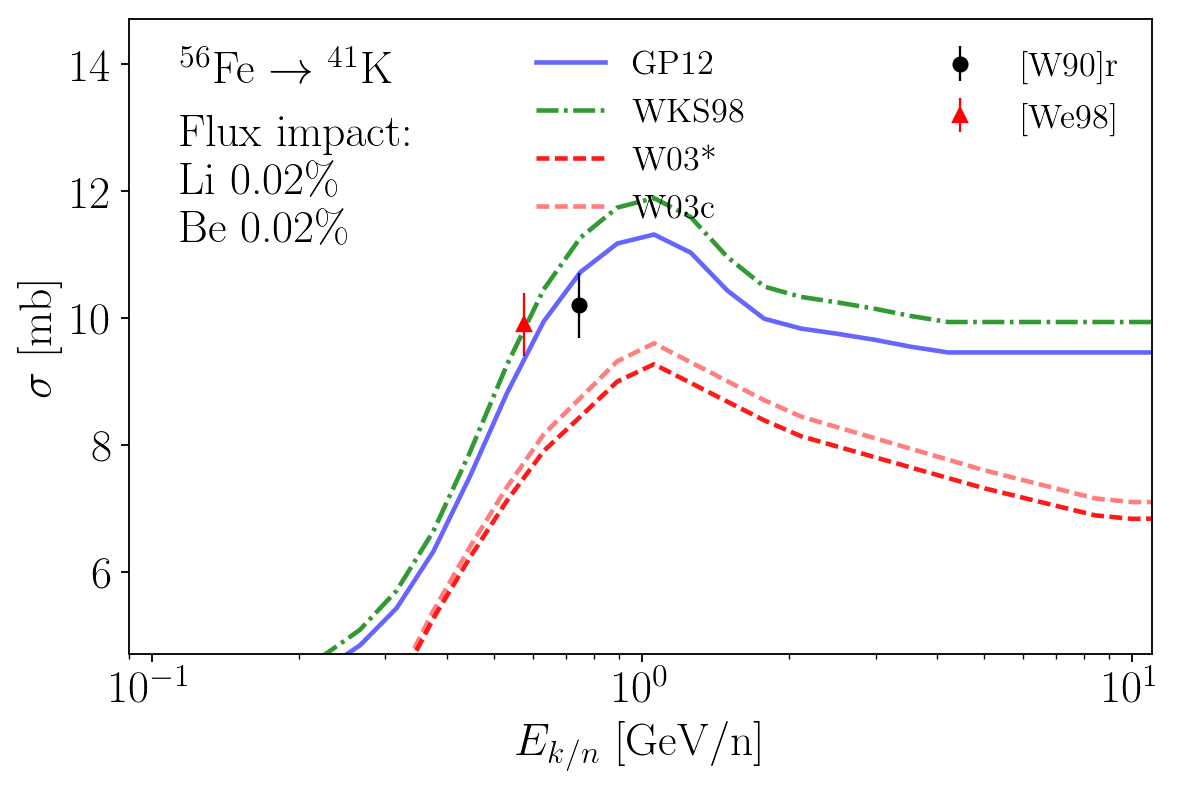}  \\ 
\includegraphics[width=0.32\textwidth]{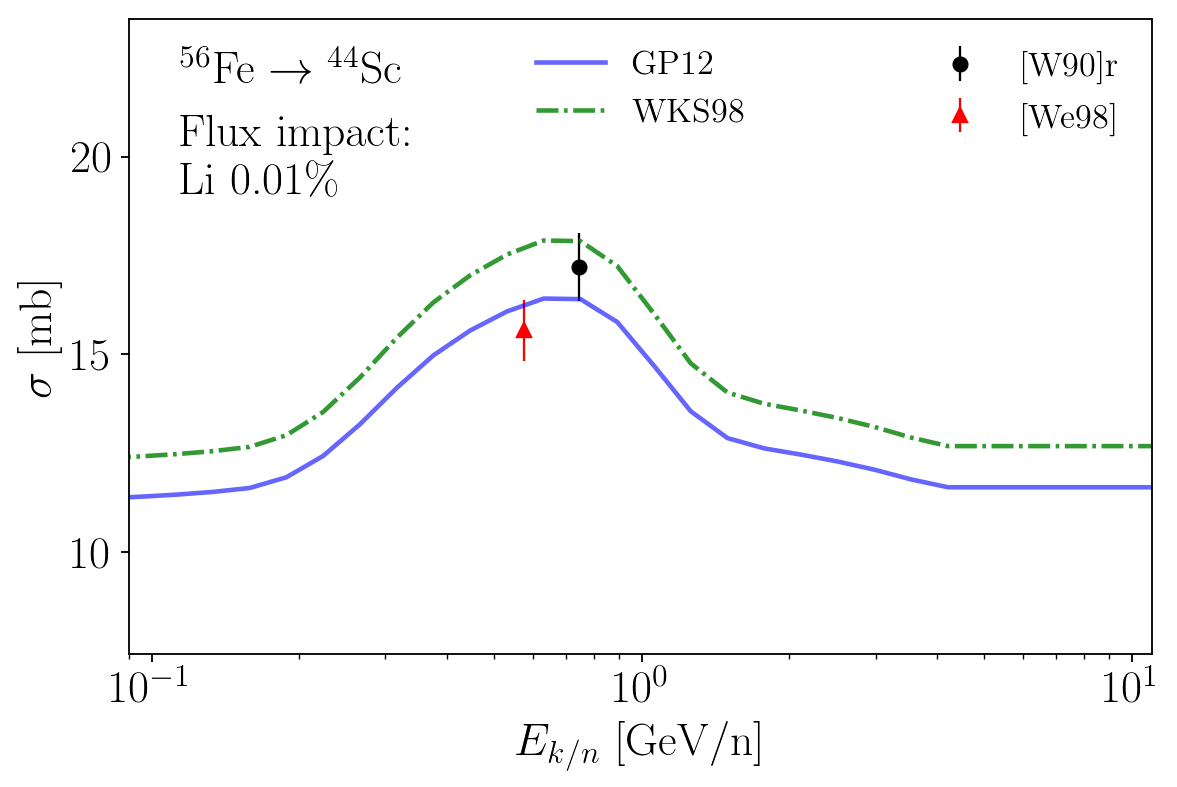}  &  
\includegraphics[width=0.32\textwidth]{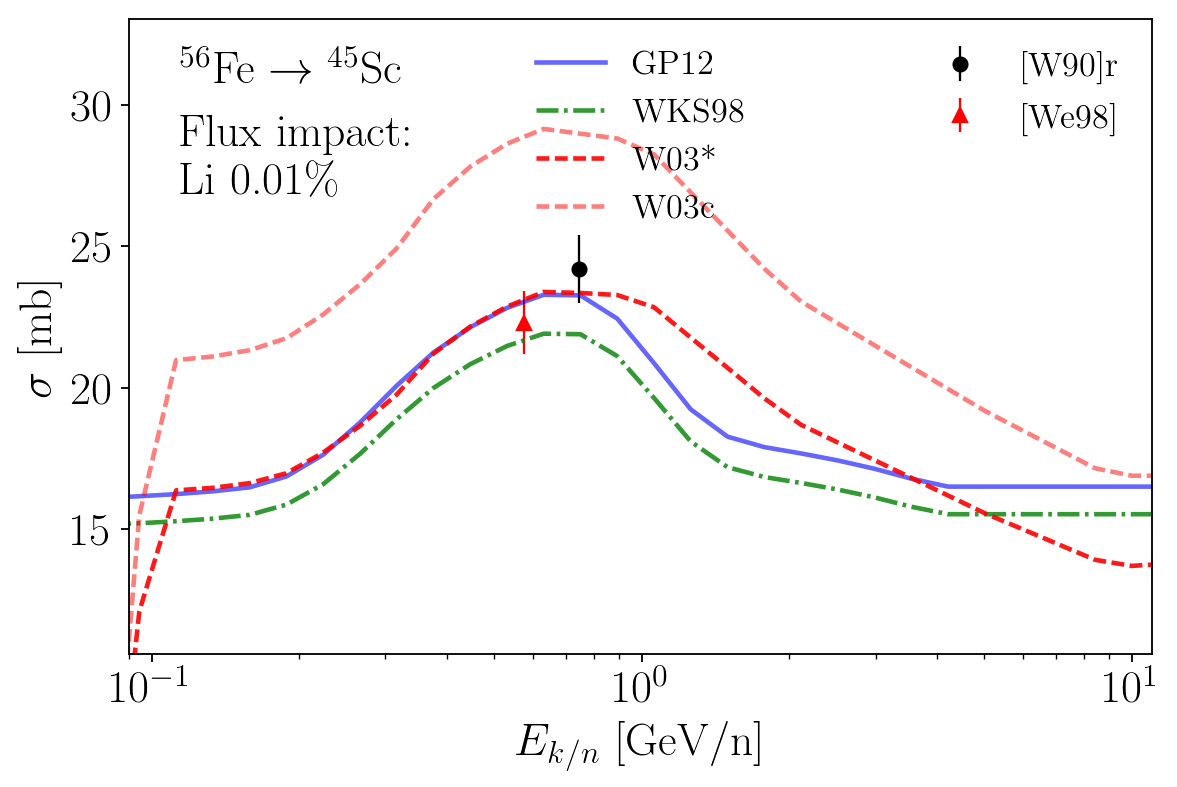} \end{longtable}
\clearpage
\newpage
\bibliography{nuc_channels,biblio_SigFrag}


\end{document}